\documentclass{ws-ijmpa-preprint}
\usepackage{amssymb}

\DeclareMathAlphabet   {\mathsc}{OT1}{cmr}{m}{sc}

\newcommand{\nnn}{ \nonumber \\ }
\newcommand{\myref}[1]{(\ref{#1})}

\newcommand{\beq}{\begin{equation}}
\newcommand{\eeq}{\end{equation}}
\newcommand{\bea}{\begin{eqnarray}}
\newcommand{\eea}{\end{eqnarray}}
\def\[{\left [}
\def\]{\right ]}
\def\({\left (}
\def\){\right )}
\newcommand{\lang}{\left\langle}
\newcommand{\rang}{\right\rangle}
\newcommand{\bigvev}[1]{{\left\langle #1 \right\rangle}}
\newcommand{\lbr}{\left\{}
\newcommand{\rbr}{\right\}}
\newcommand{\oline}[1]{\overline{#1}}
\newcommand{\jbar}{\bar{\jmath}}
\newcommand{\ibar}{\bar{\imath}}

\newcommand{\wtd}[1]{\widetilde{#1}}
\newcommand{\Lag}{\mathcal{L}}
\newcommand{\D}{\mathcal{D}}

\newcommand{\wh}[1]{\widehat{#1}}
\newcommand{\h}[1]{\hat{#1}}

\newcommand{\notD}{\not{\hspace{-.05in}D}}

\newcommand{\eV}       {~\mathrm{eV}}
\newcommand{\MeV}      {~\mathrm{MeV}}
\newcommand{\GeV}      {~\mathrm{GeV}}
\newcommand{\TeV}      {~\mathrm{TeV}}

\newcommand{\SM}       {\mathsc{sm}}

\newcommand{\VY}       {\mathsc{vy}}
\newcommand{\UV}       {\mathsc{uv}}
\newcommand{\GS}       {\mathsc{gs}}
\newcommand{\PL}       {\mathsc{pl}}
\newcommand{\PV}       {\mathsc{pv}}

\newcommand{\STR}      {\mathsc{str}}
\newcommand{\SUSY}     {\mathsc{susy}}

\newcommand{\superint}{\int\diff^{4}\theta}

\newcommand{\ww}{W^\alpha W_\alpha}
\newcommand{\bww}{\overline{W}_{\dot\alpha} \overline{W}^{\dot\alpha}}

\newcommand{\DaDa}{{\D}^{\alpha}{\D}_{\alpha}}
\newcommand{\DbDb}{{\D}_{\dot{\alpha}}{\D}^{\dot{\alpha}}}
\newcommand{\Da}{{\D}_{\alpha}}

\newcommand{\Dc}{{\D}^{\alpha}}

\newcommand{\Wa}{W_{\alpha}}

\newcommand{\Wc}{W^{\alpha}}

\newcommand{\lowest}{|_{\theta =\bar{\theta}=0}}
\newcommand{\blowest}{\bigg|_{\theta =\bar{\theta}=0}}
\newcommand{\slz}{SL(2,\mathbf{Z})}
\def\Eisen{G_{2}\(t,\bar{t}\)}

\newcommand{\baal}{b_{a}^{\alpha}}
\newcommand{\baaleff}{\(\baal\)_{\rm eff}}
\newcommand{\bpaleff}{\(b_{+}^{\alpha}\)_{\rm eff}}
\newcommand{\caleff}{\({c_{\alpha}}\)_{\rm eff}}
\newcommand{\half}{\frac{1}{2}}

\newcommand{\hc}       {\mathrm{\; h.c. \;}}

\newcommand{\hs}{\hspace{0.2cm}}
\newcommand{\pp}{\partial}

\newcommand{\STr}{{\rm STr}}
\newcommand{\Tr}{{\rm Tr}}

\newcommand{\diff}{\mbox{d}}
\newcommand{\order}{\mathcal{O}}
\newcommand{\re}{{\rm Re}}
\newcommand{\im}{{\rm Im}}
\newcommand{\Zbf}{{{\bf Z}}}

\newcommand{\gappeq}{\mathrel{\rlap {\raise.5ex\hbox{$>$}}
{\lower.5ex\hbox{$\sim$}}}}
\newcommand{\lappeq}{\mathrel{\rlap{\raise.5ex\hbox{$<$}}
{\lower.5ex\hbox{$\sim$}}}}

\def\vev{$vev\;$}
\def\vevs{$vevs\;$}
\newcommand{\chiproj}{(\oline{\mathcal{D}}^2 - 8R)}
\def\hel{\hat\ell}
\def\hK{\widehat{K}}
\def\hL{\widehat{L}}
\def\hS{\widehat{S}}

\def\vx{V_X}
\def\bpi{\mathbf{\Pi}}
\def\bfa{\mathbf{a}}
\def\bfP{\mathbf{P}}
\def\bff{\mathbf{F}}
\def\bfbf{\mathbf{\oline{F}}}
\def\F{\mathcal{F}}
\def\r{\right|}
\def\l{\left.}

\begin{document}
\markboth{Mary K. Gaillard \& Brent D. Nelson}{K\"ahler Stabilized,
Modular Invariant Heterotic String Models} \catchline{}{}{}{}{}
\title{K\"ahler Stabilized, Modular Invariant Heterotic String Models}

\author{\textbf{Mary K. Gaillard}} \address{Department of Physics,
University of California and Theoretical Physics Group, Lawrence
Berkeley National Laboratory, Berkeley, CA 94720, USA}

\author{\textbf{Brent D. Nelson}} \address{Department of Physics,
Northeastern University, Boston, MA 02115, USA}

\maketitle

\begin{abstract}
We review the theory and phenomenology of effective supergravity
theories based on orbifold compactifications of the weakly-coupled
heterotic string. In particular, we consider theories in which the
four-dimensional theory displays target space modular invariance and
where the dilatonic mode undergoes K\"ahler stabilization. A
self-contained exposition of effective Lagrangian approaches to
gaugino condensation and heterotic string theory is presented,
leading to the development of the models of Bin\'etruy, Gaillard and
Wu. Various aspects of the phenomenology of this class of models are
considered. These include issues of supersymmetry breaking and
superpartner spectra, the role of anomalous $U(1)$ factors, issues
of flavor and R-parity conservation, collider signatures, axion
physics, and early universe cosmology. For the vast majority of
phenomenological considerations the theories reviewed here compare
quite favorably to other string-derived models in the literature.
Theoretical objections to the framework and directions for further
research are identified and discussed.
\end{abstract}

\section*{Introduction}
\label{intro}

Why should we be inclined to believe in string theory? Unless we are
remarkably fortunate, and the scale at which string resonances
appear is accessible to forthcoming experiments, there will not be
-- indeed there cannot be -- any {\em direct} evidence that some
particular construction of string theory is correct. Wherefore,
then, the great interest of the high energy community in this
subject? Clearly the answer is in what string theory is (uniquely)
capable of {\em explaining}.

To some the most salient feature of string theory is its great
promise as a consistent theory of quantum gravity. But for those
whose interest lies in understanding the phenomena relevant at
energies closer to the electroweak scale it is rather string
theory's capability to explain such manifest properties of particle
physics as the presence of three generations, the gauge group of the
Standard Model (SM), the representations of the various matter
fields, and the Yukawa interactions that do such things as give mass
to the fermions but do not do such things as make the proton decay.
Consistent string constructions which address these issues generally
have supersymmetry present on the string worldsheet (base space), as
well as in the spectrum of the effective field theory description
(target space). This highly desirable phenomenological property has
not been observed in nature. But if supersymmetry is at all relevant
in understanding electroweak-scale physics new states and
interactions will likely be observed at the CERN Large Hadron
Collider (LHC).

There is, therefore, a great opportunity for string theory to make
contact with particle physics observations at accessible energies.
The elucidation of these points of contact is the purpose of the
field of string phenomenology. It is sometimes useful to divide this
field into the study of two logically distinct problems. The first
we might refer to as the ``problem of initial conditions.'' This is
the determination of the massless spectrum of gauge bosons and
matter representations -- as well as the allowed superpotential
couplings amongst these fields -- upon compactification of the
ten-dimensional (10D) theory to four dimensions (4D) on a manifold
which preserves $N=1$ supersymmetry. The ideal compactification need
not necessarily be the one which provides solely the fields of the
Minimal Supersymmetric Standard Model (MSSM), but this has often
been taken to be the supreme goal of those who study superstring
model-building. While some approaches have come close to producing
exclusively the MSSM field content, gauge group and renormalizable
superpotential it is not unfair to say that none can yet truly claim
success in this regard.

Furthermore, different choices of string compactification imply
different sets of {\em moduli} fields in the low-energy effective
Lagrangian. Very roughly speaking, these moduli parameterize the
geometry of the compact space. Their role in the low-energy
Lagrangian is to determine certain dimensionless parameters (such as
gauge and Yukawa coupling constants), yet they have no potential at
all at the classical level. That is to say, they are truly flat
directions in the scalar potential of the theory. There is therefore
a second class of problems for the string phenomenologist: the
``problem of dynamics.'' This latter class of problems includes the
issues of anomalous $U(1)_X$ cancelation, supersymmetry breaking,
dynamical electroweak symmetry breaking, and so forth. The two
problems can to a large degree be separated (at least formally), and
it is the second class of questions which we wish to address in this
review.

Over the years there has been considerable progress in understanding
the structure of effective actions describing the low-energy
dynamics of massless fields in 4D superstring theory. This success
began in the context of (weakly-coupled) heterotic string theory,
where the basic program was to extract the relevant terms in the
field-theoretical Lagrangian from the S-matrix elements computed
within the full-fledged superstring theory.\cite{Dixon:1989fj} Many
important quantities were determined at the classical level,
including the K\"ahler potentials and the gauge and Yukawa couplings
for orbifold compactifications of heterotic superstrings. Later, the
program of reconstructing effective Lagrangians from string
amplitudes was pursued to higher genus in the string loop
expansion.\cite{Dixon:1990pc,Antoniadis:1991fh} An important
development was the observation that the duality
symmetry\cite{Kikkawa:1984cp,Sakai:1985cs} between small and large
radius toroidal compactifications extends to a much larger symmetry
group of the so-called target space modular transformations acting
on the moduli fields.\cite{Shapere:1988zv} This symmetry can be very
helpful when studying the moduli-dependence of the effective actions
for orbifold compactifications.\cite{Ferrara:1989bc,Ferrara:1989qb}

\vspace{0.2in}

In this work we will review the theoretical construction and
high-energy motivation for a particular class of weakly-coupled
heterotic string models. Much of the machinery, and a great deal of
the resulting model-building, will certainly touch on issues of
low-energy string phenomenology common to other string theory
starting points as well. As the nucleus of this construction was
presented in a sequence of papers by Bin\'etruy, Gaillard and Wu we
will sometimes refer to the class as the ``BGW~model,'' (or models).
They are generally characterized by the presence of {\em K\"ahler
stabilization} to provide a realistic minimum for the dilaton
modulus. We will present this mechanism in detail in this review,
and study the resulting phenomenology at length. We choose this
class not because we necessarily believe it to be the correct model
of Nature -- though we will see that it has many excellent
phenomenological features -- but because it is, to date, the most
complete model of supersymmetric particle physics arising from
string theory in the literature. By this we mean that the
phenomenology of this class has been studied in great depth from
multiple angles: spectrum and superpotential selection rules,
effective Lagrangian construction, gaugino condensation and moduli
stabilization, supersymmetry breaking and transmission to the
observable sector, anomaly cancelation, superpartner spectra,
flavor-changing and rare processes, collider signatures, cosmology,
and so on.

In certain phenomenological areas this class of models possesses
remarkably favorable characteristics; in other areas less so. There
are also certain theoretical objections that can be raised to the
treatment presented here and we will give voice to these objections
in our concluding chapter. But the goal of this review is not to
promote the K\"ahler stabilized, modular invariant models of the
heterotic string; but rather to promote the sometimes arduous act of
building complete models of 4D~superstring dynamics generally. The
most significant property of the class of models considered here,
therefore, is that the analysis has been performed in all of these
phenomenological areas -- for a {\em single} class of theories. To
our knowledge, no other class of models (whether of string-theoretic
motivation or otherwise) has been pushed as far in as many different
directions as this class. As such, the BGW~model provides a paradigm
for what a ``complete'' theory looks like. The importance of this
idea of synthesis and completeness has been stressed recently as
crucial to any effort in effectively making contact between
superstring models and the forthcoming data era we are about to
enter.\cite{Binetruy:2003cy,Binetruy:2005ez}

\vspace{0.2in}

Understanding the phenomenology of any string-derived effective
theory begins and ends with the dynamics of the moduli in the
theory. We will therefore develop the theory behind the BGW~class of
heterotic string models by studying the types of moduli present in
the theory, their stabilization through nonperturbative dynamics,
and their role in transmitting supersymmetry breaking to the
Standard Model fields of the observable sector. In
Section~\ref{sec1} we consider theories of a single modulus, whose
vacuum expectation value ($vev$) determines the gauge coupling of a
super Yang-Mills (YM) theory. This will lead us into a discussion of
gaugino condensation from multiple directions. Much of
Section~\ref{sec1} reviews material that is more fully treated
elsewhere. We conclude the first section with a description of the
linear multiplet treatment of the dilaton and the implementation of
K\"ahler stabilization in the effective theory. In
Section~\ref{sec2} we introduce the interaction of moduli with the
chiral superfields of the MSSM matter sector and discuss the
importance of duality symmetries that relate these moduli to
themselves. Section~\ref{sec2} concludes with the full effective
Lagrangian for the BGW~class of theories and examines the sorts of
minima that can arise for the effective scalar potential of the
moduli. These minima generically break supersymmetry, and
Section~\ref{sec3} is devoted to the understanding of how
supersymmetry breaking is communicated to the other sectors of the
theory -- particularly the states of the MSSM. These results are
revisited in Section~\ref{sec4} in the presence of anomalous $U(1)$
factors which are generically present in string constructions. In
Section~\ref{sec5} we describe a series of phenomenological topics
and how they can be addressed in the context of the K\"ahler
stabilized heterotic models. In the final chapter we offer our
thoughts on where the model can be improved and how studies such as
ours can be used to construct a meaningful string phenomenology.

\section{Moduli and their stabilization}
\label{sec1}

The key to understanding the low-energy manifestation of string
physics is understanding moduli and their dynamics. The precise
definition of a ``modulus'' sometimes depends on the context, but
one property is universal: a modulus is a scalar field for which
there is no potential at the classical level. In other words, there
is no preference for any particular vacuum expectation value ($vev$)
for the scalar field over any other. Moduli are the {\em sine qua
non} of string phenomenology: all string theories, when compactified
to four dimensions, possess moduli whose \vevs determine the size of
certain dimensionless constants in the low-energy Lagrangian. An
operational definition of a 4D~string model is therefore a gauge
theory with couplings determined by the scalar components of some
set of superfields. It is hard to claim that a particular theory is
``stringy'' or ``string motivated'' if it does not contain such
fields.

Moduli are also in many ways the ``engine'' that drive any
string-derived 4D~effective Lagrangian. The primary focus of any
such effective theory is to incorporate various nonperturbative
effects -- whether of a field theoretic or string theoretic origin
-- to generate a scalar potential for these fields. A good model
would be one in which a nontrivial, finite minimum of this scalar
potential exists for all the moduli considered. But this is merely
the beginning. The values of the scalar components of these moduli
at the minimum of their potential will determine a number of
important properties of the theory, such as the gauge and Yukawa
couplings. The auxiliary $F$-term components of these chiral
superfields will generally take nonvanishing \vevs at this minimum,
implying a breakdown of supersymmetry (SUSY). The size of these
\vevs will determine the general scale of supersymmetry breaking, as
well as the potential size of any nonzero vacuum energy at the
minimum. Since the moduli couple to matter fields in a well-defined
way, the size of the soft supersymmetry breaking in the observable
sector can also be calculated. From here a wide array of
phenomenological properties of the theory can be considered.

What is more, classes of four-dimensional, effective supergravity
theories derived from string constructions can be classified by
answering the following questions: \begin{enumerate} \item What
moduli are present in the low-energy theory? \item What symmetries
(if any) relate these moduli amongst themselves? \item How do these
moduli couple to the fields of the observable sector?
\end{enumerate}
A supergravity theory built to describe the low-energy dynamics of
weakly-coupled heterotic string theory on an orbifold will have
different answers to these questions than one built to describe
strongly-coupled heterotic string theory on a Calabi-Yau manifold --
and both will be different from the theory describing Type~IIA
string theory on an orientifold with $D$-branes at intersections.
This is a powerful and under-appreciated fact, suggesting that the
phenomenology of these theories (driven as they are by the dynamics
of their moduli) will be {\em different} -- perhaps sufficiently
different to distinguish them through low-energy observations. This
connection is the heart of string phenomenology.

In this work we will be reviewing a class of supergravity models
designed to capture the physics of a large class of heterotic string
models at weak coupling. Most of the time we will be imagining the
compactification of this theory on an orbifold, though much of our
discussion is applicable to compactification on more general
manifolds. In particular we will {\em not} be considering theories
with nonperturbative structures such as $D$-branes, the positions
and orientations of which would be moduli in the effective
low-energy theory. Nor will we be considering moduli associated with
the different ways in which one can define a vector bundle $V$ on
some compact Calabi-Yau space such that it admits a connection which
satisfies the Hermitian Yang-Mills equations. We will rather be
concerned with the orbifold limit of such theories, and therefore in
a very simple set of moduli: those associated with the fields of the
ten-dimensional supergravity Lagrangian.\footnote{Additional twisted
sector moduli associated with orbifold or orientifold
compactification, such as Wilson line moduli and ``blowing-up''
moduli, are captured in our treatment to the extent that they can be
considered as twisted sector gauge-charged matter in the low-energy
effective Lagrangian.} The relevant bosonic 10D~fields are the
metric $g_{MN}$ ($M,N = 0,\dots,9$), an antisymmetric tensor
$b_{MN}$ and the real dilaton scalar $\phi$. These fields must be
dimensionally reduced to an effective four-dimensional theory. It is
common to package these degrees of freedom into chiral superfields.
The dimensional reduction of the 10D~supergravity Lagrangian and the
field redefinitions required to obtain the chiral superfields of the
4D~supergravity theory have been reviewed
elsewhere.\cite{Binetruy:1988re,Binetruy:1989hg,Bailin:1999nk} Here
we wish to provide only those facts that are relevant to our
subsequent discussion and necessary to establish our conventions
with respect to other authors.

The most important modulus, and the one that will be of central
focus in the class of theories considered here, is the dilaton field
$s$ whose real part determines the (universal) gauge coupling. By
matching the dimensionally-reduced Yang-Mills action to the
canonical form\cite{Witten:1985xb} one immediately identifies the
gauge coupling as
\begin{equation}
\frac{1}{g_{\STR}^2} = e^{3\sigma}\phi^{-3/4}
\label{realdil} \end{equation}
where $\phi$ is the ten-dimensional dilaton and $\sigma$ is the real
scalar which arises from the dimensional reduction of the graviton.
More specifically, $\sigma$ is often referred to as the ``breathing
mode'' associated to the re-scaling $g_{mn} \to e^\sigma g_{mn}$,
with $m,n=4,\dots,9$ being the coordinates for the compact space.
This allows us to set the scale of the Planck mass.
We denote the gauge-coupling as $g_{\STR}$ to indicate (a) that it
is understood to be the coupling at the string scale and (b) that it
is universal to all gauge groups in the low energy
theory.\cite{Kaplunovsky:1987rp} Note that the string scale, which
is inversely related to the string tension via $M_{\STR} =
1/\sqrt{\alpha'}$, is in turn related to the Planck scale via
\begin{equation} M_{\STR} = g_{\STR} M_{\PL} \, . \label{Mstrdef}
\end{equation}

In the effective supergravity theory the dimensional reduction of
the antisymmetric two form $b_{\mu\nu}$ ($\mu, \nu=0,\dots,3$)
appears only through its field strength $H_{\mu\nu\rho} =
\partial_{[\rho} b_{\mu\nu]}$. We may identify~(\ref{realdil}) as the real
part of a scalar field $s$, and $H_{\mu\nu\rho}$ as its pseudoscalar
partner $a$ provided we make the duality transformation
\begin{equation} \epsilon_{\mu\nu\rho\eta} \partial^{\eta} a =
\phi^{-3/2}e^{6\sigma}H_{\mu\nu\rho} \label{adef} \end{equation}
The field $a$ is the so-called ``model-independent'' axion field.

In orbifold compactifications in which the six-dimensional compact
space is factorizable into three two-torii it is possible to define
the K\"ahler and complex structure moduli in terms of the elements
of $g_{mn}$ and $b_{mn}$ (with $m,n=1,2$) for each of three internal
torii. In particular, for the important case of the K\"ahler moduli
we have in this limit the definition
\begin{equation} t^I = \phi^{3/4}\sqrt{\det(g_{mn}^I)} + \frac{i b_{12}^I}{\sqrt{2}} = (R^I)^2
+ \frac{i b_{12}^I}{\sqrt{2}} \label{Tdef} \end{equation}
where $I=1,2,3$ labels the complex plane associated with each torus
and $R^I$ is radius of the compactified subspace in string units.
The final equality in~(\ref{Tdef}) is strictly true only in the
vacuum. We therefore can identify $M_{\rm comp} \equiv 1/R^I = \lang
{\rm Re}\; t^I \rang^{-1/2} M_{\STR}$.

\subsection{Basics of gaugino condensation} \label{sec11}

Past developments in string phenomenology have by now suggested a
number of mechanisms which can be employed to generate potentials
for moduli at the nonperturbative level. Yet one source of these
effects stands out among the others for its ubiquity and generality:
that of gaugino condensation for some non-Abelian gauge group. This
phenomenon is, in principle, a purely field-theoretic effect and can
be considered without regard to some underlying string theory
construction. Before discussing moduli dynamics it is therefore
instructive to consider the physics of gaugino condensation in a
general manner. The subject has a long history in the literature and
we here intend only to motivate the effective Lagrangian approach to
appear in Section~\ref{sec2}. More detailed reviews can be found
elsewhere.\cite{Nilles:1990zd,Quevedo:1995nq}

\subsubsection{Strong coupling and confinement}

Non-Abelian gauge groups with weak coupling in some high-energy
regime are known to reach strong coupling at low-energies through
renormalization group (RG) effects, provided the beta-function
coefficient for the gauge coupling takes the correct sign. The sign
itself is determined by conventions. We will choose conventions such
that the RG equation for the gauge couplings is given by
\begin{equation} {\pp g_a(\mu)\over\pp t} = -
{3b_a\over2}g^3_a(\mu) \; , \label{RGg}\end{equation}
with
\begin{equation}
b_a = \frac{1}{8\pi^2} \( C_a - \frac{1}{3} \sum_i C_a^i \) .
\label{ba}
\end{equation}
In~(\ref{ba}) $C_a$ and $C_a^i$ are the quadratic Casimir operators
for the gauge group $\mathcal{G}_a$ in the adjoint representation
and in the representation of the matter fields $Z^i$ charged under
that group, respectively. Note that these conventions imply that a
group $\mathcal{G}_a$ with $b_a > 0$ will flow to strong coupling in
the infrared.\footnote{The choice of normalization in~(\ref{RGg})
and~(\ref{ba}) was made for future convenience when constructing the
effective superspace Lagrangian. To recover the `standard'
conventions of (for example) Martin and Vaughn\cite{Martin:1993zk}
one must take $b_a \to -(2/3) b_a|_{\rm MV}$.} To estimate the
energy scale at which strong coupling occurs one can simply solve
the RG equation for the gauge coupling. Using the one loop
beta-function, the answer is given by
\begin{equation}
\Lambda_a = \mu e^{-1/3 b_a g_a^2(\mu)}\, . \label{Lambda}
\end{equation}
This expression involves the renormalization scale $\mu$. In our
thought experiment we imagine this scale being some high-energy
scale where string theory sets the 4D effective Lagrangian; thus it
is natural to take $\mu \simeq M_{\PL}$ and $g_a^2(\mu) = g_{\STR}$.

What happens physically at this scale? Our experience with QCD
suggests that confinement can occur. This can be characterized by
nonvanishing \vevs for certain composite objects made up of fermions
charged under the strong group. In a pure super-Yang-Mills (SYM)
theory (a theory with only gauge supermultiplets and no
gauge-charged chiral superfields) the only candidate for such
`condensates' are the fermionic partners of the gauge fields -- the
gauginos. Thus one might naively expect a nonvanishing $\lang
\lambda \lambda \rang$ to develop at or near the scale $\Lambda$. By
dimensional analysis we expect
\begin{equation} \lang \lambda \lambda \rang \sim \Lambda^3
\end{equation}
and using~(\ref{Lambda}) this would seem to suggest
\begin{equation} \lang \lambda \lambda \rang \sim M^3_{\PL}
e^{-1/b_a g_a^2(M_{\PL})}\; . \label{lamlam} \end{equation}

This picture is modified in a number of ways when gauge-charged
matter is present. For starters, the beta-function itself changes to
reflect the presence of these states in the relevant loop diagrams.
More significantly we might expect the chiral matter to confine and
form composite operators. We may treat these as dynamical objects or
assume them to be very massive `matter condensates.' Modifications
to~(\ref{lamlam}) are known and have been computed by explicit
instanton calculations and/or by arguments of anomaly matching and
holomorphy.\cite{Taylor:1982bp,Affleck:1983mk,Affleck:1984xz,Amati:1988ft}
In global superspace a large number of non-Abelian gauge theories,
with and without gauge-charged matter, are known to confine and the
analogs to~(\ref{lamlam}) are
known.\cite{Shifman:1995ua,Skiba:1997sr,Thomas:1998jf,Terning:2003th}
We will discuss the presence of matter when we come to the effective
Lagrangian approach in Section~\ref{sec12} below. For the remainder
of this section we return to the pure SYM theory to make some
additional remarks.

Why do we expect such an object to break supersymmetry?
Consider the equation of motion for the auxiliary field of an
arbitrary chiral superfield in supergravity\footnote{Here and
throughout we use K\"ahler $U(1)$ superspace when discussing
supergravity as it is the most convenient for string-derived
supergravity
Lagrangians.\cite{Binetruy:1987qw,Binetruy:1991sz,Binetruy:2000zx}
Most intuition from the formalism of Wess \& Bagger\cite{WB}
continues to hold with only a few modifications, particularly in the
way expressions are written in superspace notation.
For those unfamiliar with K\"ahler $U(1)$ superspace we have
provided some introductory material on the formalism in~\ref{appA}.
}
\begin{equation} \oline{F}^{\jbar} = -e^{K/2}K^{i\jbar}(W_i + K_iW) -
\frac{1}{4}K^{i\jbar}\frac{\partial f_{a}}{\partial
\varphi_i}(\lambda_a \lambda_a)\; , \label{FEQM} \end{equation}
where $W_i = \partial W/\partial \varphi_i$, $K_{i\jbar}=
\partial^2 K / \partial \varphi_i \partial \oline{\varphi}^{\jbar}$
is the K\"ahler metric and we have set the reduced Planck mass
$m_{\PL}$ to unity, where $m_{\PL}=1/\sqrt{8\pi G} = 2.44 \times
10^{18} \GeV$. The function $f_a$ is the gauge kinetic function
which appears in the Yang-Mills kinetic part of the action
\begin{equation}
\Lag_{\rm YM} = \frac{1}{8} \sum_a \int \diff^4\theta \frac{E}{R}
f_a (\ww)_a + \hc \; . \label{LYM} \end{equation}
In this case we see that {\em provided the gauge kinetic function is
field-dependent} there will generally be a breakdown of
supersymmetry via $\lang F \rang \neq 0$ should some bilinear
$\lambda_a \lambda_a$ acquire a vacuum expectation
value.\cite{Dine:1981za,Witten:1982df,Nilles:1982ik,Ferrara:1982qs}

\subsubsection{Gauge coupling as a dynamical field}

In order to proceed further and discuss moduli stabilization we must
be somewhat more specific. Therefore, consider a case in which the
gauge coupling is determined by the \vev for the lowest component of
some chiral superfield. In anticipation of weakly coupled heterotic
strings we will refer to this chiral superfield $S$ as the dilaton
from~(\ref{realdil}). In the language of superspace effective
Lagrangians, this is equivalent to the statement that the gauge
kinetic function $f_a$ in the expression~(\ref{LYM}) is simply given
by $f_a = S$ for all gauge groups $\mathcal{G}_a$. Note that the
operator $S W^{\alpha}W_{\alpha}$ is formally of mass dimension
five, so there must be some scale $\Lambda_{\UV}$ which
enters~(\ref{LYM}) to restore the canonical mass dimension for
$\Lag_{\rm YM}$. The \vev of the lowest component $s =
S|_{\theta=0}$ determines the (universal) gauge coupling constant
$g$ at this scale $\Lambda_{\UV}$ via the relation
\begin{equation}
\frac{1}{g_a^2} = k_a \frac{\lang {\rm Re}\, s\rang}{\Lambda_{\UV}}
. \label{gbare}
\end{equation}
The integer $k_a$ is the affine level of the gauge group
$\mathcal{G}_a$, as determined by the conformal field theory of the
underlying string construction. For most purposes we will take the
simplest case of $k_a = 1$.
In what follows we will set $\Lambda_{\UV} = m_{\PL} = 1$ and
consider the \vevs of all moduli be to be given in units of this
scale.

Making the naive replacement suggested by~(\ref{gbare})
into~(\ref{lamlam}) leads to the hypothesis that gaugino
condensation will generically generate a potential for the
superfield $S$ (or at least its real part). Let us therefore take
the superpotential generated by the gaugino condensation to be (for
$k_a =1$)
\begin{equation} W_{\rm np}(S) = e^{-\frac{S}{b_a}} \,. \label{WS}
\end{equation}
Note that in writing~(\ref{lamlam}) in this way we have essentially
integrated the gaugino condensates out of the theory, replacing them
with a (holomorphic) function of the moduli. To compute the scalar
potential for the field $S$ we may use the result familiar from
supergravity
\begin{equation}
V= K_{i\jbar} F^i  \oline{F}^{\jbar} - \frac{1}{3} M \oline{M}\; ,
\label{pot}
\end{equation}
where $F^i$ is the auxiliary field associated with the chiral
superfield $\varphi_i$ and $M$ is the auxiliary field of
supergravity. The auxiliary fields can be identified by their
equations of motion
\begin{equation}
F^i = - e^{K/2} K^{i\jbar} \left(\oline{W}_{\jbar} + K_{\jbar}
\oline{W} \right), \; \; \oline{M} = -3e^{K/2} \oline{W} \;
\label{EQM}
\end{equation}
with the gravitino mass given by $m_{3/2} = -\frac{1}{3}\lang
\oline{M}\rang$. The potential~(\ref{pot}) can be
written\footnote{Note that we are here assuming that the
superpotential $W$ depends {\em only} on the chiral dilaton and not
on any other modulus. The case of more general $W(S,T)$ will be
considered in Section~\ref{sec:WST} below.}
\begin{equation}
V(s,\bar{s}) = K_{s\bar{s}} |F^S|^{2} -3e^{K}|W|^{2} = e^{K}
K^{s\bar{s}} |W_s + K_sW|^{2} -3e^{K}|W|^2 . \label{sdilpot}
\end{equation}
%

\begin{figure}
\centerline{\psfig{file=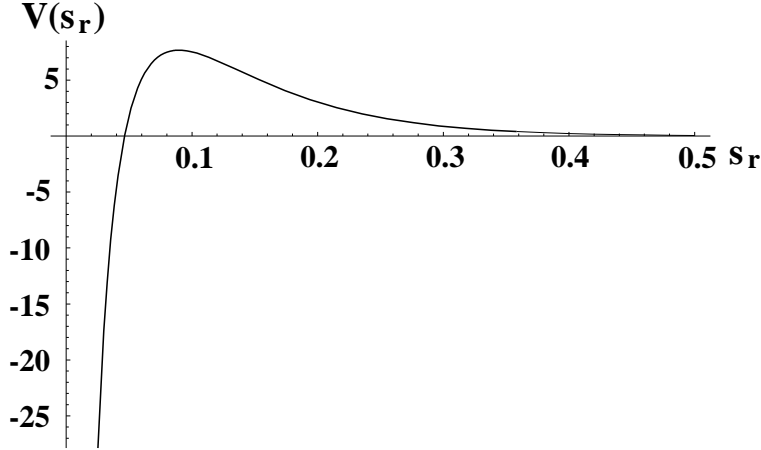,width=10cm}} \vspace*{8pt}
\caption{\textbf{Plot of dilaton potential $\mathbf{V(s+\bar{s})}$
for the superpotential $\mathbf{W_{\rm np}(S) = e^{-\frac{S}{b_a}}}$
with $\mathbf{b_a = 5/8\pi^2}$}. The shape of the potential as a
function of $s_r = (s+\bar{s}) = 2\,{\rm Re}\,s$ is given with the
height of the potential in arbitrary units ($m_{\PL} = 1$). The
asymptotic behavior as $s_r \to \infty$ is the so-called dilaton
``runaway'' problem.} \label{fig:Vs1}
\end{figure}

At the classical level in string theory, the K\"ahler potential for
the field $S$ has the form
\begin{equation}
K=-\ln(S+\oline{S})\; , \label{KStree}\end{equation}
and the nonperturbatively generated superpotential in~(\ref{WS})
gives $W_s = -(1/b_{a}) W$. The resulting dilaton
potential~(\ref{sdilpot}) is plotted in Figure~\ref{fig:Vs1} for
$b_a = 5/8\pi^2$, as would be the case for condensation of pure
$SU(5)$ Yang-Mills fields.
In the limit ${\rm Re}\,s \to 0$ (which corresponds to strong gauge
coupling) the potential is unbounded from below. There is an
extremum that depends weakly on the precise value of $b_a$ chosen
but is generally at a value of ${\rm Re}\;s$ where $\alpha =
g^2/4\pi \simeq 1$. Finally there is the `runaway' solution where
${\rm Re}\,s \to \infty$ or $g^2 \to 0$. In this limit supersymmetry
is restored as both the superpotential and its covariant derivatives
vanish.

The potential determined by~(\ref{WS}) and~(\ref{KStree}) is
incapable of breaking supersymmetry and providing a realistic
minimum for the dilaton.
A number of mechanisms were quickly suggested to correct this
behavior, and we will address some of them in subsequent sections.
For now we continue in the spirit of simplicity ({\em i.e.}
one-modulus models) and here consider the case of multiple
condensates.\cite{Krasnikov:1987jj,Casas:1990qi}
A single, simple gauge group $\mathcal{G}$ in the hidden sector was
motivated by some of the earliest Calabi-Yau constructions where the
hidden sector was an entire factor of
$E_8$.\cite{Candelas:1985en,Witten:1985xc} In general, however, (and
particularly for orbifold compactifications) we expect a hidden
sector gauge group that is given by a product of simple groups:
$\mathcal{G} = \prod_a \mathcal{G}_a$, or perhaps more generally
\begin{equation}
\mathcal{G}_{\rm hidden} = \prod_{a=1}^n\mathcal{G}_a\otimes U(1)^m
\; .\end{equation}
Some subset of these $\mathcal{G}_a$ will be asymptotically free and
can therefore form a gaugino condensate. Taking the simplest case of
just two gaugino condensates in the hidden sector we would expect a
superpotential of the form\cite{Krasnikov:1987jj,Dixon:1990ds}
\begin{equation}
W(S) = \Lambda^{3} \[d_{1}e^{-k_1 S/b_{1}} + d_{2}e^{-k_2 S/b_{2}}\]
. \label{raceW}
\end{equation}
Here $b_{1}$ and $b_{2}$ are the beta-function coefficients, defined
by~(\ref{ba}), for the two condensing groups ${\cal G}_{1}$ and
${\cal G}_{2}$. The quantities $d_{1}$ and $d_{2}$ parameterize the
presence of possible matter in the hidden sector, while the
parameters $k_1$ and $k_2$ might represent differing affine-level
for the gauge groups. For dilaton stabilization with a realistic
minimum to occur, we must require that the scalar potential $V(S)$
given in~(\ref{sdilpot}) give rise to a minimum such that $\lang
s+\bar{s}\rang/2 = 1/g_{\STR}^{2} \simeq 2$ while generating a
gravitino mass $m_{3/2} = \lang e^{K/2} W\rang$ of $\order(1 \TeV)$.
Each of the parameters in the set $\lbr b_{1},\, b_{2},\,
d_{1},\,d_{2},\,k_1,\,k_2\rbr$ are not continuously variable, but
depend upon particulars of the compactification in a calculable way.
Superpotentials of the form~(\ref{raceW}) are often referred to as
``racetrack'' models in that two exponential functions must be
balanced against one another in a delicate way to achieve the
desired minimum.

\begin{figure}
\centerline{\psfig{file=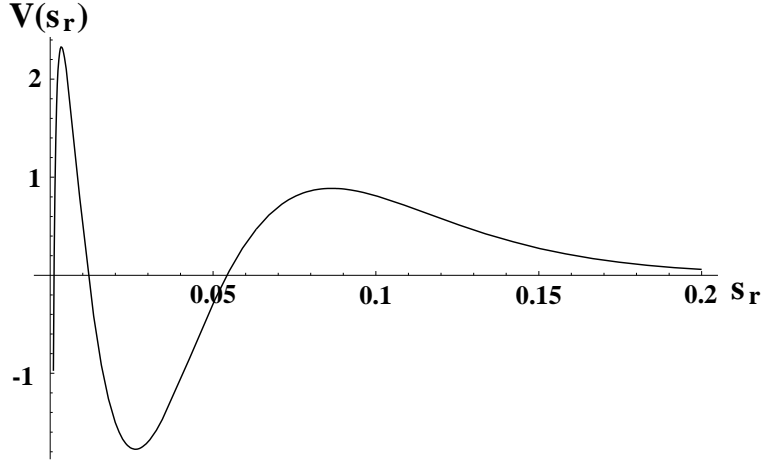,width=10cm}} \vspace*{8pt}
\caption{\textbf{Plot of dilaton potential $\mathbf{V(s+\bar{s})}$
for the superpotential of~(\ref{raceW})}. The shape of the potential
as a function of $s_r = (s+\bar{s}) = 2\,{\rm Re}\,s$ is given with
the height of the potential in arbitrary units ($m_{\PL} = 1$).
Values of the parameter set are given in the text below. }
\label{fig:racetrack}
\end{figure}

If we take the simplest possible
case\cite{Taylor:1985fz,Ferrara:1990ei,Taylor:1990wr} in which $d_a
= -b_a/6e$ and take a hidden sector comprising of ${\cal G}_{1} =
SU(7)$ with 8 $\mathbf{7}+\mathbf{\bar{7}}$'s and ${\cal G}_{2} =
SU(8)$ with 15 $\mathbf{8}+\mathbf{\bar{8}}$'s then a solution
exists for $k_1 = 2$ and $k_2 = 1$. The balancing of two $SU(N)$
groups with similar beta-function coefficients is a common feature
of solutions in the racetrack scenario. The resulting scalar
potential is plotted in Figure~\ref{fig:racetrack} above. The
asymptotic runaway behavior as $s_r \to \infty$ is still present, as
is the unbounded from below direction near the origin. But now a
nontrivial minimum develops for the real part of the dilaton scalar.
This is a major improvement over the single condensate case, though
there are two properties that make this minimum unrealistic. First,
the value of the dilaton field at this minimum suggests far too
strong a gauge coupling $g^2_{\STR} \simeq 30$. Second, the minimum
does not occur with vanishing value of $\lang V({\rm Re}\,s) \rang$
but rather with a large negative value.

The former problem of the dilaton \vev can be rectified by a number
of mechanisms. The relationship between the beta-function parameters
$b_a$ and the coefficients $d_a$ can be generalized. For example,
one might consider the effect of threshold corrections on the
beta-functions for the couplings of groups ${\cal G}_{1}$ and ${\cal
G}_{2}$ (as well as on the scales $\Lambda$ which appear in
equation~\ref{raceW}). These effects might arise from integrating
out heavy vector-like matter charged under those groups, with masses
above the condensation scale (such as states charged under some
anomalous $U(1)$ factor).\cite{Casas:1990qi} If the hidden sector
matter is to be integrated out below the scale of gaugino
condensation then a nontrivial $d_{a}$ is generated for each
condensing group whose form depends on whether the matter is
vector-like in nature\cite{deCarlos:1992da} or only forms
condensates of dimension three or higher.\cite{Binetruy:1996nx}
Another solution (often invoked in conjunction with the above) is to
allow the nonperturbative superpotential to depend on more than one
modulus -- in particular, the K\"ahler modulus
of~(\ref{Tdef}).\cite{Casas:1990qi,Cvetic:1991qm,delaMacorra:1992we,deCarlos:1992da}
Such a dependence is required in modular invariant treatments of
effective Lagrangians from heterotic string theory. We will discuss
this further in Section~\ref{sec2}.

The success of multi-condensate models in achieving supersymmetry
breaking with realistic values of the gauge coupling have made them
the preferred starting-point for most phenomenological treatments of
string constructions. Yet even in the multi-modulus (and modular
invariant) treatments they suffer from a common problem: the minimum
of the potential continues to imply negative vacuum energy ({\em
i.e.} the solution is one with anti-de Sitter spacetime). This is a
phenomenological disaster that requires the invocation of some
additional physics that enters to ``rescue'' the theory and set the
vacuum energy to zero (or very nearly zero).\footnote{It might be
hoped that quantum corrections to the vacuum energy of the theory
might conspire to ``lift'' the negative minimum to a positive or
vanishing one, without invoking any additional fields or
interactions. This is not impossible, but we argue that such
corrections are generally insufficient to rectify the problem, and
may even make it worse.\cite{Gaillard:2005cw}} Today the most
effective and robust such method is the inclusion of flux -- that
is, vacuum expectation values for the field strengths of fields from
the Ramond-Ramond sector of the string theory.\cite{Grana:2005jc}
But such mechanisms do not readily present themselves in the weakly
coupled heterotic string context. We are thus left to consider other
field-theoretic and/or string-theoretic effects which must be
included in the effective Lagrangian to allow for a realistic
minimum. For the class of models being discussed in this work the
mechanism will be that of {\em K\"ahler stabilization}, and it will
be the focus of Section~\ref{sec13} below. But before we can
consider the implementation of K\"ahler stabilization we must
discuss the effective Lagrangian approach to gaugino condensation.

\subsection{Condensates as effective operators}
\label{sec12}

The treatment of condensates in supergravity is greatly simplified
by using an effective operator
approach.\cite{Veneziano:1982ah,Nilles:1982ik} The starting point
for our analysis is the definition of the composite field operator
$U_a$ which represents a $\mathcal{G}_a$-charged gauge condensate
chiral superfield
\begin{equation} U_a\simeq {\Wc_a\Wa^a} \; . \label{Ucondef} \end{equation}
Note that the lowest component of $u_a = U_a\lowest$ involves the
gaugino bilinear $\lambda_a \lambda_a$. We wish to write an
effective Lagrangian describing the dynamics of this new chiral
superfield such that the behavior of the Lagrangian under symmetry
transformation is matched by the behavior of the underlying
supersymmetric gauge theory (with unconfined gauginos). For global
SUSY the resulting effective Lagrangian (for a single condensate) is
that of Veneziano and Yankielowicz~(VY)\cite{Veneziano:1982ah}
\begin{equation}
\Lag_{\rm eff} = \int \diff^4 \theta (U U^{\dagger})^{1/3} -
\frac{1}{4}\[ \int \diff^2 \theta \( b' \, U \ln \frac{U}{\mu^3} +
\frac{1}{g^2}U\) + \hc
\] \; . \label{VY1}
\end{equation}
Here $\mu$ is the renormalization group invariant scale of the
condensing group. The logarithmic part of the second term
in~(\ref{VY1}) can be combined with the standard Yang-Mills term
in~(\ref{LYM}) and is re\-pre\-sent\-ed by the following
su\-per\-po\-ten\-tial
ex\-pres\-sion\cite{Taylor:1988vt,Binetruy:1995hq,Binetruy:1995ta}
\begin{equation}
W(U,S)_{\VY} = \frac{1}{4}U\[ S +  b' U \ln(U/\mu^3) \] \, ,
\label{CasasW}
\end{equation}
where $b'$ is a constant coefficient which we will determine
presently. If there are multiple condensates labeled by the index
$a$ as in~(\ref{Ucondef}), then the final term in~(\ref{CasasW})
will generally involve a different coefficient $b'_a$ for each
condensate $U_a$.

To the Lagrangian in~(\ref{VY1}) we wish to add the possibility of
matter charged under the condensing group to obtain the Veneziano,
Yankielowicz and Taylor (VYT)
Lagrangian.\cite{Veneziano:1982ah,Taylor:1982bp,Lust:1990zi} From
our experience in QCD we generally expect states charged under the
strong group to experience confinement and form composites. We will
represent these by the composite field operators
\begin{equation} \Pi_a^\alpha \simeq
\prod_{i} \(\Phi_i^{(a)}\)^{n_i^{\alpha,(a)}},\label{Picondef}
\end{equation}
where the product involves only those fields $\Phi_i^{(a)}$ charged
under the confined group $\mathcal{G}_a$. In~(\ref{Picondef}) the
label $\alpha$ is a species index for the matter condensates, each
of which may consist of different component fields labeled by the
integers $n_i^{\alpha,(a)}$. Note that the canonical mass dimension
of this operator $\Pi_a^\alpha$ is given by
\begin{equation}
{\rm dim}\(\Pi_a^\alpha\)\equiv d_a^{\alpha} = \sum_{i}
n_i^{\alpha,(a)} \label{dimpi}
\end{equation}
The generalization to
supergravity\cite{Taylor:1985fz,Binetruy:1989dk} of the VYT
effective action (that is, the generalization of~(\ref{VY1}) to
include matter in K\"ahler $U(1)$ superspace) has the following
general form
\begin{equation} \Lag_{\rm VYT} = {1\over8}\superint\,{E\over R}\sum_a
U_a\[b'_a\ln(e^{-K/2}U_a) + \sum_\alpha b^\alpha_a\ln\Pi_a^\alpha\]
+ {\rm h.c.}\; , \label{VYT} \end{equation}
where there are now two separate coefficients $b'_a$ and
$b_a^{\alpha}$ which must be determined for each condensing group
$\mathcal{G}_a$. These are obtained by matching the anomalies of the
effective theory to those of the underlying theory. The
Lagrangian~(\ref{VYT}) has the correct anomaly structure under
K\"ahler $U(1)$, R-symmetry,
conformal transformations,
and modular (T-duality) transformations
provided the conditions
\begin{equation} b'_a = {1\over8\pi^2}\(C_a - \sum_iC_a^i\) ,\;\;\;\; \qquad
b^\alpha_a = \sum_{i\in\alpha}{C^i_a\over4\pi^2d_a^\alpha}\; ,
\label{cond} \end{equation}
are satisfied.  The composite chiral superfields $\Pi^\alpha_a$ are
invariant under the nonanomalous symmetries, and may be used to
construct an in\-var\-i\-ant
su\-per\-po\-ten\-tial.\cite{Taylor:1982bp,Gaillard:2003gt} Provided
there are no invariant chiral fields of dimension two, and no
additional global symmetries (such as chiral flavor symmetries), the
dynamical degrees of freedom associated with the composite
fields~(\ref{Ucondef}) and~(\ref{Picondef}) acquire
masses\cite{Wu:1996en} larger than the condensation scale
$\Lambda_a$, and may be integrated out. This results in an effective
theory constructed as described in~(\ref{VYT}) with the composite
fields taken to be nonpropagating; that is, they do not appear in
the K\"ahler potential.

We note that the conditions~(\ref{cond}) are not a unique solution
to the set of anomaly constraints. However they are the most
straightforward solution, intuitively plausible, and hold in the
presence of additional anomalous symmetries, since the weights of
the condensates $\Pi$ are just the products of the weights of their
constituents.  As an explicit example of a fully constrained model,
consider the VYT action for $SU(N_c)$ with chiral flavor symmetry.
We have N ``quark'' and N ``anti-quark'' chiral supermultiplets
$Q^A$ and $Q^c_A$, respectively.  We take the quark condensates to
be the matrix-valued ``meson'' superfield $\bpi^A_B = Q^A Q^c_B$. We
do not assume {\it a priori} that these are static fields.  If
$\mathcal{G}_a = SU(N_c)\equiv\mathcal{G}_Q$ we take
\begin{equation} \Lag^Q_{\rm VYT} = {1\over8}\superint{E\over
R} U_Q \[b'_Q\ln(e^{-K/2}U_Q) + b^\alpha_Q \ln(\det\bpi)\] + \hc \;
, \label{lvyt} \end{equation}
where here the form of the matter condensate is dictated by
invariance under flavor $SU(N)_L\otimes SU(N)_R$. For the elementary
fields we have $C_Q = N_c$, $C^i_Q = \frac{1}{2}$ and $\sum_i C^i_Q
= N$. Under K\"ahler $U(1)$ R-symmetry, anomaly matching requires
\begin{equation} b'_Q = {1\over8\pi^2}\(N_c - N\)\,
,\label{bpq}\end{equation}
and under the conformal transformation
\begin{equation} \begin{array}{ll}
\lambda_a \to e^{3\sigma/2}\lambda_a & \qquad \Phi^i\to e^{\sigma}\Phi^i \\
U_a \to e^{3\sigma}U_a & \qquad \Pi^\alpha_a \to
e^{d_a^\alpha\sigma}\Pi^\alpha_a \end{array}
\label{conft}\end{equation}
with $\bpi \to e^{2\sigma}\bpi$, we require
\begin{equation} 3b'_Q + 2N b^\alpha_Q = {1\over8\pi^2}\(3N_c - N\) = 3
b_Q,\label{conf}\end{equation}
where $b_Q$ is the $\beta$-func\-tion co\-ef\-fi\-cient defined
by~(\ref{ba}). Put\-ting these to\-geth\-er gives~(\ref{bpq}) and
\begin{equation} b^\alpha_Q = {1\over8\pi^2}, \qquad b_Q = b'_Q +
{2N\over3}b^\alpha_Q\, ,\label{baq}\end{equation}
in agreement with the general result~(\ref{cond}) for $d_Q^\alpha =
2N$ and $\pi_Q^\alpha = \det\bpi$.
It is easy to see that the anomaly matching condition
\begin{equation} 2N b^\alpha_Q =
\sum_i{C^i_Q\over4\pi^2}\end{equation}
under chiral $U(1)$ transformations $Q\to e^{i\beta}Q$, $Q^c\to
e^{i\beta}Q^c$ is also satisfied by~(\ref{bpq})
and~(\ref{baq}).\footnote{Though we have reserved the issue of
modular invariance for the next section, we remark here that it is
also very easy to see that should $Q,Q^c$ have nontrivial modular
weights under T-duality the modular anomaly matching condition is
also satisfied by~(\ref{bpq}) and~(\ref{baq}).}

It is instructive to compare the effective theory defined
by~(\ref{VYT}) with results\cite{Davis:1983mz,Seiberg:1994bz} based
on holomorphy of the superpotential by going to the rigid SUSY
limit, and neglecting the moduli and the dilaton; $s\to g_0^{-2}=$
constant. Then the superpotential reduces to the standard VYT one:
\begin{equation} W(U_Q) = {1\over4}U_Q\[g_0^{-2} + b'_Q\ln(U_Q) +
b^\alpha_Q\ln(\det\bpi)\] \; ,\label{wvy}\end{equation}
which is the analog of~(\ref{CasasW}).
Keeping $U_Q$ static and imposing the equation of motion for the
auxiliary field $F^Q$ gives the potential
\begin{eqnarray} - V &=& \Tr\[\mathbf{\bar{F}}^\pi\mathbf{K}''\mathbf{F}^\pi
+ \lbr\mathbf{F}^\pi\(\mathbf{M} + {1\over4}b^\alpha_Q
u_Q\bpi^{-1}\) + \rm h.c.\rbr\], \nonumber\\ u_Q &=&
e^{-1}\({\Lambda_Q^{3N_c-N}\over\det\bpi}\)^{1/(N_c -
N)}\label{susyv}\end{eqnarray}
where $\mathbf{K}''$ is the (tensor-valued) K\"ahler metric for
$\bpi$. The potential~(\ref{susyv}) is derivable from the following
superpotential for the dynamical superfield $\bpi$
\begin{equation} W_{\bpi} = \Tr(\mathbf{M}\bpi) - {(N_c -   N)\over32\pi^2e}
\({\Lambda_Q^{3N_c-N}\over(\det\bpi)}\)^{1/(N_c -
N)},\label{wnp}\end{equation}
which, up to a factor\footnote{The factor $e$ comes from the fact
that we take the derivative of $\int U\ln U$, while Davis {\it et
al.} start with $\int <\lambda\lambda>\ln\Lambda$ and determine
$<\lambda\lambda>$ from threshold matching.\cite{Davis:1983mz} The
minus sign comes from the convention of
Ref.~\refcite{Binetruy:2000zx}: $u\sim\left. \Wc\Wa\right| = -
\lambda\lambda$.}  $-2/e$, is the superpotential found by Davis {\it
et al.}\cite{Davis:1983mz}

\subsection{K\"ahler stabilization and the linear multiplet}
\label{sec13}

To adopt the form of~(\ref{CasasW}) as an effective superpotential
term requires that the gauge kinetic function be written in terms of
a linear combination of holomorphic objects -- that is, of {\em
chiral} superfields. Most treatments of gaugino condensation utilize
the formalism of the previous section -- namely the description of
dilaton-like objects as chiral superfields. This has the convenience
of familiarity, but almost all extensions of the theory beyond that
presented in Sections~\ref{sec11} and~\ref{sec12} quickly reveal the
limitations of the chiral superfield treatment. In this section we
will consider an alternative and dual formulation of the Yang-Mills
action in terms of {\em real} dilatonic superfields. This will prove
to be convenient in implementing the K\"ahler stabilization
mechanism for solving the vacuum energy problem described in
Section~\ref{sec11}.

\subsubsection{The linear multiplet}

It is worth recalling the properties of the fundamental degrees of
freedom obtained from direct dimensional reduction of the massless
string spectrum. For each ``dilaton'' (that is, for each field whose
\vev determines a gauge coupling) the 4D massless modes are a {\em
real} scalar $\ell$, an antisymmetric two-index tensor $b_{\mu\nu}$
%
%
and a (Majorana) Weyl fermion $\chi_{\ell}$. These are precisely the
degrees of freedom of the linear multiplet.\cite{Siegel:1979ai} In
particular there is no need to perform the duality
transformation~(\ref{adef}) in order to generate a pseudoscalar
``partner'' for the real dilaton. This is particularly important
when one considers higher-genus corrections to the effective
Lagrangian, where the duality relation~(\ref{adef}) is replaced by a
much less straightforward field identification.\cite{Giedt:2003ap}
Note also the absence of an auxiliary field in the massless spectrum
for this multiplet.

A linear multiplet $\widehat{L}$ is essentially a special case of a
real vector superfield defined by the
requirement\cite{Ferrara:1974ac,Ferrara:1986jz,Binetruy:1987eu,Cecotti:1987nw}
\begin{equation} -(\DaDa-8R^{\dagger})\widehat{L}=0, \quad -(\DbDb-8R)\widehat{L}=0 \; .
\label{lindef} \end{equation}
The requirement that the chiral projection of $\widehat{L}$ vanish
already ensures that the vector component $v_{\mu}$ of the multiplet
is Hodge-dual to the field strength of a two-index antisymmetric
tensor -- precisely the field that appears
in~(\ref{adef}).\cite{Cecotti:1987nw}
In other words, the definition~(\ref{lindef}) automatically enforces
the Bianchi identity
\begin{equation} \partial^{\mu} v_{\mu} = 0 \label{Bianchi1}
\end{equation}
for the vector component of $\widehat{L}$, which we identify as
\begin{equation} v_{\mu} = \epsilon_{\mu\nu\rho\sigma} \partial^\nu
b^{\rho\sigma} = \frac{1}{2}\epsilon_{\mu\nu\rho\sigma}
H^{\nu\rho\sigma}\, . \label{vmudef}
\end{equation}
The lowest component $\ell$ of the superfield $\widehat{L}$ is then
the dilaton and the relation~(\ref{gbare}) is replaced by
\begin{equation} \frac{g_{\STR}^2}{2} = \lang \ell \rang \; . \label{gbare2}
\end{equation}
This quantity represents the string loop expansion parameter.
Therefore string-theory information from higher loops is more
naturally encoded in terms of this set of component fields. At the
leading order the chiral and linear formalisms are related by the
simple superfield identification
\begin{equation} \widehat{L} = \frac{1}{S+\oline{S}} \label{treeLSdual}
\end{equation}
though this identification fails to be satisfactory at higher loop
level.

The antisymmetric tensor field of superstring theories undergoes
Yang-Mills gauge transformations, which implies that the superfield
$\widehat{L}$ which contains this degree of freedom is not gauge
invariant. It is possible to define a {\em modified linear
multiplet} $L$ which recovers gauge invariance by introducing
Yang-Mills Chern-Simons forms to the definition
in~(\ref{lindef}).\cite{Binetruy:1987eu} We define the YM
Chern-Simons form as
\begin{equation} \Omega_{\mu\nu\rho} = A^a_{[\mu}F_{\nu\rho]\,a} -
\frac{1}{3}f_{abc}A_{\mu}^a A_{\nu}^b A_{\rho}^c \, , \label{CSdef}
\end{equation}
where $f_{abc}$ are the structure constants of the group, and then
promote this to a real superfield $\Omega$. The new field $L$ will
obey the modified linearity conditions
\begin{eqnarray}
-(\DbDb-8R)L\,&=&\,(\DbDb-8R)\Omega\,=\,\sum_{a}(\ww)_a,
\nonumber \\
-(\DaDa-8R^{\dagger})L\,&=&\,(\DaDa-8R^{\dagger})\Omega\,=\,
\sum_{a} (\bww)_a\, . \label{modlin} \end{eqnarray}
Neither $L$ nor $\Omega$ are gauge-invariant individually, but the
combination $\widehat{L} = L + \Omega$ now is.\footnote{The choice
of signs in the relations~(\ref{modlin}) is one of conventions and
several exist in the literature. For example, the conventions used
here are those of Gaillard and Taylor.\cite{Gaillard:1992bt} They
differ by the presence of the minus sign on the first terms from
those of earlier work.\cite{Binetruy:1987eu,Binetruy:1991sz}}
For those subgroups $\mathcal{G}_{a'}$ which experience confinement
it is natural to identify the chiral projection of the modified
linear multiplet as the chiral superfield $U_{a'}$
in~(\ref{Ucondef}). We therefore have
\begin{eqnarray}
-(\DbDb-8R)L\,&=& \sum_{a}(\ww)_a + \sum_{a'} U_{a'},
\nonumber \\
-(\DaDa-8R^{\dagger})L\,&=& \sum_{a} (\bww)_a + \sum_{a'}
\overline{U}_{a'}\, . \label{Uproject} \end{eqnarray}
We will see below that this identification is crucial to a correct
implementation of certain Bianchi identities in the Yang-Mills
sector of the theory.

The generic Lagrangian describing the coupling of the modified
linear multiplet to supergravity and chiral superfields, in the
presence of Yang-Mills Chern-Simons superforms,
is\cite{Binetruy:1987eu,Binetruy:1991sz,Gaillard:1992bt}
\begin{equation}
K = k(L) + K(\Phi,\oline{\Phi}), \quad  \Lag =
-3\superint\,E\,F(\Phi,\oline{\Phi},L) \label{Klinear}
\end{equation}
where the quantity $K(\Phi,\oline{\Phi})$ represents the
contribution of chiral superfields (matter and/or additional moduli
fields) to the K\"ahler potential. We will deal with chiral matter
more thoroughly in Section~\ref{sec2}; in the discussion here they
will play only a trivial role. Note that the K\"ahler potential does
not appear explicitly in the superspace Lagrangian, but rather
implicitly through the supervierbein $E$. The component expansion
of~(\ref{Klinear}) contains the kinetic terms for the supergravity
multiplet as well as the Yang-Mills fields and the linear multiplet
itself. The two functions $k(L)$ and $F(L)$ are not entirely
arbitrary; they are constrained by the requirement that the
Einstein-Hilbert term in the component expansion of~(\ref{Klinear})
have canonical normalization. Under this constraint $k(L)$ and
$F(L)$ are related to each other by the following first-order
differential equation\cite{Binetruy:1987eu,Binetruy:1991sz}
\begin{equation}
F- L\frac{\partial F}{\partial L} = 1- \frac{1}{3}L \frac{\partial
k}{\partial L} \, . \label{relation} \end{equation}
The general solution to~(\ref{relation}) reads\cite{Binetruy:1991sz}
\begin{equation}
F(\Phi,\oline{\Phi},L) = 1 + \frac{1}{3} L V(\Phi,\oline{\Phi}) +
\frac{1}{3}L \int \frac{\diff L}{L} \frac{\partial k(L)}{\partial
L}\, . \label{FofLsol} \end{equation}
Once the functional form of $k(L)$ is specified the last term
in~(\ref{FofLsol}) is fixed. This leaves only the second term for a
nontrivial interaction between the modified linear multiplet and
matter within the function $F(\Phi,\oline{\Phi},L)$. This term is a
form of ``integration constant'' for the differential equation
in~(\ref{relation}). Such a term will play a very important role in
the implementation of the Green-Schwarz mechanism\cite{Green:1984sg}
and the inclusion of nonholomorphic threshold corrections to gauge
couplings in Section~\ref{sec2} below. Its natural emergence here is
one of the benefits of using the linear multiplet in string-derived
supergravity models. At tree level we expect from~(\ref{treeLSdual})
that the K\"ahler potential for the linear dilaton should be simply
$k(L) = \ln L$. This implies the $V=0$ solution to~(\ref{FofLsol})
is a constant $F(L) = 2/3$, and thus
\begin{equation} \Lag_{\rm kin} = -2 \superint \, E \, . \label{treeLkin}
\end{equation}

For the general form~(\ref{Klinear}) we must also generalize the
duality relationship in~(\ref{treeLSdual}). Consider the
Lagrangian\cite{Binetruy:1991sz,Gaida:1994wn}
\begin{equation} \Lag_{\rm lin} = -3\superint \, E \[
F(\Phi,\oline{\Phi},L) + \frac{1}{3} (L + \Omega)(S+\oline{S})\]
\label{Laglin} \end{equation}
where $L$ is now an unconstrained superfield. After eliminating $S$
by using its classical equation of motion one
obtains~(\ref{Klinear}). By varying with respect to $L$ and
demanding canonical Einstein term we arrive at~(\ref{FofLsol}) and
the new duality relation
\begin{equation} F(\Phi,\oline{\Phi},L) + \frac{1}{3}L(S+\oline{S}) =
1 \, . \label{newdual} \end{equation}
For the simple tree-level case with $F=2/3 + LV/3$ we have
\begin{equation} L=\frac{1}{(S+ \oline{S} + V)} \end{equation}
and we once again recover~(\ref{treeLSdual}) in the limit of
vanishing integration constant.

With the introduction of Yang-Mills Chern-Simons forms, the Bianchi
identity satisfied by the vector component of the modified linear
multiplet is no longer~(\ref{Bianchi1}) but is
instead\cite{Binetruy:1995hq,Binetruy:1996xw}
\begin{equation} \partial^{\mu} v_{\mu} = \frac{1}{8} ^{*}\Phi
\label{Bianchi2} \end{equation}
where $^{*}\Phi$ is related to the field strength of a rank-3
antisymmetric tensor field $\Gamma^{\nu\rho\sigma}$
\begin{equation}
^{*}\Phi = \frac{1}{3!} \epsilon_{\mu\nu\rho\sigma}\partial^{\mu}
\Gamma^{\nu\rho\sigma} \, . \label{4form} \end{equation}
The expression in~(\ref{4form}) is the analog to the axionic duality
relation in~(\ref{adef}).\footnote{The presence of additional terms
in the new Bianchi identity~(\ref{Bianchi2}) implies that the
would-be classical shift symmetry ({\em i.e.} a Peccei-Quinn type
symmetry) enjoyed by the model-dependent axion of the dilaton
multiplet has been broken to a restricted class of shifts by
nonperturbative effects.} The three-form supermultiplet can be
described in flat superspace by a chiral superfield $Y$ and
anti-chiral superfield $\oline{Y}$ such that\cite{Gates:1980ay}
\begin{equation} D^2 Y - \oline{D}^2 \oline{Y} = \frac{8i}{3}
\epsilon^{\mu\nu\rho\sigma} \Sigma_{\mu\nu\rho\sigma} \, ,
\label{D2Y} \end{equation}
where $\Sigma_{\mu\nu\rho\sigma}$ is the gauge-invariant field
strength of the rank-three gauge potential superfield
$\Gamma_{\nu\rho\sigma}$.
Applying~(\ref{D2Y}) to the specific case of our modified linear
multiplet in curved superspace we find the constraint
\begin{eqnarray}
(\DaDa-24R^{\dagger})U\,-\,(\DbDb-24R)\overline{U} &=& \mbox{total
derivative} \nonumber \\
 &=& \frac{i}{3}\epsilon_{\mu\nu\rho\sigma}\partial^{\mu}
\Gamma^{\nu\rho\sigma} \label{Uconstraint} \\
 &=& 2i ^{*}\Phi \nonumber
 \end{eqnarray}
with a similar constraint for the unconfined YM fields via the
replacement $U \to \ww$. Indeed, the composite operator $\ww$ can be
interpreted as the degrees of freedom of the three-form field
strength.\cite{Gates:1980ay,Binetruy:1996xw} The general solution to
the constraint equation~(\ref{Uconstraint}) is a field $U$ which is
identified with the chiral projection of a real superfield --
precisely as in~(\ref{Uproject}).

The traditional chiral formulation of gaugino condensation is
incorrect in that it treats the interpolating field $U=e^{K/2} H^3$
with $H^3 = \ww$ as an ordinary chiral superfield of K\"ahler chiral
weight $w=2$. But this is inconsistent
with~(\ref{Uconstraint}).\cite{Burgess:1995kp,Burgess:1995aa,Binetruy:1995hq,Binetruy:1995ta}
In the general formulation of duality transformations, couplings of
the dilaton to matter entail duality invariance of the corresponding
terms in the Lagrangian, as opposed to their couplings to gauge
fields, which are only an invariance of the equations of
motion.\cite{Binetruy:1995ta} One must either
apply~(\ref{Uconstraint}) religiously everywhere, or begin by using
the modified linear multiplet in the first place.

This is not a purely academic concern: failure to properly
incorporate the constrained Yang-Mills geometry inherited from the
underlying string dynamics can have demonstrable effects on the
low-energy phenomenology inferred from the component Lagrangian. For
example, in the present formalism the condensate superfields $U_a$
of~(\ref{Ucondef}) are introduced as static chiral superfields.
Their highest components $F_{U_a}$, defined by
\begin{eqnarray} F_{U_a} &=& -\frac{1}{4}\DaDa U_a \lowest
\equiv -\frac{1}{4} \D^2 U_a\, ,  \nonumber \\
\oline{F}_{\oline{U}_a} &=& -\frac{1}{4}\DbDb \oline{U}_a \lowest
\equiv -\frac{1}{4} \oline{\D}^2 \oline{U}_a\, , \label{FU}
\end{eqnarray}
therefore appear only linearly in the component Lagrangian. One
might believe that they can thus be removed from the theory by
solving their equations of motion $F_{U_a} = \oline{F}_{\oline{U}_a}
= 0$. However a subtlety arises in deriving the corresponding
equations of motion for these auxiliary fields as a result
of~(\ref{Uconstraint}),\cite{Binetruy:1996xj,Giedt:2002ku} for which
we must require\footnote{One way to ensure this constraint is to
first rewrite $F_{U_a}$ as
\begin{equation} F_{U_a} = \frac{1}{2} \(F_{U_a} + \oline{F}_{\oline{U}_a}\)
+ 2i \nabla^{\mu}v_{\mu}^a + \frac{1}{2} \(u_a\oline{M} - \bar{u}_a
M \)\, , \label{jjhe} \end{equation}
(and the conjugate expression for $\oline{F}_{\oline{U}_a}$), and
then vary the Lagrangian with respect to the {\em unconstrained}
combination $F_{U_a} + \oline{F}_{\oline{U}_a}$.}
\begin{equation}
-\frac{1}{4}\(\D^2 U_a - \oline{\D}^2 \oline{U}_a\) = F_{U_a} -
\oline{F}_{\oline{U}_a} = 4i \nabla^{\mu}v_{\mu}^a + u_a\oline{M} -
\oline{u}_a M \, . \label{Fconstraint} \end{equation}
The first term is the ``total derivative'' of~(\ref{Uconstraint}),
while the last two terms arise from the lowest components of the
superfield terms $24R^{\dagger} U_a-24R\oline{U}_a$. At a
supersymmetry-breaking minimum of the scalar potential we generally
expect nonvanishing \vevs for both the condensates ($\lang u_a \rang
\sim \lang \lambda_a \lambda_a \rang$) and the auxiliary field of
supergravity ($\lang M \rang \sim m_{3/2}$, see equation~\ref{EQM}).
Therefore~(\ref{Fconstraint}) is a nontrivial constraint that can
affect the soft supersymmetry breaking of the observable sector as
well as the physics of the axion sector (associated with
$v_{\mu}^a$).\cite{Giedt:2003ap} When the condensate field $U_a$ is
treated as an ordinary chiral superfield these last two terms do not
arise automatically (as they do in the linear multiplet treatment)
but must be included in the effective theory by hand. These terms
also can be shown to vanish in the $M_{\PL} \to \infty$ limit of
flat superspace, indicating the importance of a proper supergravity
treatment for an accurate phenomenology.


\subsubsection{K\"ahler stabilization}

In the previous subsection we have argued that it is the linear
multiplet formulation which hews most closely to the underlying
string theory. The fundamental degrees of freedom of the dilaton
multiplet, particularly in the presence of Yang-Mills Chern-Simons
forms, are easily incorporated into its structure. K\"ahler $U(1)$
superspace then provides a framework for naturally including this
constrained Yang-Mills geometry into four-dimensional supergravity.
In Section~\ref{sec22} we will also see how the integration constant
associated with~(\ref{FofLsol}) allows a beautiful implementation of
the Green-Schwarz mechanism of anomaly cancelation. Before
introducing that complication, however, we wish to recast the
familiar language of gaugino condensation from Sections~\ref{sec11}
and~\ref{sec12} in terms of our less familiar linear multiplet.
Along the way we will consider possible nonperturbative corrections
to the effective action which will allow us to achieve moduli
stabilization.

On general grounds we expect nonperturbative effects to alter the
form of any and all functions which determine the low-energy
effective supergravity Lagrangian. These may be of field-theoretic
or string-theoretic origin. An example is the gaugino condensation
of the previous sections, which we imagine to be a nonperturbative
effect from field theory which generates corrections to the
holomorphic superpotential for the moduli fields. It is certainly
not unreasonable to believe that instanton effects -- whether of the
world-sheet or of target space -- will correct the K\"ahler
potential as well. Indeed, such stringy effects were conjectured
some time ago by Shenker\cite{Shenker:1990uf} and have since been
explicitly demonstrated to exist in certain string
contexts.\cite{Polchinski:1994fq,Silverstein:1996xp,Antoniadis:1997zt}

We expect corrections to the K\"ahler potential to involve the
confinement scale $\Lambda_a$ of~(\ref{Lambda}), and on dimensional
grounds we expect corrections of field-theoretic origin which scale
like
\begin{equation} L^{-m} e^{-n/6b_aL}/M_{\PL}^{n-2} \label{dimcorr}
\end{equation}
where $n\geq 2$ and $m\geq
0$.\cite{Banks:1994sg,Burgess:1995aa,Giedt:2003ap} The simplest
example of such an effect would be to consider the leading-order
nonperturbative contribution ($n=2$ and $m=0$) to the K\"ahler
potential
\begin{equation} f(L)\,=\,A\,e^{-\,{1}/{3bL}}, \label{FTNP} \end{equation}
where $A$ is a constant to be determined by the nonperturbative
dynamics and $b$ is some effective beta-function coefficient.
Another possibility is to consider instanton contributions from the
string world-sheet, in which the function $f\(L\)$ derived
from~(\ref{dimcorr}) is slightly modified to\cite{Shenker:1990uf}
\begin{equation}
f(L) = \sum_{n} A_n (\sqrt{L})^{-n} e^{-B/\sqrt{L}}\; .
\label{nonpertsum}
\end{equation}
It is an important feature of~(\ref{nonpertsum}) that these string
instanton effects scale like $e^{-1/g}$ (when we use $\ell \sim
g^2$) and are thus stronger than analogous nonperturbative effects
in field theory which have the form $e^{-1/g^2}$. Thus they can be
of significance even in cases where the effective four-dimensional
gauge coupling at the string scale is weak.\cite{Banks:1994sg}

The corrections described above are, strictly speaking, not
corrections to the K\"ahler potential of the dilaton but to its {\em
action}, which is best investigated in component form. While these
corrections can be written in terms of modifications to the
effective four-dimensional K\"ahler potential, it is simpler to
implement the changes directly at the superfield level by modifying
the kinetic energy part of the superspace Lagrangian. We thus follow
the form of~(\ref{Klinear}) and introduce two functions of the
effective superfield $L$ which parameterize these nonperturbative
corrections arising from instanton effects as follows
\begin{equation}
\Lag_{\rm KE} = \superint\,E \[-2 + f(L)\], \quad k(L) = \ln\,L +
g(L),\label{lke}
\end{equation}
where~(\ref{relation}) implies
\begin{equation}
L\frac{\diff g(L)}{\diff L}\,=\, -L\frac{\diff f(L)}{\diff
L}\,+\,f(L) .\label{lcond} \end{equation}
In addition, we wish to obtain the classical limit $f=g=0$ at weak
coupling, so we must demand a further boundary condition at
vanishing coupling
\begin{equation} g(L=0)\,=\,0 \;\;\;\mbox{and}\;\;\; f(L=0)\,=\,0.
\end{equation}
In the presence of these nonperturbative effects the relationship
between the dilaton and the effective field theory gauge coupling is
modified from the relation in~(\ref{gbare2}) in a manner dictated by
the duality relation in~(\ref{newdual})
\begin{equation} \frac{g_{\STR}^2}{2}=\lang \frac{\ell}{1+f\(\ell\)} \rang \, .
\label{gnew} \end{equation}

We continue to use the form of~(\ref{VYT}) to describe gaugino
condensation. For simplicity we will work in this section only with
a single gaugino condensate. We will thus take the hidden sector to
be a pure Yang-Mills theory with no chiral matter. Then we have
\begin{equation} \Lag_{\VY} = \frac{b}{8} \superint\,\frac{E}{R}\,
 U\ln(e^{-K/2}U/\mu^{3}) + \hc \, ,
\end{equation}
and from~(\ref{cond}) we have $b = C/8\pi^2$ with $C$ being the
quadratic Casimir operator for the condensing group in the adjoint
representation. Using superspace integration by parts, and the
property~(\ref{Uproject}), it is possible to write the complete
Lagrangian for this simple system as
\begin{equation}
\Lag_{\rm eff}=\superint\,E\,\{\,-2 \,+\, f(L) \,+\,
bL\ln(e^{-K}\oline{U}U/\mu^{6})\,\} \, . \label{LagSec1}
\end{equation}

The method for obtaining the component field Lagrangian from
D-density expressions in K\"ahler $U(1)$ superspace is outlined in
the Appendix. Recalling the discussion surrounding the auxiliary
fields $F_U$ and $\oline{F}_{\oline{U}}$ in~(\ref{Fconstraint}), we
are careful to solve for the equations of motion for the combination
($F_U + \oline{F}_{\oline{U}}$) from which we obtain
\begin{equation}
f\,+\,1\,+\,b\ell\ln(e^{-k}\bar{u}u/\mu^{6})\,+\, 2b\ell\,=\,0 \, ,
\label{EQMfU} \end{equation}
where $k=k(\ell)$ is the dilaton K\"ahler potential in~(\ref{lke}).
This equation is easily solved to provide an expression for the
magnitude of the gaugino condensate
\begin{equation}
\bar{u}u\,=\,\frac{1}{e^{2}}\ell\mu^{6} e^{g\,-\,({f+1})/{b\ell}}\,
, \label{uu} \end{equation}
where the basic form of~(\ref{lamlam}) is recovered in the limit
$f(\ell)=g(\ell)=0$, as it should. The equation of motion for the
supergravity auxiliary field gives
\begin{equation} M=\frac{3}{4}bu \, ; \quad \quad \oline{M} =
\frac{3}{4}b \oline{u} \, , \label{Mu} \end{equation}
from which the gravitino mass can be computed. Eliminating all
auxiliary fields via their equations of motion generates the scalar
potential for the dilaton
\begin{eqnarray} V(\ell)&=&\frac{1}{16\ell^2}\[
\(1+\ell \frac{\diff g}{\diff \ell}\)
(1+b\ell)^{2}\,-\,3b^{2}\ell^{2}\,\] u\oline{u},
  \nonumber \\
  &=& \frac{1}{16e^{2}\ell}\[ \(1+\ell \frac{\diff g}{\diff \ell}\)
(1+b\ell)^{2}\,-\,3b^{2}\ell^{2}\,\]
\mu^{6}e^{g\,-\,({f+1})/{b\ell}} \label{ellpot} \end{eqnarray}
which depends only on the dilaton $\ell$. In the case of multiple
condensates each combination ($F_{U_a} + \oline{F}_{\oline{U}_a}$)
gives rise to an equation~(\ref{EQMfU}) for the individual
condensates $u_a$, and the solutions~(\ref{uu}), (\ref{Mu})
and~(\ref{ellpot}) are generalized to
\begin{equation}
\bar{u}_a u_a\,=\,\frac{1}{e^{2}}\ell\mu^{6}
e^{g\,-\,({f+1})/{b_a\ell}}\, , \label{uusol} \end{equation}
\begin{equation}
m_{3/2}=\frac{1}{3}\lang\left|M\right|\rang=\frac{1}{4}\lang
\left|{\sum_a}{b_a}{u_a} \right|\rang\, , \label{gravmasssum}
\end{equation}
\begin{equation}
V\(\ell\)=\frac{1}{16{\ell}^2}\(1+\ell \frac{\diff
  g}{\diff \ell}\) \left|{\sum_a}\(1+{b_a}\ell\)
  {u_a}\right|^{2}- \frac{3}{16}\left|{\sum_a} {b_a}{u_a}\right|^2.
\label{dilpot}
\end{equation}

\begin{figure}
\centerline{\psfig{file=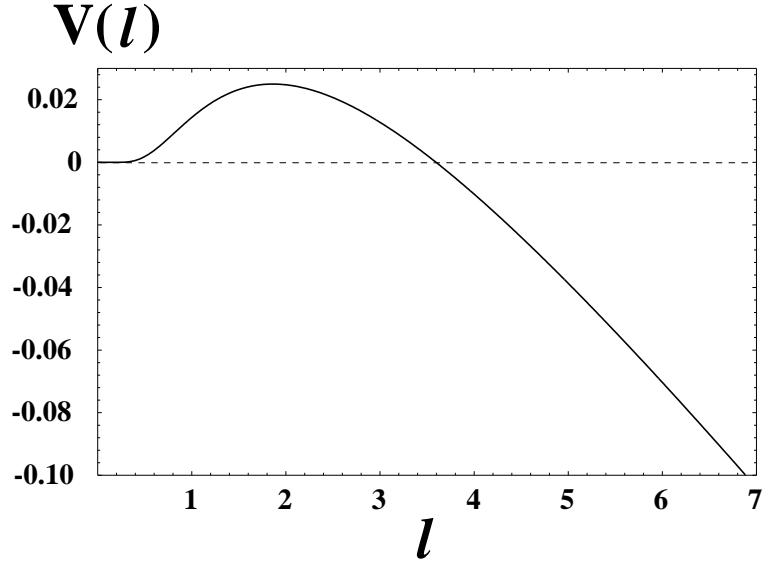,width=10cm}} \vspace*{8pt}
\caption{\textbf{Plot of dilaton potential $\mathbf{V(\ell)}$ for
the potential of~(\ref{ellpot}) without nonperturbative effects.}
The shape of the potential as a function of $\ell = L\lowest$ is
given with the height of the potential in arbitrary units ($m_{\PL}
= 1$). For this case we have taken the condensing group to be $E_8$.
} \label{fig:bgw1a}
\end{figure}

It is instructive to return to the case of Section~\ref{sec1} in
which all nonperturbative corrections are vanishing and see how the
run-away behavior is manifest in the linear multiplet formulation.
Taking~(\ref{ellpot}) for $f=g=0$ and $b = b_{E_8} = 30/8\pi^2$ we
plot the behavior of the dilaton scalar potential in Planck units in
Figure~\ref{fig:bgw1a}. Note that the general behavior is that of
Figure~\ref{fig:Vs1}, but now the asymptotic approach to vanishing
coupling occurs for $\ell \to 0$ while the unbounded-from-below
direction is for $\ell \to \infty$, which implies strong coupling.
From the identity~(\ref{lcond}) we can derive the necessary and
sufficient condition on the function $f(\ell)$ such that $V(\ell)$
in~(\ref{ellpot}) or~(\ref{dilpot}) is bounded from below:
\begin{eqnarray} f-\ell\ \frac{\diff f(\ell)}{\diff \ell}&\geq& -\order(\ell e^{{1}/{b_a \ell}})
\;\;\;\;\;\mbox{for}\;\;\;
\ell\,\rightarrow\, 0, \label{conditionA} \\
f-\ell\ \frac{\diff f(\ell)}{\diff \ell} &\geq&
\hspace{0.34cm}2\;\;\;\;\; \hspace{1.1cm}\mbox{for}\;\;\;\;\;\;\;
\ell\,\rightarrow\,\infty \, . \label{conditionB}
\end{eqnarray}
It is clear that condition~(\ref{conditionA}) is not at all
restrictive, and therefore has no nontrivial implication. On the
other hand, condition~(\ref{conditionB}) is quite restrictive; in
particular the simple tree-level model with $f=g=0$ violates this
condition -- hence the runaway solution. According to our assumption
of boundedness for $g(\ell)$ and $f(\ell)$ it must be that $\ell$=0
is the only pole of $\:g\,-\,({f+1})/{b\ell}.\:$ We therefore
recognize a relation between $\langle\bar{u}u\rangle$ and
$\langle\ell\rangle$: gauginos condense (i.e.,
$\langle\bar{u}u\rangle\neq 0$) if and only if the dilaton is
stabilized (i.e., $\langle\ell\rangle\neq 0$).

\begin{figure}
\centerline{\psfig{file=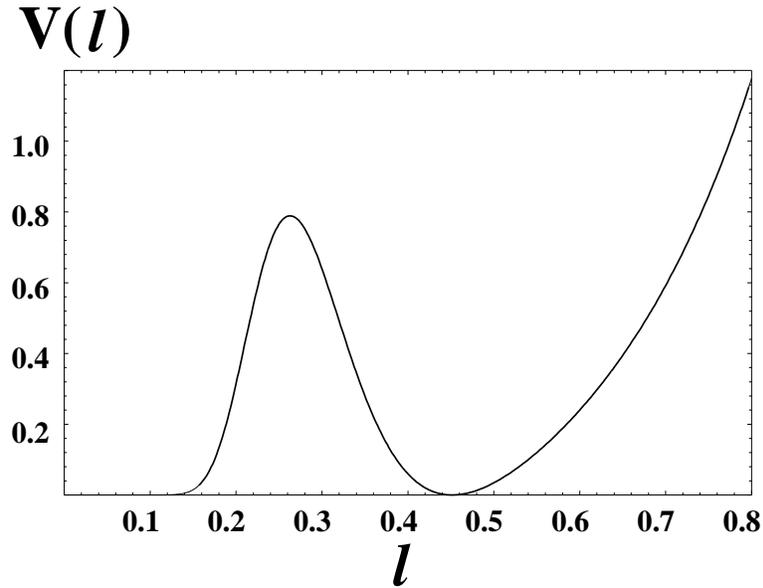,width=10cm}} \vspace*{8pt}
\caption{\textbf{Plot of dilaton potential $\mathbf{V(\ell)}$ for
the potential of~(\ref{ellpot}) with nonperturbative effects
parameterized by~(\ref{FTNP}).} The shape of the potential as a
function of $\ell = L\lowest$ is given with the height of the
potential in arbitrary units ($m_{\PL} = 1$). For this case we have
taken the condensing group to be $E_8$ and chosen the parameter
$A=6.92$ in~(\ref{FTNP}). } \label{fig:bgw1b}
\end{figure}

Let us demonstrate the K\"ahler stabilization effect by
using~(\ref{FTNP}) as our ansatz for $f(\ell)$. Note that the
boundedness condition~(\ref{conditionB}) requires $A\geq 2$. We will
therefore take $b = b_{E_8}$ and choose $A = 6.92$. The resulting
scalar potential for the dilaton is plotted in
Figure~\ref{fig:bgw1b}. Again, the plots give the value of the
dilaton potential in arbitrary units; the absolute size of the
potential values have been rescaled to more easily exhibit the shape
of the potential. For large values of $\ell$ the potential is now
bounded. A minimum occurs for $\lang \ell \rang \simeq 0.45$ for
which $f(\ell=0.45) \simeq 1$ and thus $g_{\STR}^2 \simeq 1/2$
from~(\ref{gnew}). The choice $A = 6.92$ was made so as to ensure
$\lang V(\ell) \rang =0$ at the minimum -- the other properties of
the low-energy phenomenology are generally insensitive to the
precise choice.

It has occasionally been argued that utilizing such nonperturbative
corrections in the K\"ahler potential should be avoided on the
grounds that the absence of holomorphy arguments implies that we
have (as yet) no real theoretical control over the magnitude of
these effects.
We would like to emphasize that no matter what mechanism generates
the corrections to the K\"ahler potential,
conditions~(\ref{conditionA}) and~(\ref{conditionB}) tell us
precisely how to modify the theory so as to have a stabilized
dilaton (and therefore broken supersymmetry by the argument of
above). These arguments are quite general.
Provided~(\ref{conditionB}) holds the potential of the modified
model in the strong-coupling regime is always bounded from below,
and in most cases rises as $\ell$ increases. Joining the
weak-coupling behavior of the modified model to its strong-coupling
behavior therefore strongly suggests that its potential has a
nontrivial minimum (at $\ell\neq 0$). Furthermore, if this
nontrivial minimum is global, then the dilaton is stabilized. Though
we are unable to study the exact K\"ahler potential at present, it
is nevertheless interesting to study models with reasonable K\"ahler
potentials for the purpose of illustrating the significance of
conditions~(\ref{conditionA}) and~(\ref{conditionB}) as well as
displaying explicit examples with supersymmetry breaking.
Furthermore, as we will see in Sections~\ref{sec3}~-~\ref{sec5},
these corrections give rise to a distinctive phenomenology that is
in many respects beneficial.

\section{The interaction of moduli with matter}
\label{sec2}
\setcounter{footnote}{0}

In the previous section we considered a very simple system of
moduli, focusing on those closed-string moduli (or geometric moduli)
which determine the gauge couplings of the low-energy effective
theory. We saw that the driving mechanism for moduli stabilization
and supersymmetry breaking is gaugino condensation. We briefly
considered some of the classes of mechanisms available to the
effective field theorist for achieving a realistic
supersymmetry-breaking minimum. Looking at the theory from the
four-dimensional effective field theory point of view allows us to
treat many possible constructions at once. However, not all the
mechanisms discussed in Section~\ref{sec1} can be operative in the
same theory at the same time. These issues depend on the nature of
the underlying string construction chosen. Given such a specific
construction, however, it is usually possible to do better than the
crude picture in Section~\ref{sec1} would indicate. This is because
much is known about the moduli that appear in the four-dimensional
effective field theory for most specific constructions -- including
such important quantities as the moduli dependence of Yukawa
couplings and kinetic terms, threshold corrections to gauge
couplings, and the existence of certain field theory counterterms
descended from the string theory. In this section we will focus on
the case of orbifold compactification of weakly-coupled heterotic
(WCH) string theory as this is an area in which some of the most
concrete examples can be constructed.

\subsection{Introducing the K\"ahler moduli}

\subsubsection{Modular symmetries}

Orbifold compactifications of heterotic string theory are
essentially toroidal backgrounds. Early constructions generally took
the compact six-dimensional space to factorize into three two-torii
$\mathcal{M} = T^2 \times T^2 \times T^2$, though the current
renaissance in orbifold model building has often exploited
nonfactorizable
compactifications.\cite{Kobayashi:2004ud,Buchmuller:2005jr,Buchmuller:2006ik,Lebedev:2006kn}
As mentioned in Section~\ref{sec1}, the presence of toroidal
geometry implies moduli which dictate the sizes and relative
orientations of the torii. These moduli will enjoy various symmetry
transformations that leave the spectrum and equations of motion for
the low-energy effective theory unchanged. We will refer to these
symmetries collectively as {\em modular symmetries}, though this
term is more precisely used in the context of string theory to refer
to symmetries of the string worldsheet which relate equivalent
surfaces. The symmetry group of target space modular transformations
depends on the particular orbifold (or orientifold) background.

Recall from Section~\ref{sec1} that for orbifolds we expect at
minimum three untwisted K\"ahler moduli $T^I$ which describe the
size of the three complex planes. Depending on the orbifold action
imposed there may be additional K\"ahler moduli as well as some
number of degrees of freedom $U^I$ related to deformations of the
complex structure. For example, one of the most studied orbifolds is
the $\mathbf{Z}_2 \times \mathbf{Z}_2$
orbifold\cite{Faraggi:1993pr,Lopez:1994ej,Petropoulos:1996rr} which
has three K\"ahler moduli $T^I,~I=1,2,3$ and three complex structure
moduli $U^I,~I=1,2,3$. Another commonly studied example is the $Z_3$
orbifold,\cite{Ibanez:1986tp,Ibanez:1987sn,Casas:1989wu} where there
are nine K\"ahler moduli fields labeled by $T^{IJ}$, with $I,\,J
=1,2,3$. It is common to make the assumption that the off-diagonal
components of this matrix are fixed at the string scale and do not
correspond to dynamical degrees of freedom in the low energy theory.
One thus works with only the diagonal entries $T^I \equiv T^{II}$.
Finally, there exist cases such as the (2,2) Abelian
orbifolds\cite{Dixon:1989fj} which gives rise to the low-energy
gauge group $E_8\otimes E_6\otimes U(1)^2$ in which there are
precisely the minimal number of three untwisted K\"ahler moduli
$T^I,~I=1,2,3$.

Both complex structure and K\"ahler moduli will generally transform
under some set of $SL(2,\mathbf{Z})$ symmetry groups (or their
subgroups).\cite{Kikkawa:1984cp,Sakai:1985cs} The $SL(2,\mathbf{Z})$
group acts as follows on a generic single-index modulus $M^I$ as
\begin{equation}
M^I\rightarrow\frac{a^I M^I - ib^I}{ic^I M^I + d^I }~,~~~~~a^I d^I -
b^I c^I =1~,~~~~~~a^I,b^I,c^I,d^I\in\mathbf{Z}~,
\label{mod}\end{equation}
with an analogous matrix equation for moduli which carry two
complex-plane indices. This set of transformations can be generated
from the two transformations $M^I \to 1/M^I$ and $M^I \to M^I + i$.
From the standard field definition of~(\ref{Tdef}) we note that for
K\"ahler moduli these transformations are generated by the
well-known duality transformation $R^I \to 1/R^I$ as well as by
discrete shifts of the axionic field $b_{mn}^I$ associated with each
of the three complex planes. The K\"ahler potentials describing
these untwisted moduli can be inferred from the dimensional
truncation of the ten-dimensional supergravity Lagrangian. At
leading order this K\"ahler potential is simply $K = -\ln V$ where
$V$ is the volume of the compact space. Thus
\begin{equation}
K = \sum_{I}g^I \; ; \quad \quad g^I=-\ln (T^I+\oline{T}^I) \; \;
{\rm or} \; \; g^I=-\ln (U^I+\oline{U}^I) \label{Kt} \end{equation}
for single-index fields, and
\begin{equation}
K = -\ln \det (T^{IJ}+\oline{T}^{IJ})\end{equation}
for the more general case.\cite{Ibanez:1992hc} For more details and
examples, the reader is referred to the review of Bailin and
Love,\cite{Bailin:1999nk} and references therein. For the remainder
of this section we will choose the simple case of three diagonal
K\"ahler moduli as this is sufficient to illustrate the types of
structures one expects in this class of theories.

Note that under an $\slz$ transformation~(\ref{mod}) we have
\begin{equation}
g^I \to g^I + F^I + \oline{F}^I\; ; \quad \quad F^I = \ln(icT^I + d)
\label{Ktmod} \end{equation}
and since the fields $T^I$ have no (classical) superpotential we
immediately recognize~(\ref{Ktmod}) as a K\"ahler transformation.
The classical effective supergravity action is therefore invariant
under~(\ref{mod}). This is welcome, since modular invariance is
known to be (perturbatively) preserved in string
theory;\cite{Alvarez:1989ad,Giveon:1989yf} that is, the set of
transformations~(\ref{mod}) on the various $T^I$ and $U^I$ should be
symmetries of the low-energy effective Lagrangian to all-loop order
in string perturbation theory.\footnote{When nonperturbative
objects, such as $D$-branes, are present in the construction these
$SL(2,\mathbf{Z})$ symmetries can be absent in the low-energy
theory. See, for example, Reference~\refcite{Aldazabal:1998ja}.}

Let us see how this picture is modified by the inclusion of matter
fields. We denote chiral superfields of gauge-charged matter by
$Z^i$, with lower-case Latin indices. In orbifold models the
K\"ahler metric for the matter fields arising from untwisted sectors
is precisely known. It continues to be given by the volume of the
compact space, but in this case the relation in~(\ref{Tdef}) is
modified\cite{Witten:1985xb,Derendinger:1985cv,Ferrara:1986qn} to
\begin{equation} 2(R^I)^2 = T^I + \oline{T}^I - \sum_i|(Z^i)^I|^2\; ,
\label{Runtw} \end{equation}
where the fields $(Z^i)^I$ are identified (upon dimensional
reduction) with the components of the 10D gauge fields that project
into the compact directions associated with each of the $T^I$.
Therefore the K\"ahler potential for this sector can be immediately
identified as
\begin{equation}
K = -\sum_I\ln\(T^I + \oline{T}^I - \sum_i|(Z^i)^I|^2\) \; .
\label{Kuntw} \end{equation}

For twisted sector matter, the corresponding expressions are not
known exactly but only to leading order in the matter
fields,\cite{Dixon:1989fj} though in some cases additional
information on higher-order terms can be inferred. Continuing to
work in the approximation of only three K\"ahler moduli (and no
complex structure moduli), the various matter K\"ahler metrics can
be summarized in one form
\begin{equation}
K_{i \jbar} = \kappa_i(T,\oline{T}) \delta_{i\jbar} + \order(|Z|^2)
\; \quad {\rm with} \quad  \kappa_i(T,\oline{T}) = \prod_I (T^I +
\oline{T}^I)^{-q_{i}^{I}}\;. \label{kappaZ}
\end{equation}
The parameter $q_i^I$ is the {\em modular weight} associated with
the field $Z^i$. For untwisted matter with K\"ahler
potential~(\ref{Kuntw}), for example, we have $q_i^I = 1$ for one of
the three values for $I$. It is common in phenomenological studies
to treat the three diagonal K\"ahler moduli as one common (overall)
size modulus $T$. Under this simplification the combined
matter/modulus K\"ahler potential is given by
\begin{equation}
K(T,\oline{T};\,Z,\oline{Z}) = -3\ln(T+\oline{T}) +
\sum_i\frac{|Z^i|^2}{(T+\oline{T})^{q_{i}}} \; , \label{KmatterT}
\end{equation}
where $q_{i} = \sum_{I}q_{i}^{I}$. The modular weights $q_i^I$ of
the twisted sector fields can readily be computed in Abelian
orbifold theories.\cite{Ibanez:1992hc} They are generally fractional
numbers, but the quantity $q_i$ (when $q_i^I$ are summed over all
three complex planes) will yield an $\order(1)$~integer.

A matter field $Z^{i}$ of modular weight $q_{i}^I$ transforms
under~(\ref{mod}) as
\begin{equation} Z^i \to
\prod_I(ic^I T^I + d^I)^{-q_i^I} Z^i = \exp (-\sum_I q^I_i F^I) Z^i
\; , \label{mattertrans}
\end{equation}
or simply $Z^i \to (ic T + d)^{-q_i} Z^i$ for one overall K\"ahler
modulus.\footnote{While~(\ref{mattertrans}) is strictly true for
untwisted fields, it is possible for fields in twisted sectors with
the same modular weight to mix amongst themselves under
$SL(2,\mathbf{Z})$ transformations.\cite{Lauer:1990tm}} Writing the
K\"ahler potential~(\ref{KmatterT}) as
\begin{equation}
K=\sum_{I}g^I + \sum_{i} \exp(\sum_{I}q^I_i g^I) |Z^i|^2
+\order(Z^4) \label{gtree}\end{equation}
it is easy to see that the total K\"ahler potential continues to
transform as $K \to K + \sum_I (F^I + \oline{F}^I)$
under~(\ref{mod}), or $K \to K + 3(F+\oline{F})$ with $F=\ln(icT+d)$
in the overall modulus case. Therefore the classical symmetry will
be preserved provided the superpotential for gauge-charged matter
transforms as~\cite{Binetruy:2000zx}
\begin{equation}
W \to W \(icT + d\)^{-3} . \label{pottrans}
\end{equation}
To ensure this transformation property the superpotential of
string-derived models has a moduli dependence of the
form\cite{Chun:1989se,Ferrara:1989qb,Ferrara:1990uu}
\begin{equation} W_{ijk} = w_{ijk} \left[
\eta(T) \right]^{-2(3-q_i-q_j-q_k)}. \label{WT}
\end{equation}
where $W_{ijk} = \partial^3W(Z^N)/\partial Z^i\partial Z^j\partial
Z^k$ and $w_{ijk}$ is a constant independent of the moduli.
The function $\eta(T)$ is the classical Dedekind eta function
\begin{equation}
\eta(T) = e^{-\pi T /12} \prod_{n=1}^{\infty} (1-e^{-2\pi nT})
\label{Deddef} \end{equation} and it has a well-defined
transformation under~(\ref{mod}) given by
\begin{equation}
\eta(T) \to \(icT + d\)^{1/2} \eta(T) \, . \label{etatrans}
\end{equation}
The form of~(\ref{WT}) can be readily inferred from the requirement
of modular invariance under~(\ref{mod}) for the classical
supergravity Lagrangian.\footnote{To actually see this dependence of
the superpotential on the K\"ahler moduli and Dedekind function
emerge from the underlying string theory is not trivial. It involves
factoring the level-one Euler characters for $SU(3)$ from the
three-point vertex amplitude calculation in string
theory.\cite{Dixon:1986qv,Chun:1989se,Giedt:2000es}} But this
requirement does not prevent the multiplication of the effective
Yukawa couplings by any function that is truly invariant under
modular transformations.\footnote{Similarly, in the Yang-Mills
sector, there may be modular invariant holomorphic functions of the
K\"ahler moduli that appear as a universal threshold correction to
the Yang-Mills kinetic function in models with $N=2$
sectors.\cite{Kiritsis:1996dn}} Such functions have often been
considered, in particular in conjunction with gaugino condensation
in a hidden sector.\cite{Cvetic:1991qm} In our study we will not
consider the presence of such functions of the K\"ahler moduli;
since they are modular invariant we do not expect them to shift the
eventual minimum of the potential away from the self-dual points. We
refer the interested reader to the existing
literature.\cite{Bailin:1997iz,Khalil:2001dr}

Note that the leading (large $T$ or large radius) behavior
of~(\ref{Deddef}) is $\exp(-\pi T/12)$. For three untwisted fields,
the Yukawa interaction (if allowed by string selection rules) has no
modulus-dependence, as can be seen from~(\ref{WT}) taking $q_i = 1$
for all fields. Yukawa interactions involving twisted fields where
$q_i > 1$, however, will instead come with an exponential {\em
suppression} involving the \vev of some modulus. The origin of this
suppression factor is readily understood: the Yukawa coupling in the
twisted field case is the result of stringy nonperturbative effects,
specifically world-sheet instantons, which depend on the size of the
compact space.
Such behavior is common to all string theory models in which chiral
fermions are localized at certain fixed points in the compact space
and have been useful in efforts toward building a string-theoretic
understanding of
flavor.\cite{Binetruy:1994bn,Cvetic:2002wh,Cremades:2003qj}
In the BGW model, however, the K\"ahler moduli are stabilized at
self-dual points where ${\rm Re}, t \sim 1$ and the exponential
factors are therefore not important in the weakly-coupled vacuum.

We have chosen to write the effect of a modular
transformation~(\ref{mod}) in terms of a rotation at the chiral
superfield level in~(\ref{mattertrans}). In fact, however, the
K\"ahler transformation of~(\ref{Ktmod}) is a continuous {\em
R-transformation} which affects scalars and fermions differently.
This can be accounted for by assigning the transformation property
\begin{equation} \theta\to e^{-{\frac{i}{2}}\sum_I{\rm Im}F^I}\theta
\label{thetatrans} \end{equation}
to the Grassmanian theta-parameter. All fermionic modes (modes with
unit chiral weight) in the supersymmetric Lagrangian receive such a
chiral rotation, including the gaugino fields of the gauge
supermultiplet. These transformations can be written as
\begin{equation}
\lambda_a \to e^{-{\frac{i}{2}}\sum_I{\rm Im}\,F^I}\lambda_a,\quad
\chi_A\to e^{{\frac{1}{2}}\sum_I(i{\rm Im}\,F^I - 2q_A^IF^I)}\chi_A
\; . \label{fermslz} \end{equation}
Note also that the composite operator $U_a = \Tr(\ww)_a$ introduced
in~(\ref{Ucondef}) for the confined gauginos will also have a
transformation deriving from~(\ref{fermslz}) given by
\begin{equation} U_a \to e^{-i\sum_I{\rm Im}F^I}U_a \; .
\label{Utrans} \end{equation}
That~(\ref{thetatrans}), (\ref{fermslz}) and~(\ref{Utrans}) involve
phase rotations suggests its interpretation as a type of $U(1)_R$
gauge transformation. This is the guiding principle behind
K\"ahler~$U(1)$ supergravity, which was briefly discussed in
Section~\ref{sec1} and is outlined in~\ref{appA}.

\subsubsection{General considerations of $W(S,T)$} \label{sec:WST}

Having introduced the classical symmetries of K\"ahler moduli in
weakly-coupled heterotic string theory, we are in position to
describe the nature of multi-modulus condensation models. We review
them here in part for the sake of historical completeness, but also
to provide a foil for the K\"ahler stabilization case which
concludes Section~\ref{sec2}. For the treatment of this subsection
we will use the chiral formulation of the dilaton.

Prior to the recognition that the set of transformations~(\ref{mod})
were perturbatively good symmetries of the low-energy effective
Lagrangian, it was common to assume that both the tree-level
superpotential of the observable sector $W_O$ and the effective
scalar potential generated by gaugino condensation~(\ref{WS}) would
continue to be independent of the K\"ahler moduli. In such a
situation the effective supergravity model is determined by the
functions
\begin{eqnarray} K&=& -\ln (S + \oline{S}) - 3\ln(T + \oline{T})\, ,
\nonumber \\
W &=& W_0 + W(S) = W_0 + e^{-S/b_a} \, ,\label{noscale}
\end{eqnarray}
where $W_O$ is independent of all moduli and we are taking the case
of one overall K\"ahler modulus $T^1 = T^2 = T^3 = T$. The factor of
three in the K\"ahler potential (or equivalently, the presence of
three diagonal K\"ahler moduli) is significant, in that it generates
a {\em no-scale} model for the T-moduli. In particular
\begin{equation} K_{T\oline{T}} F^T \oline{F}^{\oline{T}} = 3e^K
|W|^2 \end{equation}
and the final term in~(\ref{sdilpot}) is canceled. The potential
continues to have a run-away solution to weak-coupling ($\lang s
\rang \to \infty$), but in this case the potential is
positive-definite and monotonically decreasing.

A Lagrangian described by~(\ref{noscale}) is not invariant
under~(\ref{mod}). Classically the invariance of the Lagrangian can
be restored simply by the replacement $W \to W\eta^{-6}(T)$, and
this was the approach taken in the earliest treatments of
modular-invariant gaugino
condensation.\cite{Ferrara:1990ei,Nilles:1990jv,Binetruy:1990ck}
Consider a gaugino condensate-induced effective superpotential of
the general form
\begin{equation}
W(S,T) = \Omega(S)\eta^{-6}(T)\, , \quad {\rm with} \quad \lang
\Omega \rang \neq 0 \, . \label{WST}
\end{equation}
Minimization of the potential with respect to the field $s$ gives
rise to two possible solutions\cite{Cvetic:1991qm}
\begin{eqnarray}
{\rm Solution \; (1):} & \quad& (\Omega_s + K_s \Omega)=0
\label{FILQ1}
\\
{\rm Solution \; (2):} & \quad& (s+\bar{s})^2\Omega_{ss} =
e^{2i\gamma} \Omega^{*} \[ 2 - 3 |(t+\bar{t})\Eisen|^2\]
\label{FILQ2}
\end{eqnarray}
where $\gamma = \arg(\Omega_s + K_s \Omega)$ and we have introduced
the modified Eisenstein function
\begin{equation} \Eisen = \(2 \zeta(t) +
\frac{1}{t+\bar{t}}\)\, ; \quad \zeta(T) = \frac{1}{\eta(T)}
\frac{d\eta(T)}{dT} \, . \label{Eisen} \end{equation}
These two solutions imply very different properties for the
resulting ground state -- and hence for how supersymmetry breaking
will ultimately be transmitted to the observable sector. In the
first solution it is clear that $\lang F^S \rang =0$, and it can be
demonstrated that $\lang F^T \rang \neq 0$. Such an outcome is
typically called {\em moduli domination} (or more specifically
K\"ahler moduli domination). The value of the (overall) K\"ahler
modulus \vev can be computed in these two regimes for an arbitrary
function $\Omega(S)$. In the first case the minimum can easily be
shown\cite{Font:1990nt} to be $\lang {\rm Re} \; t \rang \simeq
1.23$ and the self-dual points are local maxima. For the second
solution the zeroes of the Eisenstein function are indeed minima and
$\lang {\rm Re} \; t \rang =1$.\footnote{More generally, the
K\"ahler moduli will be stabilized at one of two self-dual points
$\lang t_{I} \rang =1$ or $\lang t_{I} \rang = \exp{\(i\pi/6\)}$
where the Eisenstein function vanishes.}

\begin{figure}
\centerline{\psfig{file=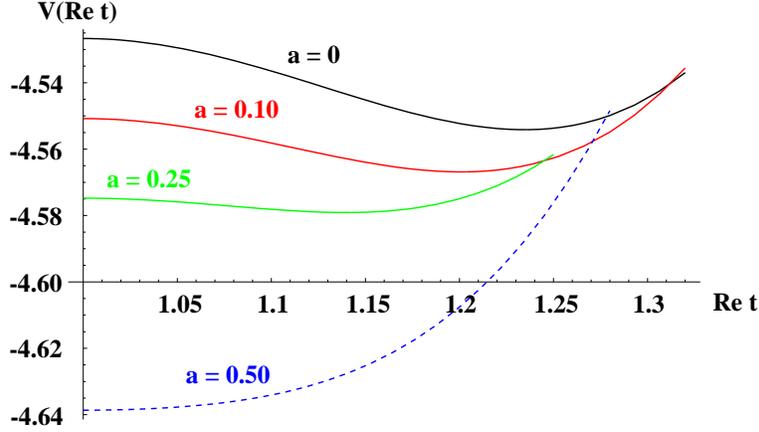,width=10cm}} \vspace*{8pt}
\caption{\textbf{Shape of $\mathbf{V(t,\bar{t})}$ in the
$\mathbf{{\rm Re}\, t}$ direction for fixed $\mathbf{\lang s+\bar{s}
\rang}$.} The value of the potential in~(\ref{FILQpot}) is given for
${\rm Re}\, t$ in the neighborhood of the fixed point at $t = 1$.
The dilatonic $F$-term is taken to have a value parameterized by
$|(s+\bar{s})\Omega_s - \Omega|^2 = a |\Omega|^2$ with four
different values of the parameter $a$. For the case $a=0$ we recover
the result $\lang {\rm Re} \; t \rang \simeq 1.23$. As in earlier
plots we have taken $m_{\PL} = 1$. The value of $V(t,\bar{t})$ for
each value of $a$ has been re-scaled independently in order to place
each plot on the same figure. } \label{fig:FILQplot}
\end{figure}

The situation determined by~(\ref{FILQ1}) with $\lang {\rm Re} \; t
\rang \simeq 1.23$ has been studied frequently in the context of
weakly-coupled heterotic string models -- so much so, in fact, that
this minimum is sometimes stated to be the {\em only} outcome
possible in modular-invariant treatments of gaugino condensation.
This statement, though erroneous, is understandable: achieving
solution~(\ref{FILQ2}) requires a deviation of the K\"ahler
potential for the dilaton from its tree-level value. If this is
achieved, the potential for the K\"ahler moduli must be re-examined.
Consider the scalar potential for~(\ref{WST})\cite{Font:1990nt}
\begin{equation} V = \frac{1}{(s+\bar{s})(t+\bar{t})^3
|\eta(t)|^{12}}\lbr |(s+\bar{s})\Omega_s - \Omega|^2 +
3|\Omega|^2\[(t+\bar{t})^2|\Eisen|^2 -1\]\rbr \, . \label{FILQpot}
\end{equation}
When~(\ref{FILQ1}) holds the first term in braces vanishes. The
potential in the ${\rm Re} \; t$ direction then has a minimum at
$\lang {\rm Re} \; t \rang \simeq 1.23$, as can be seen in the top
curve in Figure~\ref{fig:FILQplot}. But let us now imagine that
$|(s+\bar{s})\Omega_s - \Omega|^2 = K_{s\bar{s}}^{-1}
F^s\oline{F}^{\bar{s}}$ is not vanishing, but instead is given by
some coefficient $a$ times $|\Omega|^2$. The case of~(\ref{FILQ1})
is then simply the case of $a=0$. As the size of the (nonvanishing)
dilaton $F$-term increases, the potential for ${\rm Re} \; t$
becomes increasingly shallow, as is demonstrated in
Figure~\ref{fig:FILQplot}. More importantly, the minimum occurs for
values of ${\rm Re} \; t$ approaching the self-dual point of $\lang
{\rm Re} \; t \rang =1$. For some critical value (here at $a=0.5$)
the K\"ahler moduli are fixed at their self-dual points. This is the
outcome of~(\ref{FILQ2}) and, as we will see in Section~\ref{sec23},
it is precisely the outcome that the BGW model with K\"ahler
stabilization is designed to achieve. Before considering this point,
we must look at the transformations~(\ref{mod}) beyond the classical
level.

\subsection{Anomalies and counterterms}
\label{sec22}

While the diagonal modular transformations of~(\ref{mod}) leave the
classical effective supergravity Lagrangian invariant, the form
of~(\ref{fermslz}) leads us to expect the symmetry to be anomalous
at the quantum level. This is indeed the case. The variation of the
one loop Lagrangian under a modular transformation
is\cite{Louis:1991vh,Derendinger:1991hq}
\begin{equation} \delta \Lag_{\rm 1-loop} = \frac{1}{64\pi^2}
\sum_I \int \, \diff^2 \theta \frac{E}{R} \sum_a \alpha_a^I (\ww)_a
F^I + \hc\; , \label{Lagshift} \end{equation}
where $F^I = \ln(icT^I + d)$ and the coefficient $\alpha_a^I$ can be
computed from the effective field
theory\cite{Louis:1991vh,Gaillard:1992bt}
\begin{equation}
\alpha^I_a = - C_a + \sum_i\(1 - 2q^I_i\)C^i_a \;. \label{alphaIdef}
\end{equation}
Here $C_a$ and $C_a^i$ are quadratic Casimir operators in the
adjoint and matter representations, respectively, and $q^I_i$ are
the modular weights of the matter superfields $Z^i$ charged under
the group $\mathcal{G}_a$.

Since the set of $SL(2,\mathbf{Z})$ transformations should remain
good symmetries to all loop order, we expect some mechanism should
exist to cancel~(\ref{Lagshift}) and leave the resulting effective
Lagrangian invariant once again. Indeed, counterterms must be added
to the effective theory which correspond to massive modes of the
string theory which have been integrated out. From the point of view
of the effective supergravity Lagrangian, as well as the underlying
string theory, it is convenient to separate these contributions into
two types of counterterms: (1) operators which we may interpret as
threshold corrections to gauge kinetic functions and (2)
Green-Schwarz (GS) type counterterms which are analogous to the GS
anomaly cancelation mechanism in ten-dimensional string theory.
While the former are easy to interpret in the chiral formulation of
the dilaton, the latter find their natural interpretation in terms
of a linear multiplet formalism. We will stress the linear multiplet
in what follows.

The GS counterterm is universal in weakly-coupled heterotic string
models.\cite{Derendinger:1991hq,LopesCardoso:1991zt,LopesCardoso:1992yd}
In the simplest case we have
\begin{equation}
\Lag_{\GS} = b_{\GS} \int \diff^4 \theta \, E L \sum_I g^I
\label{GSlinear}
\end{equation}
where the GS coefficient $b_{\GS} = C_{\GS}/8\pi^2$ can be computed
from knowledge of the string spectrum. Note that this Green-Schwarz
coefficient $b_{\GS}$ is truly universal; it does not depend on the
gauge group $a$ or complex plane $I$ which labels~(\ref{alphaIdef}).
This is remarkable, since the anomaly coefficients
in~(\ref{alphaIdef}) seem to depend very strongly on the matter
content of each sector of the theory. Nevertheless, the anomaly
cancelation condition $b_{\GS} = \alpha_a^I \; \forall a,I$ holds
for a wide variety of orbifold constructions, including the
$\mathbf{Z}_3$ and $\mathbf{Z}_7$
orbifolds.\cite{Antoniadis:1991fh,Giedt:2001zw} In some sense this
universality was to be expected, in that the object involved in
anomaly cancelation for the K\"ahler $U(1)$ symmetry is the
antisymmetric two-index $b_{\mu\nu}$ of the universal dilaton field.
This is not the case in the open string models where multiple
closed-string moduli play this
role.\cite{Blumenhagen:2001te,Blumenhagen:2005mu}

Even in cases where the universal term is not sufficient to cancel
the entire anomaly (that is, where $b_{\GS} \neq \alpha_a^I$ for
some set of $\lbr a, I \rbr$), there is still always some part which
is universal to all sectors. We therefore will write
\begin{equation} \alpha_a^I = -C_{\GS} + b^I_a \label{alphaIdef2} \end{equation}
with
\begin{equation}
b^I_a = C_{\GS} - C_a + \sum_i\(1 - 2q^I_i\)C^i_a\; , \label{baIdef}
\end{equation}
thereby separating out the universal contribution. The
model-dependent contribution is given by the string threshold
corrections.\cite{Kaplunovsky:1995jw} These corrections vanish
completely unless the field $T^I$ corresponds to an internal plane
which is left invariant under some orbifold group transformations.
This only occurs if an $N=2$ supersymmetric twisted sector is
present. Specifically,
\begin{equation}
\Lag_{\rm th} =
-\sum_{I}\frac{1}{64\pi^2}\superint\,\frac{E}{R}\[\sum_a b_a^I
(\ww)_a + \sum_{a'} b_{a'}^I U_{a'}\] \ln\eta^2(T^I)  + {\rm h.c.}
\label{Lthresh} \end{equation}
where the two gauge sums run over unconfined and confined gauge
groups, respectively. The parameters $b_a^I$ vanish for orbifold
compactifications with no $N=2$ supersymmetry
sector,\cite{Antoniadis:1991fh} such as in the $\mathbf{Z}_3$ and
$\mathbf{Z}_7$ orbifolds mentioned above..

The form of~(\ref{Lthresh}) is suggestive of a correction to the
gauge kinetic functions of the chiral formulation. This is how such
terms are normally
presented.\cite{Derendinger:1991kr,Lust:1991yi,deCarlos:1992pd,deCarlos:1992da,Brignole:1993dj,Burgess:1995aa}
For example, modifying the universal gauge kinetic function $f_a =S$
of~(\ref{LYM}) by the addition of the group-dependent term
\begin{equation} \delta f_a = - \sum_I \frac{b_a^I}{8\pi^2} \ln
\eta^2 (T^I) \end{equation}
produces the necessary effect to cancel the anomalies, {\em
provided} the K\"ahler potential for the chiral dilaton is suitably
modified. In particular, the universal contribution associated
with~(\ref{GSlinear}) is incorporated via
\begin{equation} K(S,\oline{S}) \to -\ln\[(S+\oline{S}) +
b_{\GS}G\] \; , \label{KSmod} \end{equation}
where $G=-\sum_I \ln(T^I + \oline{T}^I)$. Finally, for overall
modular invariance to hold, the formerly invariant dilaton must now
be made to transform under~(\ref{mod}) as
\begin{equation} S \to S -b_{\GS}\sum_I F^I . \label{Strans} \end{equation}
This manner of implementing the modular invariance requirements in
the effective supergravity Lagrangian will produce the same results
(to leading order in $g_{\STR}$) as the linear multiplet treatment
we are about to use. Yet it has only one virtue: retaining the
familiar chiral formulation for the dilaton and the gauge kinetic
terms. But we feel that the downside of forcing an unphysical
K\"ahler transformation upon the dilaton -- and the resulting
kinetic mixing between dilaton and K\"ahler moduli -- make the
linear multiplet formulation superior.

Furthermore, the actual string calculation prefers the linear
multiplet treatment. To see this, consider the actual correction to
the gauge coupling (not the gauge kinetic function) as computed from
the underlying conformal field theory.\cite{Dixon:1990pc} The
resulting correction is given by
\begin{equation} \delta\(\frac{1}{g_a^2}\) =
-\sum_I \frac{b_a^I}{8\pi^2}\ln\[(T^I +
\oline{T}^I)|\eta^2(T^I)|^2\] \; ; \label{deltag} \end{equation}
which is not the real part of a holomorphic quantity. Attempting to
identify a holomorphic quantity in~(\ref{deltag}) to then factor out
and place in the gauge kinetic function leads to the complications
of~(\ref{KSmod}) and~(\ref{Strans}), as well as additional confusion
that requires considering the meaning of the Wilsonian versus 2PI
effective action for the dilaton.\cite{Burgess:1995aa}

Finally, the virtue of using K\"ahler $U(1)$ superspace in this
context is also apparent. Recall the general treatment of the
coupling of linear multiplets to supergravity in
Section~\ref{sec13}. There it was pointed out that consistency
restricts the form of the kinetic Lagrangian in~(\ref{Klinear}) only
up to an integration constant in~(\ref{FofLsol}) once the K\"ahler
potential $k(L)$ is specified. The Green-Schwarz mechanism utilizes
this integration constant with $V= \sum_I g^I$
in~(\ref{GSlinear}).\footnote{In Section~\ref{sec4} we will use this
integration constant to implement a Green-Schwarz term for anomalous
$U(1)$ gauge factors as well.} The presence of this constant implies
an additional invariance of the theory under which the function $V$
transforms as $V(\Phi,\oline{\Phi}) \to V(\Phi,\oline{\Phi}) +
H(\Phi) + \oline{H}(\oline{\Phi})$. The presence of a term
$\superint L\, V$ in the Lagrangian implies, upon integration by
parts in superspace, that terms of the form $H(\Phi) \ww + \hc$ are
produced when this transformation symmetry is applied. In the
specific case of~(\ref{GSlinear}) we see how the anomalous terms
in~(\ref{Lagshift}) are thereby naturally canceled with this
mechanism.

Since the linear multiplet is itself real the correction
in~(\ref{deltag}) is simple to implement. Taking~(\ref{GSlinear})
together with the (tree-level) kinetic Lagrangian and gaugino
condensate terms of~(\ref{LagSec1}) we arrive at the following
combined Lagrangian for models with $b_a^I =0$
\begin{equation}
\Lag_{\rm eff}=\superint\,E\,\{\,-2 \,+ f(L)\, + \, b_{\GS} LG \,+\,
b'L\ln(e^{-K}\oline{U}U/\mu^{6})\,\} \, .  \label{LagGSlin}
\end{equation}
In the above form the modular anomaly cancelation by the
Green-Schwarz counterterm is transparent. The second and third terms
in~(\ref{LagGSlin}) are {\em not} modular invariant separately, but
their sum is modular invariant, which ensures the modular invariance
of the full theory. Note that if we consider a situation in which
the various threshold corrections $b_a^I$ vanish then anomaly
cancelation implies $b'=b_{\GS}$. If we now consider nonvanishing
$b_a^I$ then using~(\ref{Uproject}) it is clear that~(\ref{Lthresh})
can also be written, upon integration by parts in superspace, as a
$D$-term superspace integral involving the modified linear
multiplet. Combining the expression in~(\ref{LagGSlin})
with~(\ref{Lthresh}) we obtain
\begin{eqnarray}
\Lag_{\rm eff} &=& \superint\,E\, \lbr \,-2 \,+\, f(L) \right.
\nonumber \\
 & & \quad \left. + L \( b'\ln(e^{-K}\oline{U}U/\mu^{6})\, -\sum_I
\frac{b_a^I}{8\pi^2}\ln\[(T^I + \oline{T}^I)|\eta^2(T^I)|^2\] \)\rbr
\, , \label{LagGSall}
\end{eqnarray}
and the correction of~(\ref{deltag}) is naturally implemented in the
effective theory.

\subsection{The BGW stabilization model}
\label{sec23}

We now have the basic pieces which make up the BGW model of moduli
stabilization. In this section we will gather all the parts to the
effective Lagrangian and exhibit the stabilization of the moduli. We
do so for the multi-condensate case, and we will allow for the
presence of matter fields in both the hidden and observable sectors.
For notational convenience we will introduce for each gaugino
condensate a vector superfield $L_a$ such that the gaugino
condensate superfields $U_{a}\simeq (\ww)_{a}$ are then identified
as the (anti-)chiral projections of the vector superfields:
\begin{eqnarray}
U_{a}=-\(\DbDb-8R\)L_{a}\, ,& \quad
{\overline{U}}_{a}=-\(\DaDa-8\overline{R}\)L_{a} \, . \label{LAdef}
\end{eqnarray}
This is a trivial notational generalization of the fundamental
definition in~(\ref{Uproject}). Its justification lies in the fact
that the individual ${L_a}\lowest$ turn out to be nonpropagating
degrees of freedom and none will appear in the effective theory
component Lagrangian. In other words, the dilaton field itself is
the lowest component of the field $L={\sum_{a}}L_{a}$. Similarly
only one antisymmetric tensor field (also associated with $L=\sum_a
L_a$) is dynamical.

\subsubsection{Elements of the effective Lagrangian}

We continue to allow for only three untwisted $(1,1)$ K\"ahler
moduli $T^I$ as in the previous sections. We also continue to make
the assumption that the K\"ahler potential for the
charged-matter/moduli system can be written as
\begin{eqnarray}
K=k\(L\)+\sum_{I}g^{I} +{\sum_i}e^{{\sum_I}{q_{i}^{I}}{g^{I}}}
\left| Z^i \right|^{2} + \order \( Z^4 \), & \; \; g^{I}=-\ln(T^I +
\oline{T}^I), \label{Kall}
\end{eqnarray}
where $k(L)$ is assumed to be of the form given in~(\ref{lke}).
The relevant part of the complete effective Lagrangian is then
\begin{equation}
{\Lag}_{{\rm eff}}={\Lag}_{{\rm KE}}  + {\Lag}_{{\rm GS}} +
{\Lag}_{{\rm th}}+  {\sum_{a}}{\Lag}_{{\rm a}} + {\Lag}_{{\rm VYT}}
+ {\Lag}_{{\rm pot}} \, , \label{Lag}
\end{equation}
with the kinetic terms given by~(\ref{lke}). The minimal form of the
Green-Schwarz counterterm in~(\ref{GSlinear}) is sufficient for
canceling anomalies associated with modular
transformations~(\ref{mod}). Any modular-invariant function of the
matter chiral superfields may also appear in the integration
constant of~(\ref{FofLsol}), however. As the exact form of the
Green-Schwarz counterterm has not been computed from the underlying
string theory, we will allow the possibility that the constant $V$
in~(\ref{FofLsol}) involves the entire K\"ahler potential for chiral
fields. We therefore take $\Lag_{\GS}$ to be of the form
\begin{eqnarray}
{\Lag}_{\rm GS}&=&{\superint}\, E\, L\, {V_{\GS}},\\
{V_{\GS}}&=& b_{\GS} {\sum_I} g^{I} + {\sum_{i}}p_{i}
e^{{\sum_I}{q_{i}^{I}}{g^{I}}} \left|Z^i \right|^{2} +
  \order\( Z^4 \) \, ,
\label{GS}
\end{eqnarray}
with the coefficients $p_i$ being as yet undetermined. A string
computation of axionic vertices in the presence of nonzero
backgrounds for twisted moduli and matter fields is needed to impose
further restrictions on $V_{\GS}$ (such as the values of the
constants $p_i$).\footnote{The computation which found the ``minimal
form'' of~(\ref{GSlinear}) set such background fields to
zero.\cite{Gaillard:1992bt}} The contribution from possible
threshold effects in $\Lag_{\rm th}$ is given in~(\ref{Lthresh}).

The terms $\Lag_{\GS}$ and $\Lag_{\rm th}$ are contributions to the
effective supergravity theory that are inherited from the underlying
string theory. Their role is to cancel the modular anomaly which
arises at the one-loop level~(\ref{Lagshift}). To ensure a
manifestly modular-invariant theory, however, we must also include
operators which {\em generate} the terms in~(\ref{Lagshift}) in the
first place. For the confined gauge sector this is already present
in the form of the VY superpotential of~(\ref{VYT}), but we still
lack the equivalent expression for the unconfined/light degrees of
freedom. This is the role of the $\Lag_a$ terms. Their specific form
in curved superspace is given by\cite{Gaillard:1999yb}
\begin{equation} \Lag_a = -\frac{1}{64\pi^2}\int d^4\theta \frac{E}{R}
\(W_\alpha \[P_\chi B_a \] W^\alpha\)_a + {\rm h.c.}\, ,
\label{Laga}\end{equation}
where $P_\chi$ is the chiral projection operator $P_\chi W^\alpha =
W^\alpha$, that reduces in the flat space limit to $(16\Box)^{-1}
\oline{\D}^2\D^2$. Here we understand $a$ to label {\em unconfined}
gauge groups. The functions $B_a$ are\cite{Gaillard:1992bt}
\begin{equation} B_a = -\sum_I\alpha_a^I g^I + \(C_a - \sum_iC_a^i\)k(L)
+ 2\sum_i C_a^i \ln\(1 + p_iL\) \, . \label{BaGT} \end{equation}
Due to the assumed invariance of $L$ under modular transformations
it is only the first term in~(\ref{Laga}) which contributes to the
modular anomaly and thus~(\ref{Lagshift}) is recovered. The other
terms, together with the threshold corrections~(\ref{Lthresh}) and
Green-Schwarz term, determine the renormalization of the coupling
constants $g_a$ at the string scale
\begin{eqnarray}  g^{-2}_a(\mu_{\STR}) &=& \Bigg<\frac{1+f}{2\ell}
- b'_a k(\ell) + \sum_i\frac{C_a^i}{8\pi^2}\ln(1+p_i\ell) \quad
\quad \nonumber
\\ & & \quad \quad - \sum_I\frac{b^I_a}{16\pi^2}\ln\[|\eta(t^I)|^2(t^I +
\oline{t}^I)\]\Bigg> \, , \label{gloop}
\end{eqnarray}
where $t^{I}\equiv T^{I}\lowest$.

The Lagrangian describing the condensates is of the VYT-type, as
described in Section~\ref{sec12} above and explicitly written
in~(\ref{VYT}):
\begin{equation} \Lag_{\rm VYT} = \frac{1}{8}\superint\,\frac{E}{R}\sum_a
U_a\[b'_a\ln(e^{-K/2}U_a) + \sum_\alpha
b^\alpha_a\ln\Pi_a^{\alpha}\] + {\rm h.c.}\, . \end{equation}
Recall the anomaly-matching conditions~(\ref{cond}) that the
coefficients $b_{a}'$ and $\baal$ must satisfy:
\begin{equation} b'_a = {1\over8\pi^2}\(C_a - \sum_iC_a^i\) ,\;\;\;\; \qquad
b^\alpha_a = \sum_{i\in\alpha}{C^i_a\over4\pi^2d_a^\alpha}\; ,
\label{baprimedef} \end{equation}
where $d_a^{\alpha}$ is the mass dimension of the corresponding
matter condensate superfield $\Pi_a^{\alpha}$.
Note the important property that when $d_a^{\alpha}=3$ for all
$\Pi$'s charged with respect to the condensing group $\mathcal{G}_a$
we have the identity
\begin{equation} b'_a + \sum_{\alpha} \baal = b_a \label{bID}
\end{equation}
with $b_a$ being the beta-function coefficient~(\ref{ba}) associated
with the coupling for the group $\mathcal{G}_a$. We will make the
assumption that $d_a^{\alpha}=3$ in what follows below, but the more
general case is easily analyzed by a slight modification of the
effective Lagrangian.\cite{Gaillard:1999et}

The final term in~(\ref{Lag}) is a superpotential term for the
matter condensates consistent with the symmetries of the underlying
theory
\begin{equation}
{\Lag}_{\rm pot}=\frac{1}{2}{\superint}\frac{E}{R}{e^{K/2}}W\(
\Pi^{\alpha}, \, T^{I} \) + \hc\;. \label{Lagpot}
\end{equation}
We will adopt the simplifying assumption\cite{Binetruy:1996nx} that
for fixed $\alpha$, $\baal\neq 0$ for only one value of $a$. In
other words, we assume that each matter condensate is made up fields
charged under only one of the confining groups. This is not a
necessary requirement, but it will make the phenomenological
analysis of the model much easier to perform.
%
%
We next assume that there are no unconfined matter fields charged
under the confined hidden sector gauge groups.
%
%
This allows a simple factorization of the superpotential of the form
\begin{equation}
W\( \Pi^{\alpha}, \, T^{I} \) = {\sum_{\alpha}}{W_{\alpha}}\(T\)
{\Pi}^{\alpha} ,
\end{equation}
where the functions $W_{\alpha}$ are given by
\begin{equation}
W_{\alpha}\(T\)=c_{\alpha}\prod_I \[\eta \(T^{I}\)\]^{2\(
q^{\alpha}_{I} -1\)}. \label{Walpha}
\end{equation}
Here $q^{\alpha}_{I}=\sum_{i} n_i^{\alpha}q_{i}^{I}$ is the
effective modular weight for the matter condensate and the Yukawa
coefficients $c_{\alpha}$, while {\it a priori} unknown variables,
are taken to be~$\order\(1\)$.

Just as in Section~\ref{sec13}, we can solve the equations of motion
for the auxiliary fields in the theory to eliminate them from the
Lagrangian. The equation of motion for $F_{U_a} +
\oline{F}_{\oline{U}_a}$ gives a formula for the real part of the
gaugino condensates analogous to the simple case of~(\ref{uusol}):
\begin{equation}
\rho_a^2 = e^{-2{\frac{b'_a}{b_a}}}e^{K}
e^{-\frac{\(1+f\)}{{b_a}\ell}}e^{-\frac{b_{\GS}}{b_a}
  {\sum_I}g^{I}}{\prod_I}
  \left|{\eta}\(t^{I}\)\right|^{\frac{4\(b_{\GS}-b_{a}\)}{b_a}}
{\prod_{\alpha}}\left|\baal/4c_{\alpha}\right|^{-2
{\frac{b_{a}^{\alpha}}{b_a}}}, \label{rhosq}
\end{equation}
where we have introduced the notation $u_{a} = U_{a}\lowest \equiv
{\rho}_{a}e^{i{\omega}_a}$. It is not hard to see that by
using~(\ref{gnew}) and taking $b'_a = b_a = b_{\GS}$ we obtain the
simple expression in~(\ref{uusol}) -- {\em i.e.} the expected
one-instanton form for gaugino condensation.
Expression~(\ref{rhosq}) encodes more information, however, than
simply the one-loop running of the gauge coupling which led us
to~(\ref{lamlam}) at the beginning of this review. Consider the
renormalization group invariant quantity\cite{Shifman:1986zi}
\begin{equation}
{\delta}_a =\frac{1}{g_{a}^{2}\(\mu\)} - \frac{3b_a}{2}\ln{\mu^2}
+\frac{2C_a}{16{\pi}^2}\ln{g_{a}^{2}\(\mu\)} +
\frac{2}{16{\pi}^2}\sum_{i}C_{a}^{i}\ln{Z_{a}^{i}\(\mu\)}.
\label{RGinvariant}
\end{equation}
Using the above expression it is possible to solve for the scale at
which the $1/g^{2}(\mu)$ term becomes negligible relative to the
$\ln{g^{2}(\mu)}$ term -- effectively looking for the ``all loop''
Landau pole for the coupling constant. This scale is related to the
string scale by the relation
\begin{equation}
{\mu_L}^2 \sim {\mu_{\STR}}^2 e^{-\frac{2}{3b_{a}g_{a}^{2}\(\mu\)}}
\prod_{i}\[Z_{a}^{i}\(\mu_{\STR}\)/Z_{a}^{i}\(\mu_{L}\)\]^
{\frac{C_{a}^{i}}{12{{\pi}^2}b_{a}}}. \label{Landau}
\end{equation}
Comparing the expression in~(\ref{Landau}) with that
in~(\ref{rhosq}) shows that the two agree provided we identify the
wave-function renormalization coefficients $Z_{a}^{i}$ with the
quantity $|4W_{\alpha}/\baal|^{2}$. This is precisely what is needed
to provide agreement between~(\ref{RGinvariant}) and~(\ref{gloop}),
indicating that the condensation scale represents the scale at which
the coupling becomes strong as would be computed using the so-called
``exact'' beta-function.

The equation of motion for the auxiliary field $F^{\alpha}$ of the
chiral supermultiplets $\Pi^{\alpha}$ gives
\begin{equation} 0 \, = \, \sum_a b^\alpha_a u_a \,+ \,4\pi^\alpha
e^{K/2}W_\alpha \quad\forall \;\; \alpha \, , \end{equation}
while that for the auxiliary field of supergravity gives
\begin{equation} M=\frac{3}{4} \(\sum_a b'_a u_a -
4We^{K/2}\) \, . \end{equation}
Using these rules the potential for the fields $\ell$ and $t^I$ is
\begin{eqnarray} V &=& \frac{1}{16\ell^2}\(v_1 - v_2 + v_3\),\nonumber \\
v_1 &=& \(1+\ell\frac{\diff
  g}{\diff \ell} \)\left|\sum_a\(1+b_a\ell\)u_a\right|^2,  \quad
v_2 = 3\ell^2\left|\sum_ab_au_a\right|^2,\nonumber \\
v_3 &=& {\ell^2\over(1+b\ell)}\sum_I\left|\sum_ad_a(t^I)u_a\right|^2
\, ,
\label{totalpot} \end{eqnarray}
where the quantity $d_a(t^I)$ is
\begin{equation} d_a(t^I) =
\(b_{\GS} - b_a\)\(1 + 4\zeta(t^I){\rm Re}\, t^I\)\,  \label{da}
\end{equation}
and the Riemann zeta-function is defined in~(\ref{Eisen}).
Note that $d_a(t^I)\propto F^I$ vanishes at the self-dual point
$t^I=1$, for which $\zeta(t^I) = -1/4$ and $\eta(t^I)\approx .77$.
Therefore we can immediately conclude that the $F$-terms for the
K\"ahler moduli vanish at the minimum of the potential. We will
return to this important phenomenological property in subsequent
sections. After eliminating $v_3$ in~(\ref{totalpot}) we are left
with the same potential as~(\ref{dilpot}).

\subsubsection{Minimizing the scalar potential}

In order to go further and make quantitative statements about the
scale of gaugino condensation (and hence supersymmetry breaking) it
is necessary to choose a specific form for the nonperturbative
effects characterized by $f(L)$ and $g(L)$. For this example we will
choose the form~(\ref{nonpertsum}), suggested originally by Shenker,
and include the first two terms in the summation
\begin{equation}
f\(L\)=\[{A_0}+{A_1}/\sqrt{L}\]e^{-B/\sqrt{L}}\, . \label{ourf}
\end{equation}
With this function it is possible to minimize~(\ref{totalpot}) with
vanishing cosmological constant and $\alpha_{\rm str}=0.04$ for
$A_{0}=3.25, A_{1}=-1.70$ and $B=0.4$ in~(\ref{ourf}). Other
combinations of these parameters can also be employed to stabilize
the dilaton at weak coupling with vanishing vacuum energy. In
general we would not expect such a truncation as~(\ref{ourf}) to be
valid for all values of $L$ -- it need only be a valid
parameterization in the neighborhood of $\lang \ell \rang$
consistent with the weak-coupling limits of~(\ref{conditionA})
and~(\ref{conditionB}).
%

With the choice of~(\ref{ourf}) the scale of gaugino condensation
can be obtained once the following are specified: ({\bf 1}) the
condensing subgroup(s) of the {\mbox original} hidden sector gauge
group $E_8$, ({\bf 2}) the representations of the matter fields
charged under the condensing subgroup(s), ({\bf 3}) the Yukawa
coefficients in the superpotential for the hidden sector matter
fields and ({\bf 4}) the value of the string coupling constant at
the compactification scale, which in turn constrains the
coefficients in~(\ref{ourf}) necessary to minimize the scalar
potential~(\ref{totalpot}).

The above parameter space can be simplified greatly by assuming that
all of the matter in the hidden sector which transforms under a
given subgroup $\mathcal{G}_a$ is of the same representation, such
as the fundamental representation. This is not unreasonable given
known heterotic string constructions. In this case the sum of the
coefficients $\baal$ over the number of condensates can be replaced
by one effective variable
\begin{eqnarray}
{\sum_{\alpha}}\baal\longrightarrow\baaleff\, ;  & \qquad
\baaleff={N_c}b_{a}^{\rm rep}\, . \label{baaleff}
\end{eqnarray}
In the above equation $b_{a}^{\rm rep}$ is proportional to the
quadratic Casimir operator for the matter fields in the common
representation and the number of condensates, $N_c$, can range from
zero to a maximum value determined by the condition that the gauge
group presumed to be condensing must remain asymptotically free.
The variable $\baal$ can then be eliminated in~(\ref{rhosq}) in
favor of $\baaleff$ provided the simultaneous redefinition
$c_{\alpha}\longrightarrow\(c_{\alpha}\)_{\rm eff}$ is made so as to
keep the final product in~(\ref{rhosq}) invariant. Combined with the
assumption of universal representations for the matter fields, this
implies
\begin{equation}
\caleff \equiv
{N_c}\({\prod_{\alpha=1}^{N_c}}c_{\alpha}\)^\frac{1}{N_c}
\label{caleff}
\end{equation}
which we assume to be an $\order\(1\)$ number, if not slightly
smaller.

\begin{figure}[t]
\centerline{
       \psfig{file=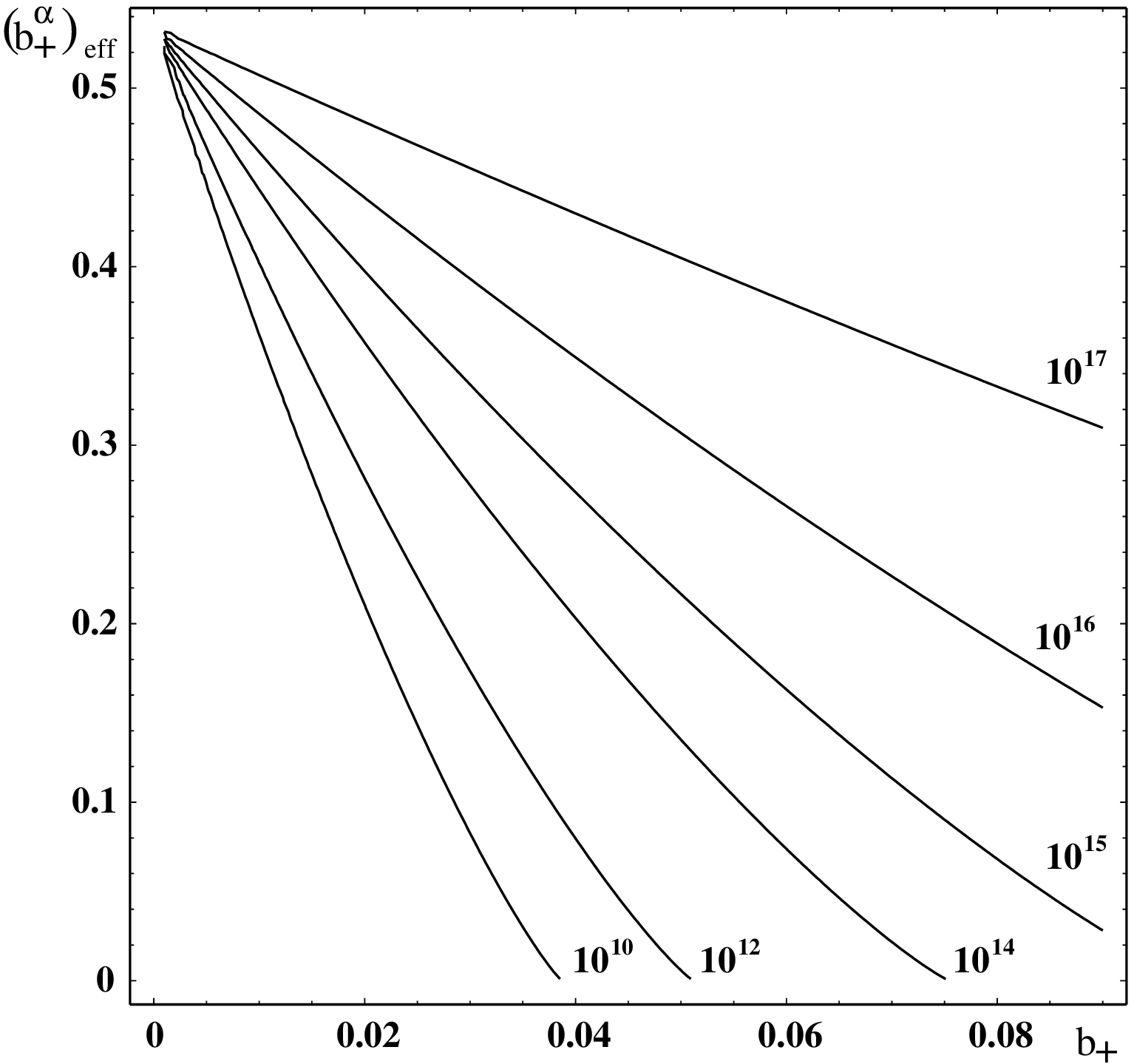,width=0.5\textwidth}
       \psfig{file=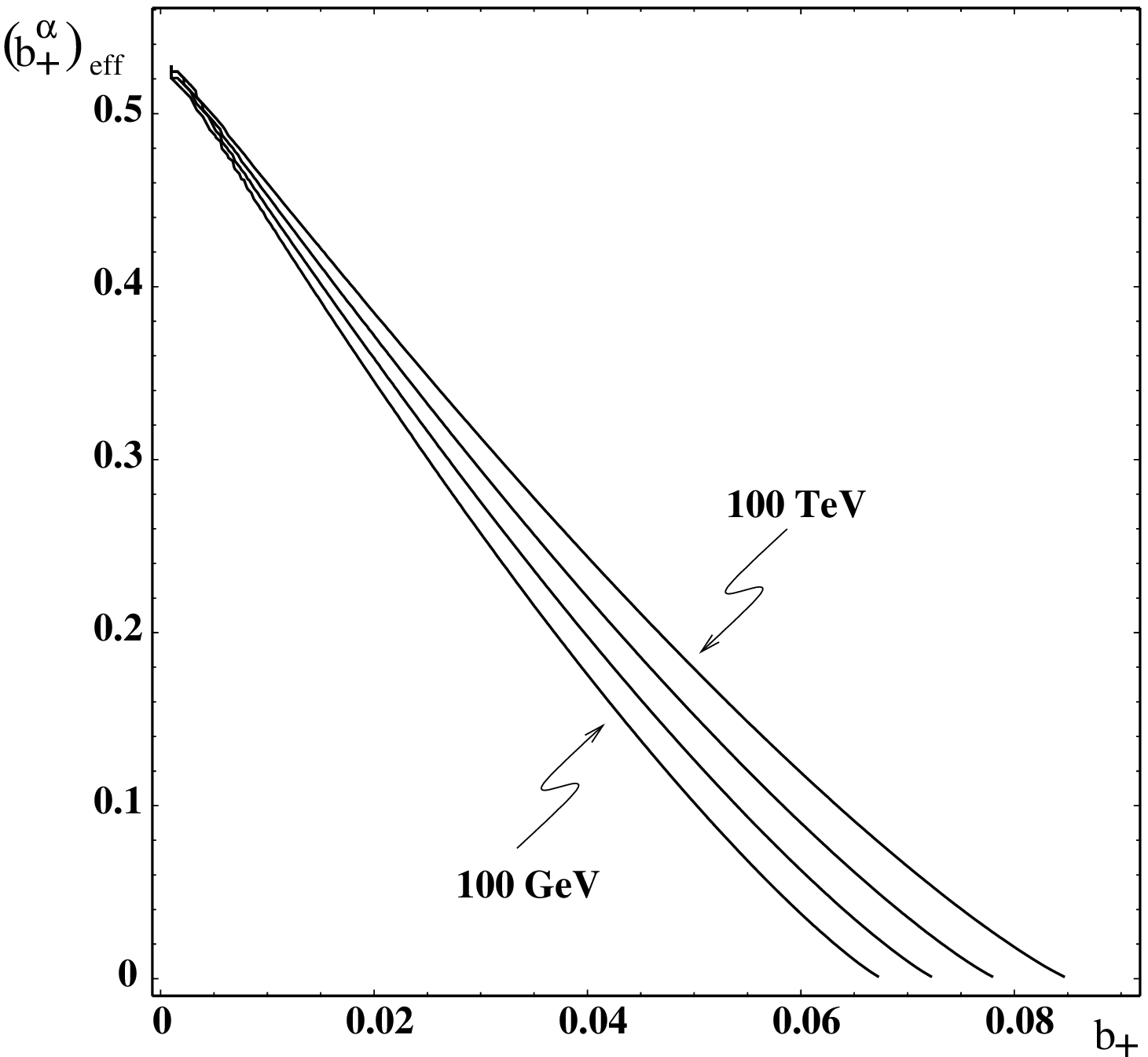,width=0.5\textwidth}}
          \caption{{\footnotesize {\bf Condensation scale and
                gravitino mass}. Contours give the
            scale of gaugino condensation in GeV in the left panel and gravitino masses
            of $10^2$ through $10^5$~GeV in the right panel for $\caleff=3$.}}
        \label{fig:conscale}
\end{figure}

From a determination of the condensate value $\rho$ the
supersymmetry-breaking scale can be found by solving for the
gravitino mass, given by
\begin{equation}
m_{3/2}=\frac{1}{3}\lang\left|M\right|\rang=\frac{1}{4}\lang
\left|{\sum_a}{b_a}{u_a} \right|\rang. \label{gravitino}
\end{equation}
In the presence of multiple gaugino condensates the low-energy
phenomenology is dominated by the condensate with the largest
one-loop beta-function coefficient. For example, the gravitino mass
for the case of pure supersymmetric Yang-Mills $SU(5)$ condensation
(no hidden sector matter fields) would be $4$~TeV. The addition of
an additional condensation of pure supersymmetric Yang-Mills $SU(4)$
gauginos would only add an additional $0.004$~GeV to the mass.
Therefore let us consider the case with just one condensate with
beta-function coefficient denoted $b_+$:
\begin{equation}
m_{3/2}=\frac{1}{4}{b_+}\lang\left|u_+\right|\rang. \label{gravmass}
\end{equation}
Now for a given choice of the effective Yukawa coupling $\caleff$
and unification-scale gauge coupling $g_{\STR}$ the condensation
scale
\begin{equation}
\Lambda_{\rm cond}=\(M_{\PL}\)\lang {\rho}^{2}_{+}\rang^{1/6}
\label{conscale}
\end{equation}
and gravitino mass can be plotted in the $\lbr b_{+}, \bpaleff\rbr$
plane. Both quantities are shown in the two panels of
Figure~\ref{fig:conscale} for $\caleff=3$. The importance of
including the possible effects of matter condensates is clear from
the left panel in Figure~\ref{fig:conscale}, in that two different
hidden sectors involving condensing groups with the same
beta-function coefficient can give rise to different scales of
gaugino condensation. This scale is a much weaker function of the
relevant parameters than that of the gravitino mass in the right
panel of Figure~\ref{fig:conscale}.

\begin{figure}
\centerline{\psfig{file=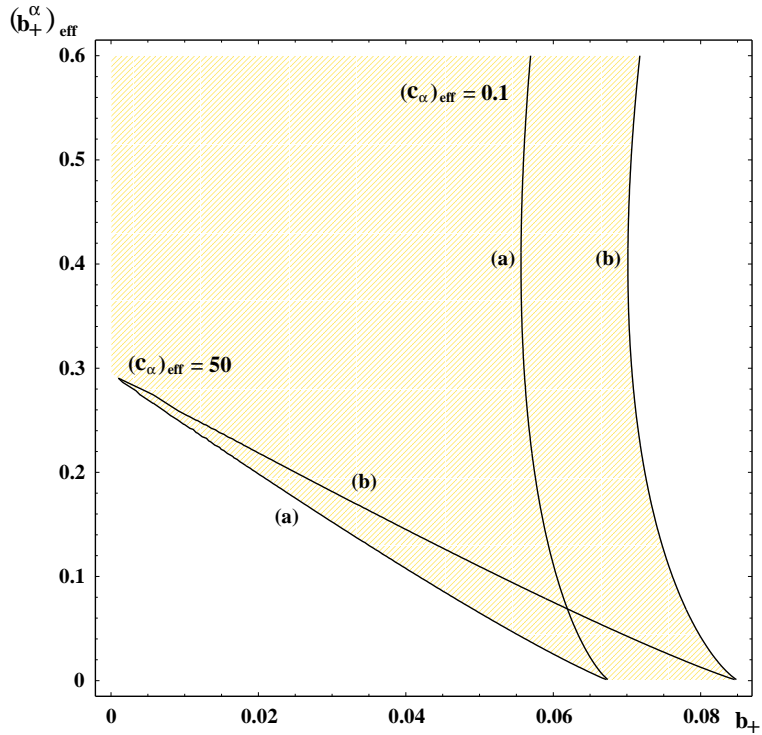,width=10cm}} \vspace*{8pt}
\caption{\textbf{Gravitino mass regions as a function of effective
Yukawa parameter.} Gravitino mass contours for (a) 100~GeV and (b)
10~TeV are shown for $\caleff=50$ and $\caleff=0.1$ with
$\alpha_{\STR}=0.04$. The region between the two sets of curves can
be considered roughly the region of phenomenological viability.}
\label{fig:gravmassvaryc}
\end{figure}

For fixed values of the unknown Yukawa coefficients $c_{\alpha}$,
the region of parameter space for which a phenomenologically
preferred value of the supersymmetry-breaking scale occurs is a
rather limited slice of the entire space available. But this is
somewhat deceptive.
The variation of the gravitino mass as a function of the Yukawa
parameters $c_{\alpha}$ is shown in Figure~\ref{fig:gravmassvaryc}.
On the horizontal axis there are no matter condensates ($\baal=0,\
\forall \alpha$) so there is no dependence on the variable
$\caleff$.
For very large values of the effective Yukawa parameter the
gravitino mass contours approach an asymptotic value very close to
the case shown here for $\caleff = 50$. We might therefore consider
the shaded region between the two sets of contours as roughly the
maximal region of viable parameter space for a given value of the
unified coupling at the string scale.

In Figure~\ref{fig:conscale} and~\ref{fig:gravmassvaryc} we have
chosen to show a range in the beta-function parameter $b_+$ for
which $b_+ \lappeq 0.09$. In principle the condensing gauge group
can be as large as $E_8$ in the weakly-coupled heterotic string, for
which $b_{E_8} = 30/8\pi^2 = 0.38$. In general we expect the hidden
sector gauge group to be a product of subgroups of $E_8$. The set of
all such possible breakings has been computed for Abelian
orbifolds:\cite{Katsuki:1989bf,KEKtable}
\begin{equation}
\lbr \begin{array}{l}
E_{7}, \,E_{6} \\
SO(16), \,SO(14), \,SO(12), \,SO(10), \,SO(8) \\
SU(9), \,SU(8), \,SU(7), \,SU(6), \,SU(5), \,SU(4), \,SU(3)
\end{array} \right. \, .
\label{KEKgroups}
\end{equation}
For each of these groups, one can define a line in the $\lbr
b_{+},\bpaleff\rbr$ plane via the relations~(\ref{ba}),
(\ref{baprimedef}) and~(\ref{bID}). These lines will all be parallel
to one another with horizontal intercepts at the beta-function
coefficient for a pure Yang-Mills theory. The vertical intercept
will then indicate the amount of matter required to prevent the
group from being asymptotically free, thereby eliminating it as a
candidate source for the supersymmetry breaking.

\begin{figure}
\centerline{\psfig{file=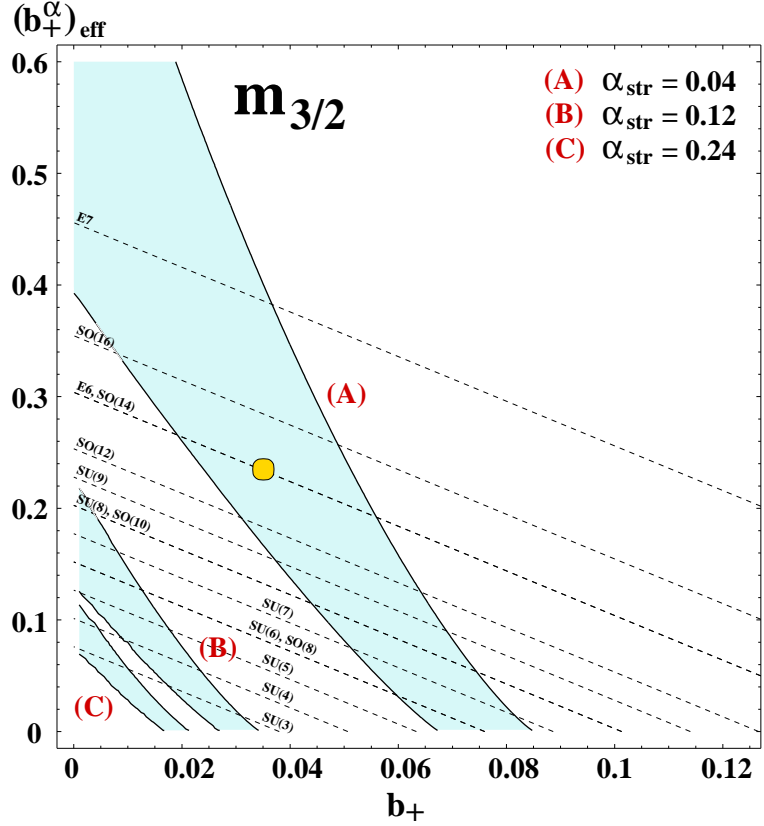,width=10cm}} \vspace*{8pt}
\caption{\textbf{Constraints on the hidden sector.} The shaded
regions give three different ``viable'' regions depending on the
value of the unified coupling strength at the string scale. The
upper limit in each case represents a 10~TeV gravitino mass contour
with $\caleff=1$, while the lower bound represents a 100~GeV
gravitino mass contour with $\caleff=10$. The dot indicates an
example point with a condensing $E_6$ gauge group and
9~$\mathbf{27}$'s of matter forming baryonic condensates.}
\label{fig:varyall}
\end{figure}

In Figure~\ref{fig:varyall} we have overlaid these gauge lines on a
plot similar to Figure~\ref{fig:gravmassvaryc}. We restrict the
Yukawa couplings of the hidden sector to the more reasonable range
of $1 \leq \caleff \leq 10$ and give three different values of the
string coupling at the string scale. A typical matter configuration
would be represented in Figure~\ref{fig:varyall} by a point on one
of the gauge group lines. As each field adds a discrete amount to
$\baaleff$ and the fields must come in gauge-invariant multiples,
the set of all such possible hidden sector configurations is
necessarily a finite one. For example, the point in
Figure~\ref{fig:varyall} with $\mathcal{G}_c = E_6$ and
9~$\mathbf{27}$'s of matter forming baryonic condensates has $b_+ =
3/8\pi^2 \simeq 0.038$ and is indicated in the plot by the circle.
Note that one cannot obtain values of $b_{+}$ arbitrarily close to
zero in practical model building. The number of possible
configurations consistent with a given choice of $\lbr
\alpha_{\STR}, \caleff \rbr$ and supersymmetry-breaking scale
$m_{3/2}$ is quite restricted. Furthermore, even moderately larger
values of the string coupling at unification become increasingly
difficult to obtain as it is necessary to postulate a hidden sector
with very small gauge group and particular combinations of matter to
force the beta-function coefficient to small values.

\section{Soft supersymmetry breaking}
\label{sec3}
\setcounter{footnote}{0}

In this section we will look in greater detail at how supersymmetry
is broken by $F$-terms from various chiral superfields in the
theory. Ultimately we wish to make contact with the formalism of
Brignole et al.\cite{Brignole:1993dj} which describes soft
supersymmetry breaking in the observable sector in terms of chiral
superfields $S$, $T^I$ and their auxiliary fields. After translating
the explicit results of the BGW model from Section~\ref{sec23} into
this notation, we will generalize to the case of arbitrary
supersymmetry breaking in models arising from weakly-coupled
heterotic strings. This will allow us to compare the K\"ahler
stabilized case to those of other moduli stabilization regimes in
the literature. It will also allow us to exhibit the one-loop
corrections to the tree-level soft Lagrangian in all generality.
Finally we will return to the specific case of K\"ahler
stabilization in the BGW model to discuss the superpartner spectrum.
Further phenomenological consequences of these results will follow
in Section~\ref{sec5}.

\subsection{Bosonic sector of the BGW model}
\label{sec31}

Returning to the superspace Lagrangian defined by~(\ref{Lag}), we
can derive the lowest component expression through the means
outlined in~\ref{appA}. The result for the bosonic part of the
component Lagrangian is given by
\bea \frac{1}{e}\Lag_{B}\,&=&\,-\,\frac{1}{2}{\cal R}
\,-\,(1+b\ell)\sum_{I}\frac{1}{(t^{I}+\bar{t}^{I})^{2}}
\(\partial^{\mu}\bar{t}^{I}\,\partial_{\mu}t^{I} - \oline{F}^{I}F^{I}\) \nonumber\\
& &\,-\,\frac{1}{16\ell^{2}}\(\ell g'+1 \)
\[4 \(\partial^{\mu}\!\ell\,
\partial_{\mu}\!\ell - B^{\mu}\!B_{\mu}\) + \bar{u} u -
4e^{K/2}\ell\(W\bar{u} +
u\bar{W}\)\]\nonumber\\
& &\,+\,\frac{1}{9}\(\ell g' -2 \)\[\oline{M}\!M - b^{\mu}b_{\mu} -
{3\over4}\lbr \oline{M}\(\sum_b b'_b u_b - 4We^{K/2}\) + \hc \rbr
\]\nonumber\\
& &\,+\,\frac{1}{8}\sum_a \Bigg\{{f+1\over\ell} +
\,b'_a\ln(e^{2-K}\bar{u}_au_a) + \sum_\alpha
b^\alpha_a\ln(\pi^\alpha\bar{\pi}^\alpha) \nonumber \\
& & \qquad + \sum_I\[bg^I -
{b_a^I\over4\pi^2}\ln|\eta(t^I)|^2\]\Bigg\}
\(F_a - u_a\oline{M} + \hc \)\nonumber\\
& &\,-\,\frac{1}{16\ell}\sum_a\[b'_a\(\ell g' + 1\)\bar{u}u_a -4\ell
u_a \(\sum_\alpha b^\alpha_a{F^\alpha\over\pi^\alpha} +
(b'_a-b){F^I\over2{\rm
Re}t^I}\) + \hc  \] \nonumber\\
& &\,+\,\frac{i}{2}\sum_a\[b'_a\ln(\frac{u_a}{\bar{u}_a}) +
\sum_\alpha
b^\alpha_a\ln(\frac{\pi^\alpha}{\bar{\pi}^\alpha})\]\nabla^{\mu}\!B_{\mu}^a
\,-\,{b\over2}\sum_{I}\frac{\partial^{\mu}{\rm Im}\,{t}^{I}}{{\rm
Re}\,{t}^{I}}B_{\mu}, \nonumber \\
&
&\,+\,\sum_{I,a}{b^I_a\over16\pi^2}\[\zeta(t^I)\(2iB_a^{\mu}\nabla_{\mu}
t^I - u_aF^I\) + \hc \] \nonumber \\
& &\,+\,e^{K/2}\[\sum_I F^I\(W_I + K_IW\) + \sum_\alpha F^\alpha
W_\alpha + \hc \], \label{Lboson1} \eea
where $\ell= L \lowest$, $g' = g'(\ell) = \diff g(L)/\diff L
\lowest$, $b_{\mu}$ and $M = (\oline{M})^{\dagger} = -6R\lowest$ are
auxiliary components of the supergravity multiplet and we have
employed the following definitions
\bea
\sigma^{\mu}_{\alpha\dot{\alpha}}B^a_{\mu}\,&=&\,
\frac{1}{2}[\,\D_{\alpha},\D_{\dot{\alpha}}\,]L_a\lowest\,+\,
\frac{2}{3}\ell_a\sigma^{\mu}_{\alpha\dot{\alpha}} b_{\mu},
\quad B^{\mu} = \sum_aB^{\mu}_a,\nonumber\\
u_a\,&=&\,U_a\lowest\,=\,-(\oline{\D}^{2}-8R)L_a\lowest,
\quad u = \sum_a u_a,\nonumber\\
\bar{u}_a\,&=&\,\oline{U}_a\lowest\,=\,-(\D^{2}-8R^{\dagger})L_a\lowest,
\quad \bar{u} = \sum_a\bar{u}_a,\nonumber \\
-4F^a\,&=&\,\D^{2}U_a\lowest, \;\;\;\;
-4\oline{F}^a\,=\,\oline{\D}^{2}\oline{U}_a\lowest,
\quad F_U = \sum_a F^a,\nonumber\\
\pi^\alpha\,&=&\,\Pi^\alpha\lowest\,
\quad \bar{\pi}^\alpha\,=\,\oline{\Pi}^\alpha\lowest\,\nonumber \\
-4F^\alpha\,&=&\,\D^{2}\Pi^\alpha\lowest, \;\;\;\;
-4\oline{F}^\alpha\,=\,\oline{\D}^{2}\oline{\Pi}^\alpha\lowest, \nonumber \\
t^{I}\,&=&\,T^{I}\lowest,\;\;\;\;
-4F^{I}\,=\,\D^{2}T^{I}\lowest,\nonumber\\
\bar{t}^{I}\,&=&\,\oline{T}^{I}\lowest,\;\;\;\;
-4\oline{F}^{I}\,=\,\oline{\D}^{2}\bar{T}^{I}\lowest \, .
 \eea
The auxiliary fields in the bosonic Lagrangian can be eliminated via
their equations of motion. For the auxiliary fields of the
supergravity multiplet these are simply
\begin{equation}
b_{\mu} \,=\,0, \quad \quad M\,=\,\frac{\, 3\,}{\,
4\,}\,\(\sum_ab'_au_a - 4We^{K/2}\)\, , \end{equation}
while those associated with the chiral superfields give
\bea F^{I}\,&=&\,{{\rm Re}\,t^I \over 2(1 + b\ell)}
\lbr\sum_a\bar{u}_a\[(b - b'_a) +
{b_a^I\over2\pi^2}\zeta(\bar{t}^I){\rm Re}\,t^I\]- 4
e^{K/2}\(2{\rm Re}\,t^I\oline{W}_I - \bar{W}\)\rbr, \nonumber \\
0\,&=&\,\sum_a b^\alpha_a u_a\,+\,4\pi^\alpha e^{K/2}W_\alpha
\quad\forall \;\; \alpha, \nonumber \\
\bar{u}_a u_a\, &=& \,\frac{\ell}{e^{2}}e^{g\,-\,({f+1})/{b'_a\ell}
-\sum_I b^I_a g^I/8\pi^2 b'_a}\prod_I|\eta(t^I)|^{b^I_a/2\pi^2b'_a}
\prod_\alpha(\pi^\alpha_r\bar{\pi}^\alpha_r)^{-b^\alpha_a/b'_a},
\label{BGWEoM} \end{eqnarray}
for $F^I$, $F^{\alpha}$ and $F^a + \oline{F}^a$, respectively. In
the above we have defined the modular invariant quantity
\begin{equation}
\pi^\alpha_r = \Pi^\alpha_r\lowest = e^{\sum_Iq^\alpha_Ig^I/2}
\Pi^\alpha \lowest \, . \end{equation}
With these~(\ref{Lboson1}) becomes simply
\begin{eqnarray}
\frac{1}{e}\Lag_{B}\,&=&\,-\,\frac{1}{2}{\cal R}
\,-\,(1+b\ell)\sum_I\frac{\partial^{\mu}\bar{t}^{I}\,\partial_{\mu}t^{I}}
{(t^{I}+\bar{t}^{I})^{2}}\,-\,\frac{1}{4\ell^{2}}\(\ell g' + 1\)
\(\partial^{\mu}\!\ell\,\partial_{\mu}\!\ell - B^{\mu}\!B_{\mu}\)  \nonumber\\
& &\,-\,\sum_a\(b'_a\omega_a + \sum_\alpha b^\alpha_a\phi^\alpha\)
\nabla^{\mu}\!B_{\mu}^a\,-\,{b\over2}\sum_{I}
\frac{\partial^{\mu}{\rm Im}\,{t}^{I}}{{\rm Re}\,{t}^{I}}B_{\mu}
\nonumber
\\ & & +\,i\sum_{I,a}{b^I_a\over8\pi^2}\[\zeta(t^I)B_a^{\mu}\nabla_{\mu} t^I
- \hc
\] - V\, , \end{eqnarray}
with the field definitions
\begin{equation} u_a = \rho_a e^{i\omega_a}, \quad \pi^\alpha = \eta^\alpha
e^{i\phi^\alpha}, \quad 2\phi^\alpha = -i\ln\(\sum_a
b^\alpha_au_a\bar{W}_\alpha\over\sum_a b^\alpha_a\bar{u}_a
W_\alpha\) \quad {\rm if} \;\; W_\alpha\ne 0 \, . \end{equation}
The final form of the scalar potential is then
\begin{eqnarray}
V\,&=&\,\frac{\(\ell g' +1\)}{16\ell^{2}}\lbr\bar{u}u +\ell
\[\bar{u}\(\sum_ab'_au_a - 4e^{K/2}W\)  + \hc \]\rbr\nonumber \\
& &\,+\,{1\over 16(1+b\ell)}\sum_I\left|\sum_au_a\(b-b'_a +
{b^I_a\over2\pi^2} \zeta(t^I){\rm Re}\, t^I\)  - 4e^{K/2}\(2{\rm
Re}\, t^I W_I - W\)\right|^2 \nonumber\\
& &\,+\,\frac{1}{16}\(\ell g' - 2 \)\left|\sum_bb'_bu_b -
4We^{K/2}\right|^2 \, ,\label{BGWpot} \end{eqnarray}
which, upon substitution of relations~(\ref{lcond}) and~(\ref{bID}),
reduces to~(\ref{totalpot}) when the form~(\ref{Walpha}) is used for
the matter condensate superpotential. The expression for the
condensate itself, as a function of model parameters, is precisely
the expression in~(\ref{rhosq}) in this case.

As indicated in the last chapter, supersymmetry is broken at the
minimum of the scalar potential in~(\ref{BGWpot}). The result should
be the appearance of new soft supersymmetry-breaking terms in the
component Lagrangian for the observable sector. These can be
identified directly by looking at the scalar potential for the
observable sector as determined from the component expansion
of~(\ref{Lag}). This direct approach was, in fact, the method
employed in the initial studies of this class of
theories.\cite{Binetruy:1997vr,Gaillard:1999et} Using the
one-condensate approximation these tree level soft terms are given
by
\begin{equation}
M_a^0 = -\frac{g^{2}_{a}\(\mu\)}{2}
\[\frac{3b_+}{1+{b_+}\ell}
+{\sum_i}\frac{p_i C_{a}^{i}}{4{{\pi}^2}{b_+}\(1+{p_i}\ell\)}
\] m_{3/2}\, ,
\label{BGWgaugino}
\end{equation}
and
\begin{eqnarray}
A_{ijk} &=& \frac{\bar{u}_+}{4} \[\frac{b_+}{1+b_+ \ell}
\frac{\(p_{i}-b_{+}\)}{\(1+{p_i}\ell\)}\] + (i\to j) + (i \to k) \,
, \label{BGWAterm} \\
(M_i^0)^2 &=& \frac{|u_+|^2}{16} \(\frac{p_i - b_+}{1+p_i \ell}\)^2
\, , \label{BGWmasssq} \end{eqnarray}
where $p_i$ is the possibly nonvanishing coupling of the chiral
matter to the Green-Schwarz term in~(\ref{GS}).
To obtain these it was necessary to use~(\ref{lcond}) and the vacuum
condition $\lang V \rang =0$ in~(\ref{totalpot}).

A great deal of simplification is possible, particularly in finding
the one-loop corrected values of these expressions, in circumstances
in which the supersymmetry breaking is associated with nonvanishing
$F$-terms for chiral superfields.\cite{Kaplunovsky:1993rd} This
simplification was exploited by Brignole et al. to systematize
possible patterns of soft supersymmetry breaking in a wide class of
string-inspired models.\cite{Brignole:1993dj,Brignole:1997dp} In
this parameterization supersymmetry-breaking effects are computed in
a context in which closed-string moduli (such as the dilaton,
K\"ahler moduli and complex structure moduli) are represented by
chiral superfields, and each is allowed to participate in
supersymmetry breaking via nonvanishing auxiliary field $vevs$.
Attempting to use this parameterization in the present case
immediately gives rise to an apparent contradiction. We are here
working in a context in which we imagine no complex structure
moduli. The K\"ahler moduli have vanishing $F^I$ at the minimum of
the scalar potential. This leaves only the dilaton to do the job of
communicating the supersymmetry breaking of the hidden sector to the
fields of the observable sector -- a situation commonly referred to
as {\em dilaton domination}. However, the dilaton multiplet in the
linear formulation has no auxiliary field! What plays the role of
nonvanishing $\lang F^S \rang$ in our case?

\subsection{Moduli as messengers of supersymmetry breaking}

The key to understand the apparent paradox can be found in the
modified linearity conditions of~(\ref{Uproject}) and the resulting
$F$-term expressions in~(\ref{FU}). The equation of motion for
$F^{U_a} + \oline{F}^{\oline{U}_a}$ generates the expression for the
condensate in~(\ref{BGWEoM}). In a sense, the lowest component of
the condensate (which is acquiring a vacuum expectation value at the
minimum of the scalar potential) plays the role of ``auxiliary
field'' for the dilaton in the modified linear multiplet. Let us see
more explicitly how this comes about.

In the presence of a (nonperturbatively induced) potential for the
dilaton, the tree level scalar Lagrangian for the condensate/moduli
sector takes the form
\begin{equation} \Lag_{\rm scalar} = - \sum_\alpha
\frac{\partial_{\mu} t^I \partial^{\mu} \bar{t}^I}{(t^I +
\bar{t}^I)^2} - \frac{k'(\ell)}{4\ell}\partial_{\mu} \ell
\partial^{\mu} \ell - \frac{\ell}{ k'(\ell)}\partial_{\mu} a \partial^{\mu} a -
V \, , \label{Lagtemp} \end{equation}
where the axion $a$ is related to the two-form $b_{\mu\nu}$ of the
linear multiplet by a duality transformation which follows
from~(\ref{realdil}), (\ref{adef}) and~(\ref{vmudef}):
\begin{equation} \frac{1}{2} \epsilon^{\mu\nu\rho\sigma}
\partial_{\nu} b_{\rho\sigma} = - \frac{2\ell}{k'(\ell)}
\partial^{\mu} a \, . \end{equation}
The scalar potential $V$ can be cast in the following suggestive
form
\begin{equation} V = \sum_\alpha \frac{1}{(t^I + \bar{t}^I)^2} F^\alpha
\oline{F}^{\alpha} + \frac{\ell}{k'(\ell)}F^2 - \frac{1}{3}M
\oline{M} \end{equation}
provided we make the identification
\begin{equation} F = \frac{k'(\ell)}{4\ell}f(\ell,t^I,z^i)
\label{Fredefined} \end{equation}
where $f(\ell,t^I,z^i)$ is a complex but {\em nonholomorphic}
function of the scalar fields.  For example in the class of models
being considered here
\begin{equation} f(\ell,t^I,z^i) = - \sum_a(1+ \ell b_a)\bar{u}_a \approx - (1+
\ell b_+)\bar{u}_+, \label{onecondapx} \end{equation}
where $\bar{u}_a(\ell,t^I,z^i)$ is the value of the gaugino
condensate for hidden gauge group $\mathcal{G}_a$. It is the
function $F$, defined via~(\ref{Fredefined}), that will play the
role of $F^S$ in this model.

To exhibit this connection, consider the variable $x(\ell) =
2g_{\STR}^{-2}(M_{\STR})$. This variable can be related to the
function $k(\ell)$ via the differential equations
\begin{equation} k'(\ell) = -\ell x'(\ell), \quad \partial\ell =
- \frac{\ell}{k'(\ell)}\partial x \, , \end{equation}
from which we derive the following relations
\begin{eqnarray} && \frac{\partial k(x)}{\partial x}
= k'(\ell)\frac{\partial \ell}{\partial x} = - \ell,
\quad \frac{\partial^2 k(x)}{\partial x^2} =
-\frac{\partial\ell}{\partial x} = \frac{\ell}{ k'(\ell)}\, ,
\nonumber \\
&& \frac{k'(\ell)}{4\ell} \partial_{\mu} \ell\partial^{\mu} \ell =
\frac{\ell}{4k'(\ell)}\partial_{\mu}x \partial^{\mu} x =
\frac{1}{4}\frac{\partial^2 k(x)}{\partial x^2}
\partial_{\mu} x \partial^{\mu} x \, .
\label{xdiffeq} \end{eqnarray}
We would like to recast~(\ref{Lagtemp}) into the standard form we
expect for a theory of only chiral superfields:
\begin{eqnarray} \Lag_{\rm scalar} &=& - \sum_N K_{N\oline{N}}
\(\partial_{\mu} z^N \partial^{\mu}\bar{z}^{\oline{N}} + F^N
\oline{F}^{\oline{N}} \) + \frac{1}{3}M \oline{M} \, , \nonumber
\\
K &=& k\(s + \bar{s}\) + K(t^I,\bar{t}^I) + \sum_i\kappa_i|z^i|^2\,
,
\end{eqnarray}
where we now let the index $N$ run over the chiral dilaton, K\"ahler
moduli and gauge-charged matter. This can be accomplished by setting
$x=s+\bar{s}$ and $a = {\rm Im}\, s$ in~(\ref{xdiffeq}) provided we
identify $F = F^S$ and $k_{s\bar{s}} = \ell/k'(\ell)$. These
relations change slightly when we include a Green-Schwarz
counterterm,\cite{Binetruy:2000md} but the basic correspondence
remains the same.

We therefore conclude that it is the dilaton that acts as the
messenger of supersymmetry breaking in this model, albeit in an
indirect way. Since the operators which connect this field to those
of the observable sector involve one inverse power of the Planck
scale (or, strictly speaking, the string scale) this is rightly
called an instance of ``gravity mediation'' as in the general
parlance. From~(\ref{xdiffeq}) we have the desired translation from
the linear multiplet to the chiral multiplet notation
\begin{equation}
\lang k_s \rang = -\ell \qquad \qquad \lang k_{s\bar{s}}\rang =
\frac{\ell^2}{1+\ell g'(\ell)}  \label{translate}
\end{equation}
which is a property of the vacuum state of the theory. The key
feature of~(\ref{translate}) is the deviation of the dilaton
K\"ahler metric from its tree level (perturbative) value.

The importance of this fact can be appreciated from another
direction. Taking the expression of~(\ref{rhosq}), let us assume the
nonperturbative superpotential generated by gaugino condensation for
the {\em chiral} dilaton would be of the form of~(\ref{WS}): $W(S)
\propto e^{-S/2b_{+}}$ with $b_{+}$ being the largest beta-function
coefficient among the condensing gauge groups of the hidden sector.
From the equations of motion~(\ref{EQM}) one immediately sees that
requiring the potential in~(\ref{sdilpot}) to vanish at the minimum
will require that
\begin{equation}
(k^{s\bar{s}})\left|k_s - \frac{1}{2b_{+}} \right|^{2} =3 \; \; \to
(k^{s\bar{s}})^{-1/2} = \sqrt{3} \frac{2b_{+}}{1-2b_{+}k_{s}}\, ,
\label{Ksstrue}
\end{equation}
where we choose to keep the precise form of the derivatives of the
dilaton K\"ahler potential unspecified. The condition
in~(\ref{Ksstrue}) is independent of the means by which the dilaton
is stabilized and is a result merely of requiring a vanishing vacuum
energy in the dilaton-dominated limit. This is sometimes referred to
as the {\em generalized} dilaton-domination
scenario.\cite{Casas:1996zi}

\begin{figure}
\centerline{\psfig{file=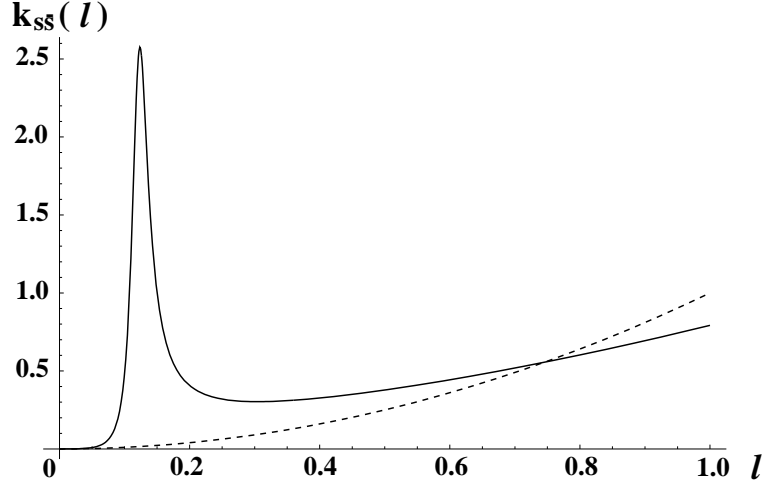,width=10cm}} \vspace*{8pt}
\caption{\textbf{Effective metric for the chiral dilaton field.} The
effective metric $k_{s\bar{s}}$ of~(\ref{translate}) is plotted as a
function of $\ell$ for the tree-level case (dashed line) and the
case with nonperturbative corrections (solid line). The
nonperturbative corrections produce a large maximum in the effective
metric near the minimum of $V(\ell)$.} \label{fig:dilmetric}
\end{figure}

The constraint of~(\ref{Ksstrue}) is precisely what is engineered by
the nonperturbative correction $f(\ell)$ and $g(\ell)$
in~(\ref{lke}). For example, the explicit case of~(\ref{ourf})
described in Section~\ref{sec23} was able to achieve~(\ref{Ksstrue})
with $\lang k_s \rang = -g_{\STR}^{2}/2$. For the parameter set
$\lbr A_0 = 3.25,\, A_1 = -1.70,\, B = 0.4 \rbr$ the tree-level
dilaton metric $k_{s\bar{s}}^{\rm tree} = \ell^2$ and corrected
dilaton metric of~(\ref{translate}) are plotted as function of
$\ell$ in Figure~\ref{fig:dilmetric}, for arbitrary units ($m_{\PL}
=1$). The effect of the nonperturbative corrections is to generate a
nontrivial feature in the metric near the minimum of the potential.
Here this peak occurs at $\lang \ell \rang \simeq 0.12$, for which
$g_{\STR}^2 \simeq 0.5$. The presence and importance of such a
feature in the dilaton metric has been noted in this scenario by
other authors.\cite{Casas:1996zi,Barreiro:1997rp}

It is helpful to parameterize the departure that~(\ref{Ksstrue})
represents from the tree level K\"ahler metric $\lang
(k_{s\bar{s}}^{\rm tree})^{1/2} \rang = \lang 1/(s+\bar{s}) \rang =
g_{\STR}^{2}/2 \simeq 1/4$ by introducing the phenomenological
parameter
\begin{equation}
a_{\rm np} \equiv \(\frac{k_{s\bar{s}}^{\rm tree}}{k_{s\bar{s}}^{\rm
true}}\)^{1/2} \label{acond}
\end{equation}
so that the auxiliary field of the dilaton chiral supermultiplet can
be expressed as
\begin{equation}
F^{S} = \sqrt{3} m_{3/2} (k_{s\bar{s}})^{-1/2} =
\sqrt{3}m_{3/2}a_{\rm np}(k_{s\bar{s}}^{\rm tree})^{-1/2} .
\label{FS}
\end{equation}
An important property of~(\ref{Ksstrue}) is to recognize that the
factor of $b_{+}$, containing as it does a loop factor, will
suppress the magnitude of the auxiliary field $F^S$ relative to that
of the supergravity auxiliary field $M$ through the
relation~(\ref{FS}). That is, provided $K_s \sim \order(1)$ so that
$K_s b_{+} \ll 1$ we can immediately see that a K\"ahler potential
which stabilizes the dilaton (while simultaneously providing zero
vacuum energy) will {\em necessarily} result in a suppressed dilaton
contribution to soft supersymmetry breaking. This is an unmistakable
property of the generalized dilaton-domination scenario when
achieved via K\"ahler stabilization. We will return to these issues
in Chapter~\ref{sec5} below.

By virtue of being able to recast the results of Section~\ref{sec31}
in terms of an effective chiral multiplet, we can employ the results
of Kaplunovsky and Louis\cite{Kaplunovsky:1993rd} to immediately
write down the soft supersymmetry-breaking parameters of the
observable sector. We will give them here to establish our notation
and conventions for what follows. The tree level gaugino mass for
canonically normalized gaugino fields is simply
\begin{equation}
M^{0}_a =  \frac{g_a^2}{2} F^n \partial_n f_{a}^{0} ,
\label{gaugtree} \end{equation}
where $f_a^0$ is the tree-level gauge-kinetic function (here taken
to be the chiral dilaton $S$) and the notation $F^n \partial_n X$
implies summation over all relevant chiral superfields. We define
our trilinear $A$-terms and scalar masses for canonically normalized
fields by
\begin{eqnarray}
V_A &=& \frac{1}{6}\sum_{ijk}A_{ijk}e^{K/2}W_{ijk}z^i z^j z^k + \hc
\nonumber \\  &=& \frac{1}{6}\sum_{ijk}A_{ijk}e^{K/2}(\kappa_i
\kappa_j \kappa_k)^{-1/2}W_{ijk}\h{z}^i \h{z}^j \h{z}^k + \hc,
\label{Apot}
\end{eqnarray}
where $\h{z}^i = \kappa_i^{{1/2}}z^i$ is a normalized scalar field,
and by
\begin{equation}
V_M = \sum_i M^2_i\kappa_i|z^i|^2= \sum_i M^2_i|\h{z}^i|^2 ,\,
\end{equation}
where the function $\kappa_i$ is the generalization of the specific
case in~(\ref{kappaZ}).

The precise form of the bilinear $B$-terms depends on how the
supersymmetric $\mu$-parameter of the Higgs potential is generated.
This remains an open problem in superstring
phenomenology,\cite{Binetruy:2005ez} as fundamental mass parameters
are generally zero for fields in the massless spectrum. For the
purposes of this section we merely give the general result. Let
$\nu_{ij}$ be a (possibly field-dependent) bilinear term in the
superpotential and let the K\"ahler potential contain a term of the
Giudice-Masiero form\cite{Giudice:1988yz}
\begin{equation} K(Z^i,\oline{Z}^{\ibar}) = \sum_i\kappa_i|Z^i|^2 +
\frac{1}{2}\sum_{ij}\[\alpha_{ij}(Z^n,\oline{Z}^{\bar{n}})Z^i Z^j +
 \hc \] + \order(|Z^i|^3) \, .\label{GMasiero} \end{equation}
Both are sources of masses for fields of the chiral supermultiplets
$Z^i$
\begin{equation} \Lag_M = - \sum_{ij}\[ \frac{1}{2} e^{K/2} \(\psi^i
\mu_{ij}\psi^j + \hc \) + e^K |z^i|^2\kappa^j|\mu_{ij}|^2\] \, ,
\end{equation}
where the effective $\mu$-parameter is given by
\begin{equation}
\mu_{ij} = \nu_{ij} - e^{-K/2}\(\frac{M}{3} \alpha_{ij} -
\oline{F}^{\bar{n}}\partial_{\bar{n}}\alpha_{ij}\) \, .
\label{muGM}\end{equation}
The $B$-term potential takes the form
\begin{equation} V_B = \frac{1}{2}\sum_{ij} B_{ij} e^{K/2} \mu_{ij} z^i z^j
+ \hc  = \frac{1}{2} \sum_{ij} B_{ij} e^{K/2}
(\kappa_i\kappa_j)^{-{1\over2}} \mu_{ij} \hat{z}^i\hat{z}^j + \hc \,
.
\end{equation}
With these conventions our tree level expressions are
\begin{eqnarray}
A^{0}_{ijk}&=&\lang F^n\partial_n\ln(\kappa_i\kappa_j\kappa_k
e^{-K}/W_{ijk})\rang \label{Atree} \\
B^{0}_{ij} &=& \lang F^n \partial_n
\ln(\kappa_i\kappa_je^{-K}/\mu_{ij}) + \frac{\oline{M}}{3}\rang
\label{Btree} \\
(M^{0}_{i})^2&=& \lang \frac{M\oline{M}}{9} - F^n
\oline{F}^{\bar{m}}\partial_n\partial_{\bar{m}}\ln\kappa_i \rang \,
. \label{scalartree}
\end{eqnarray}

If we specialize now to the case of moduli dependence given
by~(\ref{kappaZ}), (\ref{WT}) and $f_a = S$, then the tree level
gaugino masses~(\ref{gaugtree}), A-terms~(\ref{Atree}) and scalar
masses~(\ref{scalartree}) become
\begin{eqnarray}
M_{a}^{0}&=&\frac{g_{a}^{2}}{2}F^{S} \nonumber \\
A_{ijk}^{0} &=& (3-q_i - q_j - q_k)\Eisen F^{T} -k_{S}F^{S}
\nonumber \\
\(M_{i}^{0}\)^{2}&=&\frac{M\oline{M}}{9} - q_{i} \frac{|F^{T}
|^{2}}{(t + \bar{t})^{2}} \, ,
\label{treeBIM} \end{eqnarray}
where we have taken $F^{T_1} = F^{T_2} = F^{T_3} = F^T$ and dropped
the cumbersome brackets $\lang\dots\rang$. For the BGW model we have
$\lang F^T \rang = 0$ and an effective $\lang F^S \rang \neq 0$. The
tree-level soft terms are therefore simply
\begin{equation}
M_a^0 = \frac{g_{a}^{2}}{2}F^{S}, \quad A_{ijk}^{0} = -k_{S}F^{S},
\quad \(M_{i}^{0}\)^{2}= m_{3/2}^2 \, .  \label{BGWtreesoft}
\end{equation}
We can be more explicit by inserting the appropriate ``effective''
$F$-term expression for $F^S$. Starting with the definition
in~(\ref{Fredefined}) and taking the one-condensate approximation
of~(\ref{onecondapx}) we can quickly recover the expressions
in~(\ref{BGWgaugino}),~(\ref{BGWAterm}) and~(\ref{BGWmasssq}) for
$p_i = 0$.

\subsection{Loop corrections} \label{sec33}

The above expressions are insufficient to adequately describe the
superpartner spectrum of the BGW model, however. From~(\ref{acond})
we note that
\begin{equation}
a_{\rm np}=\sqrt{3}\frac{g_{\STR}^{2}b_{+}}{1- 2k_{s}b_{+}} \ll 1 \,
, \label{aBGW}
\end{equation}
and therefore from~(\ref{FS}) we see that if $g_{\STR}^2 = 1/2$ we
have
\begin{equation}
\left| \frac{F^{S}}{M} \right| = \left| \frac{F^{S}}{3m_{3/2}}
\right| = \frac{2}{g_{\STR}^2 } \frac{a_{\rm np}}{\sqrt{3}} \simeq
\frac{4a_{\rm np}}{\sqrt{3}} \ll 1 \, . \label{FtoMratio}
\end{equation}
It is evident, therefore, that quantum corrections to soft
supersymmetry-breaking terms suppressed by loop factors can, in
fact, be comparable in size to these tree-level terms for both the
gaugino mass and the trilinear $A$-term. In particular, loop
corrections arising from the conformal anomaly are proportional to
$M$ itself and receive no suppression,\footnote{We continue to use
the popular, if not particularly precise, name for these
``universal'' terms.\cite{Dine:2007me} The results presented here
were obtained through a direct one-loop computation with
supersymmetric regularization and do not appeal to notions of
superconformal transformations and their possible anomalies.} so
they can be competitive with the tree level contributions in the
presence of a nontrivial K\"ahler potential for the dilaton and
should be included.\cite{Gaillard:1999yb,Gaillard:2000fk}

In this section we aim to provide sufficient background to justify
the form of these one-loop corrections to soft terms, as well as
explain some notation we will need for our phenomenological
analysis. More complete treatments exist in the
literature.\cite{Gaillard:1998bf,Gaillard:1999yb,Gaillard:2000fk,Binetruy:2000md}
We begin with gaugino masses which can be understood as a sum of
loop-induced contributions from the field theory point of view, and
terms that can be thought of as one loop stringy corrections. The
field theory loop contribution is given
by~\cite{Gaillard:1999yb,Bagger:1999rd}
\begin{equation}
M^{1}_a|_{\rm an} = \frac{g_{a}^{2}(\mu)}{2} \left[ b_a \oline{M} -
\frac{1}{8\pi^2} \left( C_a - \sum_i C^i_a \right) F^n K_{n} -
\frac{1}{4\pi^2} \sum_i C^i_a F^n \partial_n \ln \kappa_i \right] \,
. \label{Man} \end{equation}
As described in Section~\ref{sec2}, one expects modular anomaly
cancelation to occur through a universal Green-Schwarz counterterm
with group-independent coefficient $b_{\GS}$ as well as possible
string threshold corrections with coefficient $b_a^I$. The one loop
contribution to gaugino masses from both terms are proportional to
the auxiliary fields of the K\"ahler moduli, and must vanish in the
vacuum of the BGW class of models. We are therefore left with only
the field-theory contribution of~(\ref{Man}) for $n=S$. Putting
together the tree level gaugino masses with the loop correction
gives
\begin{equation}
M_{a} = \frac{g_{a}^{2}\(\mu\)}{2} \lbr b_{a}\oline{M} +\[ 1 -
b_{a}' k_s \] F^{S} \rbr \label{Maloop}
\end{equation}
where the quantity $b_{a}'$ is defined in~(\ref{cond}).

To understand the form of the one-loop A-terms and scalar masses it
is necessary to describe how field theory loops are regulated in
supergravity, seen as an effective theory of strings. The regulation
of matter and Yang-Mills loop contributions to the matter wave
function renormalization requires the introduction of Pauli-Villars
chiral superfields $\Phi^A = \Phi^{i}$, $\wh{\Phi}^{i}$ and $\Phi^a$
which transform according to the chiral matter, anti-chiral matter
and adjoint representations of the gauge group and have signatures
$\eta_A = -1,+1,+1,$ respectively.  These fields are coupled to the
light fields $Z^{i}$ through the superpotential
\begin{equation}
W(\Phi^A,Z^i) = {1\over2}W_{ij}(Z^k)\Phi^i \Phi^j + \sqrt{2}\Phi^a
\wh{\Phi}_{i} (T_a Z)^i + \cdots, \label{PVcoup}
\end{equation}
where $T_a$ is a generator of the gauge group, and their K\"ahler
potential takes the schematic form
\begin{equation}
K_{\PV} = \sum_A \kappa_A^{\Phi}(Z^{N})|\Phi^A|^2, \label{PVKahler}
\end{equation}
where the functions $\kappa_{A}$ are {\em a priori} functions of the
hidden sector (moduli) fields. These regulator fields must be
introduced in such a way as to cancel the quadratic divergences of
the light field loops -- and thus their K\"ahler potential is
determined relative to that of the fields which they regulate.

The PV mass for each superfield $\Phi^A$ is generated by coupling it
to another field $\Pi^{A}=(\Pi^i, \widehat{\Pi}^i, \Pi^a)$ in the
representation of the gauge group conjugate to that of $\Phi^A$
through a superpotential term
\begin{equation}
W_m = \sum_{A}\mu_{A}(Z^N)\Phi^A \Pi^A, \label{PVbilinear}
\end{equation}
where $\mu_{A}(Z^N)$ can in general be a holomorphic function of the
light superfields. There is no constraint on the K\"ahler potential
for the fields $\Pi^A$ as there was for those of the $\Phi^A$.
However, if we demand that our regularization preserve modular
invariance then we can determine the moduli dependence of
$\mu_{A}(Z^N)$ as a function of the {\em a priori} unknown modular
weight of the regulator fields $\Pi^A$. Taking the case of an
overall K\"ahler modulus $T$ for simplicity we have
\begin{equation}
\lbr \begin{array}{l} \Phi^i:\;\;\kappa^\Phi_i = \kappa_i = (T +
\oline{T})^{-q_i}, \\ \wh{\Phi}^i:\;\; \h{\kappa}^\Phi_i =
\kappa^{-1}_i, \\ \Phi^a: \;\;\kappa^\Phi_a = g_a^{-2}e^K =
g_a^{-2}e^k(T + \oline{T})^{-3}\, ,
\end{array} \right.
\label{kapPhi}
\end{equation}
and the supersymmetric mass $\mu_{A}(Z^N)$ in~(\ref{PVbilinear}) is
then
\begin{equation}
\mu_{A}(Z^{N}) = \mu_{A}(S)
\[\eta(T)\]^{-2(3 - q_A - q_{A}')},
\end{equation}
with $q_A$ and $q_{A}'$ being the modular weights of the fields
$\Phi^A$ and $\Pi^A$, respectively. Furthermore, we can postulate
the form of the moduli dependence of $\kappa_A$ for the
mass-generating fields
\begin{equation}
\Pi^A:\;\;\kappa^\Pi_A = h_A(S+\oline{S})(T + \oline{T})^{-q_{A}'}.
\label{kapPi}
\end{equation}
At this point the dilaton dependence in the superpotential
term~(\ref{PVbilinear}) and the functions $h_A$, as well as the
modular weights $q_{A}'$ of the fields $\Pi^A$, are new free
parameters of the theory introduced at one loop as a consequence of
how the theory is regulated. Given~(\ref{PVbilinear}) we can extract
the Pauli-Villars masses that appear as regulator masses in the
logarithms at one loop
\begin{equation}
m_A^2 = e^K (\kappa^\Phi_A)^{-1/2} (\kappa^{\Pi}_A)^{-1/2} |\mu_A|^2
, \label{PVmass}
\end{equation}
with $m_A = (m_i, \h{m}_i, m_a)$ being the masses of the regulator
fields $\Phi^i, \wh{\Phi}^{i}, \Phi^a$, respectively.

In terms of these regulator masses, the complete one-loop correction
to the trilinear A-terms and scalar masses in a general supergravity
theory was given elsewhere.\cite{Binetruy:2000md} Here we will
simplify things to the maximum extent in order to proceed to the
phenomenology of the model. To that end, let us assume that the
functions $\mu_{A}(Z^N)$ that appear in~(\ref{PVbilinear})
and~(\ref{PVmass}) are proportional to one overall Pauli-Villars
scale $\mu_{\PV}$ so that $\h{\mu}_{i} = \mu_{a} = \mu_{i} \equiv
\mu_{\PV}$. This scale is presumed to represent some fundamental
scale in the underlying string theory. Let us further assume that
there is no dilaton dependence of the PV masses so that
$h_A(S+\oline{S})$ is trivial and $\mu_{\PV}$ is constant. With
these simplifications the complete trilinear A-term at one loop is
given by
\begin{eqnarray}
A_{ijk} &=& \frac{1}{3} A^{0}_{ijk} - \frac{1}{3}
\gamma_{i}\oline{M} - \Eisen F^T \(\sum_a\gamma_i^a p_{ia} +
\sum_{lm}\gamma_i^{lm}p_{lm}\) \nonumber \\
 & & -\ln\[(t + \bar{t})|\eta(t)|^4\]\(2
\sum_a\gamma_i^a p_{ia}M^{0}_a +
\sum_{lm}\gamma_i^{lm}p_{lm}A^{0}_{ilm}\) \nonumber \\ & & +
2\sum_a\gamma_i^a M^{0}_a \ln(\mu_{\PV}^{2}/\mu_R^2) +
\sum_{lm}\gamma_i^{lm}A^{0}_{ilm}\ln(\mu_{\PV}^{2}/\mu_R^2)
 + {\rm cyclic}(ijk) ,
\label{Atermraw}
\end{eqnarray}
where we have defined the following combinations of modular weights
from the Pauli-Villars sector
\begin{equation}
p_{ij} = 3 - \frac{1}{2} \(q_{i} + q_j + q_{i}' + q_{j}'\), \quad
p_{ia} = \frac{1}{2} \(3 - q_{a}' -\h{q}_{i}' + q_{i} \)
\label{PVmodsraw} \end{equation}
which we will refer to as ``regularization weights'' in reference to
their origin from the PV sector of the theory.

In~(\ref{Atermraw}) $M^{0}_a$ and $A^{0}_{ilm}$ are the tree level
gaugino masses and A-terms given in~(\ref{treeBIM}) and the
parameters $\gamma$ determine the chiral multiplet wave function
renormalization
\begin{eqnarray}
\gamma^j_i = \frac{1}{32\pi^2}\[4\delta^j_i\sum_a g^2_a(T^2_a)^i_i -
e^K\sum_{kl}W_{ikl}\oline{W}^{jkl}\] . \label{gam}
\end{eqnarray}
We have implicitly made the approximation that generational mixing
is unimportant and can be neglected in~(\ref{Atermraw}), and that
motivates the definitions
\begin{eqnarray}
\gamma_i^j &\approx&\gamma_i\delta^j_i, \quad \gamma_i =
\sum_{jk}\gamma_i^{jk} + \sum_a\gamma_i^a, \nonumber \\ \gamma_i^a
&=& \frac{g^2_a}{8\pi^2}(T^2_a)^i_i, \quad \gamma_i^{jk} =
-\frac{e^K}{32\pi^2}
(\kappa_i\kappa_j\kappa_k)^{-1}\left|W_{ijk}\right|^2.\label{diag}
\end{eqnarray}
The scalar masses are obtained similarly and take the form
\begin{eqnarray}
\(M_{i}\)^{2} &=& (M_{i}^{0})^{2}+ \gamma_{i}\frac{M\oline{M}}{9} -
\frac{|F^{T}|^{2}}{(t + \bar{t})^{2}} \(\sum_a\gamma^a_i p_{ai} +
\sum_{jk}\gamma^{jk}_i p_{jk}\) \nonumber \\ & & +\lbr \frac{M}{3}
\[\sum_a\gamma^a_i M^{0}_{a} +
{1\over2}\sum_{jk}\gamma^{jk}_iA^{0}_{ijk}\] + \hc \rbr \nonumber
\\ & & + \lbr  F^T \Eisen
\(\sum_a\gamma_i^a p_{ia}M^{0}_a +
\frac{1}{2}\sum_{jk}\gamma_i^{jk}p_{jk} A^{0}_{jk}\) + \hc
\rbr\nonumber \\ & & - \ln\[(t +
\bar{t})|\eta(t)|^4\]\lbr\sum_a\gamma_i^a p_{ia}
\[3(M^{0}_a)^2 - (M^{0}_i)^2\]\right. \nonumber \\
& &  \qquad\qquad + \left.\sum_{jk}\gamma_i^{jk}p_{jk}\[(M^{0}_j)^2
+ (M^{0}_k)^2 + (A^{0}_{ijk})^2\]\rbr \nonumber \\ & & +
\sum_a\gamma_i^a\[3(M^{0}_a)^2 - (M^{0}_i)^2\]
\ln(\mu_{\PV}^{2}/\mu_R^2) \nonumber \\ & & +
\sum_{jk}\gamma_i^{jk}\[(M^{0}_j)^2 + (M^{0}_k)^2 +
(A^{0}_{ijk})^2\]\ln(\mu_{\PV}^{2}/\mu_R^2), \label{massraw}
\end{eqnarray}
with $M_i^{0}$ being the tree level scalar masses
of~(\ref{treeBIM}).

To put these expressions into a less cumbersome and more suggestive
form, we will consider the case where the various regularization
weights $p_{ia}$ and $p_{jk}$ can be treated as one overall
parameter $p$. Then inserting the tree level soft
terms~(\ref{treeBIM}) into~(\ref{Atermraw}) and~(\ref{massraw})
yields
\begin{eqnarray}
A_{ijk} &=& -\frac{k_s}{3}F^S - \frac{1}{3} \gamma_{i}\oline{M} - p
\gamma_{i} \Eisen F^{T} + \tilde{\gamma}_{i} F^{S} \lbr
\ln(\mu_{\PV}^{2}/\mu_R^2) -p\ln\[(t+\bar{t}) |\eta(t)|^4\] \rbr
\nonumber \\ & & + {\rm cyclic}(ijk) \label{finalA} \\ M_{i}^{2} &=&
\lbr \frac{|M|^2}{9} -\frac{|F^T|^{2}}{(t+\bar{t})^{2}}\rbr \[ 1 +
p\gamma_i -\(\sum_{a}\gamma_{i}^{a} -2\sum_{jk}\gamma_{i}^{jk}\) \(
\ln(\mu_{\PV}^{2}/\mu_R^2) -p\ln\[(t+\bar{t}) |\eta(t)|^4\] \)
\] \nonumber \\
 & &+ (1-p)\gamma_i \frac{|M|^2}{9} + \lbr
 \wtd{\gamma}_{i}\frac{MF^S}{6}+\hc \rbr +\lbr p \wtd{\gamma}_{i}\Eisen
 \frac{\oline{F}^{T}F^S}{2} + \hc \rbr \nonumber \\
 & & +|F^{S}|^2 \[ \(\frac{3}{4}\sum_{a} \gamma_{i}^{a} g_{a}^{4}
 + k_s k_{\bar{s}} \sum_{jk} \gamma_{i}^{jk}\) \(
\ln(\mu_{\PV}^{2}/\mu_R^2) -p\ln\[(t+\bar{t}) |\eta(t)|^4\] \) \],
\label{finalscalar}
\end{eqnarray}
where $\wtd{\gamma}_{i}$ is a shorthand notation for
\begin{equation}
\wtd{\gamma}_{i} = \sum_{a} \gamma_{i}^{a} g_{a}^{2} - k_{s}
\sum_{jk} \gamma_{i}^{jk} . \label{tildegamma}
\end{equation}
The adoption of one overall regularization weight $p$ makes it
possible to identify the quantity $\ln(\mu_{\PV}^{2}/\mu_R^2)
-p\ln\[(t+\bar{t}) |\eta(t)|^4\]$ as a stringy threshold correction
to the overall PV mass scale, or effective cut-off,
$\mu_{\PV}$.\cite{Nelson:2002fk} Let us make this identification
explicit by defining
\begin{equation} \ln(\mu_{\PV}^{2}/\mu_R^2)
-p\ln\[(t+\bar{t}) |\eta(t)|^4\] \equiv
\ln(\tilde{\mu}_{\PV}^{2}/\mu_R^2) \, . \end{equation}
Specializing further to the case of generalized dilaton domination
we have\footnote{We have dropped terms of $\order\(1/(16\pi^2)^3\)$
in the scalar masses.}
\begin{eqnarray}
M_{a}&=&\frac{g_{a}^{2}\(\mu_R\)}{2} \lbr b_{a}\oline{M} +\[ 1 -
b_{a}' k_s \] F^{S} \rbr \nonumber \\ A_{ijk} &=& -\frac{k_s}{3}F^S
- \frac{1}{3} \gamma_{i}\oline{M} + \tilde{\gamma}_{i} F^{S}
\ln(\tilde{\mu}_{\PV}^{2}/\mu_R^2) + {\rm cyclic}(ijk) \nonumber
\\ M_{i}^{2} &=& \frac{|M|^2}{9}
 \[ 1 + \gamma_i
-\(\sum_{a}\gamma_{i}^{a} -2\sum_{jk}\gamma_{i}^{jk}\)
\ln(\tilde{\mu}_{\PV}^{2}/\mu_R^2) \]
\nonumber \\
 & &
\qquad  + \lbr
 \wtd{\gamma}_{i}\frac{MF^S}{6}+\hc \rbr \, . \label{BGWsoft}
\end{eqnarray}

\subsection{Superpartner spectra and fine-tuning}
\label{sec34}

With the expressions in~(\ref{BGWsoft}) we now have a starting point
for a discussion of the phenomenological implications of the
BGW~class of models. Here we will pause to look at some of the
coarse features of the model, such as the general spectrum of
superpartner masses. This will begin with a consideration of the
issue of electroweak symmetry breaking and finish, ultimately, with
the question of fine-tuning in this model class. In
Section~\ref{sec5} we will take a much more detailed look at certain
aspects of the model phenomenology, after we have considered how the
supersymmetry breaking pattern and spectrum can change in the
presence of anomalous~$U(1)$ factors in Section~\ref{sec4}.

First let us consider the issue of overall mass scales. Returning to
the expression in~(\ref{FtoMratio}) we can see that we expect a {\em
suppression of gaugino masses relative to scalar masses}. This is a
key feature of this class of models. It can be seen explicitly by
comparing~(\ref{Maloop}) to $m_{3/2}^2$ using~(\ref{FtoMratio}), for
which we obtain at tree level
\begin{equation}
\left| \frac{M_a}{M_i}\right| = \sqrt{3} a_{\rm np}
\(\frac{g_a(\mu_R)}{g_{\STR}}\)^2 \, . \label{MtoMratio}
\end{equation}
From the definition of $a_{\rm np}$ in~(\ref{aBGW}) it is clear that
gaugino masses will be suppressed by a loop factor relative to
scalar masses. If the chargino mass must be of order 100~GeV or
higher to avoid direct search constraints, this implies that the
typical size of scalars in this theory must be at least a few~TeV.
This model is therefore a manifestation of ``loop-split''
supersymmetry.\cite{Wells:2004di} We will see below that this
splitting is generally welcome phenomenologically, though it
exacerbates certain other problems. The first of these issues is
electroweak symmetry breaking (EWSB).

To adequately describe the physical masses of the superpartners in
this class of theories it is necessary to consider how electroweak
symmetry is broken. Dynamical breaking of the symmetry is possible
is the context of the MSSM provided (a) a large top Yukawa is
present, (b) there is a large range of energies between the scale of
supersymmetry breaking and the electroweak scale, and (c) a
supersymmetric Higgs mass (or $\mu$-term) can be generated of the
appropriate size. All conditions are compatible with the
requirements and constraints of the BGW class of models, but the
physics of the resulting low-energy theory depends very much on how
condition~(c) is obtained. For example, it can be
demonstrated\cite{Casas:1996wj} that if the $\mu$-parameter arises
as a fundamental parameter in the superpotential and is independent
of the moduli, then it is extremely difficult to achieve EWSB and a
sufficiently large top quark mass in this class of theories. This is
a result of the form of the supersymmetry-breaking bilinear
$B$-term, which takes the value $B=2m_{3/2}$ in this scenario.

Such a fundamental $\mu$ term generally would not arise from string
theory, however, as the Higgs doublets are part of the massless
spectrum of the theory. Instead we anticipate that the
$\mu$-parameter is dynamically generated, either via the
Giudice-Masiero mechanism\cite{Giudice:1988yz} or through the
spontaneous breaking of certain additional symmetries via the \vev
of some field which is neutral under the Standard
Model.\cite{Cvetic:1995rj,Cvetic:1997ky} In these cases the
$B$-parameter may take a variety of values, depending on the model.
In particular, for the case of dynamical generation of the
$\mu$-parameter via singlet \vevs with $W_{\mu} = \lambda_X \lang X
\rang H_u H_d$, the effective $B$-term is in fact a trilinear
$A$-term. The constraints on the model arising from electroweak
symmetry breaking are weaker in these more realistic cases.

For the sake of the present discussion, let us use the {\em
requirement} of proper electroweak symmetry breaking to {\em
determine} the values of $\mu$ and $B$ at the electroweak scale in
the usual manner. In other words, we compute the one-loop corrected
effective potential $V_{\rm 1-loop}=V_{\rm tree} + \Delta V_{\rm
rad}$ at the electroweak scale and determine the effective mu-term
$\bar{\mu}$ via
\begin{equation}
{\bar{\mu}}^{2}=\frac{\(m_{H_d}^{2}+\delta m_{H_d}^{2}\) -
  \(m_{H_u}^{2}+\delta m_{H_u}^{2}\) \tan{\beta}}{\tan^{2}{\beta}-1}
-\frac{1}{2} M_{Z}^{2} \, . \label{radmuterm}
\end{equation}
In equation~(\ref{radmuterm}) the quantities $\delta m_{H_u}$ and
$\delta m_{H_d}$ are the second derivatives of the radiative
corrections $ \Delta V_{\rm rad}$ with respect to the up-type and
down-type Higgs scalar fields, respectively. For the numerical
results which follow we will include the effects of all
third-generation particles. If the right hand side of
equation~(\ref{radmuterm}) is positive then there exists some
initial value of $\mu$ at the high-scale which results in correct
electroweak symmetry breaking with $M_{Z} = 91.187\GeV$.

\begin{figure}
\centerline{\psfig{file=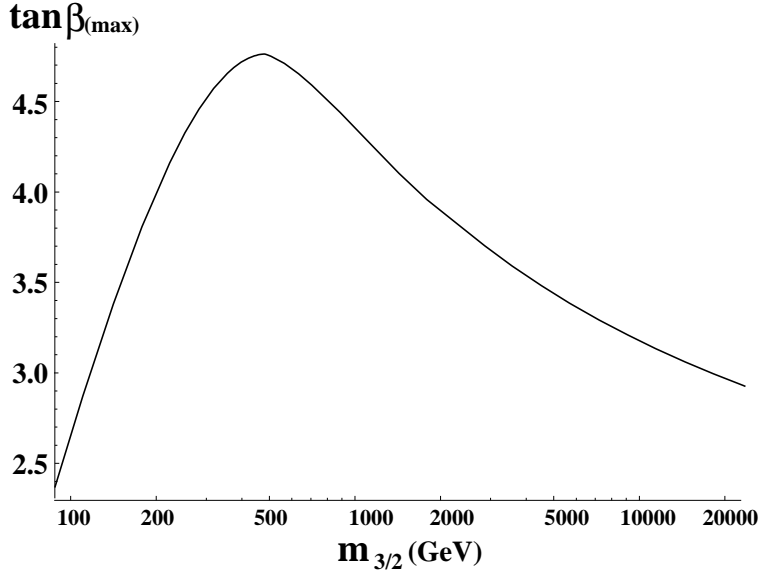,width=10cm}} \vspace*{8pt}
\caption{\textbf{A typical constraint on $\mathbf{\tan\beta}$ from
the requirement of proper EWSB.} The maximum value of $\tan{\beta}$
consistent with EWSB and positive squark masses is displayed as a
function of the gravitino mass. For this figure we have taken $b_+ =
0.08$ and $\caleff =1$.} \label{fig:lowtanbeta}
\end{figure}

Though all scalar masses (including those of the two Higgs doublets)
in the BGW~class will generally have large masses at the boundary
condition scale, it is well known\cite{Feng:1999mn} that the Yukawa
structure of the MSSM is such that the up-type Higgs soft
mass-squared $m_{H_u}^2$ is typically driven to negative values
nevertheless. Dynamical EWSB is therefore robust in this model.
However, the large scalars induce large corrections $\delta
m_{H_u}^{2}$ and $\delta m_{H_d}^{2}$ which can destabilize the EWSB
minimum and make ${\bar{\mu}}^{2} < 0$. The BGW model (with the GS
coupling of the matter fields $p_i$ in~(\ref{GS}) taken to vanish)
is therefore a ``focus-point'' model\cite{Feng:1999zg} over most of
its viable parameter space. The various requirements outlined above
tend to push the value of $\tan\beta$ required to
solve~(\ref{radmuterm}) to small values. This is shown in
Figure~\ref{fig:lowtanbeta} for the case of $b_{+}=0.08$ and
$\caleff =1$ (which we will assume from here onwards). There is some
mild dependence of the maximum value of $\tan\beta$ on these
parameters, as well as on the choice of top quark pole mass assumed.
Generally speaking, though, $\tan\beta \lappeq 10$ for most of the
model parameter space.

Returning to the soft supersymmetry-breaking parameters, the
relation~(\ref{MtoMratio}) implies that tree-level contributions to
gaugino masses arising from the effective dilaton auxiliary
field~$\lang F^S \rang$ will be of roughly the same size as those
one-loop corrections proportional to the gravitino mass. There is
one unique contribution proportional to the auxiliary field of
supergravity, which is the first term in the gaugino mass expression
in~(\ref{BGWsoft}). These terms imply a splitting in the gaugino
soft masses which will depend on the relative sizes of the
beta-function coefficient $b_a$ for the Standard Model gauge group
$\mathcal{G}_a$ and that of the largest beta-function coefficient
$b_+$ for the condensing product group. Such an outcome is not in
conflict with the possibility of gauge coupling unification in this
theory. Schematically, the gauge kinetic function in the superspace
Lagrangian density is replaced at one loop by the expression
\begin{equation}
{\cal L} \sim \int\diff^{2}\theta f_{a} \(W^{\alpha}
W_{\alpha}\)_{a} \to \int\diff^{2}\theta \( S + \frac{1}{16\pi^2}X_a
\) \(W^{\alpha} W_{\alpha}\)_{a}\, , \label{freplace}
\end{equation}
where both objects $S$ and $X_a$ obtain $\order(1)$ \vevs for their
lowest scalar components (thereby producing only small corrections
to gauge coupling unification). But only $X_a$ receives an
$\order(1)m_{3/2}$ auxiliary field $vev$; for the auxiliary field of
the dilaton this auxiliary field \vev is suppressed by a loop factor
relative to the gravitino mass. The scenario we have just described,
with tree-level and loop-level contributions to gaugino masses of
about the same magnitude, has recently re-appeared in a number of
guises.\cite{Choi:2005uz,Kitano:2006gv,Baer:2006id} Many of these
examples have been engineered to provide~(\ref{MtoMratio}), and
nonuniversal gaugino masses. In the BGW~class of heterotic models it
is an automatic feature of K\"ahler stabilization. When this
property arises many virtuous phenomenological properties follow,
which we will investigate below.

\begin{figure}[t]
    \begin{center}
\centerline{
       \epsfig{file=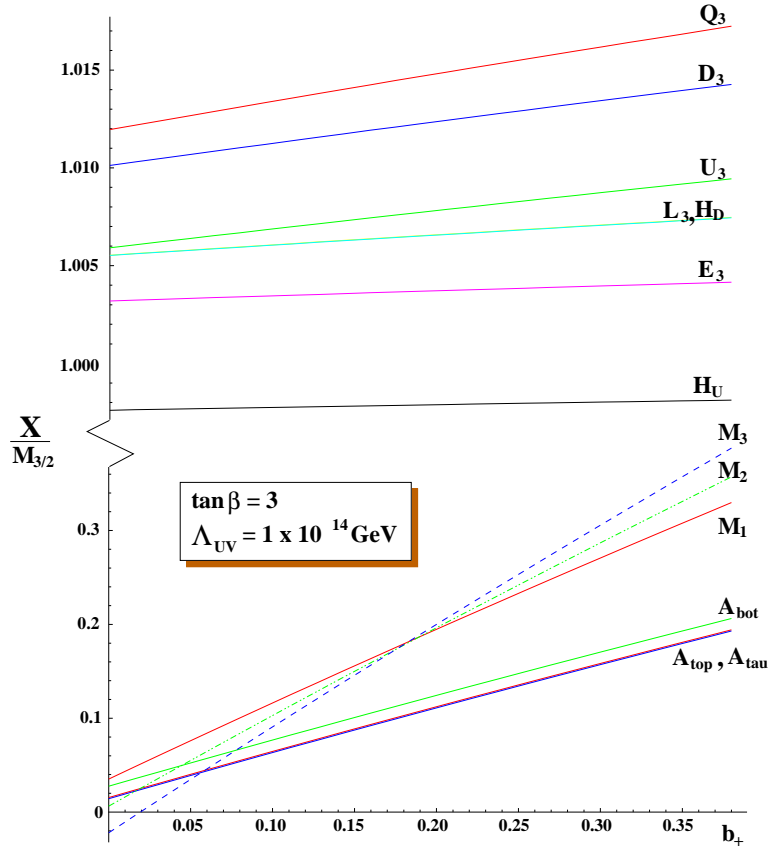,width=0.8\textwidth}}
          \caption{{\footnotesize {\bf Spectrum of soft supersymmetry
                breaking terms in BGW model.} All values are given
              relative to the gravitino mass $m_{3/2}$ at a scale
              $\mu_{\UV}= 1 \times 10^{14}$ GeV as a function
              of the condensing group beta function coefficient $b_{+}$.}}
        \label{fig:bgwspectrum}
    \end{center}
\end{figure}

To study the parameter space of the BGW model it is inconvenient to
work in the space of hidden sector parameters $\lbr \caleff
,\bpaleff, b_{+} \rbr$ as this defines a very narrow region in
Figures~\ref{fig:conscale},~\ref{fig:gravmassvaryc}
and~\ref{fig:varyall}. But we note that the values of $\caleff$ and
$\baaleff$ appear only indirectly through the determination of the
value of the condensate $\lang \rho_{+}^{2} \rang$ in~(\ref{rhosq}).
It is thus convenient to cast all soft supersymmetry-breaking
parameters in terms of the values of $b_{+}$ and $m_{3/2}$ using
equation~(\ref{gravmass}).\cite{Binetruy:2003ad} While the gravitino
mass itself is not strictly independent of $b_{+}$, it is clear from
Figure~\ref{fig:gravmassvaryc} that we are guaranteed of finding a
reasonable set of values for $\lbr \caleff ,\bpaleff\rbr$ consistent
with the choice of $b_{+}$ and $m_{3/2}$ provided we scan only over
values $b_{+} \lappeq 0.1$ for weak string coupling. This
transformation of variables allows the slice of parameter space
represented by the contours of Figure~\ref{fig:varyall} to be recast
as a two-dimensional plane for a given value of $\tan{\beta}$ and
${\rm sgn}(\mu)$.

In Figure~\ref{fig:bgwspectrum} we thus exhibit the soft
supersymmetry breaking parameters of~(\ref{BGWsoft}) as a function
of condensing group beta-function coefficient $b_+$. All masses are
given relative to the gravitino mass, and we have taken $g_{\STR}^2
= 0.5$ and chosen $\tilde{\mu}_{\PV}^{2} = \mu_{\UV}^2$, where
$\mu_{\UV}$ is the boundary condition scale. All values tend to
increase with $b_+$ as the relative size of $\lang F^S \rang$ to
$m_{3/2}$ increases. But note the importance of the anomaly
contributions to gaugino masses in changing the {\em relative} sizes
of the three Standard Model gaugino mass parameters. The values of
the one-loop soft terms are sensitive to the choice of boundary
condition scale through the logarithmic terms
$\ln(\tilde{\mu}_{\PV}^{2}/\mu_R^2)$ in the A-terms and scalar
masses, and through the dependence on the renormalized gauge
couplings $g_a^2(\mu_R)$ in the gaugino masses. This dependence is
most pronounced for the gaugino soft masses, but is milder in the
regions of phenomenological viability $0 \leq b_+ \leq 0.09$
established in Section~\ref{sec23}.

\begin{figure}
\centerline{\psfig{file=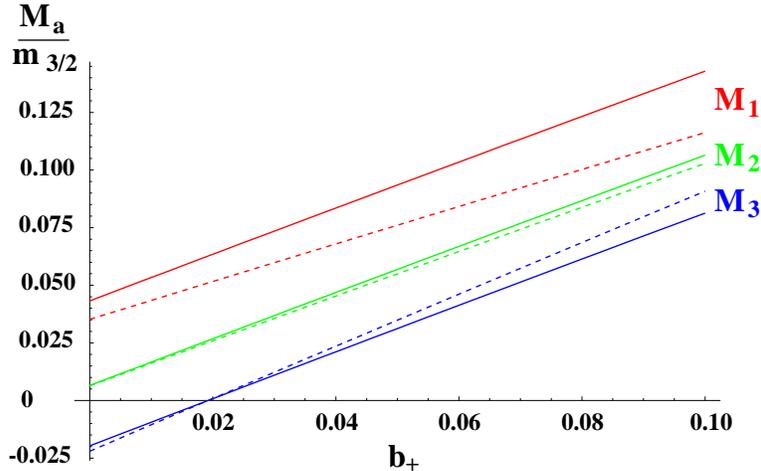,width=10cm}} \vspace*{8pt}
\caption{\textbf{Gaugino masses in the BGW model.} Gaugino masses
$M_{1}$ (top), $M_{2}$ and $M_{3}$ (bottom) are given as a function
of the condensing group beta function coefficient $b_{+}$ at a scale
$\mu_{\UV} = 2 \times 10^{16} \GeV$ (solid lines) and $\mu_{\UV} = 1
\times 10^{14} \GeV$ (dashed lines).} \label{fig:bgwgaug}
\end{figure}

This dependence is demonstrated in Figure~\ref{fig:bgwgaug}, where
we give the gaugino soft masses as a function of $b_+$ over the
range $0 \leq b_+ \leq 0.1$ for two choices of boundary condition
scale $\mu_{\UV} = 2 \times 10^{16} \GeV$ (solid lines) and
$\mu_{\UV} = 1 \times 10^{14} \GeV$ (dashed lines). The convergence
of gaugino masses as a function of $b_+$ occurs at lower values for
lower boundary condition scales. But over the range of
phenomenological interest the hierarchy of gaugino masses is
inverted relative to what occurs at low energies in unified models
such as the minimal supergravity model: the gluino mass $M_3$ is the
smallest, with the $B$-ino mass $M_1$ being the largest. Of course,
these are statements that hold at the boundary condition scale --
what is of interest is the hierarchy of gaugino masses at the low
energy (electroweak) scale. Evolving the gaugino masses to the
electroweak scale using the two-loop RG
equations\cite{Martin:1993zk} the gluino increases in mass, but
remains lighter relative to the neutralino/chargino sector than in
the minimal supergravity model.

\begin{figure}[t]
\centerline{
       \psfig{file=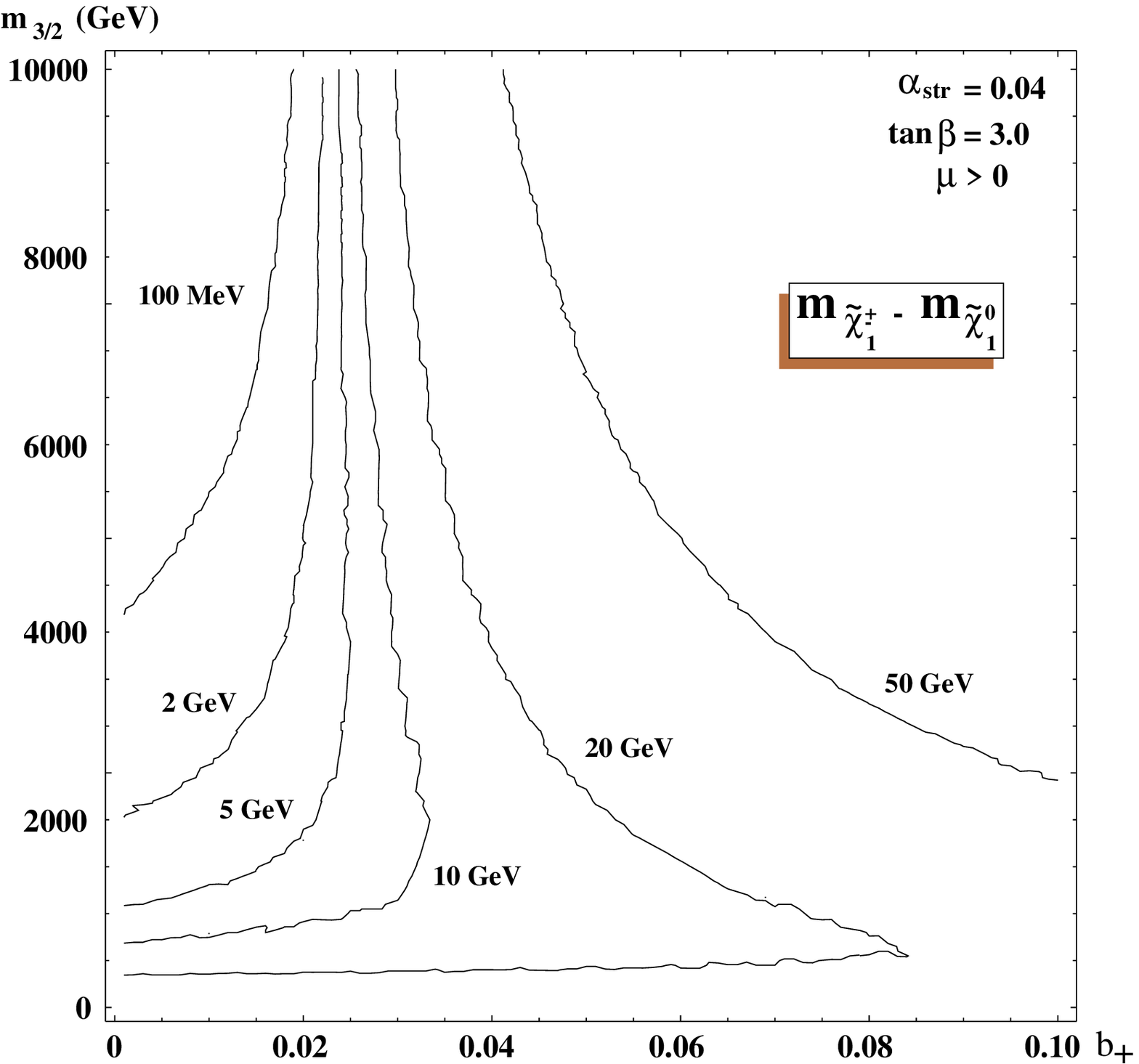,width=0.5\textwidth}
       \psfig{file=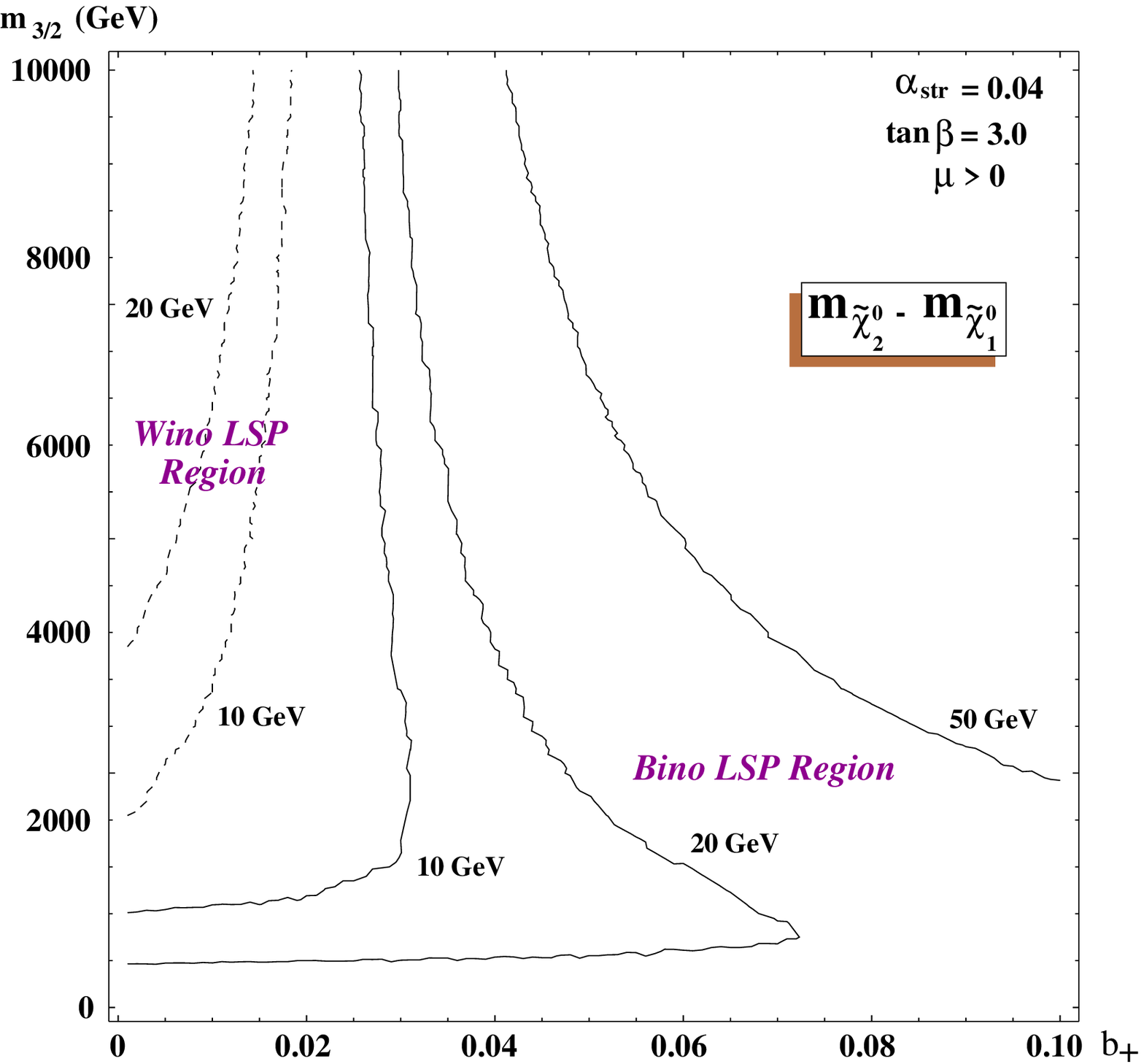,width=0.5\textwidth}}
          \caption{{\footnotesize {\bf The physical gaugino
                sector of the BGW model}. The left panel gives the mass difference
              between the $\chi^{\pm}_{1}$ and the $\chi^{0}_1$ in
              GeV. The right
              panel gives the difference in mass between the two
              lightest neutralinos $\chi^{0}_2$ and $\chi^{0}_1$. Note
              that a level crossing occurs and there exists a region in which
              the $W$-ino $W_{0}$ becomes the LSP.}}
        \label{fig:gauginosec}
\end{figure}

After evolving the gaugino masses $M_1$ and $M_2$ to the electroweak
scale, and imposing the electroweak symmetry breaking
condition~(\ref{radmuterm}), we can compute the physical neutralino
and chargino masses. Over most of the allowed range in $b_+$ the
$B$-ino is the lightest supersymmetric particle (LSP) -- just as in
minimal unified models -- but there will be a significant component
of the $SU(2)$ neutral $W$-ino in the wavefunction of the LSP. In
fact, when the condensing group beta-function coefficient $b_{+}$
becomes relatively small (i.e. similar in size to the MSSM
hypercharge value of $b_{U(1)}=0.028$) the pieces of the gaugino
mass arising from the superconformal anomaly can become equal in
magnitude to part arising from the dilaton auxiliary field. Here
there is a level crossing in the neutral gaugino sector. The LSP
becomes predominately $W$-ino like and the mass difference between
the lightest chargino and lightest neutralino becomes negligible.
This effect is displayed in Figure~\ref{fig:gauginosec}. The
phenomenology of the gaugino sector in this limit is similar to that
of the minimal anomaly-mediated supersymmetry breaking (AMSB)
model.\cite{Feng:1999hg,Gherghetta:1999sw,Paige:1999ui}

One other important implication of the hierarchy in
Figure~\ref{fig:bgwspectrum} is the issue of fine-tuning in the
electroweak sector. A widely held rule of thumb about fine-tuning
states that it generally increases with the scale of the mass of
squarks and sleptons. But this is only roughly true. Consider the
tree-level analog of the EWSB condition in~(\ref{radmuterm}). This
tree level relation can, in turn, be written in the following
way\cite{Kane:1998im}
\begin{equation}
M_{Z}^{2}= \sum_{i} C_i m_i^2(\UV) + \sum_{ij}C_{ij}m_i(\UV)
m_j(\UV) \, , \label{zsum}
\end{equation}
where $m_i$ represents a generic parameter of the softly broken
supersymmetric Lagrangian at an initial high scale $\mu_{\UV}$ with
mass dimension one, such as gaugino masses, scalar masses, trilinear
A-terms and the $\mu$ parameter. The coefficients $C_i$ and $C_{ij}$
depend on the scale $\mu_{\UV}$ and quantities such as the top mass
and $\tan\beta$ in a calculable way through solving the
renormalization group equations. For example, taking the running
mass for the top quark at the Z-mass scale to be $m_{\rm top}(M_Z) =
170 \GeV$, the starting scale to be the grand-unified scale, and
$\tan\beta=5$ we have for the leading terms
in~(\ref{zsum})\cite{Kane:2002ap}
\begin{eqnarray}
M_{Z}^{2}&=&-1.8 \mu^{2}(\UV) + 5.9 M_{3}^{2}(\UV) - 0.4
M_{2}^{2}(\UV) -1.2 m^{2}_{H_U}(\UV) \nonumber
\\ & & +0.9 m^{2}_{Q_3}(\UV) + 0.7 m^{2}_{U_3}(\UV)
-0.6 A_{t}(\UV) M_{3}(\UV) \nonumber \\ & & -0.1
A_{t}(\UV)M_{2}(\UV) +0.2 A_{t}^{2}(\UV) +0.4 M_{2}(\UV)M_{3}(\UV) +
\dots  \label{ztune}
\end{eqnarray}
where the ellipsis in~(\ref{ztune}) indicate terms that are less
important quantitatively and for our purposes. Note that the most
important parameter is, in fact, the gluino mass as evidenced by the
large value of $C_3$. In contrast, the scalar masses are far less
important. Indeed, replacing the scalars with a universal value
$m_0$ the relation in~(\ref{ztune}) becomes
\begin{equation}
M_Z^2 = -1.8 \mu^2 (\UV) + 0.4 m_{0}^{2} + 5.9 M_{3}^{2} - 0.4
M_{2}^{2} + \dots \, .\label{mSUGRAtune}
\end{equation}
Thus, models in which gluinos are light relative to the electroweak
gauginos are more likely to provide the needed cancelation of soft
parameters against the supersymmetric mass term ($\mu$) with a
minimum of fine-tuning required. This feature was the driving
motivation for the models of Ref.~\refcite{Kitano:2006gv}, which led
to construction of effective theories very similar to the BGW model.
We will pursue other phenomenological implications of the model in
Section~\ref{sec5}, but first we must consider how the pattern of
soft supersymmetry breaking can change when the model contains an
anomalous $U(1)_X$ factor.

\section{Inclusion of anomalous U(1)'s}
\label{sec4}
\setcounter{footnote}{0}

An anomalous $U(1)$, commonly denoted $U(1)_X$, is generic to
effective supergravity in string theory.  For example the
study\cite{Giedt:2001zw} of a large class of standard-like heterotic
$\mathbf{Z}_3$ orbifold models showed that 168~of 175~models had an
anomalous $U(1)_X$.  The underlying theory is anomaly free, and the
apparent anomaly is
canceled\cite{Dine:1987xk,Atick:1987gy,Dine:1987gj} by a
four-dimensional version of the Green-Schwarz term similar to the
one in~(\ref{GSlinear}) used to cancel the modular anomaly. This
leads to a Fayet-Illiopoulos (FI) term in the effective supergravity
Lagrangian. Ignoring nonperturbative corrections to the dilaton
K\"ahler potential, the $D$-term for $U(1)_X$\ takes the
form\footnote{In this section there will be a growing number of
``species'' indices with which the reader must be concerned. To
minimize future confusion, we here summarize the index conventions.
In this section we will label generic chiral superfields with
capital Latin indices (as opposed to the lower-case indices of the
previous sections). Gauge groups are labeled by lower-case Latin
indices from the beginning of the alphabet. At times it will be
convenient to single-out an anomalous $U(1)$ factor for which the
label $a$ will be replaced with $X$. Matter condensates continue to
be labeled by Greek letters and internal complex planes/K\"ahler
moduli by the capital index $I$. Repeated indices do not imply
summation -- all index summations will be explicitly shown.}
\beq D_X = - 2\(\sum_A K_A q^X_A \phi^A + \xi\),\quad K_A =
{\partial K \over
\partial\phi^A} \quad \xi = {g_{\STR}^2 \, \Tr \, Q_X \over 192 \pi^2}
m_{\PL}^2, \label{eq1} \eeq
where $K$ is the K\"ahler potential, $q_A^X$ is the $U(1)_X$\ charge
of the (complex) scalar matter field $\phi^A$, $\xi$ is the FI term,
$Q_X$ is the charge generator of $U(1)_X$, $g_{\STR}$ is the gauge
coupling at the string scale, and $m_{\PL}=1/\sqrt{8\pi G} = 2.44
\times 10^{18} \GeV$ is the reduced Planck mass that we set to unity
in the remainder of this section. Provided there are $D$-flat and
$F$-flat directions, which is also generically the case, some number
$n$ of fields $\phi^A$ acquire \vevs that break some number $m\le n$
of gauge symmetries. The corresponding gauge supermultiplets get
masses through the supersymmetric generalization of the Higgs
mechanism, but local supersymmetry remains unbroken.

In renormalizable globally supersymmetric gauge theories with a
spontaneously broken gauge theory, the associated vector multiplet
``eats'' a chiral and antichiral supermultiplet to form a massive
vector supermultiplet.  When these are integrated out, the effective
theory below the gauge symmetry breaking scale retains manifest
global supersymmetry.  In the case of supergravity, we can maintain
manifest local supersymmetry by promoting\cite{Gaillard:2002mj} the
condition
\beq D_X = 0\label{dxzero} \eeq
to a superfield condition of which~\myref{dxzero} is the lowest
component.  In effective supergravity from string theory, the chiral
supermultiplets that get \vevs have modular weights, while the
vector fields are modular invariant.  On the other hand the
condition~\myref{dxzero} does not break T-duality; it fixes the
\vevs of a linear combination of the invariant fields $\sum_A
K_A\phi^A$. Maintaining manifest T-duality in the effective theory
below the $U(1)_X$\ breaking scale requires a
generalization\cite{Gaillard:2002mj} of the usual gauge
transformation used to remove the unphysical chiral superfields to
go to unitary gauge.  Finally, in supergravity from string theory,
the parameter $\xi$ in~\myref{eq1} is not a constant; in the
classical limit $g^2_{\STR} = \lang 1/\re\, s \rang$ should be
replaced by the dynamical variable $1/\re\, s$.  In other words, the
scalar components of the eaten chiral multiplets acquire \vevs that
are functions of the dilaton and the K\"ahler moduli which
themselves remain massless and are not stabilized at the scale of
$U(1)_X$\ breaking.  When the massive modes are integrated out in a
way that preserves manifest local supergravity and T-duality, the
K\"ahler potential of the light matter fields that are charged under
the broken gauge symmetries is modified. In particular their modular
weights are shifted in a manner which depends on their charges under
the broken gauge group. In addition the GS term~(\ref{GSlinear})
that cancels the modular anomaly is modified in a parallel manner
such that modular anomaly cancelation is also manifest in the
effective theory below the gauge symmetry breaking scale. The
modified modular weights provide interesting possibilities for
generating a source for the R-parity of the MSSM -- or an even more
restrictive discrete symmetry -- that will be described in
Section~\ref{sec5}.

When supersymmetry is broken by condensation in a strongly coupled
hidden gauge sector with group $\mathcal{G}_c$, dilaton
stabilization is assured by the presence of $D$-terms -- in contrast
to the models without an anomalous $U(1)$ considered in
Section~\ref{sec3} -- but corrections to the dilaton K\"ahler
potential are still needed to stabilize the dilaton at weak
coupling.  Several promising features of the previous models
persist: enhancement of the dilaton and T-moduli masses relative to
the gravitino mass, masslessness of the universal axion, and
dilaton-mediated supersymmetry breaking that avoids potential
problems with flavor changing neutral currents. However there are
some challenges to be overcome in finding a viable scenario with an
anomalous $U(1)_X$.

There is generally a large degeneracy of the vacuum associated with
$U(1)$\ breaking,\cite{Buccella:1982nx,Gaillard:2002mj} resulting in
many massless chiral multiplets, or ``D-moduli,'' between the
$U(1)$-breaking and supersymmetry-breaking scales.  Moreover, in the
absence of superpotential couplings a number of these remain
massless even after supersymmetry breaking.  It turns out that
couplings of the D-moduli to the matter condensates in the
superpotential are sufficient to lift the degeneracy and give masses
to the {\em real} parts of the D-moduli scalars as well as the
fermions, while the imaginary parts of the scalars (``D-axions'')
remain massless in the absence of other superpotential couplings.
This remaining degeneracy may be at least partially lifted by
D-moduli couplings to other unconfined, $\mathcal{G}_c$-neutral
chiral supermultiplets.

As has been noted by a number of
authors,\cite{Kawamura:1996wn,Binetruy:1996bd,Dvali:1996rj,Barreiro:1998nd,Gaillard:2003gt}
that there is considerable tension in maintaining a vanishing
cosmological constant, a positive dilaton metric and positive and
acceptably small scalar masses in the observable sector on the one
hand, while requiring weak coupling and acceptably large
D-moduli/fermion masses on the other hand.

In Section~\ref{toy1} we illustrate the procedure for integrating
out the massive vector supermultiplet while preserving manifest
local supersymmetry and T-duality in the lower energy effective
theory using a toy model\cite{Gaillard:2002mj} with just one broken
gauge symmetry, $U(1)_X$, and just one complex scalar $vev$; in
Section~\ref{toy2} we include hidden sector gaugino condensation in
this toy model.  Since the linear multiplet formalism for the
dilaton is by far better suited to addressing these questions, we
will use it in these two sections, and impose further that the
modified linearity condition~(\ref{modlin}) remain true in the
effective theory below the $U(1)_X$\ breaking scale. The cases with
any number $m$ of broken $U(1)$'s and $n\ge m$ scalar \vevs have
been worked out in detail;\cite{Gaillard:2002ew,Gaillard:2003gt}
here we simply state the results.  In Section~\ref{vacmod} we
discuss the vacuum and the moduli sector, and in
Sections~\ref{obsmass} and~\ref{dmod} we address observable sector
and D-moduli masses, respectively, and discuss the requirements for
a viable model.

\subsection{The effective theory below the $U(1)_X$ breaking scale: a toy
model}\label{toy1}

When an anomalous $U(1)_X$ is present, the effective Lagrangian at
the string scale is defined by
\beq \Lag = \Lag_{\rm KE} + \Lag_{\rm GS} + \Lag_{\rm th} +
\sum_a\Lag_a, \label{str}\eeq
where $\Lag_{\rm th}$ is the string loop threshold correction given
in~(\ref{Lthresh}).  It is convenient here to write the kinetic
term~\myref{lke} in the form
\beq \Lag_{\rm KE} = \superint\,E\[- 3 + Ls(L)\],\label{newlkl}\eeq
such that the first term contains the usual gravity and matter
kinetic terms of K\"ahler $U(1)$ superspace, and the second term,
with  $2 L s(L) = 1 + f(L)$, includes the gauge kinetic term of the
more conventional supergravity formulation, with the vacuum value
\beq\lang s(\ell)\rang = g_{\STR}^{-2}\label{stg}\eeq
determining the gauge coupling $g_{\STR}$ at the string
scale.\footnote{In the dual chiral formulation for the dilaton
$2s(L)\to S + \bar S - bG + \vx\delta_x$.}  The GS term
(\ref{GSlinear}) now includes the FI term:
\beq \Lag_{\GS} = \superint\,E L\(b\, G - \delta_X \vx \),
\label{oit} \eeq
where $V_X$ is the $U(1)_X$\ vector superfield, $G$ is defined
by\footnote{In this section, for simplicity, we set the parameters
$p_A$ introduced as $p_i$'s in (\ref{GS}) to zero, as was done in
Ref.~\refcite{Gaillard:2003gt}.  We comment on the case with
$p^A\ne0$ in Section~\ref{obsmass}.  These parameters should not be
confused with the similarly labeled parameters introduced
in~(\ref{PVmodsraw}) above, nor those in~(\ref{pdefs6}) below.}
\beq G = \sum_I g^I, \qquad g^I = -\ln(T^I + \oline{T}^I) \, .
\label{gId} \eeq
The two parameters in~(\ref{oit}) are the coefficient $b$,
identified with $b_{\GS}$ of equation~(\ref{GSlinear}), and
\beq \delta_X = - {1\over48\pi^2}\Tr \, Q_X = -{1\over2\pi^2}\sum_A
C^A_{a\ne X}\, q_A^X = - {1\over6\pi^2}\Tr \, Q^3_X,\label{dxnorm}
\eeq
where the last two equalities are constraints on the gauge charges
of the spectrum that follow from the fact that the underlying string
theory is anomaly free.  The field theoretic quantum corrections
$\Lag_a$ take the form
\bea \Lag_a &=& - \int d^4\theta {E\over 8R} \, W_a^\alpha P_\chi
B_a W_\alpha^a + \hc \, ,
\label{lqa} \\
B_a(L,V_X,g^I) &=& \sum_I (b-b^I_a) g^I - \delta_X\vx + f_a(L) \, .
\label{lqb} \eea
The full K\"ahler potential is
\bea K &=& k(L) + G + \sum_A e^{G_A + 2 q_A^X V_X}|\Phi_A|^2 +
\order(|\Phi_A|^3) \, , \nnn k(L) &=& \ln L + g(L), \qquad G_A
\equiv \sum_I q_A^I \, g^I \, , \label{kpt} \eea
where $q_A^X$ and $q_A^I$ are $U(1)_X$\ charges and modular weights,
respectively.  Up until now we have been working in K\"ahler $U(1)$\
and Yang-Mills superspace, in which the Yang-Mills vector fields do
not appear explicitly and chiral superfields are {\em covariantly}
chiral, with the gauge connections implicit in all covariant
derivatives. When there is an anomalous $U(1)$\ the vector field
appears explicitly in the GS term needed to cancel the anomaly. The
vector field must also be introduced explicitly for the
supersymmetric regularization\cite{Gaillard:1998bf} of ultraviolet
divergences associated with an anomalous $U(1)$, which accounts for
its appearance in the expression~\myref{lqb}.  When a gauge symmetry
is broken, and the associated vector multiplet acquires a large
mass, it is appropriate to remove its gauge connection from the
covariant derivatives. In this case the vector field appears
explicitly in the K\"ahler potential as in~\myref{kpt}, and the
heavy degrees of freedom can easily be integrated out at the
superfield level, as we outline below.

Suppose that a chiral multiplet $\Phi$ with $U(1)_X$\ charge $q$ and
modular weight $q^I$ acquires a $vev$.  Setting all other matter
fields to zero, the K\"ahler potential~\myref{kpt} reduces to
\beq K = k(L) + G + e^{G_q + 2 q V_X}|\Phi|^2, \qquad G_q = \sum_I
q^I g^I, \label{simplek} \eeq
and (in Wess-Zumino gauge) the D-term~\myref{eq1} is given by
\beq D_X = - {2\over s}\(q e^{G_q (t)}|\phi|^2 - {\ell\delta_X
\over2} \),\eeq
which vanishes at the minimum of the potential if $\phi\ne 0$ is a
flat direction.  Since the scalar component $\phi$ of $\Phi$ must
get a \vev to cancel the FI term, it is consistent to write $\phi =
e^\theta$, where $\theta$ is a complex scalar field.  Promoting this
to a superfield expression, we define a chiral superfield $\Theta$
such that
\beq \Phi = e^\Theta \, . \eeq
The Lagrangian is invariant under $U(1)_X$\ gauge transformations
\beq \vx \to  \vx' = \vx + \half\(\Lambda + \oline{\Lambda}\),
\qquad \Phi^A \to \Phi'^A = e^{-q_A^X \Lambda}\Phi^A \, , \eeq
where $\Lambda$ is a chiral supermultiplet, so
we can remove the chiral supermultiplet $\Theta$ by a gauge
transformation with $\Lambda = \Theta/q$:
\beq V_X \to V' = V_X + {1\over2q}\(\Theta + \oline{\Theta} \),
\qquad \Phi \to \Phi' = e^{-\Theta}\Phi = 1 \, , \label{gaugetr}
\eeq
The field $V'$ describes a massive vector multiplet with the same
number of components as a massless vector multiplet and a (complex)
chiral multiplet, and gauge invariance assures that
\beq \Lag\(V_X,\Phi,\oline{\Phi}\) \to \Lag\(V',1,1\) \, . \eeq
The expressions for $K$ and $\Lag$ become
\bea
K & = & k(L) + G + e^{G_q + 2q V'},  \label{5_2} \\
\Lag & = &\superint E \[ -3 + 2Ls(L) + L ( b\,G - \delta_X V' ) \] +
\Lag_{\rm th} + \sum_a \Lag_a. \label{5_1} \eea

But note the important property that $\Theta$ -- and therefore $V'$
-- is not invariant under modular transformations defined
by~(\ref{mod}) and~(\ref{mattertrans}):
\beq V'\to V' - {1\over2q}\sum_I q^I \(F^I + \oline{F}^I\)\; ; \quad
\quad F^I = \ln(icT^I + d) \, .\eeq
In order to integrate out the heavy degrees of freedom in a way that
preserves the unbroken symmetries, therefore, we promote the vacuum
condition~\myref{dxzero} to a superfield condition. That is, we
require that the {\em superfield}
\beq q\, e^{G_q(T) + 2q V'} - {\delta_X\over2}L \, , \eeq
which {\em is} modular invariant, vanish in the vacuum. To this end
we introduce a vector superfield $U$ with vanishing \vev ({\em i.e.}
all component fields are defined to vanish in the vacuum) such that
this condition is maintained
\beq q\, e^{G_q + 2 q V'} =  e^{2qU} \, {\delta_X \over 2} L \, ,
\qquad \lang U \rang =0 \, . \eeq
This corresponds to the redefinition
\beq V' = U + {1 \over 2q} \( \ln {\delta_X L \over 2q} - G_q \) .
\label{aad} \eeq
In terms of the new set of independent fields $(L,\,U,\,T^I)$, the
K\"ahler potential and the Lagrangian take the form
\bea K & = & k(L) + G + e^{2q U} \, {\delta_X L \over 2q} \equiv
K(L,U) + G\, ,  \label{5_16} \\
\Lag_{\rm KE} + \Lag_{\rm GS} & = & \superint E \[ -3 + 2Ls(L) + L
\( bG - \delta_X U - {\delta_X \over 2q} \ln {\delta_X L \over 2q}
+ {\delta_X \over 2q} G_q \) \]\nonumber\\
& \equiv & \superint E \[ -3 + 2L \, S(L,U) + L\sum_I b^I g^I\]\, ,
\label{5_17} \eea
where we have introduced the new coefficient
\begin{equation} b^I = b + {\delta_X \over2q} q^I \, . \label{bIdef}
\end{equation}

When the superfield $U$ is set to zero, the net effect is a
modification of the functions $k(L)$ and $s(L)$ which are subject to
the constraint~\myref{lcond}, or, equivalently
\beq k'(L) + 2Ls'(L) = 0\, . \label{canon} \eeq
These are now replaced by the functions
\beq \tilde{k}(L) = K(L,0) = k(L) + {\delta_X\over2q}\, L\, , \qquad
\tilde{s} = S(L,0) = s(L) - {\delta_X\over4q}\ln\({\delta_X
L\over2q}\)\, ,\label{tkts}\eeq
that satisfy
\beq \tilde{k}'(L) + 2L\tilde{s}'(L) = k'(L) + 2Ls'(L) = 0\, ,
\label{canoneh}\eeq
and the Einstein-Hilbert term remains canonically normalized. This
is not the case, however, for $U\ne0$.  Moreover, there are terms
linear in $U$ that contribute to the tree-level action that must be
taken into account.  This can be done while maintaining $\lang U
\rang =0$ by a redefinition of $L$, but an arbitrary redefinition
would destroy the modified linearity condition~(\ref{modlin}) that
we wish to maintain below the $U(1)_X$\ breaking scale.  This can be
achieved by a superfield Weyl transformation\cite{Binetruy:2000zx}
that leaves the product $E\, L$ invariant but modifies the K\"ahler
potential:
\bea K(L,U) &=& \hK + \Delta k, \quad E = e^{-\Delta k/3} \wh{E},
\quad L = e^{\Delta k/3}\hL\, , \nnn 2\hL\, \hS(\hL,U) &=& 2 \hL\,
S(L,U) + 3\(1 - e^{-\Delta k/3}\)\, .\label{wyl}\eea
The condition for a canonical Einstein term in the new Weyl basis is
\beq 0 = \( {\pp \hK \over \pp \hL}\)_{U} + 2 \hL \( {\pp \hS \over
\pp \hL}\)_{U},\label{wcond}\eeq
where the subscript on derivatives indicates that $U$ is held fixed.
The $\hL$ derivatives of $\Delta k$ drop out except in an overall
factor $\pp L/\pp\hL$, giving the condition
\bea L - \hL &=& \hL\(e^{\Delta k/3} - 1\) = L^2{K'(L) + 2L\,
S'(L)\over 3 + 2L^2\, S'(L)}\nonumber \\ &=& L^2{\delta_X\(e^{2q U}
- 1\)\over2q\[ 3 - L\tilde{k}'(L)\]}
= {\delta_X L^2 \, U\over3 - L\,\tilde{k}'(L)} + \order(U^2) \nonumber \\
&=& {\delta_X \hL^2\, U \over3 - \hL\,\tilde{k}'(\hL)} +
\order(U^2)\, .\label{wyl2}\eea
We may then expand the expressions
\bea \hK(\hL) &=& \tilde{k}(L) + {\delta_X\over2q} L\(e^{2q U} - 1\)
- \Delta k,\nonumber \\ \hS(\hL) &=& \tilde{s}(L) + {3\over2L\hL}\(L
- \hL\) - {\delta_X\over2}U \, ,\label{newks}\eea
in powers of $U$. This gives
\bea \Delta k &=& 3(L - \hL)/\hL + \order(U^2),\qquad \tilde{k}(L) =
\tilde{k}(\hL) + (L - \hL)\tilde{k}'(\hL) + \order(U^2), \nonumber\\
\tilde{s}(L) &=& \tilde{s}(\hL) + (L - \hL)\tilde{s}'(\hL) +
\order(U^2) \, ,\eea
and, using~\myref{canoneh}, the terms linear in $U$ drop out
of~\myref{newks}. This means that we can just set $U= 0$ to get the
effective low energy theory in the hatted basis that has a canonical
Einstein-Hilbert term.\footnote{The terms quadratic in the $U(1)_X$\
field strengths $\Wc_V = \Wc_{V'}$ generate terms linear in $\Wc_U$;
these involve couplings to the chiral projections of spinorial
derivatives of the moduli.\cite{Gaillard:2000na} They do not
contribute to the potential as long as supersymmetry is unbroken;
see \myref{qw4} below.}

Once the heavy modes have been integrated out, the Lagrangian takes
the form~\myref{str} with
\bea \Lag_{\rm KE} &=& \superint\,E\[- 3 + L\, \tilde{s}(L)\]\, ,\label{newlkl2}\\
\Lag_{\rm GS} &=& \superint\,E L\sum_I b^I g^I\, ,\label{oit2}\eea
and the operator $B_a$ in~\myref{lqb} becomes
\beq B_a(L,\,V_X,\,g^I) = \sum_I (b^I - b^I_a) g^I +
\tilde{f}_a(L)\, , \quad\tilde{f}_a(L) = f_a(L) -
{1\over2q}\ln{\delta_X L\over2q}\, . \label{lqb2}\eeq
The shift in $f_a$ exactly cancels the effect of the
shift~\myref{tkts} in $s$ on the kinetic terms for the Yang-Mills
fields of the unbroken gauge group, so the coupling is the same as
the string scale coupling \myref{stg}, but the dilaton kinetic
energy term is modified below the $U(1)$\ breaking scale. The shift
$b\to b^I$ matches that in $\Lag_{\rm GS}$ so that the anomaly is
still canceled.  The shift in $\Lag_a$ precisely corresponds to the
shift in the metric for matter fields that occurs when the change in
variables~\myref{aad} is made.  The K\"ahler potential is now
\bea K &=& \tilde{k}(L) + G + \sum_A e^{G'_A}\({\delta_X L\over 2
q}\)^{q_A^X/q}|\Phi_A'|^2, \nonumber\\ G'_A &=& \sum_I\(q_A^I -
\frac{q_A^X}{q}q^I\)g^I \equiv \sum_I q'^I_A g^I\, , \label{kpt2}
\eea
where it is convenient to define the new ``effective'' modular
weight
\begin{equation} q'^I_A = q_A^I - \frac{q_A^X}{q}q^I \, .
\label{qprimedef} \end{equation}
The modified K\"ahler metric implies a modification of the fermion
connections that induce a shift in the functions $B_a$;
using~\myref{dxnorm},
\beq \delta B_a = {1\over4\pi^2}\sum_A C^A_a\, {q_A^X\over
q}\(\ln{\delta_X\over2q} - \sum_I q^I g^I\) =
{\delta_X\over2q}\(\ln{\delta_X\over2q} - \sum_I q^I  g^I \)\, ,
\label{delB}\eeq
giving the correction to~\myref{lqb} that appears in~\myref{lqb2}.
These modifications of the effective theory can have consequences
for phenomenology.

Finally, the right hand side of the modified linearity condition
\myref{modlin} contains the term $(W^{\alpha}W_{\alpha})_V$ where
\bea  \Wc_V & = & - {1\over4} \chiproj \D^\alpha V
=  - {1\over4} \chiproj \D^\alpha V' \nonumber\\ & = &
 - {1\over4} \chiproj \D^\alpha U - {1 \over 8q} \chiproj
\( {\D^\alpha L \over L} - \D^\alpha G_q \) \nonumber \\ &=&
\Wc_U  - {1 \over 8q} \chiproj
\( {\D^\alpha L \over L} - \D^\alpha G_q \).
\label{qw3}\eea
In the hatted basis defined by the Weyl transformation \myref{wyl},
\myref{qw3} reads
\bea  \Wc_V & = &\(1+{1 \over 2qL(\hL,U)}
{\pp L(\hL,U)\over\pp U}\)\Wc_U + W'^\alpha,\label{qw2}\\
\superint{E\over R}(W^{\alpha}W_{\alpha})_V &=&
\superint{E\over R}(W^{\alpha}W_{\alpha})_U
\[\(1+{\delta_X\hL\over 2q(3 - \hL\tilde k'(\hL)}\)^2 + O(U)\]
\nonumber \\ &&\qquad
+ \cdots,\label{qw4}\eea
where the ellipsis represents terms involving $W'^\alpha$ that are
quartic in auxiliary fields and/or derivatives. This implies a
renormalization of the $U$ wave function such that its mass is given
by\cite{Gaillard:2002mj}
\bea m^2_U &=&
\l{1\over s(\hel)}\(1+{\delta_X\hL\over 2q(3 - \hL\tilde k'(\hL)}\)^{-2}
{\pp^2\tilde K(\hL,U)\over\pp U^2}\right|_{U = 0,\hL =
 \langle\hel\rangle} \nonumber \\ &=&
 \(1+{\delta_X\langle\hel\rangle\over 2q(3 - \langle\hel\rangle
\tilde k'(\langle\hel\rangle)}\)^{-1}{q\delta_X\langle\hel\rangle\over
s\langle\hel\rangle},\label{umass}\eea
where
\beq \tilde{K} = \hK + 2\hL\hS\label{deftk}\eeq
is the ``effective'' K\"ahler potential.

\subsection{The effective theory below the condensation scale: a toy
model}\label{toy2}

Here we consider the case where the toy model described in
Section~\ref{toy1} has a strongly-coupled gauge group
$\mathcal{G}_c$ in a hidden sector.  The effective Lagrangian for
the condensates~\myref{Ucondef} and~(\ref{Picondef}) can be
constructed by anomaly matching as described in Section~\ref{sec2}.
The strongly coupled Yang-Mills sector also possesses a residual
global $U(1)_X$ invariance that is broken only by superpotential
couplings that involve the chiral superfield $\Phi$ that gets a \vev
at the $U(1)_X$-breaking scale.\footnote{Some of these coupling
could generate masses for a subset of fields that are charged under
$\mathcal{G}_c$; we will comment on that case at the end of this
section.}  These couplings enter the RGE for the $\mathcal{G}_c$
gauge coupling only through chiral field wave function
renormalization, which is a two-loop effect that is encoded in the
expression~(\ref{rhosq}) for the gaugino condensate through the
appearance of the superpotential coefficients $W_\alpha$ as
discussed in Section~\ref{sec23}.  We therefore impose the $U(1)_X$\
anomaly matching condition
\beq \sum_{\alpha,B} b^\alpha_c \, n^B_\alpha \, q_B^a =
\sum_B{C^B_c\over4\pi^2}\, q_B^a = -\half\delta_X\delta^a_{X}\,
,\label{uacond}\eeq
which is also satisfied by~\myref{cond}. Note that in the last
expression of~(\ref{uacond}) the quantity $\delta_X$ is the GS
coefficient from~(\ref{oit}) while $\delta_{X}^a$ is a Kronecker
delta function which enforces $U(1)_a = U(1)_X$. The condensate
superpotential now takes the form
\beq W(\Pi) = \sum_\alpha W_\alpha(T^I,\Phi)\, \Pi^\alpha \, ,
\label{pipot}\eeq
where T-duality and $U(1)_X$\ invariance require
\beq W_\alpha(T^I,\Phi) = c_\alpha\prod_I [\, \eta(t^I)]^{2\(q'^I_{\alpha} -
1\)} \phi^{-q^X_\alpha/q} \, , \label{pdefs} \eeq
with $q'^I_{\alpha}$ defined as in~(\ref{qprimedef}) for the matter
condensate $\Pi^{\alpha}$.

When supersymmetry is broken by condensation, we can no longer
assume {\it a priori} that $\lang D_X \rang =0$, and an $F$-term
associated with the chiral field that gets a \vev may also be
generated. To include these we modify the field
redefinitions~\myref{gaugetr} and~\myref{aad} as follows:
\beq \Phi' = e^{-\Theta + \Delta_\Phi}\Phi = e^{\Delta_\Phi}, \qquad
V' = U + {1 \over 2q} \( \ln {\delta_X L \over 2q} - G_q \) +
\Delta_X, \label{Vshift}\eeq
with $V'$ defined as in~\myref{gaugetr}. The objects $\Delta_X$ and
$\Delta_\Phi$ are constant vector and chiral superfields,
respectively, with vanishing fermionic components. We can make a
$U(1)_X$\ gauge transformation with a constant chiral superfield
$\Lambda$ to eliminate the scalar and auxiliary components from
$\Delta_X$, which just redefines the corresponding components of
$\Delta_\Phi$:
\bea 0 &=& \l\Delta_X\r = -{1\over4}\l D^2\Delta_X\r,
\qquad D_X = {1\over8}\l\Dc\(\oline{D}^2 - 8R\)\Da\Delta_X\r, \nonumber\\
\delta &=&\l\Delta_\Phi\r, \quad F = -{1\over4}\l D^2\Delta_\Phi\r,
\label{delaux}\eea
and we take
\beq \lang U\rang = 0\eeq
as before. However we cannot set $U=0$ before taking into account
linear couplings of $U$ that arise in the presence of the
supersymmetry-breaking \vevs \myref{delaux}, the matter fields
$\Phi^M$ and the now nonvanishing auxiliary fields represented by
the ellipsis in~\myref{qw4}. As before, we go to the basis where the
Einstein-Hilbert term is canonical, starting with the Lagrangian
defined by
\bea K & = & k(L) + G + e^{2q(\Delta_X + U) + \Delta_\Phi +
\oline{\Delta}_\Phi}{\delta_X L
\over 2q} + \sum_A e^{G'_A}\({\delta_X L\over 2
q}\)^{q_A^X/q}|\Phi_A'|^2\nonumber\\&\equiv& K(L,M,U) + G, \label{kdel} \\
\Lag_{\rm KE} + \Lag_{\rm GS} & = & \superint E \[ -3 + 2Ls(L) + L
\( b\,G - \delta_X(\Delta_X + U) - {\delta_X \over 2q} \ln {\delta_X L
\over 2q}
+ {\delta_X \over 2q} G_q \) \]\nonumber\\
& \equiv & \superint E \[ -3 + 2L\, S(L,M,U) + L\sum_I b^I g^I\],
\label{ldel} \eea
where now
\bea K(L,M,U) &=& \tilde{k}(L) + {\delta_X L\over2q}\(e^{2q(\Delta_X + U)
+ \Delta_\Phi + \oline{\Delta}_\Phi} - 1\) + \sum_A x^A,
\qquad  M=\Delta,\Phi,T,\nonumber \\ S(L,M,U) &=&
\tilde{s}(L) - \delta_X(U + \Delta_X)\qquad
x^A =  e^{G'_A}\({\delta_X L\over 2
q}\)^{q_A^X/q}|\Phi_A'|^2\, .\label{defKS}\eea
We make the superfield Weyl transformation~\myref{wyl} and impose the
condition~\myref{wcond} with $U\to U,M$ held fixed
in~\myref{wcond} to obtain
\beq L - \hL = L^2\lbr {\delta_X\(e^{2q(\Delta_X + U) + \Delta_\Phi +
\oline{\Delta}_\Phi} - 1\)
+ 2q\sum_A x'^A\over2q\[ 3 - L\,\tilde{k}'(L)\]}\rbr \,
.\label{LhL}\eeq
The functions $\hK(\hL,M,U)$ and $\hS(\hL,M,U)$ can be expanded as
before to obtain the component Lagrangian which now includes $F$-terms
for $\Phi,\Phi^M$ and the $U(1)_X$\ $D$-term.  Here we will just
describe the modifications with respect to the BGW case of
Section~\ref{sec3}.

The couplings of $\Phi$ in the condensate
superpotential~\myref{pipot} introduce additional parameters in the
potential.  We define
\beq p^I = - {q^I\over q}\sum_\alpha b_c^\alpha q^X_{\alpha} \equiv
p\, q^I \qquad p = - {1\over q}\sum_\alpha b_c^{\alpha} q^X_\alpha\,
. \label{pdefs2}\eeq
In this toy model it follows from the $U(1)_X$\ anomaly matching
condition~\myref{uacond} that
\beq p^I = b^I - b,\qquad p  = {\delta_X\over2q}\, .
\label{pdefs3}\eeq
Neglecting corrections of order $\Delta$, the \vev of $|\phi|^2$ is
given by
\beq \lang |\phi|^2 \rang = e^{-\sum_I q^I g^I -
2q\vx}{\delta_X\ell\over2q} = p\, \ell\, e^{-\sum_I p^I g^I -
2q\vx}>0\, ,\eeq
so $p$ and $p^I$ are positive since the modular weights, as defined
here, are generally positive.\footnote{A rare exception to this
statement can be found in Ref.~\refcite{Ibanez:1992hc}.}  Once the
matter condensates have been eliminated by the equations of motion for
their auxiliary fields, the equation of motion for the condensate
auxiliary field $F^c$ determines $|u_c|^2$; in particular there is a
factor
\beq \prod_\alpha|W_\alpha|^{2b^\alpha_c/b_c}\, .\eeq
The relation~\myref{pdefs3} holds for the $p^I$ defined
in~\myref{pdefs2} in terms of the original modular weights at the
string scale, in which case
\bea \sum_\alpha b_c^\alpha q^I_\alpha &=& \sum_B b_c^\alpha
n^B_\alpha q^I_B = \sum_B {C_c^B\over4\pi^2}q^I_B = b - b'_c - b^I_c
, \quad \sum_\alpha b_c^\alpha = b_c - b'_c, \label{qconds}\eea
giving
\beq \sum_\alpha b_c^\alpha\(q'^I_\alpha  - 1\) = b^I - b_c -
b_c^I\, ,\label{bqconds}\eeq
which is just what one would have gotten using the modular weights
obtained in unitary gauge with $q'=0$.
Now setting $\phi\to\phi'=1$ in unitary gauge and proceeding as in
Section~\ref{sec3}, we obtain\footnote{There are some errors in
(3.33)--(3.39) of Ref.~\refcite{Gaillard:2003gt}.}
\beq \bar u_c u_c = e^{-2b'_c/b_c}e^{\kappa -2\hS/b_c}
\prod_\alpha\left|\frac{b_c^\alpha}{4c_\alpha}\right|^{-2b_c^\alpha/b_c}
\prod_I\[\,2\re\, t^I|\eta(t^I)|^4\]^{(b^I - b_c)/b_c} +
\order(\delta), \label{solu}\eeq
where $\kappa = \hK - G$ is the modular invariant part of the
K\"ahler potential in the new basis. The K\"ahler moduli $F$-terms
are now given by
\bea F^I &=& - {2\re \,t^I\over1 + b^I\hel}{\bar u\over4}\(b_c - b^I
\)\[1 + 4\re \, t^I\bar{\zeta(t^I)}\].\label{solt}\eea
Note that the exponential factor in~\myref{solu} contains $\hS =
\tilde{s} + \order(\Delta)$.  When re-expressed in terms of the
actual string coupling that expression acquires a factor
\beq e^{2(s - \tilde{s})/b_c} = (p\,\ell)^{p/b_c}\,
,\label{ufactor}\eeq
which, in the weak coupling limit with $p\,\ell<<1$, amounts to a
decrease in the scale of supersymmetry breaking, allowing for
somewhat larger values of $b_c$.

To incorporate the effects of terms linear in $U$ we need to retain
terms quadratic in the components of $U$ in the Lagrangian.
Referring to~\myref{qw2} we have
\bea D_V &=& -{1\over2}\D^\alpha W_\alpha^V\blowest =
-\half\D^\alpha\[\(1+{1 \over 2qL(\hL,U,M)}{\pp L(\hL,U,M)\over\pp U}\)
W_\alpha^U + W'_\alpha\]\blowest \nonumber \\ &=&
\[1 + {1\over2q\ell(\hel,u,m)}{\pp\ell(\hel,u.m)\over\pp u}\]D +
D',\qquad u = U\lowest\qquad m = M\lowest,\eea
where the bosonic, nonderivative part of $D'$ is quadratic in
auxiliary fields:
\beq D' \sim |u_c|^2,\eeq
since the \vev $u_c$ of the hidden sector gaugino condensate sets the
scale of supersymmetry breaking.
The part of the Lagrangian that contains the auxiliary fields $D$
is
\beq \Lag(D) = {s(\hel)\over2}D_V^2 + D{\pp\wtd{K}\over\pp D} = -
V(D) \, .\eeq
Eliminating $D$ by its equation of motion gives
\bea V(D) &=& {1\over2s(\hel)}\[A^2(\hel,u,m) + 2s(\hel)A(\hel,u,m)D'\]
 \nonumber \\ &=& {1\over2s(\hel)}\[A(\hel,u,m) + s(\hel)D'\]^2 -
{s(\hel)\over2}D'^2,\label{LD}\\
A(\hel,u,m) &=& \(1 + {1\over2q\ell(\hel,u,m)}
{\pp\ell(\hel,u,m)\over\pp u}\)^{-1}{\pp\tilde K\over\pp D}.
\eea
The last term in~\myref{LD} is $\order(|u_c|^4)$ and we may ignore
it. Expanding
\beq A(\hel,u,m) + s(\hel)D' = A_0(\hel,m) + u A_1(\hel,m) +
u^2A_2(\hel,m) \, , \eeq
we note that $A_0$ is the $u$-independent part of $A + s D' = A +
\order(|u_c|^2)$. But $\wtd{K}$ has no term linear in $U$ when we
neglect $\epsilon\equiv  \delta,x^m$. So $\pp A/\pp u\sim\epsilon,u$
and $A_0\sim\epsilon,|u_c|^2$, giving
\bea V(D) &=& {1\over2s(\hel)}\[A_0 + A_1 u + A_2u^2\]^2 +
\order(|u_c|^4)
 \nonumber \\ &=& {1\over2s(\hel)}\(A_0^2 + 2A_0A_1 u +  [A^2_1 + O(\epsilon,|u_c^2|)]u^2\)
+ \order(|u_c|^4)\, .\label{VD}\eea
When order $\epsilon$ terms are included, in addition to the
$\epsilon$-dependence of $\kappa= \hK - G$ and $\hS$, the
expression~\myref{solu} contains a factor
\beq e^{-(\delta + \bar\delta)\sum_\alpha b^\alpha_c q^X_\alpha/q \,
b_c} = e^{(\delta + \bar\delta) p/b_c},\eeq
and the $F$-component of $\Phi'$ is given by\cite{Gaillard:2003gt}
\bea F &=& - {\bar
u\over4}\({\pp\tilde{K}\over\pp\delta\pp\bar\delta}\)^{-1}
\[2{\pp\hS\over\pp\bar\delta} - p - b_c{\pp\hK\over\pp\bar\delta}\],
\label{solfa}\eea
where the derivatives are taken with $\hel$ held fixed.

In this toy model, $F$ is of order $\delta$. To see this, write
\beq 2{\pp\hS\over\pp\bar\delta} - p - b_c{\pp\hK\over\pp\bar\delta}
= 2\(1 + b_c\hel\){\pp\hS\over\pp\bar\delta} - p -
b_c{\pp\tilde{K}\over\pp\bar\delta}\,\label{odelterm}\eeq
Taking the lowest component of the relation
\beq \({\pp \wtd{K}\over\pp\Delta}\)_{\hL} = \({\pp
K\over\pp\Delta}\)_L + 2\hL\({\pp
S\over\pp\Delta}\)_L,\label{tkderiv}\eeq
that follows\cite{Gaillard:2003gt} from~\myref{wcond} with
$U\to\Delta$, gives
\beq \({\pp\wtd{K}\over\pp\bar\delta}\)_{\hel} =
\({\pp{K}\over\pp\bar\delta}\)_{\ell} = p\,\ell\[1 +
\order(\delta,u)\] = p\hel + \order(\epsilon,u)\label{dtkdd}\, .
\eeq
The derivatives of the lowest component of~\myref{LhL}
satisfy\cite{Gaillard:2003gt}
\beq {\pp\ell\over\pp\bar\delta} = {\pp\ell\over\pp\hel}
{p\,\ell^2\over3 - \ell\tilde k'(\ell)}\[1 + O(\delta,u)\] =
{p\,\ell^2\over3 - \ell\tilde k'(\ell)} + \order(\epsilon,u)\,
,\label{dl}\eeq
and, referring to~\myref{canoneh} and~\myref{newks},
\beq {\pp\hS\over\pp\bar\delta} =
{\pp\ell\over\pp\bar\delta}\(\tilde{s}'(\ell) + {3\over2\ell^2}\)
 = \frac{1}{2} p\, + \order(\epsilon,u), \eeq
so the expression in~\myref{odelterm} is of order $\epsilon,u$.

The full potential is given by
\begin{equation} V = V(D) + \frac{|u|^2}{16}v(\hel,\epsilon,u) +
\({\pp\tilde{K}\over\pp\delta\pp\bar\delta}\)^{-1}F\oline{F} +
\sum_I{1 + b^I\hel\over(t^I + \bar{t}^I)^2} \oline{F}^I F^I\, ,
\label{pot1}\end{equation}
with
\beq{\pp\tilde{K}\over\pp\delta\pp\bar\delta} = p\,
\(\ell + {\pp\ell\over\pp\bar\delta}\)\[1 + O(\delta,u)\],\eeq
and the function
\beq v(\hel,\epsilon,u) = {\pp_{\hel}\hK\over\hel}(1 + b_c\hel)^2 -
3b_c^2\label{minv}\eeq
is the same as in the BGW case with the replacement $k \to \hK$.  It
follows immediately from~\myref{deftk}, \myref{dtkdd} and~\myref{dl}
that $\pp\hK/\pp\delta$ is of order $\epsilon,u$, that is, $\hK(\hel,m,u) =
\tilde{k}(\hel) + \order(x^M,\delta\epsilon,u\epsilon,u^2)$.
This is just a consequence of the fact that $\Delta$ appears in $\hK$
in a linear combination with $U+\Delta_X$, and as shown in
Section~\ref{toy1} there is no term linear in $U\to U+\Delta_X$ in either
$\hK$ or $\hS$.  $\hS$ contains a term linear in $\Delta$ that is
independent of $\hL$, so all terms linear in $\Delta$ and $U+\Delta_X$
drop out of the condition~\myref{tkderiv}.  Expanding the potential
in powers of $u$, it takes the form
\beq V = V_0(\hel,m) + V_1(\hel,m)u + V_2(\hel,m)u^2 + \order(u^3),
\label{upot}\eeq
where $V_1$ and $V_2$ are dominated by the $D$-term contribution
\myref{VD}:
\beq V_1 = {1\over s(\hel)}A_0A_1 +
\order(\epsilon|u_c|^2)\sim\epsilon, \qquad V_2 =
{1\over2s(\hel)}A^2_1 + \order(\epsilon,|u_c|^2)\sim 1.
\label{delmass}\eeq
When rewritten in terms of the canonically normalized scalar $v$,
which, to leading order, is related to $u$ by the normalization
factor found in \myref{qw4}, the potential \myref{upot} takes the
form
\bea V &=& V_0(\hel,m) + {1\over\sqrt{s(\hel)}}\[A_0 m_U +
\order(\epsilon|u_c|^2)\]v + \half\[m^2_U +
\order(\epsilon,|u_c|^2)\]v^2 + \order(|u_c|^4) \nonumber \\ &=&
V_0(\hel,m) - {1\over2s(\hel)}A^2_0+ \half\[m^2_U
+\order(\epsilon,|u_c|^2)\]v'^2\label{upot2}\\ v' &=& v +
{A_0\over\sqrt{s(\hel)}m_U}\[1 + \order(\epsilon,|u_c|^2)\] +
\order(\epsilon|u_c|^2) + \order(|u_c|^4), \eea
where $m^2_U$ is given in~\myref{umass}.  The potential that results
from integrating out $u$ at tree level is obtained by setting $v'=0$
in \myref{upot2}.  The term proportional to $A_0^2$, which includes
the conventional $D$-term, cancels out and the potential $V(\ell,m)$
and is just given by the last three terms in~\myref{pot1} evaluated
at $u=0$, up to order $|u_c|^4$ corrections. At the vacuum where
$\langle x^M\rangle = \langle F^I\rangle = 0$,
\beq V(\hel,\delta) = {|u_c|^2\over16}\[v(\hel) + v_1(\hel)\delta^2\]
+ \order(|u_c|^4),\eeq
and minimization of the potential with respect to $\delta$ gives
$\delta\sim |u_c|^2$, so the potential for $\hel$ is just the BGW
potential $v(\hel)$ up to order $|u_c|^4$ corrections.

In more realistic models with $n$ scalar \vevs and $m\le n$ broken
$U(1)$'s, the functions introduced in~\myref{defKS} are replaced by
\bea K(L,M,U) &=& \tilde{k}(L) + \sum_A
k^A(L)\(e^{\Delta_{\Theta_A} + \oline{\Delta}_{\oline{\Theta}_{A}} +
2\sum_a q^a_A(U_a + \Delta_a)} - 1\) + \sum_M x^M,
\nonumber \\
S(L,M,U) &=& \tilde{s}(L) - \delta_X\(U_X + \Delta_X\) \, .
\label{defKS2}\eea
The functions
\beq k^A = |C_A|^2 \, e^{2\sum_a q^a_A h_a(L)}\eeq
are the modular invariant \vevs generated at the $U(1)_X$\ breaking
 scale, with $h_a$ the $L$-dependent part of the shifts in the
$U(1)_a$\ vector superfields analogous to~\myref{Vshift}.  The $k^A$
are subject to the constraints
\beq 2q\sum_A q^a_A k^A = \delta^a_{X}\delta_X L\,
,\label{kcond}\eeq
that ensures vanishing $D$-terms and unbroken supersymmetry at that
scale.  It also ensures that the functions
\beq \tilde{k}(L) = k(t) + \sum_A k^A(L),\qquad \tilde{s}(L) = s -
\frac{1}{2}\delta_X h_X, \label{tkts2}\eeq
satisfy the Einstein-Hilbert normalization
condition~\myref{canoneh}. The constant superfields
$\Delta_{\Theta_A}$ and $\Delta_a$ are the potential deviations of
the \vevs from their supersymmetric values.

The superpotential~\myref{pipot} is also modified, with now
\beq W_\alpha(T,\Phi) = c_\alpha\prod_I[\,\eta(T^I)]^{2(q^I_\alpha +
p^I_\alpha - 1)}\prod_A\(\Phi_A\)^{q^A_\alpha} \label{pdefs4} \eeq
where the parameters $q^A_\alpha$ and $p_\alpha^I$ satisfy the
constraints
\beq \sum_A q^a_A q_\alpha^A = - q_\alpha^a, \qquad p^I_\alpha =
\sum_A q^I_A q_\alpha^A\, . \label{pdefs5} \eeq
The anomaly matching constraints in~\myref{qconds} are unchanged,
and~\myref{uacond} and~\myref{bqconds} are replaced by
\beq \sum_\alpha b_c^\alpha q^a_\alpha = \sum_{\alpha,B} b_c^\alpha
n^B_\alpha q^a_B = \sum_B {C_c^B\over4\pi^2}q^a_B =
-{1\over2}\delta_X\delta^a_{X},\label{uacond2} \eeq
and
\beq \sum_\alpha b_c^\alpha\(q^I_\alpha + p^I_\alpha - 1\) = b - p^I
- b_c - b_c^I,\label{bqconds2}\eeq
where
\beq p^I = \sum_\alpha b_c^\alpha p^I_\alpha\, .\label{pidef}\eeq
Evaluating the \vevs of $\phi_A\to\phi'_A$ in unitary gauge,
\myref{solu} and~\myref{solt} become
\beq \bar u u = e^{-2b'_c/b_c}e^{\kappa -2\hS/b_c}
\prod_\alpha\left|\frac{b_c^\alpha}{4c_\alpha}\right|^{-2b_c^\alpha/b_c}
\prod_I\[2\re\, t^I|\eta(t^I)|^4\]^{(b + p^I - b_c)/b_c} +
\order(\delta), \label{solu2}\eeq
and
\bea F^I &=& - {2\re \, t^I\over1 + b^I\hel}{\bar u\over4}\(b_c - b
- p^I \)\[1 + 4\re \, t^I\bar{\zeta(t^I)}\].\label{solt2}\eea
Finally, the $F$-component~\myref{solfa} of $\Phi'$ takes the form
\beq F^A = - {\bar u\over4}\sum_B\widetilde{K}^{A\oline{B}}
\[2\hS_{\bar B} - p_B - \hK_{\oline{B}}b_c\],
\label{solfa2}\eeq
where as usual the indices denote derivatives with respect to chiral
and anti-chiral superfields, $\widetilde{K}^{A\oline{B}}$ is the
inverse of the metric derived from the ``effective'' K\"ahler
potential~\myref{deftk}, and the parameters\footnote{See the second
footnote in~\ref{toy1}.}
\beq p^A = \sum_\alpha b_c^\alpha q^A_\alpha,\label{pdefs6}\eeq
are constrained by the anomaly matching condition~\myref{uacond2}
which implies
\beq 2\sum_A q_A^a p^A = 2\sum_{\alpha,A}b^\alpha_c q_\alpha^A q^a_A
= - 2\sum_\alpha b^\alpha_c q_\alpha^a =
\delta_X\delta_{X}^a.\label{pcond} \eeq
There are two distinct cases:
\begin{itemize}
\item Minimal models: if $n = m$, the matrix $q^a_A$ has an inverse
$Q^A_a$, since if $m$ $U(1)$'s are broken, by definition the $q$'s
form a set of $m$ linearly independent $m$-component vectors.  The
$m$ vector superfields $V_a$ ``eat'' $m$ modular invariant
combinations
\beq -\frac{1}{2}\sum_A Q^A_a\(\Theta_A + \oline{\Theta}_A +
G_A\)\eeq
of chiral superfields, with $\Phi^A = C_Ae^{\Theta_A}$, resulting in
a redefinition of the modular weights of the chiral superfields
$\Phi^M$:
\beq q'^I_M = q^I - \sum_{a,A} q^a_M Q^A_a q_A^I, \label{minmodwt}
\eeq
and a corresponding shift in the coefficient of the GS term:
\beq b^I = b + \frac{1}{2}\delta_X\sum_A Q^A_a
q_A^I.\label{newbi}\eeq
The constraint~\myref{pdefs5} on $q^A_\alpha$ can be inverted:
\beq q^A_\alpha = -\sum_a Q^A_a q^A_\alpha,\qquad p_\alpha^I = -
\sum_{A, a} q_A^I Q^A_a q^a_\alpha \, .\eeq
Using the anomaly matching condition~\myref{uacond2} in the
definition~\myref{pidef} gives
\beq p^I = b^I - b \label{minpi}\eeq
as before, because in unitary gauge $\Phi'^A= C_A$ is again modular
invariant. Contracting the constraints~\myref{kcond}
and~\myref{pcond} with $Q^B_a$ gives
\beq k^B = \ell p^B = \frac{1}{2}\delta_X Q^B_X.\label{minkp}\eeq
Then, proceeding as for the toy model, it is straightforward to
show\footnote{Some terms linear in $\delta$ in the F-term where
inadvertently dropped in Ref.~\refcite{Gaillard:2003gt}} that
$F^A$ and $\pp v/\pp\delta_A$ are of order $\epsilon,u$ as in the toy
model.
\item Nonminimal models:  if $n > m$, the matrix $q^a_A$ is not invertible
and in general $p^I\ne b^I$, because the transformation to unitary
gauge does not remove the full modular weight from $\Phi^A$;
$\Theta'_A$ is a linear combination of the $n-m$ D-moduli with net
modular weight $q'^I_A\ne0$. In addition, $k^A \ne \ell p^A$ in
general, and $F^A$ does not vanish when $\delta = 0$. An exception
is the case $n = N m$ with $N$ replicas of a minimal set of
$\phi^A$, with identical $U(1)_a$\ charges and modular weights, that
get \vevs. They are all made modular invariant by the same
transformation so $p^I = b^I$, and
\beq k^A = \ell p^A = {1\over2N}\delta_X Q^A_X\, .\eeq
The effective Lagrangian for the moduli sector is the same as in
the minimal case, but there are $m(N-1)$ D-moduli.
\end{itemize}

Before discussing the phenomenology of these models, we note that in
typical orbifold compactifications with Wilson lines there is no
asymptotically free gauge group above the scale of $U(1)$\ breaking.
One or more asymptotically free gauge sector will emerge below that
scale, provided a sufficient number of gauge-charged chiral
multiplets get masses through their superpotential couplings to the
$\Phi^A$ that acquire large $vevs$.  In their contribution to the
quantum Lagrangian~\myref{lqb} the effective IR cut-off is their
mass, rather than the condensate scale, as in the
Lagrangian~\myref{wvy2} introduced later in Section~\ref{axion} for
supersymmetric QCD.  The mass terms in the superpotential are
$U(1)$\ and modular invariant, containing factors analogous to the
$W_\alpha$ in~\myref{pdefs4} that assure that $\Lag_a$ has the
correct anomaly structure at the string scale.  The net result is
that the potential is identical to that given above, with
$p^I,\,p^A$ determined by the full massless spectrum at the string
scale, except that the $\beta$-function coefficient $b_c$ is the one
that controls the RGE running below the $U(1)_a$\ breaking scale,
without the contributions from the heavy
modes.\cite{Gaillard:2003gt}

\subsection{The vacuum and the moduli sector}\label{vacmod}

The potential is modular invariant, with a similar $t$-dependence as
is found in the BGW model described in Section~\ref{sec3}, so the
moduli are still stabilized at self-dual points $t_1 = 1, \; t_2 =
e^{i\pi/6}$, with $\lang F^I\rang =0$. The T-moduli masses are
determined by the coefficients of $F^I$ in~\myref{solt2}. In models
that satisfy~\myref{minkp}, the squared masses are positive since
$b\ge b_c$ and, using~\myref{newbi}, \myref{minpi}
and~\myref{minkp},
\beq p^I = \sum_A p^A \equiv p >0\label{newp},\eeq
because $k\sim |\Phi|^2$ and $\ell$ are positive.
For example in a $\mathbf{Z}_3$ orbifold model of Font et al. that
we will refer to as the FIQS model,\cite{Font:1989aj} the three
K\"ahler moduli are approximately degenerate with $m_{\re \,t}(t_i)
\approx 5(6)m_{3/2}, \; m_{\im\, t}(t_i)\approx 1(2)m_{3/2}$ for $i
= 1(2)$.

If we set the moduli at self dual points, the potential for the
dilaton becomes, after integrating out the scalar components $u_a$
of $U_a$ as in \myref{upot}\footnote{In nonminimal models with
$F^A,\pp V(\hel)/\pp\delta_A\sim |u_c|^2$ instead of
$\delta|u_c|^2$, it is necessary to keep order $\delta^3$ terms in
the $u_a$ expansion of the D-terms analogous to \myref{LD}.  Then
minimization of the potential with respect to $\delta$ gives
$\delta^2\sim |u_c|^2,$ $\delta\sim |u_c|\ll 1$, $\delta^3\sim
|u_c|^3,$ so the potential for $\hel$ is still dominated by the
explicit terms in \myref{newV}, and $\delta$ can be ignored in the
analysis of these models discussed briefly below.}
\beq V(\hel) = {|u_c|^2\over16}\[w(\hel) +
v(\hel)\] + \order(\epsilon|u_c|^2,\delta^3,|u_c|^4),\label{newV}\eeq
where $v(\hel) = v(\hel,0)$ is given in \myref{minv}, and the $F$
contribution has
\beq  w = w_0 - \delta_X h'_X(\hel)(1 + b_c\hel)^2,\qquad w_0 =
\sum_A{(p^A + b_c k^A)^2\over k^A}.\label{fpot}\eeq
For the minimal case characterized by~\myref{minkp}, $w=0,\;V =
V(\hel) + \order(\delta^2|u_c|^2$), the dilaton potential is the
same as in the BGW case discussed in Sections~\ref{sec2}
and~\ref{sec3}, except for the shift $k\to\tilde{k}$ in the dilaton
metric in both the potential and in the kinetic energy term, so that
the constraint~(\ref{Ksstrue}) on the dilaton metric is unchanged.
In the general case, imposing vanishing vacuum energy at leading
order in $\delta\sim|u_c|^2$ gives
\beq \hel^{-1}\(1 + b_c \hel\)^2\tilde{k}'(\hel) = 3b_c^2 - w, \eeq
and positivity\footnote{If the dilaton metric goes through zero, one
should rewrite the theory in terms of the canonically normalized
field, in terms of which the zero of the metric becomes a
singularity in the potential.  It is not clear that there might not
be some viable region of parameter space in this case.} of the
dilaton metric requires $w< 3b_c^2.$ The viable region of parameter
space can be explored\cite{Gaillard:2003gt} by separating out a
minimal subset of the $n>m$ $\Phi^A$ with nonvanishing $p^A$:
\beq k^A,\;\; A = 1,\ldots n,\;\;\to\;\; (k^P,k^M), \quad P =
1,\ldots, m,\quad M = 1,\ldots, n-m,\label{kakm}\eeq
and writing
\beq k^A = \ell p^A + y^A\, , \qquad \sum_A q^a_A y^A = 0 \,
,\label{ycond}\eeq
where the last equality ensures that if~\myref{pcond} is satisfied,
so is~\myref{kcond}.  The vacuum is degenerate at the $U(1)_a$\
breaking scale; at the condensation scale, the (approximate) vacuum
values will be those that minimize $v$ with respect to the $y^A$
subject to the condition in~\myref{ycond}.  If $y^A=0$, for all $A$,
the dilaton potential is identical to the minimal case. Defining
$Q^P_a$ as the inverse of the matrix $q^a_P$, the constraint
in~\myref{ycond} reads
\beq y^P = - \sum_M\zeta^P_M y^M, \qquad \zeta^P_M = \sum_a Q^P_a
q_M^a,\eeq
and we may take the $y^M$ as independent variables.  Since
\beq \tilde{k}'(\hel) = k'(\hel) + \sum_A k'^A(\hel) = k'(\hel) +
\delta_X\hel h'_X(\hel),\eeq
minimizing $V\propto v+w$ with respect to $y^M$ amounts to minimizing
$w_0$, giving the conditions
\beq 0 = {\pp w_0\over\pp k^M} - \sum_P\zeta^P_M {\pp w_0\over\pp
k^P}= - (p^M/k^M)^2 + b_c^2 + \sum_P\zeta^P_M
\[(p^P/k^P)^2 - b_c^2\].\label{miny}\eeq
If $p^M = 0$, we require $k^M= y^M\ge0$, and
\beq \l{\pp w_0\over\pp y^M}\r_{y=0} = b_c^2 +
\sum_P\zeta^P_M\(\hel^{-2} - b_c^2\).\label{pm0}\eeq
If $\hel b^2_c<1$ and $\sum_P\zeta^P_M\ge 0$, the minimum indeed
corresponds to $y=0$.  However if $\sum_P\zeta^P_M<0$, the minimum
corresponds to a smaller $w_0$ with $y>0$.  Larger $y$ results in
smaller $h'_X$, so the net effect is to increase $w$, and positivity
of the dilaton metric becomes difficult to
maintain.\cite{Gaillard:2003gt}

If $p^M\ne0$ we have instead of~\myref{pm0} the expression
\beq \l{\pp w_0\over\pp y^M}\r_{y=0} =
\(\sum_P\zeta^P_M - 1\)\(\hel^{-2} - b_c^2\).\label{pm1}\eeq
In this case $y=0$ is the minimum if and only if $\sum_P\zeta^P_M =
1$.  If $\sum_P\zeta^P_M>0$, the minimum will in general shift
slightly from $y=0$.  The dilaton potential for these cases is not
substantially different from the minimal case.  On the other hand if
$\sum_P\zeta^P_M<0$, the situation is similar to the case with
$p^M=0$: the minimum occurs for larger $k^A$ and larger $w$, such
that positivity of the dilaton metric is difficult to maintain. For
example in the FIQS model there are minimal flat directions in which
one, two or three charge-degenerate sets of six chiral
supermultiplets acquire \vevs and break six $U(1)$'s.  There are
additional $F$-flat and $D$-flat directions associated with
``invariant blocks'' ${\cal B}$ of fields such that
\beq \sum_{P\in\cal B}q^a_P + \sum_{M\in\cal B}q_M^a = 0.\eeq
It is clear that if we choose $\Phi^M$ that form invariant blocks
with the $\Phi^P$, at least some $\zeta^P_M<0$.  The numerical
solution to the minimization equations for one such
choice\cite{Gaillard:2003gt} gave $\lang \tilde{k}'\rang <0$, and
this result is likely to be generic.

The vacuum conditions $v(\hel) = v'(\hel) = 0$ require
$\tilde{k}'\sim \hel \tilde{k}'' \sim \hel b_c^2$ in the vacuum, but
there is no {\it a priori} reason to expect $\tilde{k}'''$ to be
similarly suppressed if $\hel\sim 1$, in which case the dilaton mass
is enhanced as will be discussed in Section~\ref{inflate}. If there
is just one hidden sector condensate the universal axion is massless
at the scale of supersymmetry breaking, and is a candidate for the
QCD axion, as will be discussed in Section~\ref{axion}.  The
fermionic superpartners of the moduli have masses similar to those
of the K\"ahler moduli. Using the FIQS model again as an example, it
can be shown that for a minimal set of \vevs there are two linear
combinations of the three fermions $\chi^I$ that have no mixing with
the dilatino, while one linear combination that is approximately an
equal admixture of the $\chi^I$, mixes with the
dilatino.\cite{Gaillard:2003gt} All four masses are roughly the same
as $m_{\re\, t}(t_i) \approx 5$--$6m_{3/2}$.


\subsection{Observable sector masses: conditions for a viable
model}\label{obsmass}

In the minimal models as defined above, the tree level masses of the
observable sector gauginos are identical to those in the BGW case;
they are suppressed due to a factor of the dilaton metric
$\tilde{k}' \sim 3b_c^2$. In nonminimal models they will be
suppressed even further since the metric is suppressed further, as
discussed in~\ref{vacmod}. When we consider chiral fields $\Phi^M$
with no couplings to the condensates we have to include their
F-terms which is the same as those for the $\Phi^A$ except that $p^M
= 0$:
\bea V&\ni& \sum_M F^M F_M = {|u_c|^2\over16}\sum_M{\[x^M b_c
 - \(1 + b_c\hel\)x'^M\]^2\over
x^M}\[1 + \order(x^2)\] \nonumber \\ &=& {|u_c|^2\over16}\sum_M\[x^M b^2_c
 - 2b_c\(1 + b_c\hel\)x'^M + 2\(1 + b_c\hel\)^2\sum_a q^a_M h'_a x'^M\]
.\label{sums2}\eea
For fields with vanishing $vevs$, $\pp V/\pp\phi \propto \phi$
vanishes in the vacuum, the mass matrix is diagonal:
\bea m^2_M &=& {\pp V\over\pp x^M} =  {|u|^2\over16\hel}(1 +
b_c\hel)^2{\pp\hK'(\hel)\over\pp x^M} + {\pp\over\pp
x^M}(\wtd{K}_{M\bar N}F^M\bar F^{\bar N}) + {\pp\over\pp
x^M}(\wtd{K}_{A\bar B}F^A\bar F^{\bar B}) \nonumber \\ &=&
{|u|^2\over16}\[b^2_c - 4b_c(1 + b_c\hel)\sum_a h'_a q^a_M - 2(1 +
b_c\hel)^2\sum_a h''_a q^a_M\] \nonumber \\ && + {\pp\over\pp
x^M}(\wtd{K}_{A\bar B}F^A\bar F^{\bar B}) + \order(x^M)\, .
\label{m2m} \eea
In minimal models with
\beq h'_a = -\hel h''_a = \sum_A
Q^A_a/2\hel\, , \qquad p^A = k^A/\hel = \delta_X Q^A_X/2 \, ,
\label{mincond}\eeq
the last term in~(\ref{m2m}) is of order $\delta|u_c|^2\sim|u_c|^4$, giving
\beq m_M=m_{3/2}\[1 + {\zeta_M\over b_c^2\hel^2}\(1 - b_c^2\hel^2\)\]
+ \order(|u_c|^4),
\qquad \zeta_M = \sum_{a,A}Q^A_a q_M^a.\label{m2min}\eeq
The first term in~(\ref{m2min}) is just the $F$-term contribution of
the BGM model; the second term is expected to dominate because of
the factor $b^{-2}_c>>1$, and can give large (and in some cases
negative) masses to squarks and sleptons.  As discussed in
Section~\ref{sec31}, the gravity/dilaton mediated $F$-term
contribution to scalar masses can be enhanced if they have couplings
in the GS term. So, for example, if $p_i = b_{\GS} \gg b_+$
in~(\ref{BGWmasssq}) it might be possible to arrange for all masses
to be positive (except possibly the Higgs) at the condensation
scale, but this would involve some measure of fine tuning.  The
possibility of including higher order terms in the gauge-charged
chiral superfields $|\Phi|$ in the K\"ahler potential was considered
in Ref.~\refcite{Gaillard:2003gt} and found to make little
difference.  The problem can be attenuated somewhat if $b_c$ and/or
$\hel$ is larger than expected.\footnote{The point $b_c^2\hel^2 = 1$
where the charge-dependent contribution to~\myref{m2min} actually
vanishes is also the point where~\myref{pm1} vanishes identically,
which is the condition for a minimal solution.} The
factor~\myref{ufactor} that now appears in the expression for the
condensate \vev may allow for the former, and alternate
parameterizations\cite{Gaillard:2003gt} of nonperturbative string
effects, based on corrections to the action rather that to the
function $f(L)$ in~\myref{nonpertsum} as well as
perturbative\cite{Gaillard:2005cw} field theoretic quantum
corrections to $k(\ell)$ can allow for a value of $\hel$ larger than
one while preserving weak coupling $s(\hel)\le1$.  Another
difficulty with large charge-dependent contributions to observable
sector scalar masses is that they can generate flavor-changing
neutral currents. This can be avoided if MSSM states come in sets of
three with all the same quantum numbers ({\em i.e.}  $U(1)$ charges
and modular weights as well Standard Model quantum numbers), as can
happen for example in $\mathbf{Z}_3$ orbifold compactifications. The
simplest solution to these problems would be to find a
compactification such that $\zeta_M = 0$ for observable sector
chiral multiplets, which implies strong constraints on the various
$U(1)$ charges.

In nonminimal models that do not involve just charge-degenerate
minimal sets, the expressions for observable sector masses are more
complicated, but they are always of the form $m^2_M = m^2_{3/2}\(1 +
\zeta_M f(b_c\hel)/b_c^2\hel^2\)$. The parameter space for these
nonminimal models is constrained to be rather small because of the
positivity constraint on the dilaton metric, the need to avoid an
overly large scalar/gaugino mass ratio and a very large axion
coupling constant.  Although it has recently been
shown\cite{Fox:2004kb} that the original cosmological
bounds\cite{Preskill:1982cy,Abbott:1982af,Dine:1982ah} on the axion
coupling can be evaded, increasing the coupling at the string scale
leads to a proportional increase at the QCD scale that might further
restrict the class of compactifications that allow the universal
axion to be the QCD axion. We will return to this issue in
Section~\ref{axion}.  A possible advantage of more general
nonminimal models might be that when more fields get $vevs$, more
unwanted states are removed from the spectrum. On the other hand,
there are other scales where masses could be generated for
additional fields, such as the condensation scale itself,
$\Lambda_c\sim 10^{13-14}\GeV$, and a scale $\Lambda_\nu\sim
10^{11-12}\GeV$ suggested by observed neutrino masses.  Indeed
masses generated at one or more intermediate scales can be useful in
reconciling the data with string scale
{unification}.\cite{Dienes:1996du}

\subsection{Lifting the vacuum degeneracy: D-moduli masses}\label{dmod}

In minimal models all complex scalars that vanish in the vacuum and
have no couplings to matter condensates acquire masses $m = m_{3/2}$
(or somewhat larger masses if they couple to the GS term), while
their fermionic superpartners remain massless, just as for the
observable sector of the MSSM.  In nonminimal models where the
number $n$ of scalar fields that get large \vevs is larger than the
number $m$ of broken $U(1)$'s, the situation is very different.  The
K\"ahler potential in~\myref{defKS2} for the chiral multiplets with
nonvanishing \vevs is replaced by
\beq K(L,\Delta,\widehat{\Sigma}) = \tilde{k}(L) + \sum_A
k^A(L)\(e^{\widehat{\Sigma}_A + \Delta_A + \oline{\Delta}_{A} +
2\sum_a q^a_A(U_a + \Delta_a)} - 1\),\eeq
where the modular invariant fields $\widehat{\Sigma}_A$ are defined
by\cite{Gaillard:2003gt}
\beq \widehat{\Sigma}_A = \Sigma_A - \sum_{a,b,B}q_A^a x^B q^b_B
N^{-1}_{ab}\Sigma_B, \qquad \Sigma_A = \Theta_A + \oline{\Theta}_A +
G_A ,\eeq
and satisfy
\beq \sum_A q^a_Ax^A(L)\widehat{\Sigma}_A(L) = 0, \qquad x^A = k^A
e^{\Delta_A + \oline{\Delta}_{A} + 2\sum_a q^a_A\Delta_a}.
\label{sigcond}\eeq
They have vanishing \vevs and are the $n-m$ uneaten Goldstone
supermultiplets above the scale of supersymmetry
breaking.\cite{Buccella:1982nx,Gaillard:2000na} But when
$\widehat{\Sigma}\ne0$ we have to include them among the superfields
$M$ that appear in the superfield Weyl transformation~\myref{LhL} to
the basis where~\myref{wcond} is satisfied. If we restrict ourselves
to the class of ``minimal'' models with $n = N m$ where $N$
charge-degenerate sets of scalars get $vevs$, the
constraints~\myref{sigcond}, that insure the absence of a linear
coupling of $\widehat{\Sigma}$ to $U_a$, imply the additional
constraint
\beq \sum_A x^A(L)\widehat{\Sigma}_A(L) = 0, \label{sigcond2}\eeq
with
\beq x^A = k^A = L p^A = L k'^A = L\delta_X
Q^A_X/2N,\label{vecs}\eeq
so we may drop contractions of $\widehat{\Sigma}_A$ with any of the
vectors in~\myref{vecs}, as well as higher $\ell$-derivatives of
$k^A$. Then the expressions
\bea L - \hat L &=& {L^2\sum_A k'^A(e^{\widehat{\Sigma}_A} -1)\over3
- L\tilde{k}'(L)}, \nonumber\\ \widehat{K}(\widehat{L}) &=&
\tilde{k}(L) + \sum_A k^A(e^{\widehat{\Sigma}_A} -1)
+ 3\ln(\hL/L), \nonumber\\
\hS(\hL) &=& \tilde{s}(L) + {3\over2\hL L}(L - \hL),\eea
have no terms linear in $\widehat{\Sigma}$.  Expanding as before, we
obtain for the effective K\"ahler potential
\beq \tilde{k} = \hK + 2\hL\hS = \tilde{k}(\hL) + 2\hL\tilde{s}(\hL)
+ \half\sum_A k^A(\widehat{\Sigma}_A)^2 +
\order(\widehat{\Sigma}^3),\eeq
and, using the constraint~\myref{sigcond}--\myref{vecs}
\beq \hK(\hel) = \tilde{k}(\hel) + \order[(\ell - \hel)^2] =
\tilde{k}(\hel) + \order(\widehat{\Sigma}^4), \eeq
where we have dropped corrections of order $\delta\sim|u_c|^2$.
The D-moduli mass comes from the potential term
\beq \tilde{K}_{A\oline{B}}F^A\oline{F}^{\oline{B}} =
{|u|^2\over32\hel^2}\sum_A k^A(\widehat{\Sigma}_A)^2.\eeq
The modular invariant fields $\widehat{\Sigma}_A$ with
$<\widehat{\Sigma}_A> = 0$ are independent of $L$ in these models,
and may be expressed\cite{Gaillard:2002ew} in terms of chiral and
anti-chiral fields $D_A,\oline{D}_A$ by expanding the moduli about
their \vevs $\lang t^I\rang$:
\bea \widehat{\Sigma}_A &=& D_A+ \oline{D}_A +
\order\(\[\widehat{T}^I + \widehat{\oline{T}}^I\]^2/\lang t^I +
\bar{t}^I
\rang^2\),\nonumber\\
D_A &=& {\Theta}_A + \half\bigvev{G_A} + \bigvev{\pp G_A\over\pp
t^I} \widehat{T}^I - {1\over N}\sum_{B; q^a_B=q^a_A}\({\Theta}_B +
\half\bigvev{G_B} + \bigvev{\pp G_B\over\pp t^I}\widehat{T}^I\),
\nonumber \\
T^I &=& \lang t^I\rang + \widehat{T}^I, \qquad \bigvev{D_A +
\oline{D}_A} = 0\, . \label{ddef}\eea
When we re-express the chiral multiplets $D_A$ in terms of an
orthonormal set $D_i$ subject to the constraint~\myref{sigcond2}:
\beq D_A = \sum_{i=1}^{n-m}c_A^i D_i\, , \qquad \sum_A k^A c_A^i =
0\, , \label{dadi} \eeq
the K\"ahler metric and the squared mass matrix
\beq K_{i\jbar} = \sum_A C^A_i k^A C^A_{j}, \qquad \mu_{ij}^2 =
{m_{3/2}^2\over b_c^2\ell^2}\sum_A C^A_i k^A C^A_{j}\, ,\eeq
with $C^A_i$ the inverse of $c_A^i$, are diagonalized by the same
unitary transformation:
\beq D_i\to D_i = U_i^j D_j\, , \qquad d_i = \l D_i\r = N_d(\sigma_i
+ i a_i)\, ,\label{diprime}\eeq
where the normalization constant $N_d$ is chosen to make the kinetic
energy term canonically normalized. Then the Lagrangian quadratic in
the scalar D-moduli reads
\beq \Lag_D = \half\sum_i\[\pp_\mu\sigma_i\pp^\mu\sigma_i + \pp_\mu
a_i\pp^\mu a_i - 2{m_{3/2}^2\over b_c^2\hel^2}(\sigma_i)^2\].
\label{ld}\eeq
The Yukawa couplings for condensation models of the type considered
here are given in Ref.~\refcite{Giedt:2002ku}; they generate squared
masses for the D-moduli fermions $(\chi_{i\,L})_\alpha = \l\D_\alpha
D_i\r/\sqrt{2}$ that are equal to half the scalar squared masses up
to corrections of order $b_c\hel=
m_{3/2}/m_\chi$.\cite{Gaillard:2003gt} If $\hel\sim1$ and $b_c\sim
.03-.04$ as suggested by the dark matter analyses to be discussed in
Section~\ref{dmatter} below, the scalar and fermion masses are much
larger than the gravitino mass:
\beq m_\chi\approx m_\sigma/\sqrt{2} \approx (24-30)m_{3/2}\, .\eeq
However, as discussed in Section~\ref{obsmass}, this hierarchy
requires constraints on observable scalar $U(1)$ charges for a
viable phenomenology.  On the other hand, the D-axions remain
massless because the phase of $\theta = \l\Theta\r$ is undetermined.
These axions might play an interesting role in cosmology with
possible contributions to dark energy or dark matter.  Just as in
the minimal models, in these restricted nonminimal ones there are
generically many fields with nonvanishing \vevs at the $U(1)$
breaking scale for which the complex scalars acquire masses $m =
m_{3/2}$, and the fermions remain massless.  This would lead to a
disastrous
cosmology,\cite{Coughlan:1983ci,Goncharov:1984qm,Ellis:1986zt} but
presumably many of these fields acquire supersymmetric masses
through superpotential terms coupling them to the $\Phi^A$ that do
get $vevs$, as well as to other fields that might acquire \vevs at
some intermediate scales as mentioned at the end of
Section~\ref{obsmass}.

\section{Particle physics phenomenology}
\label{sec5}
\setcounter{footnote}{0}

\subsection{R-parity and modular invariance}\label{rparity}
One of the challenges of string theory is to provide a mechanism for
forbidding operators that violate lepton and baryon number in the
low-energy effective theory.  In the MSSM this is achieved by
imposing a discrete symmetry called R-parity such that the unwanted
operators are forbidden, while those that give masses to quarks and
leptons are allowed, as is the Higgs mass term ($\mu$-term) that is
needed to produce the correct electroweak symmetry breaking pattern.
We saw in Section~\ref{sec23} that the T-duality, or modular
invariance, of the weakly-coupled heterotic string ensures that when
SUSY is broken by gaugino condensation in a hidden sector the
K\"ahler moduli are stabilized at self-dual points where their
$F$-terms vanish.  As a result, in the absence of other sources of
supersymmetry breaking, transmission from the hidden sector to the
observable sector is dilaton-mediated. Provided that the matter
couplings $p_i$ to the Green-Schwarz term~(\ref{GS}) vanish (or are
degenerate), the resulting scalar masses will be flavor-diagonal at
tree level. The loop corrections are then merely RG-induced, the
off-diagonal scalar masses are small and scalar-mediated flavor
changing neutral current (FCNC) effects are therefore sufficiently
small.
We saw however in Section~\ref{sec4} that charge-dependent scalar masses
may be generated when there is $U(1)$ anomaly cancelation by a second
Green-Schwarz term. Such a term amounts to an FI term that triggers
the breaking of some number $m$ of $U(1)$'s by $n$ nonvanishing scalar
\vevs $\lang \Phi_A^{\alpha} \rang$; in this case the theory remains
FCNC-free if observable sector fields that are degenerate under Standard
Model charges are also degenerate under the broken $U(1)$ charges.
A second consequence of modular invariance is that there is generally a
residual discrete symmetry that might play the role of R-parity in
the MSSM.\cite{Gaillard:2004aa}

The modular transformations given in~\myref{mod}
and~\myref{mattertrans} are those of the minimal subgroup of a
generally larger group of modular transformations.  For example in
$\mathbf{Z}_3$ orbifolds with just three diagonal K\"ahler moduli,
the largest possible symmetry group is $[SL(2,\Zbf)]^3$
with\footnote{Here we are again neglecting
mixing\cite{Chun:1989se,Lauer:1990tm} among twisted sector fields of
the same modular weights $q_A^I$ with mixing parameters that depend
on the integers $a^I,b^I,c^I,d^I$.}
\bea &&T^I \to {a^I T^I - i b^I \over ic^I T^I + d^I}, \qquad \Phi^A
\to e^{i\delta_A -\sum_I q^I_A F^I} \Phi^A, \qquad F^I = \ln \( i
c^I T^I + d^I \),\nnn &&a^I d^I - b^I c^I = 1, \qquad
a^I,b^I,c^I,d^I \in \Zbf \qquad \forall \; I=1,2,3, \label{mdtr}
\eea
under which the K\"ahler potential and superpotential transform as
\beq K\to K + F + \oline{F}, \qquad W\to e^{- F}W, \qquad F = \sum_I
F^I.\eeq
The $T^I$ are trivially invariant under \myref{mdtr} with
\beq a^I = d^I = \pm 1,\qquad b^I = c^I =0,\qquad e^{F^I} =e^{i
n\pi}\, . \label{Z2}\eeq
The self-dual vacua $T_{s\, d}$ are further invariant
under~\myref{mdtr} with
\beq b^I = - c^I = \pm 1 \, , \eeq
and for $\lang t^I\rang = 1$
\beq a^I= d^I = 0,\qquad e^{F^I} =  e^{i{n\over2}\pi},\label{Z4}\eeq
or for $\lang t^I\rang = e^{i\pi/6}$
\beq \cases{a^I = b^I,\; d^I = 0,\cr d^I = c^I,\;a^I = 0\cr},\qquad
e^{F^I} = e^{i{n\over3}\pi}\label{Z6}.\eeq
Thus for three moduli at self-dual points the residual symmetry
group is $G_R = \mathbf{Z}_4^m\otimes Z_\mathbf{6}^{m'},\; m + m' =
3$.

The gaugino condensate \vevs $\lang u\rang \ne0$ break this further
to a subgroup with
\beq e^{i\,\im F} = e^F = e^{2n i\pi},\label{subg}\eeq
under which $\lambda_L\to e^{-{i\over2}\im F}\lambda_L = \pm
\lambda_L$. It is natural to identify the case with a minus sign
with R-parity.  This subgroup also leaves invariant the soft
supersymmetry-breaking terms in the observable sector, if no other
field gets a \vev that breaks it.  Consider, for example a scenario
in which the $\mu$-term comes from a superpotential term $X H_u
H_d$, with the \vev $\lang x \rang = \lang \left. X \right| \rang
\ne 0$ generated at the TeV scale. Then the symmetry could be broken
further to a subgroup $R\in G_R$ such that $R\, X = X$, unless there
is a concomitant breaking of a $U(1)$ gauge factor such that $X$ is
invariant under redefined $G_R$ transformations that include global
transformations under the broken $U(1)$.  On the other hand if the
$\mu$-term comes from a K\"ahler potential term generated by
invariant \vevs above the scale where the moduli are fixed, there
would be no further breaking until the Higgs get $vevs$. Since an
invariant $\mu$-term implies that $H_{u}$ and $H_{d}$ have opposite
$G_R$ charges as well as opposite weak hypercharge, the $G_R$
transformations can be redefined to include global hypercharge
transformations such that they are both invariant, and no further
breaking of the discrete symmetry occurs at the electroweak scale.

The transformation property~\myref{etatrans} of the Dedekind
$\eta$-function is not fully general. More precisely it reads
\beq \eta(T^I)\to e^{i\delta_I}e^{\frac{1}{2} F(T^I)}\eta(T^I),\quad
F(T^I) = F^I,\quad \delta_I = \delta_I(a^I,b^I,c^I,d^I)\, .
\label{phase}\eeq
The constant phases\cite{Ferrara:1989bc} $\delta_I$ are commonly
dropped in the literature because they do not appear in the scalar
potential, and it follows from T-duality that they can be
re-absorbed\cite{Ferrara:1989qb} into the transformation properties
of the twisted sector fields by removing the phases $\delta_A$
in~\myref{mdtr}. This was implicitly done in
writing~\myref{mattertrans}, following the commonly used convention.
When these phases are taken into account the constraints obtained by
imposing T-duality covariance on superpotential terms of the form
\beq W = \prod_A\Phi^A\prod_I\eta(T^I)^{2\(\sum_A q_A^I -
1\)},\label{wterm}\eeq
coincide with selection rules\cite{Hamidi:1986vh} that follow from
the discrete symmetries of orbifold compactification.

Consider for example a $\mathbf{Z}_3$ orbifold with untwisted sector
fields $U_A^{I}$, and twisted sector fields $T_A$ and $Y_A^{I}$ with
modular weights
\beq \(q_{I A}^{\, J}\)_U = \delta_I^J,\qquad \(q^I_A\)_T =
\({2\over3},{2\over3},{2\over3}\),\qquad \(q_{I A}^{\, J}\)_Y =
\({2\over3},{2\over3},{2\over3}\) + \delta_I^J\, .\label{Z3wts}\eeq
Allowed superpotential terms trilinear in matter fields are of the
form\cite{Ferrara:1989qb} $U_1U_2U_3$ and $(T)^3$. When the
appropriate Dedekind $\eta$ factors in~\myref{wterm} are included,
all such terms are covariant provided
$\delta_U = 0$, $\delta_T = - {2\over3}\delta = -
{2\over3}\sum_I\delta_I$.  If we further impose $\delta_{Y^I} = -
{2\over3}\delta - 4\delta_I$, the superpotential terms~\myref{wterm}
constructed from the monomials
\beq U_1U_2U_3,\quad T^3,\quad U_I Y^I T^2,\quad U_I Y^I U_J Y^J T,
\quad U_I Y^I U_J Y^J U_K Y^K,\label{mons}\eeq
are covariant under~\myref{mdtr} and~\myref{phase}.  Higher
dimensional monomials can be constructed by adding factors of
invariant monomials. For example
\beq \eta^6\Pi, \qquad \Pi = Y^1Y^2Y^3, \qquad \eta =
\prod_I\eta_I,\label{Piinv}\eeq
is invariant. The group~\myref{mdtr} of duality transformations on
$T^I$ is generated\cite{Schellekens:1986xh} by $T^I\to1/T^I$ with
$\delta(0,1,-1,0) = \pi/4$ mod $2\pi$, and $T^I\to T^I -i$ with
$\delta(1,1,0,1)=\pi/12$ mod $2\pi$. Therefore $\delta = n\pi/12$ is
a rational number, and invariant operators can be constructed from
products of covariant operators \myref{wterm} multiplied by $\eta$:
\beq \eta^{2m}\prod_{i=1}^m W_i, \qquad 2m\delta = m'{\pi\over6} =
2\pi n\, . \label{invmons} \eeq
These potential terms are consistent with the selection
rules;\cite{Hamidi:1986vh} they are further restricted by additional
selection rules and gauge invariance. The invariant
operators~\myref{Piinv} and~\myref{invmons} can also be used to
construct terms in the K\"ahler potential. For the subgroup defined
by~\myref{Z2}~-~\myref{subg}, $i\delta = - \frac{1}{2} F = - in\pi$,
the superpotential is invariant, as are the monomials
in~\myref{mons}. Therefore any product of them could appear in the
superpotential or K\"ahler potential in the effective theory below
the scale where the T-moduli are fixed and supersymmetry is broken
({\em e.g.} through quantum corrections and/or integrating out
massive fields) with perhaps additional \vevs that are invariant
under some $G_R$ generated at that scale.

Superpotential terms of dimension three can be generated from higher
order terms when some fields acquire $vevs$.  In models with an
anomalous $U(1)_X$, with $m$ $U(1)$'s broken by $n = N m$ \vevs as
discussed above, the modular weights are modified in the same way as
in~\myref{minmodwt} for the minimal case $n = m$. Then for a term in
the superpotential~\myref{wterm} with some $\lang \Phi^A \rang\ne0$:
\bea W &=& \prod_M\Phi^M\prod_A\lang\Phi^A\rang \prod_I[\eta(
T^I)]^{2\(\sum_M q_M^I + \sum_B q_A^I - 1\)}\nnn &=&
\prod_M\Phi'^M\prod_A\lang\Phi^A\rang\prod_I[\eta( T^I)]^{2\(\sum_M
q'^I_M - 1\)}, \label{wterm2}\eea
because $W$ is also $U(1)_a$ invariant: $\sum_M q_M^a + \sum_A q_A^a
= 0$.\footnote{Recall from Section~\ref{sec4} that $q_M^a$ is the
charge of field $\Phi^M$ with respect to the Abelian factor
$U(1)_a$, and similarly for $q_A^a$.} In order to make T-duality
fully manifest below the $U(1)$-breaking scale, we have to redefine
the transformation ~\myref{mdtr} on $\Phi^A$ by including a global
$U(1)_a$ transformation such that $\Phi^A$ is fully invariant, and
\beq \Phi'^M \to e^{i\delta'_M - \sum_I q'^I_M F^I} \Phi^M, \qquad
\delta'_M = \delta_M - \sum_{A,\, a}q_M^a Q_a^A\delta_A\,
.\label{deltr2}\eeq

{\em A priori} we expect that $\lang\Phi^A\rang \sim 0.1$, so that
couplings arising from high dimension operators in the
superpotential are suppressed.\footnote{The factors multiplying
these terms can in fact be rather
large.\cite{Cvetic:1998gv,Giedt:2000es}} We would like to have one
large coupling ($Q_3,T^c,H_u$) which should correspond to one of the
dimension-three operators in~\myref{mons}.  Until recently, most
models studied were $\mathbf{Z}_3$ orbifold compactifications with
all quark doublets in the untwisted
sector.\cite{Ibanez:1987sn,Casas:1988se,Font:1988nc,Casas:1988hb,Casas:1989wu,Giedt:2000bi,Giedt:2001zw}
In this case we should take $T^c$ and $H_u$ in the untwisted sector
as well, and require $q_{Q_3}^a + q_{T^c}^a + q_{H_u}^a = 0$.  That
is, if we identify the $Q_I$ generation index with the modulus
index, we can have, {\it e.g.}, $T^c = T^c_2$, $H^u = H^u_1$.  Then
to suppress the $Q_2 C^c H^u$ and $Q_2 U^c H^u$ couplings we require
$C^c,U^c\notin U_3$, so one of these must be in the twisted sector
$T$. These requirements can be met in the FIQS
model\cite{Font:1989aj} mentioned in Section~\ref{sec4}. This model
was analyzed in Ref.~\refcite{Gaillard:2004aa} where it was found
not to produce the desired R-parity.  The subgroup $G_R$ that leaves
the self-dual points invariant has $i\delta_I = -\frac{1}{2} F^I$,
and after the redefinitions~\myref{wterm2} and~\myref{deltr2} the
MSSM chiral multiplets transform with phases $e^{n i\pi/33}$.
However these phases combine in the superpotential terms in such a
way that baryon number violating couplings cannot be forbidden.
Moreover the symmetry must be broken to a smaller group to allow
mass terms for all quarks and leptons and to generate a $\mu$-term;
once this is done all gauge invariant trilinear terms are allowed.

We can turn the question around and ask what are the constraints on
the $U(1)$ charges that allow for R-parity in a given
compactification.  Taking $\mathbf{Z}_3$ models as an example, assume
that, as in the FIQS model, $Q_I,\;T^c$, and $H^u$ are in the
untwisted sector, all other MSSM superfields are in the twisted sector
$T$, and in each sector fields degenerate in MSSM charges are also
degenerate in $U(1)$ charges. The last condition, which assures the
suppression of FCNC, implies that fields in the twisted sector $T$
that are degenerate in MSSM charges are further degenerate in
R-parity. Then for all the observed Yukawas to be generated the $Q_I$
must all have the same R-parity, which is compatible with their having
the same $U(1)$ charges only if the R-parity group is further
constrained to have\cite{Gaillard:2004aa}
\beq F^I = 2n^I i\pi\qquad\forall\;\; I.\eeq
Assuming further that there are no standard model singlets in the
untwisted sector, the R-parity transformations take the form
\bea \Phi^m &\to& e^{2i\pi\beta^m}\Phi^m,\nnn \beta^{U^m_J} &=&
\sum_I n_I\({1\over3}\sum_{a,A\in T,Y}q_m^a Q_a^A - \sum_{a,A\in
Y_I}q_m^a Q_a^{Y^A_I}\)\nnn \beta^{T^m} &=& \sum_I
n_I\[{1\over3}\(\sum_{a,A\in T,Y}q_m^a Q_a^A - 1\) - \sum_{a,A\in
Y_I}q_m^a Q_a^{Y^A_I}\].\eea
In addition, requiring that mass terms be allowed for all quarks
implies that the untwisted field $T^c\in U$ has the same R-parity as
$U^c,D^c\in T$, which entails a relation between their (different)
$U(1)$ charges.  We are left with the following set of phases:
\bea \beta^{H^u} &=& - \beta^{H^d} = \gamma, \qquad \beta^Q = \beta
+ 2\gamma, \qquad \beta^{U^c} = - \beta - 3\gamma,\nnn \beta^{D^c}
&=& - \beta - \gamma, \qquad \beta^L = \alpha - \gamma, \qquad
\beta^{E^c} = - \alpha + 2\gamma.\eea
With these phases all MSSM superpotential terms are allowed, while
$Q D^c L$ and $L L E^c$ are forbidden provided $\alpha\ne n$, and
$U^c D^c D^c$ is forbidden provided $3\alpha + 5\beta\ne n$. Below
the scale of electroweak symmetry breaking we redefine R-parity as
discussed above:
\beq \beta^m\to \beta'^m = \beta^m - 2Y_m\gamma,\eeq
so that $\beta'^{H^u} = \beta'^{H^d} = 0$.  In contrast with the
conventional definition of R-parity, higher dimension operators that
generate $B$ and $L$ violation can also be suppressed. For example,
the dimension-four superpotential operator $\tilde U^c\tilde
U^c\tilde D^c E^c$, allowed by conventional R-parity, leads to
dimension-five operators in the effective Lagrangian that may be
problematic\cite{Harnik:2004yp} even if these couplings are Planck-
or string-scale suppressed, given the current bounds on the proton
decay rate. This problem is easily evaded in the current context;
for the choice of phases considered above this requires $3\beta +
\alpha + 3\gamma\ne n.$ The stability of the lightest neutralino is
assured at the same level as proton stability since its decay
products would have to include an odd number of Standard Model
fermions and hence violate $B$ and/or $L$.

There are many other possibilities even within the context of
$\mathbf{Z}_3$ orbifolds, such as cases with all MSSM particles in
the twisted sector.\cite{Kim:1992en} The FIQS model, which is the
best studied of the $\mathbf{Z}_3$ models, cannot reproduce the
observed Standard Model Yukawa textures,\cite{Giedt:2000es} quite
apart from the issue of providing an R-parity. More generally it is
difficult to get the standard $SU(5)$ normalization of weak
hypercharge in these models,\cite{Giedt:2001zw} and different
compactifications, such as recently proposed $\mathbf{Z}_6$-based
orbifolds,\cite{Buchmuller:2005jr,Buchmuller:2006ik,Lebedev:2006kn,Lebedev:2006tr}
with the first two generations of quark doublets in the twisted
sector, would be interesting to analyze in this context.

Finally we remark that it is not actually necessary for the moduli
to be stabilized at their self-dual points for R-parity to be a
symmetry of the superpotential.  This is because the superpotential
is a sum of monomials that are products of matter fields with
coefficients that are functions of the moduli only.  These functions
are invariant at the self-dual points, which means that products of
matter fields are invariant by themselves {\em provided} $F^I(T^I)$
is replaced by $F^I(T^I_{s\,d})$ in the transformation property of
$\Phi_A$ in~\myref{mdtr}.  In other words, there is a
$(\mathbf{Z}_4\otimes \mathbf{Z}_6)^3$ subgroup of the group of
modular transformations that coincides with a subset of the orbifold
selection rules, which have also been considered a potential source
of R-parity.  The use of residual discrete symmetries from T-duality
allow these selection rules to be rephrased in terms of the explicit
symmetries of the effective supergravity theory, and combining them
with $U(1)$ symmetries to obtain an unbroken discrete symmetry at
each stage of gauge symmetry-breaking generates noninteger charges
that make it possible to exclude higher dimension operators as well
as the dimension-four operators excluded by conventional R-parity.

\subsection{General flavor changing processes} \label{FCNC}
The issue of R-parity is clearly critical to a viable low energy
phenomenology. By eliminating terms odd under R-parity, the
contribution of superpartners to FCNC processes will generally be
small if the size of the scalar masses is fairly large. But the
previous statement assumes that soft scalar masses themselves are
diagonal in the flavor basis. That is, that all flavor-changing
processes involving squarks and sleptons are proportional to Yukawa
couplings. To a good approximation this is true in the BGW model:
scalar masses are generally large and universal at tree level, with
loop corrections violating this universality only at the few percent
level. In fact, the constraints on the off-diagonal elements of the
scalar mass matrices weaken very quickly as the overall scale of the
masses exceeds 1~TeV, though the imaginary parts of these same
masses can remain tightly constrained.\cite{Gabbiani:1996hi}

Naively, therefore, we might conclude that the BGW class of theories
is among those that are largely immune to the supersymmetric FCNC
problem. But it is often said that the staring point for a
discussion of flavor in the context of supergravity should be one of
arbitrary off-diagonal scalar masses -- not the diagonal tree-level
masses we derived in Section~\ref{sec3}. The argument runs something
like this. One can imagine {\em a priori} operators of the form
\begin{equation}
\superint \frac{\oline{X} X}{m_{\PL}^{2}} \oline{Q}^{\bar{i}} Q^j,
\label{FCNCbad} \end{equation}
where $X$ is a Standard Model singlet that is presumably a hidden
sector field. If it participates in supersymmetry (SUSY) breaking,
then it will generate off-diagonal soft-masses. In the absence of a
rule for how this $X$ couples to Standard Model matter (such as via
gauge charges in gauge mediation) we must assume that different
flavors can be treated differently. In other words, there is no
symmetry argument as to why operators of the form of~(\ref{FCNCbad})
which mix flavors should be absent.

But by ``symmetry'' what is typically being considered is a {\em
gauge} symmetry. Yet the operator in~(\ref{FCNCbad}) may admit a
{\em geometrical} interpretation. Let us rewrite things to make this
more apparent
\begin{equation}
\superint \frac{R_{\bar{i}j\bar{k}\ell}}{M_{\PL}^{2}}
\oline{X}^{\bar{k}} X^{\ell} \oline{Q}^{\bar{i}} Q^j \, ,
\label{bad2} \end{equation}
where the tensor $R_{\bar{i}j\bar{k}\ell}$ is the curvature tensor
formed from the field-reparameterization connection
\begin{equation}
R^{i}_{jk\bar{m}} = D_{\bar{m}} \Gamma^{i}_{jk}; \quad
\Gamma^{i}_{jk} =  K^{i\bar{n}}\partial_{j} K_{k\bar{n}} =
K^{i\bar{n}}\partial_{j}\partial_{k}\partial_{\bar{n}}K\, ,
\label{gammas} \end{equation}
where $D_{\bar{m}} = K_{\ell\bar{m}}D^{\ell}$ is a covariant
derivative with respect to field reparameterization and
$K_{\ell\bar{m}}$ is the K\"ahler metric. Thus, while an
understanding of the form of this tensor in~(\ref{bad2}) may not be
possible in terms of the gauge quantum numbers of the fields
involved, an understanding in terms of the isometries of the
manifold defined by the chiral superfields of the theory may indeed
exist. This point has been emphasized recently for string-based
effective supergravity
theories~\cite{Chankowski:2005jh,Lebedev:2005uh}.

To address the question of what constraints are needed to avoid
experimentally excluded FCNC effects, we first note that the tree
potential of an effective supergravity theory includes a term
\beq V_{\rm tree}\ni e^K K_i K_{\jbar}K^{i\jbar}|W|^2\,
.\label{vkahl}\eeq
So prior to any discussion of large loop-induced contributions to
flavor-changing operators it is necessary to ensure their absence at
the tree level. The observed suppression of FCNC effects thus
constrains the K\"ahler potential already at the leading order -- to
a high degree of accuracy we require that\cite{Louis:1994ht}
\beq K_i K_{\jbar}K^{i\jbar} \not{\hspace{-.03in}\ni} \lang f(X,\bar
X)\rang \phi^a_f\bar\phi^{\bar a}_{f'\ne f},\eeq
where $f,f'$ are flavor indices, $a$ is a gauge index, $\phi^a_f$
any standard model squark or slepton, and $X$ is a singlet of the
Standard Model gauge group. For example, in the no-scale models that
characterize the untwisted sector of orbifold compactifications, we
have
\beq K_i K_{\jbar}K^{i\jbar} = 3 + K_S K_{\bar S} K^{S\bar S},\eeq
which is safe, since $K_S$ is a function only of the dilaton.  The
twisted sector K\"ahler potential~(\ref{KmatterT}) is flavor
diagonal and also safe.  But the higher order corrections
to~(\ref{KmatterT}) could be problematic if some $\phi^a = X^a$ have
large \vevs ({\em i.e.} within a few orders of magnitude of the
Planck scale).  Thus phenomenology requires that we forbid couplings
of the form $\phi^a_f\phi^{\bar a}_{f'\ne f} |\phi^{a'}_{f"}|^2
X^{b_1}\cdots X^{b_n}$, $n\le N$, where $N$ is chosen sufficiently
large to make the contribution $\lang X^{b_1}\cdots X^{b_n}\rang$ to
the scalar mass matrix negligible.

In addition to these higher-order terms, the quadratically divergent
one-loop corrections generate a term
\beq V_{\rm 1-loop}\ni e^K K_i K_{\jbar}R^{i\jbar}|W|^2, \qquad
R^{i\jbar} = K^{i\bar k}R_{\bar k l} K^{k\jbar}.\label{vricc}\eeq
where $R_{i\jbar}$ is the K\"ahler Ricci tensor. Since the Ricci
tensor involves a sum of K\"ahler Riemann tensor elements over all
chiral degrees of freedom, a large coefficient may be generated,
proportional to the number of chiral superfields $N_{\chi}$. For
example, for an untwisted sector $U$ with three untwisted moduli
$T^i$ and K\"ahler potential as in~(\ref{Kuntw}) we get
\beq R^n_{i\jbar} = (N_n + 2)K^n_{i\jbar}\, .\label{runtw}\eeq
While this contribution is clearly safe, since the Ricci tensor is
proportional to the K\"ahler metric, the condition that the tree
potential be FCNC safe does not by itself ensure that~(\ref{vricc})
is safe in general.  For this we require in addition the absence of
K\"ahler potential terms of the form $\phi^a_f\bar\phi_{f'\ne
f}^{\bar a}|\phi_{f''}^{a'}|^4(X^b)^{n\le N}$.  On the other hand,
if the K\"ahler metric is FCNC safe due to an {\it isometry}, the
same isometry will protect the Ricci tensor from generating FCNC.

There is a large class of models in which FCNC are suppressed
independently of the details of the structure of the K\"ahler
potential, provided the moduli $t^I$ are stabilized at self dual
points. The supersymmetric completion of the potential in any given
order in perturbation theory yields (in the absence of D-term
contributions) the scalar squared mass matrix
\beq (m^2)^i_j = \delta^i_jm_{3/2}^2 - \lang \wtd{R}^i_{j
k\bar{m}}\rang \wtd{F}^k \wtd{\oline{F}}^{\bar{m}}, \label{mtree}
\eeq
where $\wtd{R}^i_{j k\bar{m}}$ is an element of the Riemann tensor
derived from the fully renormalized K\"ahler metric, and $\wtd{F}^i$
is the auxiliary field for the chiral superfield $\Phi^i$, evaluated
by its equation of motion using the quantum corrected Lagrangian.
Since the latter is perturbatively modular invariant, the K\"ahler
moduli $t^i$ are still stabilized at self-dual points with $\lang
\wtd{F}^{t^i}\rang=0$. Classically we have $R^a_{b s\bar{s}}=0$
where the indices $a,b$ refer to gauge-charged fields in the
observable sector. This need not be true at the quantum level. For
example, if the quantum correction to the K\"ahler potential
includes a
term\cite{Binetruy:1985ap,Gaillard:1993es,Gaillard:1996hs,Gaillard:1996ms}
\beq \Delta K = \frac{1}{32\pi^2}\STr\Lambda^2_{\rm eff} \ni \frac{c
N_{\chi}}{32\pi^2}e^{\alpha K}\, ,\eeq
with $\alpha$ and $c$ coefficients which depend on the nature of the
Pauli-Villars regulating sector, we get
\beq \lang \wtd{R}^a_{b s\bar{s}} \rang = \delta^a_b\frac{c
N_{\chi}}{32\pi^2} \alpha^2 e^{\alpha K}\(K_{s\bar{s}} + \alpha K_s
K_{\bar{s}}\),\eeq
which is flavor diagonal, and therefore FCNC safe.

A more substantive comparison can be made by considering a
particular K\"ahler potential. Take the case of
\begin{equation}
K = g(M,\oline{M}) + \sum_a f_a(M,\oline{M})|\Phi^a|^2 +
\frac{1}{4}\sum_{a b}X_{a b}f_a f_b|\Phi^a|^2|\Phi^b|^2 +
\order(|\Phi|^3) , \label{Kquad} \end{equation}
\begin{equation} g(M,\oline{M}) = -\sum_i\ln(T^i +
\oline{T}^i),\qquad f_a(M,\oline{M}) = \prod_i(T^i +
\oline{T}^i)^{-q_a^i}\, . \label{orb} \end{equation}
The term proportional to $X_{ab}$ may give rise to potentially
dangerous off-diagonal elements through the second term
in~(\ref{mtree}). The form of these terms will depend on how the
quadratically divergent parts of $\wtd{R}^i_{j k\bar{m}}$ are
regularized. Returning to the discussion of loop corrections in
supergravity from Section~\ref{sec33} we recall that there were two
sets of Pauli-Villars (PV) fields necessary to regulate loops
involving light matter fields. These fields were labeled $\Phi$ and
$\Pi$ and were coupled through a supersymmetric mass term as
in~(\ref{PVbilinear}). We can relate the field-dependent effective
cut-off for the quadratic divergences to this PV mass via the
relation
\begin{equation} \Phi^I,\Pi^I \;: \quad e^{(1
-\alpha_\alpha)K}K^{i\ibar}|\beta_{\alpha}\mu|^2 \equiv
\(\beta_{\alpha}\)^2 \Lambda_\alpha^2\, , \label{PVlambda}
\end{equation}
where $i$ labels the light (observable sector) chiral superfield
$Z^i$ and $\alpha$ labels the PV regulating fields associated with
$Z^i$. Note that $\Lambda_{\PV} \simeq m_{\PL} = 1$ and $\beta$ is
assumed to be an $\order(1)$ parameter. If the supersymmetric
Pauli-Villars regulator masses are independent of the dilaton, then
in the BGW model the potential off-diagonal scalar masses take the
form
\begin{equation} (m^2)^i_j =
3m_{3/2}^2\beta^2_\alpha\Lambda^2_\alpha \delta^i_j \(\sum_k X_{j k}
+ \sum_I q_j^I\)\(1 - \alpha_\alpha\)\(2 - \alpha_\alpha\)\, .
\label{trouble} \end{equation}
Therefore, even in this particularly simple case of dilaton
domination there is a potential for sizable FCNCs since the
summation in the first term of~(\ref{trouble}) runs over all fields
which participate in the quartic coupling
of~(\ref{Kquad}).\cite{Choi:1997de} However, the presence of this
off-diagonal scalar mass contribution depends on the parameters
$\alpha_{\alpha}$, which are determined by Planck-scale physics. In
particular, the contribution vanishes completely -- independent of
the values of the modular weights or the values of $X_{ab}$ --
provided $\alpha =1 \; {\rm or} \; 2$.

The issue of whether supergravity effects induce large FCNCs
therefore depends on the physics of the UV~completion of the theory.
In this case that UV~completion is heterotic string theory itself.
The manner in which the Planck-scale theory softens ultraviolet
divergences is reflected in the K\"ahler potential factors
$\alpha_A$. As argued in Section~\ref{rparity}, there are good
reason to believe that the higher-dimension operators involving
$X_{ab}$ in~(\ref{Kquad}) {\em are not arbitrary}, but are in fact
tightly constrained. Ultimately the issue of FCNCs thus becomes an
issue of how flavor physics is encoded in string models in the first
place.\cite{Binetruy:2005ez}

\subsection{Collider signatures} \label{collider}
The relatively low masses for the gauginos in this class of models
should make them easily accessible at hadron colliders. With the
reduction in the gluino mass it is possible to greatly increase the
accessibility of the gluino at Tevatron energies. This suggests that
these models can be probed significantly in the short term even
before the LHC data taking begins. It was for this reason that the
K\"ahler-stabilized models of the weakly-coupled heterotic string
were included in a set of benchmark models for the Tevatron
constructed in Ref.~\refcite{Kane:2002qp}.

As described in Section~\ref{sec34}, a convenient parameter space
for the BGW~class of models is $\lbr \tan\beta, m_{3/2}, b_+ \rbr$
and ${\rm sgn}(\mu)$. The three benchmark points chosen in
Ref.~\refcite{Kane:2002qp} were defined by the set $\lbr \tan\beta,
m_{3/2}, a_{\rm np} \rbr$, but we can replace $a_{\rm np}$ with
$b_+$ through~(\ref{aBGW}). The boundary condition scale was taken
to be $\Lambda_{\UV} = 2 \times 10^{16} \GeV$ as this is a common
convention in the literature and makes for easier comparisons with
previous results. The specific values were
\begin{eqnarray}
{\rm Case\; A:} \qquad  \lbr \tan\beta,\; m_{3/2},\; b_+
\rbr & = & \lbr 10,\; 1500 \GeV,\;  0.152 \rbr \label{benchA} \\
{\rm Case\; B:} \qquad \lbr \tan\beta,\; m_{3/2},\; b_+
\rbr & = & \lbr 5,\; 3200 \GeV,\;  0.063 \rbr  \label{benchB} \\
{\rm Case\; C:} \qquad \lbr \tan\beta,\; m_{3/2},\; b_+ \rbr & = &
\lbr 5,\; 4300 \GeV,\;  0.038 \rbr \, . \label{benchC}
\end{eqnarray}
The last two cases correspond to beta-function coefficients $b_+ =
5/8\pi^2 = 0.063$ and $b_+ = 3/8\pi^2 \simeq 0.038$. The former
could result from a condensation of pure $SU(5)$ Yang-Mills fields
in the hidden sector. The latter case could be obtained either from
a similar condensation of pure $SU(3)$ Yang-Mills fields or from the
condensation of an $E_6$ hidden sector gauge group with 9
$\mathbf{27}$'s condensing in the hidden sector.\footnote{This is
precisely the case marked with the circle in
Figure~\ref{fig:varyall}.} The first case is just on the edge of
where realistic gravitino masses can be obtained from some pair of
values for $\lbr \caleff ,\bpaleff\rbr$. It corresponds to a
condensing group beta-function coefficient of $b_{+}= 12/8\pi^2
\simeq 0.152$, which could result from a hidden sector condensation
of pure $E_6$ Yang-Mills fields.

\begin{table}[h]
\tbl{Sample Spectra for benchmark points. All masses are in GeV.}
{\begin{tabular}{@{}lcccc@{}}
\toprule Point & A & B & C & mSUGRA \\ \colrule
$\tan\beta$ & 10 & 5 & 5 & 10 \\ \colrule
%
%
%
%
$m_{\chi^0_1}$ & 78 & 93 & 91 & 98 \\
$m_{\chi^0_{2}}$ & 122 & 132 & 110 & 182 \\
$m_{\chi_{1}^{\pm}}$ & 120 & 132 & 110 & 181 \\
$m_{\tilde{g}}$ & 471 & 427 & 329 & 582 \\
$\wtd{B} \; \% |_{\rm LSP}$ & 89.8 \% & 98.7 \% &  93.4 \% &
 99.9 \% \\
$\wtd{W}_{3} \% |_{\rm LSP}$ & 2.5 \% &  0.6 \% & 4.6 \% & 0.0 \%
\\ \colrule
$m_{h}$ & 114.3 & 114.5 & 116.4 & 112.0 \\
$m_{A}$ & 1507 & 3318 & 4400 & 381 \\
$m_{H}$ & 1510 & 3329 & 4417 & 382 \\
$\mu$ & 245 & 631 & 481 & 332 \\ \colrule
$m_{\tilde{t}_{1}}$ & 947 & 1909 & 2570 & 392 \\
$m_{\tilde{t}_{2}}$ & 1281 & 2639 & 3530 & 571 \\
$m_{\tilde{c}_{1}}$, $m_{\tilde{u}_{1}}$ & 1553 & 3254 & 4364 & 528
\\
$m_{\tilde{c}_{2}}$, $m_{\tilde{u}_{2}}$ & 1557 & 3260 & 4371 & 547
\\ \colrule
$m_{\tilde{b}_{1}}$ & 1282 & 2681 & 3614 & 501 \\
$m_{\tilde{b}_{2}}$ & 1540 & 3245 & 4353 & 528 \\
$m_{\tilde{s}_{1}}$, $m_{\tilde{d}_{1}}$ & 1552 & 3252 & 4362 & 527
\\
$m_{\tilde{s}_{2}}$, $m_{\tilde{d}_{2}}$ & 1560 & 3261 & 4372 & 553
\\ \colrule
$m_{\tilde{\tau}_{1}}$ & 1491 & 3199 & 4298 & 137 \\
$m_{\tilde{\tau}_{2}}$ & 1502 & 3207 & 4308 & 208 \\
$m_{\tilde{\mu}_{1}}$, $m_{\tilde{e}_{1}}$ & 1505 & 3207 & 4309 &
145 \\
$m_{\tilde{\mu}_{2}}$, $m_{\tilde{e}_{2}}$ & 1509 & 3211 & 4313 &
204
\\ \botrule
\end{tabular}}
\end{table}

From these values and the expressions in~(\ref{BGWsoft}) it is
possible to construct the entire superpartner spectrum, assuming the
MSSM field content and proper electroweak symmetry breaking. The
results are give in Table~1. For comparison we also give the
physical spectrum for a unified model of the minimal supergravity
type, for which the unified scalar mass is taken to be $m_0 = 100
\GeV$, the unified gaugino mass is taken to be $m_{1/2} = 250 \GeV$,
and the unified trilinear coupling $A_0$ is taken to vanish. This is
model point~B of Battaglia et al.\cite{Battaglia:2001zp}, which is
very nearly the Snowmass point~1A.\cite{Allanach:2002nj} A few
initial comments are in order. First we note that the
$\mu$-parameter for the three BGW~points is quite small relative to
typical scalar masses. This is the focus-point effect at work,
alluded to earlier. Note also the departure of these models from the
typical mSUGRA relation $m_{\chi^0_1} \simeq 0.5 m_{\chi^{\pm}_1}$,
with the significant $W$-ino component of the LSP (indicated by the
size of $\wtd{W}_{3} \% |_{\rm LSP}$ versus $\wtd{B} \; \% |_{\rm
LSP}$). Finally, the three BGW~points have scalars which are much
larger than those mSUGRA models (such as Snowmass point~1A) which
are typically used for collider studies. This impacts on both the
production of superpartners as well as the branching fractions of
these superpartners to ``typical'' SUSY-indicating final states.

\begin{figure}
\centerline{\psfig{file=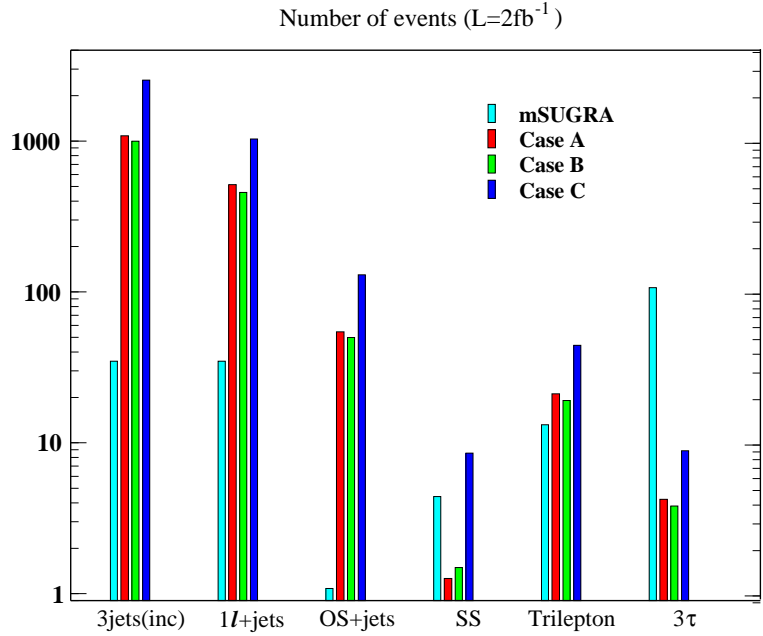,width=10cm}} \vspace*{8pt}
\caption{\textbf{Number of superpartner events of different
signatures for models of Table~1 at the Tevatron.} These numbers are
based on counting topologies from {\tt PYTHIA} at the parton level
with no kinematic or geometric cuts. Note the logarithmic scale in
numbers of events. Descriptions of each signature type are given in
the text.} \label{fig:signatures}
\end{figure}

Figure~\ref{fig:signatures} shows naive estimates of numbers of
events of various signatures in 2 fb$^{-1}$ integrated luminosity at
Tevatron center-of-mass energies for the models of Table~1.
The signature of these models was calculated using {\tt
PYTHIA},\cite{Sjostrand:2006za} but only at the generator level. No
geometric or kinematic cuts, no triggering efficiencies are applied,
no jet clustering is performed, tau leptons are not decayed, etc.
The event numbers are only meant to illustrate the generic features
of each model and demonstrate the experimental challenges. All
events have missing transverse energy as we assume an intact
R-parity. The signature set is as follows
\begin{enumerate}
\item Inclusive multi-jets $n_{\rm jets}\ge 3$,
\item One lepton plus $n_{\rm jets}\ge 2$,
\item Opposite sign (OS) dileptons plus $n_{\rm jets}\ge 2$,
\item Same-sign (SS) dileptons plus any number of jets,
\item Trileptons plus any number of jets,
\item Three taus plus any number of jets [before decaying the taus].
\end{enumerate}
The very large number of jet events with missing energy, relative to
the Snowmass benchmark, is the result of the much lighter gluino for
this model. The signatures are most pronounced for point~C with the
lightest gluino mass. In general, the SUSY signature space is
dominated by gluino production and decay for the BGW~class of
models.\cite{Nelson:2003tk}

This is only the beginning of a meaningful analysis. A study of the
backgrounds must of course be done to be sure any given channel is
detectable, but models with hundreds of events are presumably
detectable for the first two signatures, and models with tens of
events for the rest. The same-sign dilepton channel has smaller
backgrounds: even a handful of clean events may constitute a signal.
Furthermore, despite the fact that the event numbers in
Figure~\ref{fig:signatures} are based on unsophisticated estimates,
taking ratios of such inclusive signatures should be robust under
more detailed analyses. For example, the ratio of OS dilepton + jets
events to trilepton events should be roughly $5:2$ in the BGW model,
independent of the other parameters, as a result of the predominance
of gluino pair production over squark production.\cite{Kane:2002qp}

The potential utility of using combinations of inclusive observables
to separate classes of models was explored in
Ref.~\refcite{Binetruy:2003cy}, where point~C of~(\ref{benchC}) was
studied in much greater detail for the LHC environment. Consider the
situation after 10 fb$^{-1}$ of integrated luminosity ({\em i.e.}
one year at $10^{33} {\rm cm}^{-2} {\rm sec} ^{-1}$) at the LHC.
Looking at just the most inclusive signatures -- those of
Figure~\ref{fig:signatures} -- plus crude kinematic information, can
separate the K\"ahler-stabilized models from alternative
SUSY-breaking paradigms. For example, in Table~2 we collect the
predictions for numbers of events with missing transverse energy in
excess of 100 GeV, at least two jets (each with transverse energy
above 100 GeV) and various final state topologies for Standard Model
processes as well as various paradigms of new supersymmetric
physics. We also include the peak in the effective mass
distribution, where $m_{\rm eff}$ is defined as
\begin{equation}
m_{\rm eff} = E^T_{\rm miss} + \sum_{\rm jets} E^T_{\rm jet} \, .
\label{Meffdef} \end{equation}
In the table, model~A is the Snowmass point SPS~1A (only slightly
modified from the unified model in Figure~\ref{fig:signatures}) with
parameter set $m_0 = 200 \GeV$, $m_{1/2} = 250 \GeV$, $A_0 = -800
\GeV$, $\tan\beta=10$ and positive $\mu$. Model~B is also a unified
model, but in this case with very heavy scalars to achieve a ``focus
point'' model similar in nature to the BGW class. The parameter set
for this model is $m_0 = 2150 \GeV$, $m_{1/2} = 300 \GeV$, $A_0 = 0
\GeV$, $\tan\beta=10$ and positive $\mu$. Model~D is a
strongly-coupled heterotic string model without
$D_5$-branes,\cite{Horava:1995qa,Witten:1996mz,Horava:1996ma,Lukas:1997fg,Choi:1997cm}
which we include here for comparison to our K\"ahler stabilized
model. More examples were considered in the original paper of
Reference~\refcite{Binetruy:2003cy}.

\begin{table}[h]
\tbl{Number of events in excess of the Standard Model prediction for
different signatures. For each channel the Standard Model baseline
is given in the first column. Subsequent columns give the excess
beyond this baseline for selected models described in the text. }
{\begin{tabular}{@{}lccccc@{}}
\toprule Channel & SM & A & B & BGW Point~C & D \\
\colrule Jets ($\times 10^3$) & 100.0 & 59.5 & 0.7 & 31.7& 6.6
\\
$1\ell$ ($\times 10^3$) & 13.0 & 17.1 & 0.5 & 7.3 & 1.7 \\
OS ($\times 10^3$) & 7.0 & 5.7 & 0.2 & 2.0 & 0.6 \\
SS & 20 & 1332 & 99 & 504& 160 \\
$3\ell$ & 60 & 737 & 97 & 204 & 77 \\
$m_{\rm eff}^{\rm peak}$ (GeV) & - & 812 & 1140 & 838 & 1210 \\
\botrule
\end{tabular}}
\end{table}

We note that each model can be ``discovered'' above the Standard
Model background estimation in at least one channel (assuming a very
naive estimate of $\sqrt{N_{\SM}}$ for a measure of the approximate
experimental error in the observations). What is more, the models
can be distinguished from one another -- this is particularly
important for using models of string physics as interpreters of
possible LHC signals. The ability of inclusive signatures to
distinguish the K\"ahler-stabilized heterotic string model from
other SUSY models was tested using global fits to simulated
data.\cite{Binetruy:2003cy}. For example, the minimal supergravity
point given by
\begin{equation}
\tan\beta=10 \quad m_{1/2}=380 \quad m_0=600 \quad A_0 =0 \quad {\rm
sgn}(\mu)>0.
\label{sugrapoint} \end{equation}
was used to simulate 50,000 events for LHC center-of-mass energies
using {\tt PYTHIA}. The inclusive observables of Table~2 were then
computed. Also computed was the SUSY contribution to the anomalous
magnetic moment of the muon $a_{\mu}^{\SUSY}$ and to the rate for
$b\to s \gamma$ processes. In an effort to fit the resulting
``data,'' the same simulation was performed on an ensemble of
minimal supergravity models with $A_0 = 0$ but varying the
parameters $\tan\beta$, $m_0$ and $m_{1/2}$. For each resulting set
of collider + indirect observables, a $\chi^2$-fit was performed to
determine how well the test point reproduced the target data. Not
surprisingly, the best-fit $\chi^2$ corresponded to a point very
close to~(\ref{sugrapoint}) in the ensemble, namely the case
$\tan\beta=10$, $m_{1/2}=380 \GeV$ and $m_0=500 \GeV$ with a minimum
chi-squared of $(\chi^2)_{\rm min} = 1.7$ with three degrees of
freedom. But would another model fit this data equally well?

When an ensemble of experiments from the BGW model was constructed,
by simulating 50,000 events at each point in a mesh over the
three-dimensional parameter space defined by $\lbr \tan\beta,
m_{3/2}, b_+ \rbr$, the resulting best fit point had the parameters
$\lbr \tan\beta, m_{3/2}, b_+ \rbr$ = $\lbr 5\, , 2750 \GeV \, ,
b_{+} = 8/8\pi^2 \rbr$ which corresponds to $\(\chi^2_T\)_m = 2.8$
with five degrees of freedom. It would appear that the nonuniversal
model is doing a fairly adequate job of reproducing the inclusive
signatures! But when additional kinematic data is taken into account
the two can clearly be distinguished. For example, when we include
the peak in the $m_{\rm eff}$ distribution and the peak in the
invariant mass distribution $m_{\ell \ell}$ of opposite-sign
dilepton events we obtain $\lbr m_{\rm eff}^{\rm peak}, m_{\ell
\ell}^{\rm peak}\rbr = \lbr 1360 \GeV, 92 \GeV \rbr$ for the true
``data,'' which was well reproduced by the best-fit minimal
supergravity point. In contrast, the best-fit K\"ahler-stabilization
model point produced $\lbr m_{\rm eff}^{\rm peak}, m_{\ell
\ell}^{\rm peak}\rbr = \lbr 987 \GeV, 58 \GeV \rbr$. Even allowing
for relatively large uncertainties in the measurement of these
quantities these are without question measurable differences.

Is it reasonable to ask how well do the above results hold when
trying to distinguish two models across their entire parameter space
(as opposed to trying to fit a single point)? Doing so requires
expanding the list of observables to consider to include additional
kinematic information and asymmetries in the LHC event rates. Kane
et al.\cite{Kane:2006yi} have considered the parameter space of
several prominent string constructions that have been studied in the
literature, including the K\"ahler stabilization model reviewed
here. They conclude that models of the BGW~class can be
distinguished from the other string-based effective theories over
most of its parameter space. The key observables are
\begin{enumerate}
\item The number of ``clean'' multi-lepton events
\item The total rate of dilepton events with at least two jets
\item The fraction of dilepton + jet events for which at least one of the
jets is tagged as a b-jet
\item The charge asymmetry in events in which there is a single
high-$p_T$ lepton
\item The number of events with a single (hadronically-decaying) tau
\item The peak in the missing $E_T$ distribution over all events
with at least 100 GeV of missing $E_T$
\end{enumerate}

Let us consider just a couple of examples of how these observables
help to distinguish this class of models from others. Events with
high-$p_T$ leptons but with no jets over 100 GeV in transverse
energy are called clean events. They arise from direct production of
neutralino/chargino pairs, which certainly has a sizable
cross-section in the BGW class. However, the mass difference between
the lightest chargino and lightest neutralino is typically small.
The resulting leptonic decay products therefore tend to be soft and
thus the number of clean leptonic events will be very sensitive to
the $p_T$ threshold demanded of the leptons. To be included in the
data sample in the study of Kane et al. a minimum $p_T$ of 10 GeV
was demanded for electrons and muons. Therefore there are
essentially no clean dilepton/trilepton events in the BGW model when
LHC-motivated cuts are used.

Associated production of squarks with gluinos can give rise to a
charge asymmetry in events with a single high-$p_T$ lepton. At the
LHC, the produced squark is most likely to be an up-type squark as
the initial state is asymmetric between up and down quarks. These
squarks will decay preferentially to a positive chargino, which then
decays to a positively charged lepton and missing energy. When
squarks are relatively light one therefore expects to see more
events with jets and a single high-$p_T$ lepton as well as an
asymmetry in favor of positive charges for these leptons. The heavy
squarks of the BGW~model reduce this production rate and give rise
to fewer events and almost no asymmetry. By combining observations
such as these it should be possible to separate models from one
another, as well as eliminate classes of models when confronting
them with actual signals at the LHC.

\subsection{Dark matter} \label{dmatter}
Another important experimental arena in which nonuniversal gaugino
masses can play a significant role is in the thermal production of
cold dark matter in the form of stable LSP
neutralinos.\cite{Weinberg:1982tp,Goldberg:1983nd} One of the prime
virtues of supersymmetry as an explanation of the hierarchy problem
is that it tends to also provide a solution to the dark matter
problem as a nearly automatic consequence of R-parity conservation.
But models in which gaugino masses are universal at some high scale
tend to predict too large a relic neutralino density for generic
parameter choices.\cite{Ellis:2001zk,Baer:2002gm} The reason is not
hard to understand. For neutralinos of a mass of approximately $100
\GeV$, the thermally produced density of particles after reheating
is far larger than that required to account for the nonbaryonic dark
matter. But as the universe cools relic neutralinos must find one
another and annihilate into light fields (such as leptons) before
the rate of annihilation falls below the expansion rate of the
cosmos. Typically, weak-scale interaction rates are very close to
the needed annihilation rate as a first approximation. But on closer
examination it becomes clear that annihilation of neutralinos into
leptons will only be efficient if diagrams involving t-channel
exchange of sleptons contribute significantly to the total rate. For
models with slepton masses much above $100 \GeV$ this rate quickly
drops, particularly when the LSP is predominantly $B$-ino like,
resulting in far too much relic density at freeze-out.

\begin{figure}
\centerline{\psfig{file=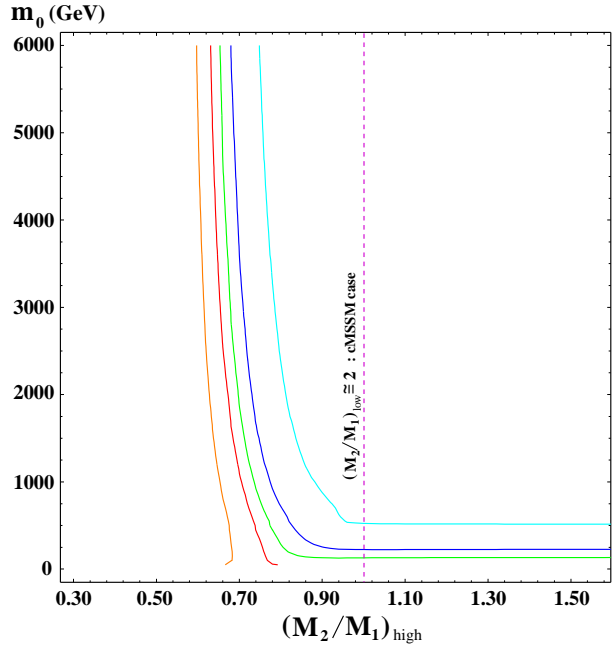,width=8cm}} \vspace*{8pt}
\caption{\textbf{Preferred dark matter region for nonuniversal
gaugino masses.} Contours of $\Omega_{\chi} {\rm h}^2$ of 0.01, 0.1,
0.3, 1.0 and 10.0 from left to right, respectively, are given as a
function of the ratio of $SU(2)$ to $U(1)$ gaugino masses
$M_{2}/M_{1}$ at the high scale. The unified model is recovered
where the two masses are equal at the high scale, as has been
indicated by the dashed line. In this plot we have set $M_3 =
m_{1/2} = 200 \GeV$.} \label{fig:E4plot}
\end{figure}

In the BGW~class of models all scalar fields are very massive, so
{\em a priori} we would expect a very significant problem for the
relic neutralino density. But the rather simplified description from
the previous paragraph is remarkably sensitive to the wave-function
of the LSP. Let us adopt the following parameterization for this
wave-function
\begin{equation}
\chi^{0}_{1} = N_{11} \tilde{B} + N_{12} \tilde{W} + N_{13}
\tilde{H}^{0}_{d} + N_{14} \tilde{H}^{0}_{u} \;, \label{LSPcontent}
\end{equation}
which is normalized to $N_{11}^2+N_{12}^2+N_{13}^2+N_{14}^2=1$. In
unified models (with low to moderate $\tan\beta$) the LSP is
overwhelmingly $B$ino-like, which is to say that $N_{11} \simeq 1$.
If we relax the GUT relationship between the gaugino masses but
still remain in the large $|\mu|$ limit (low $\tan\beta$) then we
will continue to have a predominantly {\em gaugino-like} LSP
($N_{11}^2 + N_{12}^2 \simeq 1$) with the relative values of
$N_{11}$ and $N_{12}$ governed by the relative values of $M_{1}$ and
$M_{2}$. Decreasing $M_{2}$ relative to $M_{1}$ at the electroweak
scale increases the wino content of the LSP until ultimately $M_{1}
\gg M_{2}$ and $N_{11}\simeq 0$, $N_{12} \simeq1$. The $B$-ino
component of the neutralino couples with a U(1) gauge strength
whereas the wino component couples with the larger SU(2) gauge
strength, thus enhancing its annihilation cross section and thereby
lowering its relic density. As $N_{12}$ is increased more SUSY
parameter space should open up for correct dark matter abundance
until eventually annihilation becomes too efficient and we are left
with no neutralino dark matter at all.

\begin{figure}
\centerline{\psfig{file=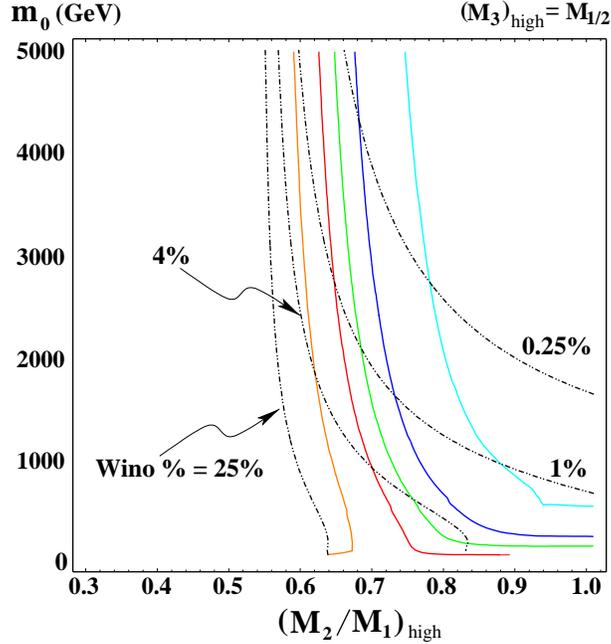,width=8cm}} \vspace*{8pt}
\caption{\textbf{$W$-ino content of the LSP for cosmologically
preferred regions.} Contours of constant relic density are given by
the solid lines for $\Omega_{\chi} {\rm h}^2 =$ 0.01, 0.1, 0.3, 1,
and 10 from left to right.  The dotted lines are curves of constant
wino content for $N_{12}^{2}$ = 0.25, 0.04, 0.01 and 0.025 from left
to right.} \label{fig:Eplots1}
\end{figure}

Another important effect of the deviation of gaugino masses from GUT
relations is the increasing importance of co-annihilations between
the LSP and other light neutralinos and charginos in the early
universe. In unified models like minimal supergravity
co-annihilation is only important in limited regions of parameter
space -- through resonant annihilation diagrams involving the CP-odd
neutral Higgs or from co-annihilation diagrams involving the
stau.\cite{Edsjo:1997bg,Ellis:1999mm,Baer:2002fv} But in the
K\"ahler-stabilized models the mass difference between the lightest
neutralino and the next-to-lightest neutral gaugino, or lightest
chargino, can be relatively small. This was demonstrated in the
plots of Figure~\ref{fig:gauginosec}. When mass differences between
gauginos reach a few GeV, additional coannihilation processes become
efficient at removing relic LSPs from the early universe. Such
processes include $\chi^{\pm}_{1} \chi^{0}_{1} \rightarrow f f'$
(such as $e^{\pm} \nu_{e}$), $W^{\pm} \gamma$, and $W^{\pm} Z$.

These additional processes were studied in the context of
nonuniversal gaugino masses in
References~\refcite{Birkedal-Hansen:2001is}
and~\refcite{Birkedal-Hansen:2002am}. There it was found that for
particular special ratios of the soft Lagrangian parameters
$M_2/M_1$ at the high-energy input scale the relic abundance of
neutralinos achieves the observationally preferred values of
$\Omega_{\chi} {\rm h}^2 \simeq 0.1-0.2$ {\em independent of the
scalar fermion masses}. More specifically, we fix an overall mass
scale for the gauginos by a value of $m_{1/2} \equiv {\rm min}(M_1,
M_2)$ and then allow the larger to vary according to the ratio
$(M_2/M_1)$. Once the gluino mass is determined in relation to
$m_{1/2}$ the parameters at the low-energy scale can be found
through RG evolution. The preferred ratio of $(M_2/M_1)$ at the high
scale for the resulting relic density of the LSP is then only a mild
function of the value of the gluino mass relative to $m_{1/2}$.

The results of scanning over a range of values $\(M_{2}/M_{1}\)_{\rm
high}$ at the high scale is given in Figure~\ref{fig:E4plot}, where
contours of constant relic density $\Omega_{\chi} {\rm h}^2$ are
given. Cosmological observations therefore suggest a preferred value
of this ratio in the range $0.6 \leq \(M_{2}/M_{1}\)_{\rm high} \leq
0.85$. In Figure~\ref{fig:Eplots1} we overlay the contours of
constant $W$-ino fraction in the LSP wavefunction. The impact of
additional coannihilation channels and increased annihilation rates
is felt even at relatively small admixtures of $SU(2)$ gaugino to
the predominantly $B$-ino LSP.

\begin{figure}
\centerline{\psfig{file=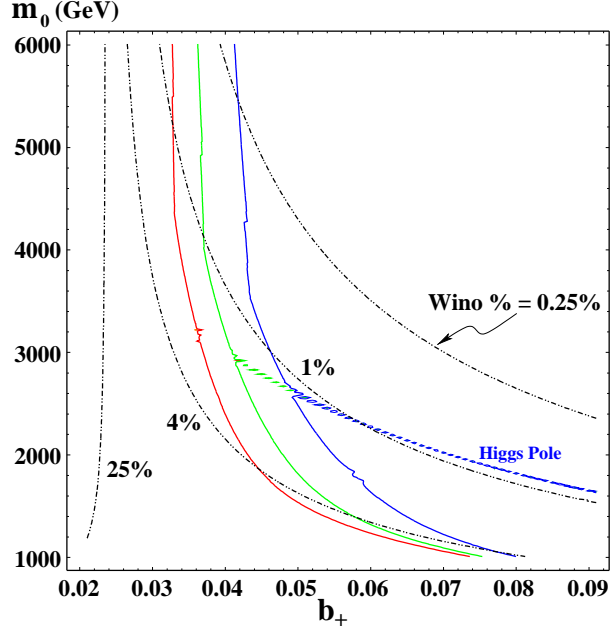,width=8cm}} \vspace*{8pt}
\caption{\textbf{Preferred dark matter region in the BGW model.}
Contours of constant relic density are given as a function of $M_{0}
= m_{3/2}$ and $b_{+}$ by the solid lines. Moving outward from the
left are contours of $\Omega_{\chi} {\rm h}^2 =$0.1, 0.3 and 1.0.
The dotted lines give the value of the LSP wino content (25\%, 4\%,
1\%, 0.25\% from left to right).} \label{fig:C2plot2}
\end{figure}

In the BGW~class of models this ratio is determined
through~(\ref{BGWsoft}) by the beta-function coefficients of the
Standard Model electroweak gauge groups, the beta-function
coefficient $b_+$ of the condensing gauge group, and the dilaton
\vev
\begin{equation}
\frac{M_{2}\(\mu_{\UV}\)}{M_{1}\(\mu_{\UV}\)} =
\frac{g_{2}^{2}\(\mu_{\UV}\)}{g_{1}^{2}
  \(\mu_{\UV}\)}\frac{\(1+b'_{2} \ell\)-(b_{2}/b_{+}) \(1+b_{+}
  \ell\)}{\(1+b'_{1} \ell\)-(b_{1}/b_{+})\(1+b_{+} \ell\)} \, .
\label{eq:bgwratio}
\end{equation}
It is therefore possible to map the results of
Figure~\ref{fig:Eplots1} directly onto the parameter space of this
class of models. The result is shown in Figure~\ref{fig:C2plot2}.
The feature along the $\wtd{W}_{3} \% |_{\rm LSP} = 1\%$ contour is
the Higgs annihilation resonance. The recent data from the WMAP
experiment favors a nonbaryonic dark matter relic density of
$\Omega_{\chi} {\rm h}^2 \simeq 0.127$,\cite{Spergel:2006hy} which
translates into an $\order(1\%)$ $W$-ino fraction in the LSP
wave-function for large scalar masses. This is a natural outcome of
the K\"ahler-stabilization mechanism for reasonable hidden sector
configurations.

\begin{figure}
\centerline{\psfig{file=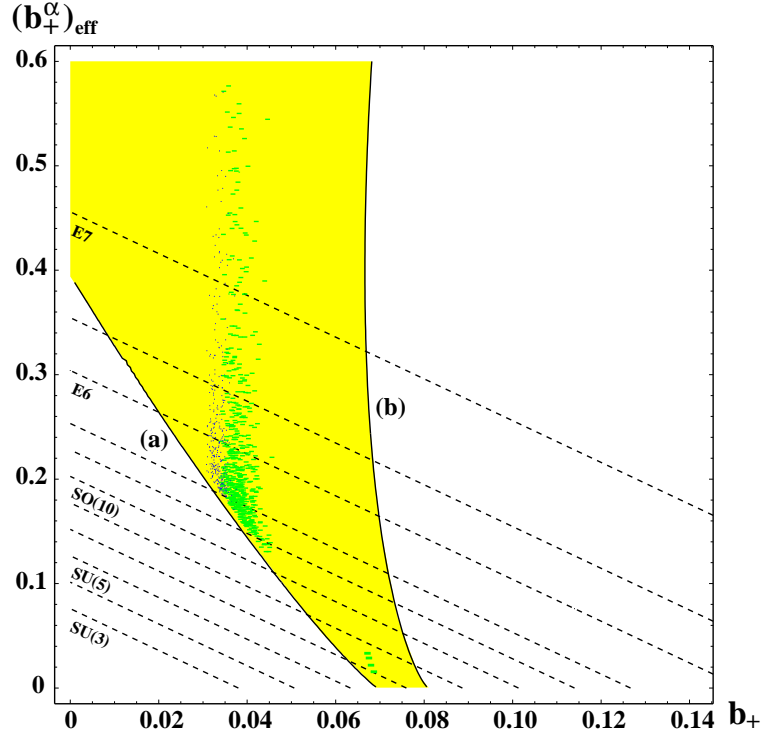,width=10cm}} \vspace*{8pt}
\caption{\textbf{Preferred dark matter region in hidden sector
configuration space.} This plot illustrates the dark matter
parameter space in terms of the gauge group and matter content
parameters of the hidden sector.  The fine points on the left have
the preferred value $0.1 \leq \Omega_{\chi} {\rm h}^2 \leq 0.3$ and
the coarse points have $0.3 < \Omega_{\chi}{\rm h}^2 \leq 1.0$. The
swath bounded by lines (a) and (b) is the region in which the $0.1
\leq \caleff \leq 10$ and the gravitino mass is between 100 GeV and
10 TeV. The dotted lines are the possible combination of gauge
parameters for different hidden sector gauge groups.}
\label{fig:c1plot}
\end{figure}

In Figure~\ref{fig:c1plot} we reproduce the parameter space of
Figure~\ref{fig:varyall}, highlighting the region in which the
Yukawa parameter takes the values $0.1 \leq \caleff \leq 10$ and the
resulting gravitino mass is between 100 GeV and 10 TeV. Achieving
the proper LSP thermal relic density singles out a very specific
subset of the model space. The dotted points in
Figure~\ref{fig:c1plot} represent the subset of parameter choices in
a scan over 25,000 possible combinations of $\{b_{+}, \baaleff,
\caleff\}$ which (a) give rise to gravitino masses between 100~GeV
and 10~TeV, (b) yield a particle spectrum consistent with
experimental bounds, and (c) yield a realistic relic density. Those
that give rise to the WMAP value for $\Omega_{\chi} {\rm h}^2$ tend
to cluster around the value of $b_+ \simeq 3/8\pi^2$ -- the
benchmark point of~(\ref{benchC}) used in the collider studies. We
will see below that this same value of $b_+$ has intriguing
implications for the axion sector of the model as well.

\subsection{The axion sector} \label{axion}
Banks and Dine\cite{Banks:1994sg} pointed out some time ago that in
a supersymmetric Yang Mills theory with a dilaton chiral superfield
that couples universally to Yang-Mills fields
\beq \Lag_{YM} = {1\over8}\sum_a\int \diff^2\theta
\,S(W^{\alpha}W_{\alpha})_a+\hc , \label{bd}\eeq
there is a residual R-symmetry in the effective theory for the
condensates of a strongly coupled gauge sector, provided that (a)
there is a single condensation scale governed by a single
$\beta$-function, (b) there is no explicit R-symmetry breaking in
the strongly coupled sector, (c) the dilaton $S$ has no
superpotential couplings, and (d) the K\"ahler potential is
independent of $\im\, S$. The latter two requirements are met in
effective supergravity obtained from the weakly-coupled heterotic
string, and explicit realizations of this scenario are found in the
BGW~model described in Sections~\ref{sec2} and~\ref{sec3}, and
generalizations thereof to include an anomalous $U(1)_X$ that are
discussed in Section~\ref{sec4}.

The R-symmetry transformations on the gauginos $\lambda_a$ and
chiral fermions $\chi^A$
\beq \lambda_a\to e^{{i\over2}\alpha}\lambda_a, \qquad \chi^A\to
e^{-{i\over2}\alpha}\chi^A,\label{rferm}\eeq
leave the classical Lagrangian \myref{bd} invariant, but are
anomalous at the quantum level:
\beq \Delta\Lag_{YM} = {i\alpha\over8}\sum_a b'_a\int
d^2\theta(W^{\alpha}W_{\alpha})_a+\hc\, , \label{delbd}\eeq
with $b'_a$ given in~(\ref{cond}). In the case that there is a
single simple condensing gauge group $\mathcal{G}_c$, the symmetry
can be restored by an axion shift
\beq a = \im\, s \to a - ib'_c\alpha\, .\eeq
If this gauge group becomes strongly coupled at a scale
\beq \Lambda_c \sim e^{-1/3b_c g^2_0}\Lambda_0 \, , \label{lambda}
\eeq
with $b_a$ as defined in~(\ref{ba}),
then the effective
theory\cite{Veneziano:1982ah,Taylor:1982bp,Lust:1990zi,Taylor:1985fz}
below that scale will have the same anomaly structure as the
underlying theory.  A potential is generated for the dilaton $\re \,
s$, but not for the axion.

If the gauge group is not simple $\mathcal{G} = \prod_a
\mathcal{G}_a$, the R-symmetry is anomalous but no mass is generated
for the axion as long as there is a single condensate.  In the
two-condensate case with $\beta$-functions $b_2\ll b_c$ in dilaton
stabilization models\cite{Binetruy:1996nx,Gaillard:2003gt} the axion
acquires a small mass
\beq m_a \sim (\Lambda_2/\Lambda_c)^{3\over2}m_{3\over2}\, .
\label{ma}\eeq
In the context of the BGW~models, we saw in Sections~\ref{sec2}
and~\ref{sec3} that a viable scenario for supersymmetry breaking
occurs if a hidden sector gauge group $\mathcal{G}_c$ condenses with
$b_c\approx .036$, $\Lambda_c\sim 10^{13} \GeV$ and $m_{3/2}\sim$
TeV. Then if there is no additional condensing gauge group other
than QCD, the universal axion is a candidate Peccei-Quinn axion. The
result~(\ref{ma}) would suggest an axion mass
\beq m_a \sim 10^{-9}{\rm eV}\, .\label{ma2}\eeq
However, the result~\myref{ma} cannot be directly applied to the QCD
axion since QCD condensation occurs far below the scale of
supersymmetry breaking and heavy modes need to be correctly
integrated out.  Moreover, the result~\myref{ma} was obtained under
the assumptions that in the hidden sector there are no additional
spontaneously-broken nonanomalous symmetries such as the chiral
flavor $SU(N)$ of QCD, and no gauge invariant dimension-two
operators such as quark mass terms.  Under these assumptions
two-condensate hidden sector models have a point of enhanced
symmetry where the condensing gauge sectors $\mathcal{G}_a$ have the
same beta-function coefficients $b_a$, and the axion mass is
proportional to $|b_1 - b_2|$.

To investigate whether the string axion can be the QCD axion, we
need to consider the case where the second condensing gauge group
$\mathcal{G}_Q$ is $SU(N_c)$ with a $U(N)$ flavor symmetry for quark
supermultiplets. In this case the point of enhanced symmetry occurs
for\cite{Gaillard:2005gj}
\beq b_c = {N_c\over8\pi^2}.\label{sympt}\eeq
As a result, the standard relation between the axion mass and its
Planck scale coupling constant is modified in this class of models
due to a contribution to the axion-gluon coupling that appears below
the scale of supersymmetry breaking when gluinos are integrated out.
Put differently, {\em the axion coupling constant is different above
and below the scale of supersymmetry breaking}.  The axion mass
vanishes at the point of enhanced symmetry; for QCD with $N_c = 3$,
this occurs for
\beq b_c = {3\over8\pi^2} = .038,\eeq
which is again in the preferred range $.3\le b_c\le .4$ found
earlier. As a consequence, the axion mass is suppressed and higher
dimension operators\cite{Banks:1994sg,Butter:2005wr} might lead to
strong CP violation.  If on the other hand the string axion acquires
a mass as in~\myref{ma} from multiple hidden sector condensates, it
cannot be the QCD axion. The possibility of detecting both types of
axions has been analyzed\cite{Kain:2005ic} following the procedures
developed by Fox, Pierce and Scott\cite{Fox:2004kb} for the case of
the string axion with the standard relation\cite{Srednicki:1985xd}
between its mass and its string-scale coupling.  In all of these
cases the string-scale axion coupling, which in these models is of
the order of the reduced Planck mass, is outside the conventional
cosmological bound.\cite{Preskill:1982cy,Abbott:1982af,Dine:1982ah}
Yet this can be evaded\cite{Fox:2004kb} by reducing relic axion
production with a sufficiently small initial misalignment angle
and/or late entropy release.  More specifically, if the classical
dilaton K\"ahler potential is used, the axion coupling constant is
approximately\cite{Fox:2004kb} $F_a \simeq 10^{16}$ GeV. When string
nonperturbative corrections are invoked to stabilize the dilaton,
they have the effect of enhancing both the dilaton mass and the
axion coupling constant:\footnote{In Ref.~\refcite{Binetruy:1997vr}
it was incorrectly stated that the axion coupling constant was
suppressed by these effects.} $F_a \sim
(1/\sqrt{3}b_c)\times10^{16}$ GeV $\sim 10^{17}$ GeV.

To understand the origin of the enhanced symmetry points, first
consider a supersymmetric model with condensing gauge group
$\mathcal{G}_c\otimes\mathcal{G}_Q$.  Although this is not a
realistic model for QCD, it has the advantages that all the
symmetries are manifest and the effective Lagrangian is highly
constrained by supersymmetry. In the absence of a superpotential
there is a classical [$U(1)$]${}^3$ symmetry defined
by~\myref{rferm}, the chiral $U(1)$ transformations
\beq Q\to e^{i\beta}Q,\qquad Q^c\to e^{i\beta}Q^c, \label{uone}\eeq
introduced in Section~\ref{sec12} and
\beq \Phi^A_c\to e^{i\gamma}\Phi^A_c\label{uonec},\eeq
all of which can be made nonanomalous by an axion shift.  The
superpotential\footnote{This may reflect a potential $W(\Phi)$ in
the classical theory and/or arise from nonperturbative QFT effects.}
$W(\Pi)$ for the hidden sector condensates $\Pi^\alpha_c$ that is
needed for condensation to occur breaks this symmetry down to
[$U(1)$]${}^2$, with the parameter $\gamma$ restricted such that
\beq \Pi^\alpha_c\to e^{i d^\alpha_c\gamma}\Pi^\alpha_c =
e^{i\alpha}\Pi^\alpha_c,\eeq
and the anomaly coefficient of the hidden sector condensate $U_a$ is
simply given by the $\beta$-function coefficient:
\beq b_c = b'_c + \sum_\alpha b^\alpha_c = {1\over8\pi^2}\(C_c -
\textstyle{1\over3}\sum_A C^A_c\),\eeq
for dimension-three matter condensates. If there is a second hidden
sector with the same condensate structure, the residual classical
symmetry is just $U(1)$, and the anomaly cannot be canceled by an
axion shift unless the beta functions are the same.  On the other
hand if the only other condensate is the SUSY QCD one and the quarks
are massless, there is a nonanomalous $U(1)$ defined
by~\myref{rferm} and~\myref{uone} with
\beq \alpha\, b_c = {1\over8\pi^2}\[\alpha\(N_c - N\) + 2\beta
N\]\label{nonan}.\eeq
If there are no massless quarks, flavor-chiral $U(1)$ symmetry is
broken, and there is no longer the freedom to choose the R-parity of
$Q$; in this case the classical R-symmetry has $\beta = \alpha/2$,
and it is anomalous at the quantum level unless
\beq b_c = {N_c\over8\pi^2} = {b'_Q N_c\over N_c -
N}.\label{sym}\eeq
The effective meson Lagrangian for this toy model has been worked
out explicitly.\cite{Gaillard:2005gj} The K\"ahler moduli sector is
essentially unchanged with respect to the BGW case, and we will
neglect it here.  In the absence of quark masses the scalars and the
flavor singlet pseudoscalar get masses of the order of the gravitino
mass, which is the same as for the BGW model if $|u_c|\gg|u_Q|$, and
the flavor adjoint pseudoscalars $\bfa$ as well as the axion $a$ are
massless.  When (flavor invariant and generally complex) quark
masses are turned on, these acquire masses in the ratio
\beq {m_a\over m_{\bfa}} \approx \sqrt{3v\over N}(8\pi^2 b_c -
N_c),\qquad v = \langle Q^c Q\rangle.\label{amass} \eeq
The axion mass vanishes at the symmetry point.

The above toy model is not a realistic model for QCD, but we can
modify it in several ways to make it more closely resemble the MSSM
while keeping manifest supersymmetry of the effective Lagrangian.
In the MSSM only $n<N$ chiral supermultiplets have masses below the
condensation scale $u_Q^{1/3}\sim\Lambda_Q$, while $m = N-n$ chiral
supermultiplets have masses $M_A$ above that scale. The latter
decouple at scales below their masses, which explicitly break the
nonanomalous $U(N)_L\otimes U(N)_R$ symmetry to a $U(n)_L\otimes
U(n)_R$ symmetry if $m = N - n$ quarks are massive. They do not
contribute to the chiral anomaly at the $SU(N_c)$ condensation
scale.  To account for these effects we replace $b'_N$
in~\myref{lvyt} by
\beq b'_n = (N_c - n)/8\pi^2, \label{defbn}\eeq
and replace the second term in~\myref{lvyt} by
\beq  {b^\alpha_Q\over8}\superint{E\over R}U_Q
\[\ln(\det\bpi_n) - \sum_{A=1}^m\ln
M_A\]+\hc,\label{lvyt3}\eeq
where $\bpi_n$ is an $n\times n$ matrix-valued composite operator
constructed only from light quarks. This result can be formally
obtained from~\myref{lvyt} by integrating out the heavy quark
condensates as follows. As the threshold $M_A$ is crossed, set
$\det\bpi_{n+A}\to \pi^A\det\bpi_{n + A -1}$ and take the condensate
$\pi^A \sim Q^A Q_A^c$ to be static: $K(\bpi_{n+A})\to K(\bpi_{n + A
-1})$. Then including the superpotential term $W(\pi^A) = -
M_A\pi^A$, the equation of motion for $F^A$ gives $\pi^A =
e^{-K/2}u_Q/32\pi^2M_A$, giving~\myref{defbn} and~\myref{lvyt3} up
to some constant threshold corrections.  The flat SUSY analog
of~\myref{wvy} is now
\beq W = {1\over4}U_Q\lbr g_0^{-2} + b'_n\ln(U_Q) +
b^\alpha_Q\[\ln(\det\bpi) - \sum_{A=1}^m\ln
M_A\]\rbr,\label{wvy2}\eeq
and we recover~\myref{susyv}--\myref{wnp} with now
\beq \Lambda_Q = e^{-1/3b_n g^{2}}\prod_{A=1}^m M_A^{b_3/3b_n},
\qquad 3b_n = {3N_c - n\over8\pi^2} = 3b'_n + 2n\, b_3\, ,\eeq
which corresponds to running $g^{-2}(\mu)$ from $g^{-2}(1)=
g_0^{-2}$ to $g^{-2}(\Lambda_Q)= 0$ using the $\beta$-function
coefficient $(3N_c - n - A)/8\pi^2$ for $m_{A}\le \mu\le m_{A+1}$,
again in agreement with the results of nonperturbative flat SUSY
analyses.\cite{Davis:1983mz}

The anomaly matching conditions in the toy model correctly
reproduce\cite{Binetruy:1996nx} the running of $g^{-2}$ from the
string scale to the condensation scale provided supersymmetry is
unbroken above that scale.  This is not the case for QCD, and the
effective ``QCD'' Lagrangian~\myref{lvyt} or~\myref{lvyt3} is not
valid below the scale $\Lambda_c$ of supersymmetry breaking.  The
arguments of the logs are effective infra-red cut-offs. For gauginos
and squarks, they should be replaced by the actual masses, as was
done in~\myref{lvyt3} for quark supermultiplets with masses above
the QCD condensation scale. The gaugino and squark mass terms are
given by
\beq \Lag_{\rm mass} = - \half {|m_{3/2}|\over F_a^2 b_c}
\(e^{i\omega_c}\bar\lambda_R\lambda_L + \hc\) -
m_{3/2}^2|\tilde{q}|^2,\label{tgtqmass}\eeq
where $F_a$ is the axion coupling constant
\beq \frac{1}{F_a^2} = 2\langle K_{s\bar s}\rangle \approx
{3\over2}b^2_c,\eeq
as defined by its coupling to the gauge fields above the
condensation scales
\beq \Lag\ni - {a\over4F_a}\sum_a(F\cdot\wtd{F})_a \, .
\label{alag}\eeq
The phase $\omega_c$ of the static condensate $u_c$
in~(\ref{tgtqmass}) is determined by the equations of motion in
terms of the axion
\beq \omega_c = - {a\over F_a b_c} + \phi \, , \eeq
where the constant phase $\phi$ includes the phases of the quark
mass and of the meson condensate.  The mass terms~\myref{tgtqmass}
are invariant under~\myref{rferm} which is spontaneously broken by
the vacuum value $u_c\ne0$, but remains an exact (nonlinearly
realized) symmetry of the Lagrangian, since the anomaly can be
canceled by an axion shift as long as QCD nonperturbative effects
can be neglected. Since $U_c$ transforms the same way as $U_Q$, an
effective theory with the correct anomaly structure
under~\myref{rferm} and~\myref{uone} is obtained by
using~\myref{lvyt3} and replacing the first term of~\myref{lvyt} by
\beq  {1\over8}\superint{E\over R}U_Q\[ b_1\ln(e^{-K/2}U_Q) +
b_2\ln(e^{-K/2}U_c)\] + \hc,\label{lvyt4}\eeq
provided
\beq b_3 = {1\over8\pi^2} = b^\alpha_Q\, , \qquad b_1 + b_2 = {N_c -
n\over8\pi^2} = b'_n \, ,\label{newb}\eeq
and we can choose $b_1$ and $b_2$ to better reflect the correct
infrared cut-offs for squarks and gauginos.  The potential is
modified, but its qualitative features are the same; in particular
the axion mass is unchanged since it depends only on $b_1 + b_2 =
b'_n$.
The scalar components of the composite chiral superfields $U_Q$ and
$\bpi$ are composed of gauginos and squarks that get large masses
proportional to $m_{3/2}$, while the true light degrees of freedom
are the quarks and gauge bosons.  The corresponding composite
operators are the $F$-components $F_Q,\, \bff$ of $U_Q,\, \bpi$;
these were eliminated by their equations of motion to obtain an
effective Lagrangian for the scalars.  We can trade the former for
the latter by inverting these equations.  Then setting everything
except the light pseudoscalars at their vacuum values, to leading
order in $1/m_{\PL}$ and the quark mass
\beq m_q = e^{i\delta}m,\eeq
one obtains the identification
\beq \bfbf_n \approx e^{i\delta}\(c_1e^{- i\bfP} - c_0\), \quad \bfP
= -{\sqrt{2}\bfa\over F_\pi} - {8\pi^2 b_c - N_c\over b_c n} {a\over
F_a}, \quad F_\pi \approx2\sqrt{v},\eeq
and the effective potential for the light pseudoscalars takes the
form
\beq V = c\,\Tr\,\bff_n + \hc,\label{pspot}\eeq
which is the standard result in QCD if $\Tr\,\bff_n$ is identified
with the quark condensate.  To check that this identification is
correct, we note that %
under the K\"ahler $U(1)$ transformation~\myref{rferm} and the
transformation~\myref{uone} on the $n$ light quark supermultiplets,
the anomalies induce a shift in the Lagrangian
\beq \delta\Lag\ni - {1\over4}\[\alpha b_c(F\cdot\wtd{F})_c +
(\alpha b'_n + 2n\beta)(F\cdot\wtd{F})_Q\],\eeq
which is canceled in the nonanomalous case \myref{nonan}  by a shift
in~\myref{alag} due to the axion shift
\beq a \to a - \alpha b_c F_a \, .\eeq
This gives
\beq \bff_n\to e^{i\alpha b_{\bff}}\bff_n, \qquad b_{\bff} ={ 8\pi^2
b_c - N_c\over n},\eeq
which matches the phase transformation of the quark condensate
\beq\chi_L\chi^c_L \to e^{i\alpha b_{\chi}}\chi_L\chi^c_L\, , \qquad
b_\chi = 2{\beta\over\alpha} - 1 = {8\pi^2b_c - N_c + n\over n} - 1
= b_{\bff}\, .\eeq

For $n=2$ we identify the factor $\exp(i\sqrt{2}\bfa/F_{\pi})$ with
the operator $\Sigma = e^{2i\pi^i T_i/F_{\pi}}$ of standard chiral
Lagrangians, where $\pi^i$ are the canonically normalized pions, and
$T_i$ is a generator of $SU(2)$. That is, we identify $a_i$ with
$\pi_i$ giving
\beq m_a = {\left|8\pi^2 b_c - N_c\r\over b_c n}{\sqrt{n}F_\pi\over
F_a\sqrt{2}}m_\pi \approx {\sqrt{3}|8\pi^2b_c -
N_c|F_\pi\over2\sqrt{n}}m_\pi\, ,\quad F_\pi \approx
93\MeV\label{fpi}\, .\label{ma3}\eeq
The result~\myref{ma3} appears to differ from the standard result by
a factor $1- N_c/8\pi^2 b_c$.  However, $F_a$ is the axion coupling
to Yang-Mills fields {\em above} the scale of supersymmetry
breaking. When the gluinos of the supersymmetric extension of the
Standard Model are integrated out, a term is generated that modifies
the axion coupling strength to $(F\wtd{F})_Q$ by precisely that
factor. This can be seen in two ways.

First, note that in the absence of the light quark mass the
Lagrangian is invariant under~\myref{rferm} together
with~\myref{uone} with $\beta = \alpha/2$ for the heavy quark
supermultiplets ($\phi^A,\chi^A$), and
\beq \beta = \half + {8\pi^2 b_c - N_c\over2n},\label{fund2}\eeq
for the light quark supermultiplets ($\phi^i,\chi^i$).  In order to
keep this approximate symmetry manifest in the low energy effective
theory, we can redefine the fields so as to remove the
$\omega_c$-dependence from all terms in the Lagrangian for the heavy
fields that do not involve the light quark mass $m$:
\bea \lambda_a &=& e^{i\omega_c/2}\lambda'_a,\qquad \phi^A=
e^{i\omega_c/2}\phi'^A, \qquad \chi^A= \chi'^A, \nonumber \\
\phi^i &=& e^{i\beta\omega_c}\phi'^i,\qquad \chi^i=
e^{i\gamma\omega_c}\chi'^i,
 \qquad \gamma = {8\pi^2 b_c - N_c\over2n}.
\label{redef}\eea
The primed fields are invariant under the nonanomalous symmetry, and
when expressed in terms of them, the potential and Yukawa couplings
have no dependence on $\omega_c$ when $m\to0$. This ensures that any
effects of integrating out the heavy fields will be suppressed by
powers of $m/M_A$, $m/m_{3/2}$ relative to the terms retained.

However, these transformations induce new terms in the effective
Lagrangian. First, because the transformation~\myref{redef} with
$\omega_c$ held fixed is anomalous, it induces a term
\beq \Lag'\ni \Delta\Lag = - {\omega_c b_c\over
4}(F\cdot\wtd{F})_Q\, .\label{shift}\eeq
Second, there are shifts in the kinetic terms. The ones that concern
us here are the shifts in the fermion axial connections $A_m$
\beq \Delta A_m^\lambda = - \half\pp_m\omega_c\, ,\qquad \Delta
A_m^{\chi^i} = - \gamma\pp_m\omega_c \, .\label{dela}\eeq
Quantum corrections induce a nonlocal operator coupling the axial
connection to $F\cdot\wtd{F}$, at scales $\mu^2\sim\Box\gg
m^2_\lambda$, through the anomalous triangle diagram
\beq \Lag_{\rm qu}\ni -
{1\over4}(F\cdot\wtd{F})_Q{1\over\Box}\({N_c\over4\pi^2}\pp^{\mu}
A_{\mu}^\lambda + {n\over2\pi^2}\pp^{\mu}
A_{\mu}^{\chi^i}\).\label{tri}\eeq
The contribution to~\myref{tri} from the shift~\myref{dela} exactly
cancels the shift~\myref{shift} in the tree level Lagrangian,
leaving the $a F\wtd{F}$ S-matrix element unchanged by the
redefinition~\myref{redef}. However at scales $\mu^2\ll
m^2_\lambda$, we replace $\Box\to m^2_\lambda$ in the first term
of~\myref{tri} because the contribution decouples, but the analogous
contribution~\myref{shift} to the tree Lagrangian $\Lag'$ remains in
the effective low-energy Lagrangian.  This is a reflection of the
fact that the classical symmetry of the unprimed variables, without
a compensating axion shift, is anomalous.  The gluino contribution
to that anomaly is not canceled by the gluino mass term, because the
gluino mass does not break the symmetry. Its phase $\omega_c$ is
undetermined above $\Lambda_{QCD}$ and transforms so as to make the
mass term invariant.

To see that the gluino contribution to the anomaly does not
decouple, we write the (unprimed) gaugino contribution to the
one-loop action as
\beq S_1 = - {i\over2}\Tr\ln(i\notD + m_\lambda) = S_A +
S_N,\label{s1}\eeq
where \beq S_A = - {i\over2}\Tr\ln(i\notD)\eeq
is mass-independent and contains the gaugino contribution to the
anomaly
\beq \delta\Lag \ni \delta S_A = - {\alpha
N_c\over32\pi^2}(F\cdot\wtd{F})_Q\, .\label{dell}\eeq
The mass-dependent piece
\beq S_N = - {i\over2}\Tr\ln(-i\notD + m_\lambda) +
{i\over2}\Tr\ln(-i\notD)\eeq
is finite and therefore nonanomalous. A constant mass term would
break the symmetry and the contribution from $S_N$ would exactly
cancel that from $S_A$ in the limit $\mu/m_\lambda\to0$. However it
clear that $S_N$ is invariant under~\myref{rferm} because the
gaugino mass is covariant. On can show\cite{Gaillard:2005gj} by
direct calculation that gaugino loops give the
contribution~\myref{dell} under a nonanomalous $U(1)$ transformation
in the limit $m_\lambda\gg\mu$, which in this limit arises only from
the phase of the mass matrix. This implies that the effective low
energy theory must contain a coupling
\beq \Lag_{\rm eff}\ni\Lag_{\rm anom}= - {\omega_c
N_c\over32\pi^2}(F\cdot\wtd{F})_Q,\label{lanom}\eeq
which is precisely the term that is generated by the redefinitions
in~\myref{redef}.

Below the scale of QCD condensation standard effective chiral
Lagrangian techniques can be used\cite{Gaillard:2005gj} to
recover~\myref{ma3}, or, taking $n=2$ and allowing for $m_u\ne m_d$,
\beq m_a \approx 2m_\pi{F_\pi\sqrt{z}\over f(1+z)}, \qquad z =
{m_u\over m_d},\eeq
where \beq {n a\over32\pi^2}f^{-1} = \(1 -
{N_c\over8\pi^2b_c}\)F_a^{-1},\eeq
in agreement\footnote{The coupling constant $f$ used by
Srednicki\cite{Srednicki:1985xd} is a factor two larger than the one
defined in~\myref{lit} and used by Fox {\it et
al.}\cite{Fox:2004kb}} with the result\cite{Srednicki:1985xd} of
Srednicki. The coupling $f$ is the low energy axion coupling
constant as defined by the convention
\beq \Lag\ni -{n a\over32\pi^2f}\sum_b(F\cdot\wtd{F})_b,\qquad f =
{n F_a \over8\pi^2},\label{lit}\eeq
used by Fox {\it et al.}\cite{Fox:2004kb}
The axion coupling to photons is important in direct axion searches.
A similar shift in the axion coupling to $SU(2)_L$ gauge fields is
generated when the $W$-inos are integrated out, with $N_c = 2$
in~\myref{lanom}. Then the couplings $f_{\gamma,g}(\mu)$ for the
canonically normalized gauge fields as measured at a scale
$\mu<m_\lambda$ are in the ratio
\beq {f_\gamma\over f_g} = {g_3^2(\mu)\over e^2(\mu)}{8\pi^2b_c - 3
\over8\pi^2b_c - 2\sin^2\theta(\mu)},\eeq
which could be very different from the value at $\mu>m_\lambda$
which is just the ratio of the fine structure constants.  In
addition there is an induced $\gamma W a$ coupling for
$\mu<m_\lambda$.

If use the preferred value $b_c = .036$, we are very close to the
symmetric point for $N_c = 3$: $8\pi^2b_c = 2.84$, so we get an
(accidental) suppression of the axion mass. In particular, if
$b_c\ell\ll1$ and $n=2$ we obtain
\beq m_a\approx 5\times 10^{-13}\eV,\label{bgn}\eeq
which raises the issue of the importance of higher dimension
operators that might contribute to the axion potential and destroy
the solution to the strong CP problem.\cite{Banks:1994sg} Indeed,
exactly at the point of enhanced symmetry $b_c = 8\pi^2N_c$, the
nonanomalous symmetry with~\myref{fund2} does not include a chiral
transformation on the quarks, and one loses the solution to the CP
problem.  The axion decouples from the quarks in the effective
Lagrangian, and its \vev cannot be adjusted to make the quark mass
matrix real in the $\theta = 0$ basis.

There is no reason to expect that nature sits at this point, but if
the axion mass is very small one should worry about other sources of
an axion potential. The contribution of higher dimension operators
was studied in Reference~\refcite{Butter:2005wr} in the context of
modular-invariant gaugino condensation models. Modular invariance
severely restricts the allowed couplings; the leading contribution
to the axion mass takes the form
\beq m'^2_a \approx {9\over4}b_c
p^3|u|^2k'\lambda|\eta^2e^{-K/2}u|^p\label{hdmass}\eeq
where $\lambda$ is a dimensionless coupling constant, and $p$ is the
smallest integer allowed by T-duality.  An orbifold compactification
model with three complex moduli and an $[SL(2,{\bf Z})]^3$ symmetry
has $p=12$, and, with the values of the various parameters used
above and $\lambda\approx1$, one finds $m'_a \approx 10^{-63}$eV,
which is completely negligible.  However if the symmetry is
restricted, for example, to just $SL(2,{\bf Z})$ one has $p=4$ and
the contribution from~\myref{hdmass} is of the order of~\myref{bgn}.
The axion potential takes the form
\bea V(a) &=& - f^2m^2_a\cos(a/f + \phi_0) - f'^2m'^2_a\cos(a/f')\, , \nonumber \\
\frac{1}{f'} &=& {p\over b_c}\frac{1}{F_a} = {p\, n\over8\pi^2b_c
-N_c}\frac{1}{f}\, ,\label{cp}\eea
where we have absorbed constant phases in $a$ and/or in $\phi_0$ so
as to make the coefficients negative. The strong CP problem is
avoided if for some value of $b_c$ the vacuum has
$\langle{\bar\theta}\rangle = \langle{n(a/f +
\phi_0)/2}\rangle<10^{-9}$ for any value of $\phi_0$.  For values of
$b_c$ in the preferred range $.3\le b_c \le .4$ this does not occur.
For example for $b_c = .036$ with $p=4$ and $f'/f\approx1/50$, this
requires $f'^2m'^2_a/f^2m^2_a<10^{-10}$, whereas
evaluating~\myref{cp} gives $f'^2m'^2_a/f^2m^2_a\approx
4\times10^{-4}$ in this case.  A numerical analysis shows that the
CP problem is avoided provided $p\ge 5$, that is, provided the
T-duality group is not the minimal one, which is in fact the case
for most orbifold compactifications of the weakly-coupled heterotic
string.

The reasoning leading to~\myref{hdmass} is similar to that used in
Section~\ref{sec5} in the discussion of R-parity.  In the language
of K\"ahler $U(1)$ supergravity, superpotential terms must have
K\"ahler $U(1)$ weight 2, where chiral fields $Z^A$ have weight 0
and the Yang-Mills superfield strength $\Wa$ has weight 1.  Thus the
following terms with at least one factor $W^{\alpha}W_{\alpha}$ are
allowed
\beq \Lag_{SP} = {1\over2}\superint F
W^{\alpha}W_{\alpha}\F(e^{-K/2}W^{\alpha}W_{\alpha},\, Z^A)
+\hc.\label{lsp}\eeq
Invariance under the T-duality transformations~\myref{mdtr}
restricts the function $\F$ to the form
\beq \F = \F(\eta^2e^{-K/2}W^{\alpha}W_{\alpha},\eta^A\Phi^A), \quad
\eta = \prod_I\eta_I,\quad \eta^A = \prod_I\eta_I^{2q^A_I}, \quad
\eta_I = \eta(T^I).\label{hsp}\eeq
Consider first terms with no $\Phi^A$-dependence; since for a
general transformation~\myref{mdtr} $\delta_I = n_I\pi/12$, the only
invariant superpotential is of the form:
\beq \Lag_{H W} = {1\over2}\superint{E\over
R}W^{\alpha}W_{\alpha}\F(\eta^2e^{-K/2}W^{\alpha}W_{\alpha})
+\hc,\quad \F(X) = \sum_{n=1}\lambda_n X^{12n}.\label{lhw}\eeq
If the $[SL(2,{\bf Z})]^3$ symmetry implied by~\myref{mdtr} were
instead restricted, say to just $SL(2,{\bf Z})$, with
$a_I,b_I,c_I,d_I,$ independent of $I$ as in~\myref{mod}
and~\myref{mattertrans}, then the phase of $\eta$ is $3\delta_I =
n\pi/4$, and lower dimension operators would be allowed: $\F(X) =
\sum_{n=1}\lambda_n X^{4n}$.  These give the estimate
in~\myref{hdmass}.

In addition to the operators in~\myref{hsp} chiral superfields with
zero chiral weight can be constructed using chiral projections of
any functions of chiral fields.  Operators of this type were
found\cite{Antoniadis:1996qg} in (2,2) orbifold compactifications of
the heterotic string theory with six dynamical moduli.  In the class
of models considered here we can construct zero-weight chiral
superfields of the form
\beq\F =
e^{-(p+n)K/2}(W^{\alpha}W_{\alpha})^p\eta^{2(p+n)}\prod_{i=1}^n\chiproj
f_i[|\eta_I|^4(t^I+\bar{t}^I)]\label{chiproj}\eeq
that are modular invariant provided $(p+n)\sum_I\delta = m\pi.$
Since $\langle{F^I}\rangle=0$, the corresponding terms in the
potential at the condensation scale are proportional to\footnote{The
coefficients of the nonpropagating condensate superfield auxiliary
fields vanish by their equations of motion.}
$|u|^p(m_{3\over2})^{n+1}$, so for fixed $p+n$ one is trading
factors of $|u|$ for factors of $m_{3/2}\sim 10^{-2}|u|$, and these
contributions to the axion mass will be smaller than those
in~\myref{hdmass}.

We may also consider operators with matter fields that have
nonvanishing $vevs$.  Since $e^{-K/2}W^{\alpha}W_{\alpha}$
transforms like the composite operators $U_1U_2U_3$ constructed from
untwisted chiral superfields, the rules for construction of a
covariant superpotential including this chiral superfield can be
directly extracted from the discussion in Section~\ref{rparity} of
modular-invariant superpotential terms in the class of
$\mathbf{Z}_3$ orbifolds considered here.  They take the form
of~\myref{hsp} with
\beq \F_{p n q} = \Pi^q(e^{-K/2}W^{\alpha}W_{\alpha})^p\eta^{2(p
+n)}\prod_{\alpha = 1}^n W_i,  \qquad (p+n)\sum_I\delta_I =
m\pi,\label{fpnq}\eeq
where $\Pi = Y^1Y^2Y^3$ is the product of twisted sector oscillator
superfields introduced in Section~\ref{rparity}, and $W_i$ is any
modular covariant zero-weight chiral superfield that is a candidate
superpotential term (subject to other constraints such as gauge
invariance).  For example, the superpotential terms for matter
condensates could contribute to this expression.  However the
equations of motion for the auxiliary fields of these condensates
give $W_i\sim m_{3/2}$ for these terms, so again they are less
important than the contribution in~\myref{hdmass}.

Most $\mathbf{Z}_3$ orbifold compactifications of the type
considered here have an anomalous $U(1)$ gauge group, with the
anomaly canceled by a Fayet-Illiopoulos D-term, as discussed in
Section~\ref{sec4}. A number $n$ of scalars $\phi^A$ acquire $vevs$
along an $F$- and $D$-flat direction such that $m\le n$ $U(1)_a$
gauge factors are broken at a scale $\Lambda_D$ that is close to the
Planck scale. {\em A priori} there might be gauge- and
modular-invariant monomials of the form~\myref{fpnq} with
considerably larger \vevs than those in~\myref{lhw}, and no modular
covariant, gauge invariant superpotential term $W_i$, so that the
direction $\phi^A\ne0$ is $F$-flat. However if $m=n$, there is no
gauge invariant monomial $\prod_A(\phi^A)^{p_A}$.  Gauge invariance
requires
\beq \sum_A p_A q^a_A = 0 \qquad \forall a,\label{gauge}\eeq
where $q^a_A$ is the $U(1)_a$, charge of $\phi^A$. If $m=n$ these
are linearly independent and form an $m\times m$ matrix with inverse
$Q^A_a$; then~\myref{gauge} implies
\beq p_A = 0 \qquad \forall A.\label{gauge2}\eeq
Similarly, for the chiral projection of a monomial
$\prod_A(\phi^A)^{p_A + q_A}(\bar\phi^{\bar A})^{q_A}$ gauge
invariance still requires~\myref{gauge} and~\myref{gauge2}, so any
such monomial can be written in the form
\beq f(T^J,\oline{T}^{\oline{J}}) \prod_A\[|\phi^A|^2\prod_I(T^I +
\oline{T}^{\oline{I}})^{-q^A_I}\]^{q_A}.\eeq
It is the modular invariant composite fields $|\phi^A|^2\prod_I(T^I
+ \oline{T}^{\bar I})^{-q^A_I}$ that acquire large $vevs$; any
coefficients of them appearing in overall modular-invariant
operators are subject to the same rules of construction as the
operators in~\myref{chiproj}. The same considerations hold if $N$
sets of fields $\phi^A_i$ with identical $U(1)_a$, charges
$(q^i_A)^a = q^a_A$, $i = 1,\ldots,N$ acquire $vevs$.

In the general case with $n>m$ one cannot rule out the above terms.
However in this case part of the modular symmetry is realized
nonlinearly on the $U(1)_a$-charged scalars after $U(1)_a$-breaking.
Monomials of the above type would generate mixing of the axion with
massless ``D-moduli'' that are Goldstone particles associated with
the degeneracy of the vacuum at the $U(1)_a$-breaking scale,
requiring a more careful analysis.

\subsection{Early universe physics} \label{inflate}

So far in this section we have focused on phenomena relevant for
low-energy observations in the late-time universe. Yet string theory
is meant to provide a single framework for understanding {\em all}
phenomena, including the physics of the early universe. Therefore, a
string-derived effective supergravity model -- to the extent that it
is a complete description of the underlying string degrees of
freedom -- should provide such things as an inflaton candidate, a
baryogenesis mechanism and a source for the dark energy in the
universe. And it should do these things while allowing for the
successful predictions of the Big Bang Nucleosynthesis (BBN) theory.
In this section we will concern ourselves with some of these topics
where a definite statement can be made in the context of the
BGW~class of models.

Scalar field inflation has long been the leading paradigm for
understanding the horizon and flatness problems of the big bang
cosmology.\cite{Olive:1989nu,Kolb} All supersymmetric theories
provide many such scalar fields -- and string-derived models have
still more. The latter include the moduli fields that have been our
focus throughout this work. These fields carry no Standard Model
quantum numbers and provide an interesting possibility to realize
the ``sterile'' field models that are common in inflation theories.
Unfortunately, it has long been appreciated that maintaining
adequate flatness of the potential to achieve slow-roll is difficult
in supergravity models of scalar fields in an expansionary
universe.\cite{Copeland:1994vg} Higher-order K\"ahler potential
terms are the most troublesome since they do not benefit from
nonrenormalization theorems.

In this section we will consider a hybrid inflation scenario
suggested by the BGW class of theories, in which some field other
than the inflaton is displaced from its vacuum value and thereby
generating most of the potential during inflation. The necessary
linear and bilinear terms in the superpotential will be the result
of some fields getting \vevs when an FI $D$-term is driven to small
values, while preserving SUSY and some flat directions. To be more
specific, consider a theory of untwisted matter having a K\"ahler
potential given by~(\ref{Kuntw}) from Section~\ref{sec2}. Let us
assume that during some period of inflation all matter from twisted
sectors have vanishing $vevs$. If we further assume that the limit
$\lang (z^i)^I \rang \ll \lang t^I \rang$ during inflation, then it
is permissible to regard the scalar fields
\begin{equation} x^I = t^I + \bar{t}^I - \sum_i |(z^i)^I|^2
\label{xscalar} \end{equation}
as the low energy degrees of freedom. Theories described in this
manner are of the flat no-scale type and enjoy a Heisenberg
symmetry.\cite{Binetruy:1987xj,Gaillard:1995az} Typically such a
model has a minimum only in the limit as $x^I \to \infty$, but the
presence of FI $D$-terms of the form~(\ref{eq1})
\begin{equation}
D_X = - 2\(\sum_i K_i q^X_i (z^i)^I + \xi\),\quad K_i = {\partial K
\over
\partial (z^i)^I } \quad \xi = {g_{\STR}^2 \, \Tr \, Q_X \over 192 \pi^2}
m_{\PL}^2, \label{newFI}
\end{equation}
can prevent this runaway behavior because of the nontrivial $x^I$
dependence of the metric $K_i$. The GS term will preserve this
Heisenberg invariance provided that it also depends on the moduli
only through the combination $x^I$. This is to say, $p_i = b_{\GS}$
in~(\ref{GS}) for the untwisted fields.

Other terms in the superpotential which involve the twisted sector
fields will explicitly break the Heisenberg invariance, but these
are small effects during inflation by assumption. For the scalar
components of the twisted sector fields we can define a quantity
\begin{equation} X_i \equiv \prod_I (x^I)^{-q_i^I} |z^i|^2
\end{equation}
such that (the scalar part of) the K\"ahler potential reads
\begin{equation} K = \ln (\ell) + g(\ell) - \sum_I \ln x^I + \sum_i
X_i \, . \end{equation}
Therefore, near the origin in scalar field space, we can write the
K\"ahler metric for arbitrary matter field $z^i$ as
\begin{equation} K_i = \prod_I \bar{z}^i
(x^I)^{-q_i^I}\, ; \quad K_{i \jbar} = \prod_I \delta_{i\jbar}
(x^I)^{-q_i^I} \, . \label{Kalpha}
\end{equation}
As for the superpotential, we assume as always that its form is
dictated by the requirements of modular invariance. Written in terms
of scalar fields it is given by
\begin{equation} W = \sum_m \lambda_m \prod_{i}
(z^i)^{n_m^i} \prod_I \eta(t^I) ^{2(\sum_{i} n_m^{i}q_{i}^I -1)}\, ,
\label{Winflate} \end{equation}
which is the equivalent to what was considered in~(\ref{Walpha}) for
the hidden sector matter condensates. The integers $n_m^{i}$ are
nonnegative.
Note that this implies
\begin{equation} \frac{\partial W}{\partial t^I} \equiv W_I = 2
\zeta(t^I) \(\sum_{i} q_{i}^I z^{i} W_{i} - W\)\,  , \label{dWdT}
\end{equation}
and $t^I$ are stabilized at one of the two self-dual points.

Now if we are concerned with cases in which $\lang V \rang \gg u^2$
-- that is, cases in which the vacuum energy is much greater than
the size of the eventual gaugino condensates -- then the terms
involving the condensates $u$ can be neglected in the scalar
potential of the BGW~model. This leaves us with two sources for the
scalar potential: possible $D$-terms from an anomalous $U(1)$ and
derivatives of the superpotential in~(\ref{Winflate}) with respect
to the chiral matter. Let us take a very simple ansatz. Assume that
each individual term in~(\ref{Winflate}) vanishes (which is to say
that at least one scalar field in each term in~(\ref{Winflate}) has
vanishing $vev$). Also assume that all $W_{i}$ vanish except for one
$i$ in the untwisted sector. Without loss of generality assume that
this field is associated with $I=3$ so that $i = C3$. Finally,
assume that all matter field values are much smaller than one in
Planck-scale units. The potential in this limit is
\begin{equation} V = \frac{\ell e^{g(\ell)}}{(1+b_{\GS}\ell)x^1 x^2}
|W_{C3}|^2 \, . \label{Vinflate} \end{equation}
Let us take $W_{C3}$ to depend only on $x^1$ and $x^2$ so that the
moduli are stabilized but a flat direction persists for $t^3$ as
well as matter fields in the $I=3$ sector. The inflaton could be
identified with a particular combination of these
fields.\footnote{In fact, the canonically normalized inflaton field
$\varphi$ can be found (up to a phase) in this simple case by
inverting the relationship $|z_C^3| = \sqrt{t^3 +
\bar{t}^3}\tanh(\varphi/\sqrt{2})$.\cite{Kain:2006nx}}

To achieve something like this simplified scenario one might imagine
that the superpotential that arises from the string compactification
has a form
\begin{equation} W = \lambda \[ \eta(t^1)\eta(t^2) \]^{-2} z^{C3}
\prod_{i\neq C3 } z^{i}\prod_I
\[\eta(t^I)\]^{-2q_{i}^I} \, . \label{Wlong} \end{equation}
Then to obtain the desired form of $W_{C3}$ we must suppose that
during inflation there are nonzero and modular-invariant \vevs for
certain $z^{i}$'s:
\begin{equation} \lang |z^{i}|^2 \prod_I (x^I)^{-q_{i}^I} \rang =
c_{i} \lang \ell^{\,d_{i}} \prod_I \[\, x^I|\eta(t^I)|^4\]^{p_{i}^I}
\rang \label{vevinflate}
\end{equation}
where $c_{i}$ is a constant. But in fact \vevs of precisely this
form are indeed induced at the anomalous~$U(1)_X$ scale, as was
discussed in Section~\ref{sec4}. Simply consider the expression
in~(\ref{newFI}) with $g_{\STR}^2 = 2\ell/1+f(\ell)$ and
use~(\ref{Kalpha}) to find the form for $K_i$. This yields a \vev of
the form in~(\ref{vevinflate}) with $p_i^I = 0$ and $d_i = 1$. Once
these \vevs are integrated out of the theory the moduli dependence
of~(\ref{Vinflate}) is then simply
\begin{equation} V \propto \prod_I \[ |\eta(t^I)|^4 x^I\]^{n^I} \, , \quad
n^{1,2} = \sum_{i} (p_{i}^{1,2} q_{i}^{1,2}) -1 \, , \quad n^3 =
\sum_{i} (p^3_{i} + q^3_{i}) \, .
\end{equation}
To achieve flat directions and a realistic inflaton it is necessary
that at least one of the $n^I$ vanish. If the remaining $n^I$ are
negative then the corresponding moduli are stabilized at $t^I =
e^{i\pi/6}$. The dilaton dependence of~(\ref{Winflate}) is then
\begin{equation} V \propto \frac{e^{g(\ell)}
\ell^d}{1+b_{\GS}\ell}\, , \quad d =1 + \sum_{i}d_{i}
\end{equation}
which slightly modifies the minimization conditions for the dilaton
when $d\neq 1$, though weak-coupling solutions with the domain of
attraction can be
found.\cite{Gaillard:1998xx,Cai:1999aj,Kain:2006nx}

Note that this scenario actually generates a precisely flat
potential. Ending inflation therefore requires a perturbation on the
scenario. These could come from (a) having some terms in $W$ that do
not vanish during inflation, (b) having additional fields which
contribute to~(\ref{Winflate}) beyond our one untwisted field,
and/or (c) assuming that the $D$-term is not forced to vanish but
instead driven to very small but nonvanishing values. The resulting
theory will be a hybrid inflation model which is a variant on that
of Ref.~\refcite{Stewart:1994ts}. Flat directions involving the
untwisted sector are lifted by mass terms with a typical size that
is $|m_{(z^i)^I}|^2 \sim m_{t_I}^2$ provided the field \vevs satisfy
$\lang |(z^{i})^I|^2 \rang \lappeq \lang {\rm Re}\, t^I \rang$. This
contribution is generally negative and much smaller than that
induced by loop effects.\cite{Gaillard:1995az} If either is the
dominant source of nonvanishing slope for the inflaton potential
then one expects the spectral index $n = 1+2m^2/V$ to be very close
to unity.

Finally, we must worry about the overall scale of the inflationary
potential, which is constrained by the Cosmic Background Observer
(COBE) normalization to have $V^{1/4} \lappeq 10^{-2} m_{\PL}$. This
will require that the superpotential~(\ref{Wlong}) that gives rise
to~(\ref{Winflate}) involve nonrenormalizable operators. The scale
of the potential is given by
\begin{equation} V = \lambda \Lambda^{-2n} \xi_D^{2(2+n)}
\label{Vscale} \end{equation}
where $\Lambda^2 = g_{\STR}^2 m_{\PL}^2$ and $\lambda$ is a ratio of
dimensionless couplings in the superpotential. The mass dimension of
the term that contains $z_{C}^3$ is $3+n$. We see, therefore, that
\begin{equation} V^{1/4} \sim \lambda^{1/4}g_{\STR} \(\frac{\Tr \,
Q_X}{192\pi^2}\)^{(2+n)/4} \, m_{\PL} \end{equation}
which implies $n=1$ or $n=2$ is necessary to match the COBE
normalization.

After inflation ends the moduli must end up at the minima of the
scalar potential. This can be problematic for moduli whose potential
is generated by nonperturbative effects. In particular, for the
dilaton the scalar potential after inflation is generated by gaugino
condensation and is quite steep for large field values (see
Figure~\ref{fig:bgw1b}, with only a small barrier separating the
nontrivial minimum from the vacuum with vanishing gauge coupling.
One might worry that the dilaton field will ``overshoot'' the
desired minimum at the end of inflation.\cite{Brustein:1992nk}
However, recent work demonstrates that this is not the case with the
nonperturbative corrections~(\ref{nonpertsum}) employed in the
BGW~model.\cite{Barreiro:1998aj,Skinner:2003bi,Kain:2006nx} If
during an expansionary period the dilaton begins at some point in
its scalar potential which corresponds to a regime of strong
coupling, then over a wide range of initial field values it will
enter a quasi-scaling regime as it evolves towards its
(weak-coupling) true minimum. This is in spite of the rather steep
potential set up by gaugino condensates (and K\"ahler
stabilization). This expansionary period must be generated by some
other field in the theory, however. In
Ref.~\refcite{Barreiro:1998aj} the minimum number of e-foldings
$N_{\rm min}$ required such that the dilaton enters the scaling
regime for $m_{3/2} = 1 \TeV$, $g_{\STR}^2 = 1/2$ and a
radiation-dominated universe was found to be $N_{\rm min} = 11$. It
was therefore claimed in that work that the dilaton in the BGW~model
should enter its scaling regime and therefore end up at the global
minimum after inflation (without overshooting) provided the dilaton
scalar is significantly more massive than the gravitino.

This is not the only instance in which the masses of the moduli
fields play an important role in the physics of the early universe.
We generally expect re-heating after inflation to produce states
whose masses are less than or on the order of the reheat temperature
$T_{\rm RH}$. Of particular interest are those scalar fields which
have no classical potential. Being flat directions, these scalars
are likely to take large field values away from their eventual
minima. As the universe cools, oscillations of these fields about
their minimum-energy configurations will generally produce too much
energy density to be consistent with the known age of the universe
unless their masses are impossibly
small.\cite{Coughlan:1983ci,Goncharov:1984qm,Ellis:1986zt} They
therefore must decay, but as these particles interact with the
observable sector only via Planck-suppressed operators, their
resulting lifetime is generally quite long. When they decay, light
elements produced via BBN will be dissociated. It is therefore
necessary that the decay of these moduli fields reheat the universe
once again, with a reheat temperature this time of order the BBN
scale of 1~MeV. A simple computation reveals that the needed $T_{\rm
RH}$ can be achieved provided the moduli have masses of
$\order(10~\TeV)$ or higher.\cite{Ellis:1986zt,Banks:1993en}

In the BGW~class of models the corrections to the dilaton metric
result in an enhancement of the dilaton scalar mass relative to the
that of the gravitino. The masses obey the relation
\begin{equation} \frac{m_{\ell}}{m_{3/2}} \simeq \frac{1}{b_+^2} \gg
1 \, . \label{dilmass} \end{equation}
For the case of the $E_6$ condensates with 9~fundamentals of matter
this implies $m_{\ell} \sim 10^3 m_{3/2} \sim 3000 \TeV$. This is
sufficient to keep the dilaton in the domain of attraction for its
post-inflation potential while simultaneously avoiding the
cosmological problem for this modulus.

More problematic are the K\"ahler moduli. Their masses can be found
from the second derivative of the scalar potential
\begin{equation} \frac{\partial^2 V}{\partial (t^I)^2} \simeq
\frac{1}{32\ell^2}\sum_{ab} \rho_a \rho_b
\[\frac{\pi^2}{9}\frac{\ell^2}{(1+b_{\GS}\ell)}(b-b_a)(b-b_b)\]
\simeq \rho_+^2\frac{\pi^2}{288}\frac{(b-b_+)^2}{1+b_{\GS}\ell} \, ,
\end{equation}
from which we extract the normalized mass-squared
\begin{equation} m_{t_I}^2 \simeq \lang \rho_+^2\frac{\pi^2}{36}
\frac{(b-b_+)^2}{1+b_{\GS}\ell} \rang \,  . \label{tmass}
\end{equation}
When $b_{\GS} > b_+$, as when $b_+ = 3/8\pi^2$ and $b_{\GS} =
b_{E_8} = 30/8\pi^2$, one can enhance the K\"ahler modulus mass by
an order of magnitude relative to the gravitino mass. But this is
roughly the largest such an enhancement can be. For more realistic
scenarios we expect $b_{\GS}$ to be similar in magnitude to the
value of $b_+$. However, when there are anomalous~$U(1)$ factors the
equation of motion for the auxiliary fields of the K\"ahler moduli
are modified to the form of~(\ref{solt}) from Section~\ref{sec4}. In
this case the K\"ahler moduli masses are enhanced relative to the
those given above for the same values of $b_{\GS}$ and $b_+$.  We
conclude that the masses of the moduli in the BGW~class of models
are likely to be sufficiently large to avoid the cosmological moduli
problem,\footnote{Recall that typical gravitino masses are already
in the multi-TeV range, so enhancement factors of 5~to~10 may be
sufficient.} but that the actual resolution of the problem is
dependent on the parameters that arise from the underlying string
construction.

\section*{Conclusion -- Where do we go from here?}

In this review we have tried to take the reader on a largely
self-contained exploration of one particularly well-understood
corner of the moduli space of M-theory. We began by introducing the
notion of string moduli, reviewing the manner in which
nonperturbative field theory effects can be employed to produce a
potential for their scalar components. As these scalar fields
determine all dimensionless parameters in the low-energy effective
supergravity Lagrangian, this is clearly the heart of any
phenomenological treatment of string theoretic models. The
well-known shortcomings of the traditional treatments of moduli
stabilization were demonstrated: the need for multiple condensates
and the difficulty in finding a minimum with vanishing vacuum
energy. The BGW class of models remedies both problems by utilizing
nonperturbative corrections to the K\"ahler potential of the
dilaton, thereby generating an acceptable level of supersymmetry
breaking with vanishing vacuum energy. As with all such methods of
supersymmetry breaking, some degree of tuning between the various
parameters of the theory must be employed.

The importance of dealing with the above issues cannot be
overstated. Without achieving a minimum for the overall scalar
potential such that the vacuum energy is negligible, no truly
meaningful statements about supersymmetry breaking, the superpartner
spectrum or phenomenology can be made. By negligible we will mean
small on the scale of particle physics experiments, {\em i.e.}
significantly smaller than the electroweak scale. Given any
particular mechanism to achieve $\lang V \rang \simeq 0$, the degree
to which we can achieve precisely vanishing vacuum energy is a
function of the amount of fine-tuning we can engineer in the model.
It is to be expected -- and is generally the case in explicit
examples -- that any such mechanism for achieving vanishing \vev for
the scalar potential will have some sort of ``back-reaction'' on the
observable sector particle physics. In other words, if one simply
assumes that $\lang V \rang \simeq 0$ by some unspecified mechanism
then it is natural to wonder whether this mysterious sector truly
plays no role in determining the low-energy phenomenology of the
model in question. The BGW class of weakly-coupled heterotic string
models has the virtue of being forthright in addressing the problem.
Since the K\"ahler stabilization has real effects on the resulting
phenomenology (mostly good ones, we hasten to add) it is perfectly
reasonable as string {\em phenomenologists} to study this mechanism.
Comparing eventual data to the resulting ``complete'' theory then in
part tests this mechanism for stabilizing the moduli and achieving
appropriate vacuum energy. We do not know, in general, how a small
vacuum energy arises from any quantum theory, let alone a
string-based one. Therefore other mechanisms should be sought out,
but a framework which envisions no such mechanism must eventually
make only empty statements about Nature.

Starting from the context of weakly-coupled heterotic string theory
an effective supergravity theory was built in Section~\ref{sec2} to
describe the dynamics of the low energy four-dimensional world. The
form of this effective Lagrangian was guided by the principle of
target space modular invariance. This symmetry acts as a classical
symmetry of the supergravity Lagrangian. The underlying string
theory informs us that it should be a good symmetry to all orders in
the string perturbation expansion. Therefore we ought to ensure that
it remains an intact symmetry to all orders in quantum field theory.
The anomalies associated with modular invariance can be remedied by
terms arising from the string theory itself: Green-Schwarz
counterterms and threshold corrections to gauge coupling constants.
We saw that these are easiest to implement when the dilaton is
packaged in the linear multiplet. This is no surprise since the
degrees of freedom of the string are precisely those of the linear
multiplet and the genus-counting parameter is the real object
$\ell$. The drawback is that the axion sector is slightly more
difficult to discuss and less familiar than the dual pseudoscalar
treatment.

Modular invariance proves to be a powerful constraint. Preserving it
when integrating out heavy matter at the anomalous $U(1)$ scale can
imply a relation between string selection rules, modular invariance
and an intact R-parity. This, in turn, has implications for stable
relics (dark matter) and the issue of rapid proton decay. In
conjunction with the K\"ahler corrections for the dilaton it led to
a vacuum solution in which K\"ahler moduli are fixed at self-dual
points where their auxiliary fields vanish. It is the dilaton,
therefore, which communicates supersymmetry breaking to the fields
of the observable sector. This is an example of the (generalized)
dilaton domination scenario. In Section~\ref{sec3} we demonstrated
that at tree-level scalar masses and trilinear A-terms are
universal. The field-theory loop corrections are small and the
higher-order terms in the K\"ahler potential may preserve the
resulting FCNC suppression provided certain well-motivated
isometries exist from the compactification. Gaugino masses are
generally an order of magnitude smaller than scalars since the same
mechanism which allows for an acceptable vacuum tends to
dramatically alter the K\"ahler metric of the dilaton (whose
effective auxiliary field determines the size of gaugino masses).
The model therefore predicts scalar masses to be in the multi-TeV
range, with implies a further suppression of superpartner-mediated
FCNC and CP-violating effects. The relatively light gluino implies
reduced fine-tuning in the electroweak sector despite the heavy
scalars. In Section~\ref{sec5} we saw that it also has profound
implications for the signature of this class of models at hadron
colliders. The fact that tree level gaugino masses are similar in
size to certain loop-induced terms gives rise to a mixed
modulus/anomaly-mediation scenario for gauginos. Achieving the right
scale of superpartner masses tended to imply a very particular set
of possible hidden sector configurations. These just so happen to be
configurations in which the relic abundance of (stable) neutralinos
is naturally in the right range to account for the non-baryonic dark
matter as suggested by the WMAP experiment. Furthermore, the range
of parameters singled out by these considerations also happens to
coincide with the range in which the model-independent axion can be
the QCD axion which solves the strong-CP problem. Finally we note
that (modulo the issues of Section~\ref{sec4}) the physical masses
of the scalar moduli tend to be significantly larger than that of
the gravitino, perhaps by as much as an order of magnitude. Since
avoiding the direct search constraints on light gauginos will tend
to imply $m_{3/2}$ of several TeV, this should significantly
mitigate the cosmological problems associated with the moduli. The
above results are promising, but require specific parameter choices
from the string model. This is not necessarily a bad thing -- it
demonstrates that low-energy phenomenology has an impact on the
viability of a certain string framework! This is the essence of
string phenomenology.

Though this class of theories is arguably the most complete string
model in the literature, there are many topics that were not
addressed. Most significantly is the issue of initial conditions,
alluded to in the Introduction. We have assumed the minimal field
content for a supersymmetric version of the Standard Model. It is by
no means clear whether this should be the goal of string
model-builders. For example, we expect on general principles that no
fields in the massless spectrum of the superstring will have a
supersymmetric mass. Therefore we expect the Higgs $\mu$-parameter
and any potential Majorana mass for right-handed neutrinos to arise
only dynamically from the \vev of some fields from {\em outside} the
MSSM field content. We have addressed neither issue within this
review, and indeed string theory is largely silent on these
important issues.\footnote{Some examples specific to the
weakly-coupled heterotic string have been
studied.\cite{Witten:1985bz,Antoniadis:1994hg,Kokorelis:1998cc,Giedt:2005vx,Langacker:2005pf,Braun:2005xp,Bouchard:2006dn,Buchmuller:2007zd}}
We also assumed that the tree level K\"ahler metric for chiral
matter can be made diagonal in the flavor basis. This is the
standard supersymmetric flavor problem which all theories, whether
string-based or otherwise, share. In some string constructions, such
an assumption can be directly examined, but in many constructions
the relation between the flavor indices of chiral matter and the
underlying compact geometry is obscure. Other important issues are
the question of CP-violating phases in the soft
supersymmetry-breaking Lagrangian, the possible presence of charge
and color-breaking minima, the mechanism for baryogenesis, and the
issue of creating an explicit model for inflation.

Let us finally address the very place we began this investigation:
the weakly-coupled $E_8 \otimes E_8$ heterotic string. Despite the
many clear phenomenological advantages of this starting point, other
string constructions have moved to the fore in recent years. In part
this is due to their ability to achieve the same outcomes of the
weak-coupling heterotic string: $N=1$ supersymmetry, chiral matter,
a natural hidden sector, anomaly cancelation, and (sometimes) gauge
coupling unification. But primarily it is because of the issue that
has so motivated this concluding section -- the issue of
supersymmetry breaking and vanishing vacuum energy. Very promising
mechanisms now exist in other corners of the M-theory landscape for
addressing these problems.\cite{Grana:2005jc} Of significance is the
fact that a very large number of solutions with gaugino condensation
and background flux are possible, each such solution giving rise to
a vacuum with a slightly different value of $\lang V \rang$ for the
geometrical moduli. The numerical coefficients in the gaugino
condensation part of the effective scalar potential are ultimately
discrete numbers, determined by the underlying string theory, just
as in the examples considered in this review. But the part of the
effective scalar potential generated by the non-trivial fluxes can
be treated as effectively continuous in these contexts. This allows
for a very finely-grained instrument for tuning the resulting
vacuum-energy -- a ``knob'' which is explicitly lacking in the
context of weakly-coupled heterotic
models.\cite{Kachru:2003aw,Ashok:2003gk,Denef:2004ze,Denef:2004dm,Giryavets:2004zr}

In the current context we have not sought to use the nonperturbative
corrections represented by the expansion in~(\ref{nonpertsum}) to
tune $\lang V \rang$ to match the comparative small value implied by
recent supernova data. Nevertheless, the results we achieve requires
tuning the various coefficients in~(\ref{nonpertsum}). Just like the
coefficients in~(\ref{cond}) that appear in the effective Lagrangian
for the condensates, we should expect that these parameters can not
be changed by infinitesimally small amounts. It remains an open
question, therefore, as to whether the nonperturbative effects
in~(\ref{nonpertsum}) can truly generate a potential with vanishing
\vev in realistic heterotic string compactifications. At the moment
it is unclear whether analogs exist in the heterotic context for
these flux compactification scenarios, though treatments of a more
phenomenological spirit have given rise to interesting
results.\cite{deCarlos:2005kh} The nonperturbative K\"ahler
potential corrections included in the BGW model have two virtues.
They are known to exist and take a well-defined functional form. Yet
it may be argued that it is inconsistent from the point of view of
effective field theory to include nonperturbative contributions
prior to examining the relevant {\em perturbative}
contributions.\cite{Louis:1994ht,Anguelova:2005jr} Consideration of
perturbative corrections to the K\"ahler potential has already
proven fruitful in other string
frameworks.\cite{Becker:2002nn,Balasubramanian:2004uy,Balasubramanian:2005zx,Berg:2005yu,Greene:2005rn}

This is an exciting time for string phenomenology. The rapid
progress being made in understanding the dynamics of string theory
in various corners of the M-theory landscape is about to be joined
by data from a number of forthcoming experiments. The domain in
which these two themes intersect is the effective supergravity
Lagrangian describing string moduli, their interaction with matter
and their stabilization. Unlike bottom-up models constructed to
describe only one set of phenomena (such as the process of
electroweak symmetry breaking), string-based models have the burden
and the opportunity to describe much more. Classes of string models
with some claim to this level of ``completeness'' are growing,
though many challenges remain ahead of us. The question of whether
the post-LHC era will be a golden age for string phenomenology will
largely depend on how well we address these challenges and the
deepening of the vertical integration between phenomenologists and
formal theorists.

\section*{Acknowledgements}

We would like to thank Joel Giedt for reviewing the manuscript and
for helpful conversations on many of the points contained in this
article. We would also like to acknowledge discussions with G.~Kane,
T.~Taylor, F.~Quevedo, B.~Kain and S.~Kachru. Finally, we would like
to thank the Kavli Institute for Theoretical Physics at U.C. Santa
Barbara for hosting the authors during a portion of the writing of
this work. This research was supported in part by the National
Science Foundation under Grant no. PHY99-07949. MKG is supported in
part under Department of Energy contract DE-AC02-05CH11231 and in
part by the National Science Foundation under grant PHY-0457315.

\appendix

\section{An overview of K\"ahler $U(1)$ superspace}
\label{appA}

The action for the coupling of matter fields to supergravity is
described in the conventional superspace approach\cite{WB} by the
quantity
\begin{equation} \Lag_{\rm kin} = -3 \int\,
E \, e^{-\frac{1}{3}K(\Phi,\bar{\Phi})} \label{WBLag}
\end{equation}
where $E$ denotes the super-determinant of the super-vielbein
$E_M^{\, A}$ and the measure is understood to be $\diff^2 \theta\,
\diff^2 \bar{\theta}^2$. The component expression of~(\ref{WBLag})
contains the kinetic terms for the supergravity multiplet as well as
the matter multiplets $\Phi$. However, this component expression
yields the correct normalization for the Einstein term of the
gravity action only after a field-dependent re-scaling of the
component fields.\cite{Cremmer:1982en} One must assign the
transformation property under K\"ahler transformations to these
re-scaled fields (not the superfields that appear in the superspace
Lagrangian).

Through a combination of classical symmetries of the supergravity
Lagrangian -- specifically K\"ahler transformations and super-Weyl
(or Howe-Tucker) re-scalings -- it is possible to absorb the
exponential factor into the super-determinant
\begin{equation} \Lag_{\rm kin} = -3 \int\,
E'\, . \label{BGGLag} \end{equation}
In other words, in this new frame the kinetic Lagrangian is nothing
more than -3 times the volume of superspace. The K\"ahler~$U(1)$
formalism incorporates these transformations into the structure
group of superspace,\cite{Binetruy:2000zx} thereby dispensing with
the need for conformal compensators and/or Weyl re-scalings of the
component field Lagrangian. K\"ahler transformations are now
understood at the {\em superfield} level and the Einstein term
automatically has the canonical normalization.

The structure group of K\"ahler~U(1) geometry contains the usual
Lorentz transformations as well as an additional chiral $U(1)$. This
additional chiral $U(1)$ acts on a superfield $\Phi$ of chiral
weight $w$ as $\Phi \to \Phi \exp\(-\frac{i}{2}\, w {\rm Im}\, F\)$,
where $F$ is a superfield. To incorporate this transformation into
the structure of superspace one constructs a connection $A_M$ with
which one forms a covariant derivative in superspace with respect to
this additional symmetry operation. The superfield $A_M$ associated
with this transformation has the following components
\begin{equation} A_{\alpha} = \frac{1}{4}\D_{\alpha} K, \quad \quad
A^{\dot{\alpha}} = -\frac{1}{4}\D^{\dot{\alpha}}K \end{equation}
\begin{equation} A_\mu = \frac{1}{4} \( K_i \partial_\mu \varphi^i
- K_{\jbar} \partial_{\mu} \oline{\varphi}^{\jbar}\) +
\frac{i}{8}K_{i\jbar}\oline{\sigma}_{\mu}^{\dot{\alpha}\alpha}\chi_{\alpha}^i
\oline{\chi}_{\dot{\alpha}}^{\jbar} \, ,
\end{equation}
where $\varphi = \Phi \lowest$ and $\chi_{\alpha} = \D_{\alpha}\Phi
\lowest$. We see that the gauge field of a K\"ahler $U(1)$
transformation is a composite object made up of the various chiral
fields in the theory. The effect of a K\"ahler~U(1) transformation
is simply to shift the vector part of the connection as
\begin{equation} A_{\mu} \to A_{\mu} -
\partial_{\mu}\(-\frac{i}{2}{\rm Im}\, F\) \end{equation} in analogy
to Abelian gauge theory.

One can now use this symmetry to remove the superpotential and
Yang-Mills kinetic terms from the F-density part of the Lagrangian
(those terms that involve integration over only half of superspace)
and recast the entire superspace Lagrangian in the form of
D-densities (integration over all of superspace). The K\"ahler
potential only appears implicitly, through the connections in the
covariant derivatives used to obtain the component-field expression.
The K\"ahler~$U(1)$-invariant kinetic terms for the matter fields
are now interpreted as the ``FI term'' for the K\"ahler potential.
The single expression~(\ref{BGGLag}) contains the kinetic terms for
the entire supergravity/matter system. In the chiral formulation of
the dilaton (or in the absence of a dilaton) the kinetic terms for
the Yang-Mills sector must be introduced through an F-density
expression as in~(\ref{LYM}). In the formalism of the modified
linear multiplet, however, these terms are also incorporated in a
single expression of the form~(\ref{BGGLag}), though the
expression~(\ref{BGGLag}) is generalized to the form
\begin{equation}
\Lag_{\rm kin} = -3\int\,E\,F(\Phi,\oline{\Phi},L) \, ,
\end{equation}
as was done in~(\ref{Klinear}) of Section~\ref{sec13}.

To obtain the component expression from any superfield Lagrangian
density in K\"ahler~$U(1)$ superspace, one may employ the chiral
density method. This is the locally supersymmetric generation of the
$F$-term construction in global supersymmetry. Consider a superspace
expression of the form $\Lag = \int E \Omega$, where $\Omega$ is a
real superfield that has weight $w_{\Omega} = 0$. Using integration
by parts in $U(1)$~superspace we can rewrite this expression as
\begin{equation} \Lag = \int \frac{E}{2R} \mathbf{r} ;
\quad r = -\frac{1}{8} \chiproj \Omega \, . \label{rrule}
\end{equation}
The component Lagrangian can now be expressed in terms of this
quantity $\mathbf{r}$ via the following rule
\begin{equation} \frac{1}{e}\Lag_{eff}=
-\frac{1}{4}\oline{\mathcal{D}}^{2}{\bf r} \lowest +
\frac{i}{2}(\bar{\psi}_{\mu}\bar{\sigma}^{\mu})^{\alpha}
\mathcal{D}_{\alpha}{\bf r} -
(\bar{\psi}_{\mu}\bar{\sigma}^{\mu\nu}\bar{\psi}_{\nu} + \oline{M})
{\bf r}\lowest \,\,+\,\, \mbox{h.c.}\, , \label{Lagr} \end{equation}
where $\oline{M} = -6R^{\dagger}\lowest$ is the supergravity
auxiliary field whose \vev determines the gravitino mass via $9
m_{3/2}^2 = \lang |M|^2 \rang$. More details on the procedure can be
found in Reference~\refcite{Giedt:2002ku} where the complete
component Lagrangian for the most general model of the BGW~class is
given.

As an example, let us consider an extension of the simple model from
Section~\ref{sec1} with Lagrangian density
\begin{equation}
\Lag_{\rm eff}=\superint\,E\,\{\,-2 \,+\, f(L) \,+\, bL\sum_I g^I
\,+\, bL\ln(e^{-K}\oline{U}U/\mu^{6})\,\} \, . \label{LagApp}
\end{equation}
That is, we consider the theory defined by~(\ref{LagSec1}) but
extended to include K\"ahler moduli and universal anomaly
cancelation. The Green-Schwarz term of~(\ref{GS}) is taken to have
vanishing $p_i$ so that $V_{\GS} = \sum_I g^I \equiv G$, and
$b_{\GS} = b$ for the single condensing gauge group. To
apply~(\ref{Lagr}) we must work with the quantities
\bea {\bf r}\,&=&\,-\,\frac{1}{8}\chiproj\{\,(\,-2\,+\,f(L)\,)
\,+\,bLG\,+\,bL\ln(e^{-K}\bar{U}U/\mu^{6})\,\}, \nonumber\\
{\bf
\bar{r}}\,&=&\,-\,\frac{1}{8}(\mathcal{D}^{2}-8R^{\dagger})\{\,(\,-2\,
+\,f(L)\,)\,+\,bLG\,+\,bL\ln(e^{-K}\bar{U}U/\mu^{6})\,\}. \eea
If we are interested in the purely bosonic part of the component
Lagrangian then we must compute the first and last terms
of~(\ref{Lagr}). Computing $\oline{M} {\bf r}\lowest + \hc$ is
straight-forward, but the first term in~(\ref{Lagr}) requires more
care. In particular we will need to concern ourselves with
quantities such as $(\mathcal{D}^2 R + \oline{\mathcal{D}}^2
R^{\dagger})$ and $(\mathcal{D}^{\alpha}X_{\alpha} +
\mathcal{D}_{\dot{\alpha}} X^{\dot{\alpha}})$, where
\begin{equation} X_{\alpha} = -\frac{1}{8} \chiproj
\mathcal{D}_{\alpha} K \, ; \quad X^{\dot{\alpha}} = -\frac{1}{8}
(\mathcal{D}^{2}-8R^{\dagger}) \mathcal{D}^{\dot{\alpha}} K \, .
\end{equation}
In our simple example these quantities are related in the following
manner
\begin{equation} \(L\frac{\diff g(L)}{\diff L} + 1\)(\mathcal{D}^2 R + \oline{\mathcal{D}}^2
R^{\dagger}) + (\mathcal{D}^{\alpha}X_{\alpha} +
\mathcal{D}_{\dot{\alpha}} X^{\dot{\alpha}}) = \Delta \end{equation}
where the bosonic components of $\Delta$ are
\bea
\Delta\lowest &=&\,-\,\frac{1}{\ell^{2}}(\ell^{2}\!g''-1)
\partial^{\mu}\!\ell\,\partial_{\mu}\!\ell
\,+\,\frac{1}{\ell^{2}}(\ell^{2}\!g''-1)v^{\mu}\!v_{\mu}
\,+\,2\nabla^{m}\!\nabla_{\!m}k
\nonumber\\
& &\,+\,4\sum_{I}\frac{1}{(t^{I}+\bar{t}^{I})^{2}}
\partial^{\mu}\bar{t}^{I}\,\partial_{\mu}t^{I}
\,-\,\frac{4}{9}(\ell^{2}\!g''-\ell g'-2)\oline{M}\!M
\nonumber\\
& &\,+\,\frac{4}{9}(\ell^{2}\!g'' +2\ell g' +1)b^{a}b_{a}
\,-\,4\sum_{I}\frac{1}{(t^{I}+\bar{t}^{I})^{2}} \oline{F}^{I}F^{I}
\nonumber\\
& &\,-\,\frac{4}{3\ell}(\ell^{2}\!g'' +\ell g') v^{\mu}e_{\mu}^{\hs
a}b_{a} \,-\,\frac{1}{2\ell}(\ell g' +1)(F_{U}+\oline{F}_{\bar{U}})
\nonumber\\
& &\,-\,\frac{1}{6\ell}(2\ell^{2}\! g'' -\ell g' -3)
(\,u\oline{M}\,+\,\bar{u}M\,) \,-\,\frac{1}{4\ell^{2}}(\ell^{2}\!g''
-1)\bar{u}u \, . \eea
In the above we have defined
\begin{equation} g' = g'(\ell) = \frac{\diff g(L)}{\diff L} \blowest
\, ; \quad g'' = g''(\ell) = \frac{\diff^2 g(L)}{\diff L^2} \blowest
\, ,
\end{equation}
the field $v_{\mu}$ is the vector component of the linear
multiplet~(\ref{vmudef}), and $b_a = -3 G_a \lowest$ is an auxiliary
field of the supergravity multiplet which is constrained to vanish
in the vacuum. The complete bosonic component Lagrangian
for~(\ref{LagApp}) is given by\cite{Binetruy:1996xj}
\bea \frac{1}{e}\Lag_{B}\,&=&\,-\,\frac{1}{2}{\cal R}
\,-\,\frac{1}{4\ell^{2}}(\ell g'+1 )
\partial^{\mu}\!\ell\,\partial_{\mu} \!\ell \,-\,(1+b\ell)
\sum_{I}\frac{1}{(t^{I}+\bar{t}^{I})^{2}}
\partial^{\mu}\bar{t}^{I}\,\partial_{\mu}t^{I}
\nonumber\\
& & +\,\frac{1}{4\ell^{2}}(\ell g' +1)v^{\mu}\!v_{\mu}
\,+\,\frac{1}{9}(\ell g' -2)\[\oline{M}\!M \,- b^{a}b_{a}\]
\,+\,(1+b\ell)\sum_{I}\frac{1}{(t^{I}+\bar{t}^{I})^{2}}
\oline{F}^{I}F^{I}
\nonumber\\
& &\,+\,\frac{1}{8\ell}\{\,f\,+\,1\,+\,b\ell\ln(e^{-k}\bar{u}u/
        \mu^{6})\,+\,2b\ell\,\}(F_{U}+\oline{F}_{\bar{U}})
\nonumber\\
& &\,-\,\frac{1}{8\ell}\{\,f\,+\,1\,+\,b\ell\ln(e^{-k}\bar{u}u/
\mu^{6})\,+\,\frac{2}{3}b\ell(\ell g' +1)\,\}
(\,u\oline{M}\,+\,\bar{u}M\,)
\nonumber\\
& &\,-\,\frac{1}{16\ell^{2}}(1+2b\ell)(\ell g' +1)\bar{u}u
\nonumber\\
& &\,-\,\frac{i}{2}b\ln(\frac{\bar{u}}{u})\partial^{\mu}\!v_{\mu}
\,-\,\frac{i}{2}b\sum_{I}\frac{1}{(t^{I}+\bar{t}^{I})}
(\,\partial^{\mu}\bar{t}^{I}\,-\,\partial^{\mu}t^{I}\,)v_{\mu}. \eea


\end{document}